\documentclass[letterpaper,11pt,vi,twoside,singlespace]{mitthesis}

\usepackage{aastex_hack}
\usepackage{aas_macros}
\usepackage{amsfonts}
\usepackage{amsmath}
\usepackage{amssymb}
\usepackage{deluxetable}
\usepackage{graphicx}
\usepackage{latexsym}
\usepackage{lscape}
\usepackage{multirow}
\usepackage{natbib}
\bibpunct[]{(}{)}{;}{a}{}{,}
\usepackage{dcolumn}
\usepackage{array}
\usepackage{calc}
\usepackage{appendix}
\providecommand{\noun}[1]{\textsc{#1}}
\providecommand{\rmn}[1]{{\mathrm{#1}}}
\providecommand{\bmath}[1]{\mathbf{#1}}
\providecommand{\mathbfss}[1]{\textbf{\textsf{#1}}}
\providecommand{\tabularnewline}{\\}
\def\elll{L}
\def\ga{\gtrsim}
\def\la{\lesssim}

\newcolumntype{f}{D{.}{.}{5.0}}
\newcommand{\lyxdot}{.}

\def\decayright#1{\kern#1em\raise1.1ex\hbox{$|$}\kern-.5em\rightarrow}

\def\Ob{\Omega_b}
\def\Oc{\Omega_{\rm c}}
\def\Ok{\Omega_k}
\def\Ol{\Omega_\Lambda}
\def\Om{\Omega_m}

\def\On{\Omega_\nu}

\def\ob{\omega_b}

\def\ocdm{\omega_{\rm c}}
\def\od{\omega_d}

\def\om{\omega_{\rm m}}
\def\on{\omega_\nu}

\def\fn{f_\nu}
\def\ns{{n_s}}
\def\nt{{n_t}}
\def\al{\alpha}
\def\Ot{\Omega_{\rm tot}}
\def\As{A_s}
\def\At{A_t}
\def\Ap{A_{\rm peak}}

\def\zeq{{z_{\rm eq}}}
\def\zrec{{z_{\rm rec}}}
\def\zion{{z_{\rm ion}}}
\def\zacc{{z_{\rm acc}}}
\def\teq{t_{\rm eq}}
\def\trec{t_{\rm req}}
\def\tion{t_{\rm ion}}
\def\tacc{t_{\rm acc}}
\def\tnow{t_{\rm now}}

\def\Th{\Theta_s}

\def\Mnu{M_\nu}

\def\Qnl{Q_{\rm nl}}

\def\rhob{\rho_{\rm b}}
\def\rhoc{\rho_{\rm c}}

\def\rhon{\rho_\nu}

\def\rhom{\rho_{\rm m}}

\def\rhohalo{\rho_{\rm halo}}

\def\xib{\xi_{\rm b}}
\def\xic{\xi_{\rm c}}

\def\xin{\xi_\nu}

\def\ng{n_\gamma}

\def\rdamp{r_{\rm damp}}
\def\rsound{r_{\rm s}}

\def\percent{\%}
\def\Pg{P_{\rm g}}
\def\spose#1{\hbox to 0pt{#1\hss}}
\def\simlt{\mathrel{\spose{\lower 3pt\hbox{$\mathchar"218$}}
     \raise 2.0pt\hbox{$\mathchar"13C$}}}
\def\simgt{\mathrel{\spose{\lower 3pt\hbox{$\mathchar"218$}}
     \raise 2.0pt\hbox{$\mathchar"13E$}}}
\def\simpropto{\mathrel{\spose{\lower 3pt\hbox{$\mathchar"218$}}
     \raise 2.0pt\hbox{$\propto$}}}

\usepackage{bm}

\IfFileExists{url.sty}{\usepackage{url}}
                      {\newcommand{\url}{\texttt}}

\begin{document}
%
%
%
%
%
%
%
\title{Particle Physics in the Sky and Astrophysics Underground: Connecting the Universe's Largest and Smallest Scales}

\author{Molly E.~C. Swanson}
\prevdegrees{BS Physics\\California Institute of Technology (2002)}
\department{Department of Physics}
\degree{Doctor of Philosophy in Physics}
\degreemonth{June}
\degreeyear{2008}
\thesisdate{May 1, 2008}


\supervisor{Max Erik Tegmark}{Associate Professor of Physics}

\chairman{Thomas J. Greytak}{Professor of Physics \\ Associate Department Head for Education}

\maketitle



\cleardoublepage
\setcounter{savepage}{\thepage}
\begin{abstractpage}
Particles have tremendous potential as astronomical messengers,
and conversely, studying the universe as a whole also teaches us
about particle physics. This thesis encompasses both of these research directions.

Many models predict a diffuse flux of high energy neutrinos
from active galactic nuclei and other astrophysical sources. 
The ``Astrophysics Underground'' portion of this thesis 
describes a search for this neutrino
flux performed by looking for extremely high energy upward-going muons 
using the Super-Kamiokande detector, and comparing
the observed flux to the expected background.
We use our results to 
to set an upper limit on the diffuse neutrino flux from astrophysical sources.

In addition to using particles to do astronomy, we can also use the universe itself as a particle physics lab.  
Cosmology provides new insights
that could never be observed in terrestrial laboratories.
The ``Particle Physics in the Sky'' portion of this thesis focuses
 on extracting cosmological information from galaxy surveys.

To overcome technical challenges faced by the latest galaxy surveys, we 
produced a comprehensive upgrade to 
{\scshape mangle}, a software package that processes the angular masks defining
the survey area on the sky.  We added dramatically faster algorithms and new useful features 
to this software that are necessary for managing complex masks of the Sloan Digital Sky
Survey (SDSS) and will be invaluable for future surveys as well. 

With this software in hand, we
utilized galaxy clustering data from SDSS to 
investigate the relation between galaxies and dark matter by studying relative bias, 
i.e., the relation between different types of galaxies. If all 
galaxies were perfect tracers of dark matter, different subpopulations would trace
each other perfectly as well. However, separating galaxies by their luminosities 
and colors reveals a complicated picture: red galaxies are clustered 
more strongly than blue galaxies, with both the brightest and the faintest 
red galaxies showing the strongest clustering. Furthermore, red and blue galaxies tend 
to occupy different regions of space, effectively introducing an element of stochasticity 
(randomness) when modeling their relative distributions. In order to make precise 
measurements from the next generation of galaxy surveys, it will be essential to 
account for this complexity.

\end{abstractpage}


\cleardoublepage

\section*{Acknowledgments}
I would like to extend my heartfelt thanks:

To my first teachers, my parents Barb and Rick Swanson.  To my dad for everything from counting Cheerios to
help with calculus homework, and for teaching me that physicists answer questions in the language of math, even when their 
8-year-old daughters can't bear to see another equation scribbled on a napkin, and to my mom for her
endlessly enthusiastic support of everything I've ever done.

To my most recent teachers, my advisors Max Tegmark and Kate Scholberg, for being such great role models, for always sticking up for me,
and for showing me two such vastly different yet equally inspiring ways to be a scientist.

To all the great teachers I've had in between.

To my late grandmother Helen Hage -- it was worth it, even though I'm not a boy.

To all of the Dr. Swanson physicists before me -- my dad, my uncle Charlie Swanson, and my grandfather Bob Swanson, thanks to whom
I always thought getting a PhD in physics was a common thing to do.  

To the rest of the Swanson(/Olson/McLaughlin) clan, for doing other things so that we are not \emph{all} huge nerds.

To the Crosby/Newman/Allen/etc. clan, especially to my grandfather-in-law Da (known to the rest of the world as Bob Newman) for taking
a genuine interest in what I do and for my favorite definition of a neutrino ever.

To the Techer-Boston crew and various other dear friends in the area,
including Wedge Cheung, Robin Friedman, Mike Fuerstman, Todd Schuman, Dave Guskin, Mick Garvey, Marlena Fecho, Brian and 
Sarah Bairstow, Craig Chu, Charley Mills, Pat Codd, Dave and Jess Tytell, Kate Jensen, and Dan Recht.
The Thursday-night-dinners-on-Tuesdays-on-Wednesdays-on-Thursdays are among my fondest memories of my time in Boston.

To all my fellow astrograds, and everyone else I've had the pleasure of knowing at MIT, especially Miriam Krauss, Jake Hartman,
Judd and Cassie Bowman, Ed Boyce, Robyn Sanderson, Ben Cain, Mike Stevens, Yi Mao, Bonna Newman, Julie Millane, Jessie Shelton, 
Tom and Liz Pasquini, Will Fox, 
and Sarah-Jane White. Having such awesome people with whom to share the pain made grad school infinitely more fun.

To my fellow 37-602 denizens, most especially Josh Carter and Ryan Lang, who were in it with me for the long haul, but also to
Judd Bowman, Sam Conner, Chris Williams, Leslie Rogers, and (albeit briefly) Sarah Vigeland. It wouldn't have been the same without 
dodging Nerf footballs, green tea, Penzey's spice rub, indoor paper baseball, cutthroat games of Facebook Scrabble and 
World Traveler, multiwing bats, SoftLips pranks, paper airplane contests, and countless other diversions, not to mention
the constant support.

To Gideon Koekoek, virtual officemate extraordinaire, who, thanks to the magic of teh intarnets, was just as distracting
and just as inspiring as his real-life counterparts all the way from Amsterdam.  Words can't do it justice, so here are some *doctoral hugs* instead.

To all of Walking Fish and Blacker Hovse, for being the most amazing friends a nerdy girl could ask for, especially to
Katie Romportl, my first partner-in-crime, to Mike Borchert and Matt Kuzma for throwing pennies at my window late at
night in order to sneak into my room to chat about quantum mechanics, to Mike Jorgenson for fractals and ensanguination,
to Joe Carroll for frisbee and folk songs, 
to Nate Austin for castles and zepplins, 
and to Katharina Kohler for always understanding exactly what I'm going through.

And most of all, to my husband Tim Crosby, for being the reason I come home every night, and for being the home I'll
always come home to.  Swarling swar swar, swarling swar.

\clearpage

The work presented in chapters \ref{chap:superk}, \ref{chap:mangle},
\ref{chap:bias} and \S\ref{sec:cosmological-parameters} was all done
in collaborations -- here I would like to acknowledge my co-authors 
and detail my own role in each of these projects.

For chapter~\ref{chap:superk}, I did the analysis and write-up
primarily by myself, with
valuable guidance from Kate Scholberg, Alec Habig, Shantanu Desai, and Jodi Cooley.
The other members of the Super-Kamiokande collaboration helped via building
the detector, keeping it running, and providing insightful comments on
drafts of the paper. I also gratefully gratefully acknowledge 
the cooperation of the Kamioka Mining and
Smelting Company.  The Super-Kamiokande experiment has been built and
operated from funding by the Japanese Ministry of Education, Culture,
Sports, Science and Technology, the United States Department of Energy,
and the US National Science Foundation.

For chapter~\ref{chap:mangle}, I added the new pixelization
algorithms described here
to the \noun{mangle} software which was originally written by Andrew Hamilton,
and did the bulk of the write-up as well.
The \emph{HEALPix} algorithms were done primarily by UROP student Colin Hill.
Colin and I also worked together to create the new \noun{mangle} website -- Colin 
deserves a special thank you for pulling an all-nighter with me the night before
the website went live.
I also thank Michael Blanton for providing the SDSS DR6 VAGC mask, Krzysztof
G\'{o}rski and collaborators for creating the \emph{HEALPix} package, and the WMAP team
for making their data public via the Legacy Archive for Microwave
Background Data Analysis (LAMBDA; \protect\url{http://lambda.gsfc.nasa.gov}).%
. Support for LAMBDA is provided by the NASA Office of Space Science.

For \S\ref{sec:cosmological-parameters}, which summarizes \citet{2006PhRvD..74l3507T},
I contributed primarily by using my new version of \noun{mangle} to create angular
masks for each of the angular subsamples used in the analysis. I also helped with 
proofreading and editing the original paper.

For chapter~\ref{chap:bias}, I did the analysis and write-up, with 
 guidance from Max Tegmark. Idit Zehavi provided 
careful editing and insightful suggestions, and
Michael Blanton provided the SDSS VAGC-DR5 data.
I also wish to thank 
Daniel Eisenstein, David Hogg, Taka Matsubara, Ryan
Scranton, Ramin Skibba, and Simon White for helpful comments. 
Funding for the SDSS has been provided by the Alfred P.~Sloan Foundation,
the Participating Institutions, the National Science Foundation, the
U.S.~Department of Energy, the National Aeronautics and Space Administration,
the Japanese Monbukagakusho, the Max Planck Society, and the Higher
Education Funding Council for England. The SDSS Web Site is \url{http://www.sdss.org}.
The SDSS is managed by the Astrophysical Research Consortium for the
Participating Institutions. The Participating Institutions are the
American Museum of Natural History, Astrophysical Institute Potsdam,
University of Basel, Cambridge University, Case Western Reserve University,
University of Chicago, Drexel University, Fermilab, the Institute
for Advanced Study, the Japan Participation Group, Johns Hopkins University,
the Joint Institute for Nuclear Astrophysics, the Kavli Institute
for Particle Astrophysics and Cosmology, the Korean Scientist Group,
the Chinese Academy of Sciences (LAMOST), Los Alamos National Laboratory,
the Max-Planck-Institute for Astronomy (MPIA), the Max-Planck-Institute
for Astrophysics (MPA), New Mexico State University, Ohio State University,
University of Pittsburgh, University of Portsmouth, Princeton University,
the United States Naval Observatory, and the University of Washington.

This work was supported by a National Defense Science and Engineering Graduate
Fellowship, a Bruno Rossi Fellowship,
NASA grant NNG06GC55G, NSF grants AST-0134999,
0607597 and 0708534, the Kavli Foundation, and fellowships from the
David and Lucile Packard Foundation and the Research Corporation.


\pagestyle{plain}

\tableofcontents
\newpage
\listoffigures
\addcontentsline{toc}{chapter}{List of Figures}
\newpage
\listoftables
\addcontentsline{toc}{chapter}{List of Tables}


\chapter{Connecting particle physics and astrophysics}
\begin{quotation}
Meg looked about.  Ahead of her was a tremendous rhythmic swirl of wind and flame, but it was wind and flame quite different from the cherubim's; this was a dance, a dance ordered and graceful, and yet giving an impression of complete and utter freedom, of ineffable joy.  As the dance progressed, the movement accelerated, and the pattern became clearer, closer, wind and fire moving together, and there was joy, and song, melody soaring, gathering together as wind and fire united. 

And then wind, flame, dance, song, cohered in a great swirling, leaping, dancing, single sphere.

Meg heard Mr. Jenkins's incredulous, ``What was that?''

Blajeny replied, ``The birth of a star.''

Mr. Jenkins protested, ``But it's so small I can hold it in the palm of my hand.''  And then an indignant snort, ``How big am I?''

``You must stop thinking about size, you know.  It is both relative and irrelevant.''

\hfill --- Madeleine L'Engle, \emph{A Wind in the Door} 
\end{quotation}

One of the most remarkable things about physics is that it can be applied over such a vast range of size scales. Some physicists focus on understanding our world on almost inconceivably tiny scales, studying theories of fundamental building blocks of matter as small as $10^{-35}$ meters, a factor of $10^{-20}$ smaller than the diameter of a proton. 
Other physicists study our world at equally inconceivably huge scales, looking at the most distant parts of our universe that it is possible to see, over $10^{26}$ meters away. These largest and smallest scales in physics are separated by over 60 orders of magnitude.

Perhaps even more astounding is the fact that these two extreme regimes of physics are deeply connected: learning about how our universe behaves on the very largest scales can teach us things about elementary particles that could never be discovered in a terrestrial laboratory, and studying particles coming from space can also teach us about properties of distant astronomical objects. Over the past several decades, physicists have gained many great insights about both the macroscopic and microscopic regimes by taking advantage of this connection.
The link between particle physics and astrophysics, which is the focus of this thesis, is but one example of a fruitful interaction between traditionally separate fields of science -- there is a tremendous amount of exciting work being done at interfaces between other fields as well. Some examples include applying statistical physics to biological systems, quantum computation research that aims to use quantum physics principles to revolutionize computer science, and neuroscience research integrating biology, psychology, and computer science. 

Doing science at these interfaces is no easy task -- integrating expertise across different fields generally leads to certain challenges. Scientists on either side of the fence will have different academic cultures, traditions, and methodologies, and they frequently do not mix well. For example, concerns have been raised that the trend towards using astronomy to study fundamental physics could ultimately be damaging to the astronomical community: \citet{2007RPPh...70..883W} notes that ``By uncritically adopting the values of an alien system [of high-energy physics traditions], astronomers risk undermining the foundations of their own current success and endangering the future vitality of their field.''
However, the scientific potential of the crossover is large enough that such obstacles are well worth overcoming. As \citet{2007RPPh...70.1583K} says, the ties between astronomy and particle physics ``can either encourage or choke creativity ... It is up to us to choose wisely.'' If these two scientific communities intelligently navigate the cultural interface, the science of both fields will be greatly enriched.

\begin{figure*}
\includegraphics[bb=0bp 275bp 575bp 536bp,clip,width=1\textwidth]{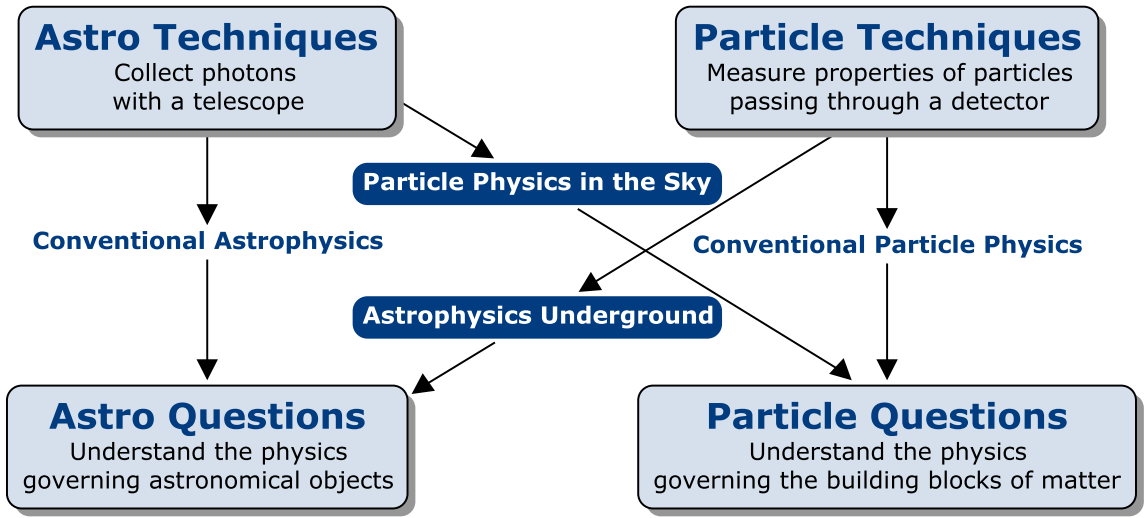}

\caption{\label{fig:connections}Flow diagram of connections between particle physics and astrophysics.}
\end{figure*}

The nature of the particle-physics--astrophysics connection used in this thesis is illustrated in Fig.~\ref{fig:connections}. The traditional means of doing astrophysics is to look at the light emitted from distant astronomical objects with a telescope and use this to infer the physics of these objects.  On the other hand, traditional particle physics is done by measuring the properties of particles passing through a detector and using this to infer the nature of the particles themselves or the interactions that produced them. 

One means of connecting astrophysics and particle physics is to use particles as astronomical messengers, i.e. to look for particles other than photons coming from astronomical objects to get a more complete picture of the physical system.  We dub this ``Astrophysics Underground'' because when the particle messenger is a neutrino, the ``telescopes'' used for this endeavor are typically large particle detectors located deep underground to shield from background radiation. Essentially this is using the tools of particle physics to do astronomy.

We can also turn this connection in the other direction and use the tools of astronomy to do particle physics, which we dub ``Particle Physics in the Sky.'' One particularly exciting area in which this can be applied is cosmology, the study of our universe as a whole.  By doing astronomical observations cosmologists have built up a consistent theory of how our universe has evolved that turns out to depend on the properties of elementary particles and has also provided evidence for physics beyond the standard model of particle physics. Thus by studying the our universe at the largest scales we can also learn about the smallest.

The rest of this thesis is organized as follows: an exploration into doing ``Astrophysics Underground'' using neutrinos is given in chapters \ref{chap:nuastro} and \ref{chap:superk}. Chapter~\ref{chap:nuastro} provides an overview of the field of neutrino astronomy, and chapter~\ref{chap:superk} describes a search for high energy neutrinos coming from astrophysical sources using the Super-Kamiokande neutrino detector.  The following three chapters turn the connection arrow in the other direction and focus on ``Particle Physics in the Sky'' by doing cosmology with galaxy surveys. Chapter~\ref{chap:cosmology} gives a general introduction to the field of cosmology with a particular focus on the tool of galaxy surveys and what this can tell us about particle physics. In chapter~\ref{chap:mangle}, we describe a set of computational techniques for managing the complex data of modern galaxy surveys, 
and in chapter~\ref{chap:bias} we use data from the Sloan Digital Sky Survey to study how the clustering of galaxies depends on and their luminosities and colors and the implications of this for using galaxies as cosmological tools.
Finally, we conclude in chapter~\ref{chap:conclusions} with a summary of what we have learned, some ideas for future research directions, and reflections on the nature of doing science at this fascinating interface between astrophysics and particle physics.

\cleardoublepage
\part{Astrophysics Underground}
\chapter{Neutrino astronomy}
\label{chap:nuastro}

\section{History of multi-messenger astronomy}

Historically, the study of astronomy has been the study of photons
coming from the sky. Early astronomers learned a tremendous amount
observing only the visible-wavelength photons that could be seen with
the naked eye. Since then, revolutions in astronomy have come about
through expanding our ability to detect different types of astronomical
messengers. 

The invention of the telescope in the early seventeenth century enabled
the collection of much larger numbers of photons, opening a window
on fainter sources. The twentieth century saw an explosion of astronomical
information as techniques were developed to observe photons all across
the electromagnetic spectrum from radio waves to gamma rays, leading
to the discovery of remarkable astrophysical phenomena such as pulsars
and gamma ray bursts.

Detecting particles other than photons from astrophysical sources
provides yet another window on the universe. The discovery of cosmic
rays in 1912 was the first step in this endeavor. Cosmic rays are
charged particles -- protons, electrons, and atomic nuclei -- originating
from cosmic sources. Studies of cosmic rays have led to insights about
astrophysical processes such as the evolution of galactic magnetic
fields and particle acceleration within remnants of supernova explosions.
High-energy cosmic rays penetrating the atmosphere also provide a
natural particle collider -- this was an instrumental tool in particle
physics as well: for example, positrons, muons, and pions were all
discovered in cosmic ray studies. Today the astrophysical origin and
acceleration mechanisms of the highest energy cosmic rays is one of
the primary open questions in particle astronomy.

\section{Neutrinos as astronomical messengers}

Another exciting step towards using particles as astronomical messengers
is to look for neutrinos coming from cosmic sources. In terms of providing
astrophysical information, neutrinos have some advantages over photons
and charged particles. Firstly, they are not readily absorbed by intervening
matter, so they can act as probes of extremely high-density environments
such as the central cores of stars, active galactic nuclei, or supernovae
from which photons could never escape. Secondly, they are not deflected
by cosmic magnetic fields, so unlike charged particles, they always
point directly back to their source.

The advantage of low absorption, however, is the biggest disadvantage
as well -- they tend to pass through our detectors unseen just as
readily as they pass through dense material at the source. Thus neutrino
astronomy demands sophisticated detection techniques, and as such,
this field is still in its infancy. So far, neutrinos have only been
detected from two astrophysical sources: the sun \citep{2003RvMP...75..985D,2003RvMP...75.1011K}
and supernova
1987a \citep{1987PhRvL..58.1490H,1987PhRvL..58.1494B}. However, the insights gained into both astronomy and
particle physics from just these two neutrino sources have been tremendous. 

The work of Raymond Davis and others on detecting neutrinos from the sun generated
the famous ``solar neutrino problem'', whereby the observed flux
of solar neutrinos was only about a third of what was expected based
on astrophysical models of energy production in the sun \citep{1996NuPhS..48..284D}. This
called both the experimental measurements and the standard solar models
into question, and remained a controversy for decades. Its ultimate
resolution actually came from a revolution in particle physics: measurements
from the Super-Kamiokande neutrino detector (Super-K; \citealt{1998PhRvL..81.1562F})
provided conclusive evidence that
neutrinos oscillate from one type to another and therefore must have
mass, in conflict with the standard model of particle physics which
treats them as massless. This explained the solar neutrino problem
via some fraction of electron neutrinos produced in the sun oscillating
to muon or tau neutrinos by the time they reach the Earth and thus
going undetected. Ultimately, observations of solar neutrinos have
both solidified our understanding of astrophysical processes in the
sun and provided the first concrete evidence for physics beyond the
standard model.

The 20 neutrinos detected from supernova 1987a by the Kamiokande-II
and IMB detectors \citep{1987PhRvL..58.1490H,1987PhRvL..58.1494B} similarly provided insights into both particle
physics and astrophysics. The flux observed is consistent with astrophysical
models of core-collapse supernovae in which nearly all of energy from
the collapse is emitted in neutrinos. Additionally, the observation
that neutrinos of different energies all arrived nearly simultaneously
provides a clue to the nature of the neutrinos themselves: it means
that the neutrinos of different energies all travel at nearly the
speed of light, which implies that the electron neutrino rest mass
must be quite small. Thus the neutrinos observed from the supernova
can be used to set an upper limit on the mass of the electron neutrino
\citep{1995NuPhB.437..243K}.

The next frontier of neutrino astronomy is to detect astrophysical
neutrinos at high energies (GeV-PeV) -- several experimental efforts
to do this are currently underway. Such high-energy neutrinos will
open yet another new window on the universe -- they will shed light
on astrophysical processes in extreme environments and will also provide
clues to the origins of the highest energy cosmic rays. (See 
\citealt{2002RPPh...65.1025H} for more details.)

\section{Astrophysical production mechanisms}

\subsection{Diffusive shock acceleration}

The general mechanism for producing high-energy neutrinos is the following:
first, protons or other nuclei are accelerated by shocks in an astrophysical
plasma. Then the accelerated hadrons interact with photons in a surrounding
radiation field and produce charged pions that decay into high-energy
neutrinos:

\begin{eqnarray}
p+\gamma\longrightarrow & \hskip-.8em\pi^{+} & \hskip-1.0em+n\nonumber \\
 & \hskip-.8em\decayright{.2} & \hskip-.8em\mu^{+}+\nu_{\mu}\nonumber \\
 &  & \hskip-.8em\decayright{.2}e^{+}+\bar{\nu}_{\mu}+\nu_{e}\label{eq:pgamma}\end{eqnarray}

\begin{eqnarray}
n+\gamma\longrightarrow & \hskip-.8em\pi^{-} & \hskip-1.0em+p\nonumber \\
 & \hskip-.8em\decayright{.2} & \hskip-.8em\mu^{-}+\bar{\nu}_{\mu}\nonumber \\
 &  & \hskip-.8em\decayright{.2}e^{-}+\nu_{\mu}+\bar{\nu}_{e}\label{eq:ngamma}\end{eqnarray}

These interactions can also produce neutral pions, which generate
high-energy gamma rays:

\begin{eqnarray}
p+\gamma\longrightarrow & \hskip-.8em\pi^{0}+p\nonumber \\
 & \hskip-.8em\decayright{.2}2\gamma\label{eq:gamma}\end{eqnarray}

\begin{figure}
\begin{centering}\includegraphics[width=0.5\textwidth]{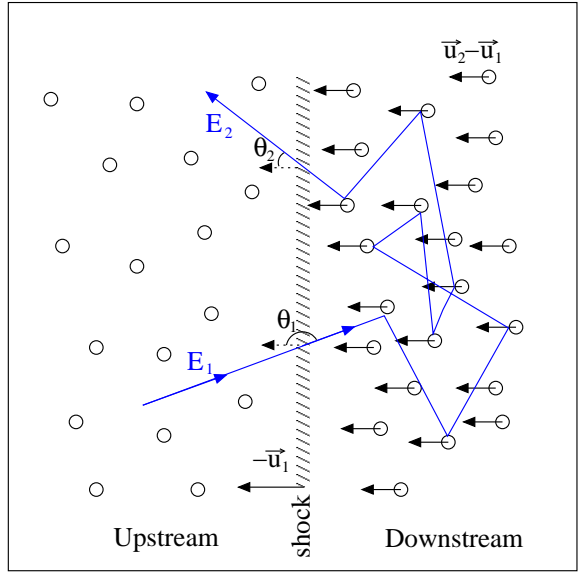}\par\end{centering}

\caption[Geometry of diffusive shock acceleration.]{\label{fig:fermi}Geometry of diffusive shock acceleration, shown
in the rest frame of the material upstream of the shock. The trajectory
of an accelerated particle is shown in blue/grey. Scattering centers
(i.e. magnetic field irregularities) are represented as circles.}
\end{figure}

The proton acceleration mechanism at work here is known as diffusive
shock acceleration, also known as the first-order Fermi mechanism \citep{fermi}.
The difference in velocities on either side of a shock in a magnetized
plasma can transfer kinetic energy to particles as they cross the
shock multiple times. The basic geometry of a plane shock is illustrated
in Fig.~\ref{fig:fermi}. This mechanism generically predicts a power-law
spectrum of particle energies with $dN/dE\propto E^{-\alpha}$ where
$\alpha\approx2$. We give a rough derivation of this below following
\citet{1990cup..book.....G}; for a more detailed treatment see \citet{2002cra..book.....S}. 

The basic equations governing the behavior of shocks can be derived
from conservation of mass, momentum, and energy across the shock boundary.
For a non-relativistic hydrodynamic plane shock, these conditions are given
by

\begin{eqnarray}
\rho_{2}u_{2}-\rho_{1}u_{1} & = & 0,\label{eq:shock1}\\
\rho_{2}u_{2}^{2}-\rho_{1}u_{1}^{2} & = & P_{1}-P_{2},\label{eq:shock2}\\
\rho_{2}u_{2}\left(w_{2}-\frac{1}{2}u_{2}^{2}\right)-\rho_{1}u_{1}\left(w_{1}-\frac{1}{2}u_{1}^{2}\right) & = & 0,\label{eq:shock3}\end{eqnarray}
where 1 subscripts denote conditions upstream of the shock and 2 subscripts
denote downstream conditions. $u$ is the component of the velocity
normal to the shock front in the rest frame of the shock, $\rho$
is the density, $P$ is the pressure, and $w$ is the specific enthalpy,
given by $w=c_{p}T$ for temperature $T$ and specific heat at constant
pressure $c_{p}$. For an ideal gas, $w=\gamma P/\left(\left(\gamma-1\right)\rho\right)$
where $\gamma$ is the ratio of specific heats at constant pressure
and constant volume; $\gamma=5/3$ for a monatomic ideal gas. In
a magnetized plasma the magnetic fields will also contribute to these
conditions -- here we assume either that the magnetic field is parallel
to the normal of the shock plane or that the magnetic fields are small
enough to be negligible for the continuity conditions.

In order for a shock to form, $u_{1}$ must be greater than the sound
speed $c_{s1}=\sqrt{\gamma P_{1}/\rho_{1}}$. The Mach number $M$
of the shock is defined as $M=u_{1}/c_{s1}$, and by using equations~\eqref{eq:shock1},
~\eqref{eq:shock2}, ~and~\eqref{eq:shock3}, and much clever algebra,
the ratio of the upstream to downstream velocities is given by \begin{equation}
\frac{u_{1}}{u_{2}}=\frac{M^{2}\left(\gamma+1\right)}{M^{2}\left(\gamma-1\right)+2}.\label{eq:velocities}\end{equation}

The energy of a charged particle will increase as it diffuses across
the shock from the upstream medium to the downstream medium and back,
as illustrated by the trajectory in Fig.~\ref{fig:fermi}. To calculate
the energy change, we start with a relativistic ($E\approx pc$) population
of particles with velocities that are isotropic in the frame of the
upstream medium. A particle with $ $energy $E_{1}$ that crosses
the shock at an angle $\theta_{1}$ in the upstream frame will have
an energy$ $\begin{equation}
E_{1}^{\prime}=\Gamma E_{1}\left(1-\beta\cos\theta_{1}\right)\label{eq:lorentz1}\end{equation}
in the downstream frame, where $\beta=\left(u_{1}-u_{2}\right)/c$
and $\Gamma$ is the corresponding Lorentz factor. Primes denote quantities
in the downstream frame; unprimed quantities are in the upstream frame.
Averaging over an isotropic population gives $\left\langle \cos\theta_{1}\right\rangle =-2/3$
for particles crossing the shock.

In the downstream medium, the particle undergoes elastic scattering
off of irregularities in the magnetic field generated by low-frequency
magnetohydrodynamic waves. As the particle diffuses, its average motion
will coincide with that of the downstream medium, so we now have an
isotropic population in the downstream frame. To a first approximation,
these scatterings do not cause the particle to lose energy, so a particle
that is scattered back into the upstream medium at an angle $\theta_{2}^{\prime}$
will have energy $E_{2}^{\prime}=E_{1}^{\prime}$. Transforming back
into the upstream frame gives$ $

\begin{equation}
E_{2}=\Gamma E_{2}^{\prime}\left(1+\beta\cos\theta_{2}^{\prime}\right),\label{eq:lorentz2}\end{equation}
and since the particles have isotropized relative to the downstream
medium, $\left\langle \cos\theta_{2}^{\prime}\right\rangle $ can
be obtained by averaging over an isotropic population as well, yielding
$\left\langle \cos\theta_{2}^{\prime}\right\rangle =2/3$. Plugging
equation~\eqref{eq:lorentz1} in for $E_{2}^{\prime}$ in equation~\eqref{eq:lorentz2}
and averaging over the angular factors yields the following formula
for the fractional energy change in one {}``encounter'' (diffusing
from upstream to downstream and back):

\begin{eqnarray}
\epsilon\equiv\frac{E_{2}-E_{1}}{E_{1}} & = & \frac{1+\frac{4}{3}\beta+\frac{4}{9}\beta^{2}}{1-\beta^{2}}\nonumber \\
 & \approx & \frac{4}{3}\beta,\label{eq:energychange}\end{eqnarray}
where the last approximation holds for a non-relativistic shock with
$\beta\ll1$.

Once in the downstream medium, the particles have some probability
to escape rather than get scattered back into the upstream medium.
We can estimate this probability by considering the flux of particles
through an imaginary plane one scattering length away from the shock
in the downstream medium. Particles that pass through this plane back
in the upstream direction will get swept up by the shock, and particles
that pass through towards the downstream direction can escape. Working
in the frame of the downstream medium, particles will pass back through
the shock if $c\cos\theta<-u_{2}$. Thus the flux upstream is\begin{eqnarray*}
F_{u} & \propto & \int_{-1}^{-u_{2}/c}\left(c\cos\theta+u_{2}\right)d\cos\theta\\
 & = & -\frac{\left(c-u_{2}\right)^{2}}{2c}\end{eqnarray*}
and the flux downstream is

\begin{eqnarray*}
F_{d} & \propto & \int_{-u_{2}/c}^{1}\left(c\cos\theta+u_{2}\right)d\cos\theta\\
 & = & \frac{\left(c+u_{2}\right)^{2}}{2c}.\end{eqnarray*}
The probability for a particle to escape is then$ $\begin{eqnarray}
P_{\mathrm{esc}} & = & 1-\left|\frac{F_{u}}{F_{d}}\right|\nonumber \\
 & = & 1-\frac{\left(c-u_{2}\right)^{2}}{\left(c+u_{2}\right)^{2}}\nonumber \\
 & \approx & \frac{4u_{2}}{c}\label{eq:escprob}\end{eqnarray}
for $u_{2}\ll c$.$ $

The energy spectrum produced in this scenario -- where a particle
gains fractional energy $\epsilon$ and has a probability $P_{\mathrm{esc}}$
of escaping with each encounter -- can be calculated as follows: if
particles initially have energy $E_{0}$, the energy after $m$ encounters
is $E_{n}=E_{0}\left(1+\epsilon\right)^{m}$, so the number of encounters
$n$ needed to reach energy $E$ is\[
n=\frac{\log\frac{E}{E_{0}}}{\log\left(1+\epsilon\right)}.\]
The probability of remaining after $m$ encounters is $\left(1-P_{\mathrm{esc}}\right)^{m}$,
so the number of particles with energy greater than $E$ is thus \begin{eqnarray*}
N\left(>E\right) & \propto & \sum_{m=n}^{\infty}\left(1-P_{\mathrm{esc}}\right)^{m}\\
 & = & \frac{\left(1-P_{\mathrm{esc}}\right)^{n}}{P_{\mathrm{esc}}}\\
 & = & \frac{1}{P_{\mathrm{esc}}}\left(\frac{E}{E_{0}}\right)^{-\left(\alpha-1\right)},\end{eqnarray*}
where\begin{equation}
\alpha-1=\frac{\log\frac{1}{1-P_{\mathrm{esc}}}}{\log\left(1+\epsilon\right)}.\label{eq:alpha}\end{equation}
The differential energy flux is thus $dN/dE\propto E^{-\alpha}$.
Plugging in the results of equations~\eqref{eq:energychange}~and~\eqref{eq:escprob}
into equation~\eqref{eq:alpha} and using the approximations that $u_{1}\;\mathrm{and\;}u_{2}\ll c$,$ $\[
\alpha\approx1+\frac{3}{\frac{u_{1}}{u_{2}}-1}.\]
Now we can insert equation~\eqref{eq:velocities} for $u_{1}/u_{2}$,
use $\gamma=5/3$ for a monatomic ideal gas, and take the limit of
a strong shock where $M\gg1$, yielding\begin{equation}
\alpha\approx2+\frac{4}{M^{2}}.\label{eq:fermipowerlaw}\end{equation}
Thus, $dN/dE\sim E^{-2}$ for first order Fermi acceleration.

Calculating the details of the high-energy neutrino spectrum of a
particular type of astrophysical source requires a more sophisicated
treatment that relaxes some of the assumptions we have made here and
accounts for appropriate energy loss mechanisms as well as how the pion energy
is distributed among the decay products in a particular context. However,
since most astrophysical models invoke this first-order Fermi mechanism,
the prediction of a spectrum roughly proportional to $E^{-2}$ is
a generic feature. As such, $d\Phi_{\nu}/dE_{\nu}\propto E_{\nu}^{-2}$
is frequently used as a fiducial model for an astrophysical neutrino
flux -- we make use of this convention in chapter~\ref{chap:superk}.

\subsection{Active galactic nuclei}
\label{sec:AGNs}

Astrophysical systems that produce high energy neutrinos via the mechanism
outlined in the previous section must have two features: a shocked,
magnetized plasma that can accelerate protons to very high energies,
and a sufficiently intense radiation field to allow for photo-pion
production. These conditions are expected to be present in a number
of astrophysical systems, including gamma ray bursts and active galactic
nuclei (AGNs). The expected signal from AGNs is the strongest for
the energy range to which Super-K is sensitive, so they are the primary
astrophysics target for the study discussed in chapter~\ref{chap:superk}.

AGNs are supermassive black holes accreting matter at the centers
of galaxies. They are the most luminous persistent sources in the
universe and emit radiation over a vast range of the electromagnetic
spectrum, from radio waves to gamma rays. There are two locations
where an AGN could produce neutrinos: in the core or in the jets.
We give a rough outline of these processes here; further details can
be found in \citet{1995PhR...258..173G}.

In models of neutrino production within AGN cores, a shock is formed
in an approximately spherical shell around the black hole, where the
ram pressure of the infalling material is balanced by the radiation
pressure of the emission. The region inside the shock is optically
thick and provides ample radiation density for $p\gamma$ collisions.
This process generates high-energy neutrinos from quite close to the
black hole, thus probing a region from which photons cannot escape.
A detailed model of neutrino production in AGN cores can be found
in \citet{Stecker:1995th}, along with a prediction for the diffuse flux of neutrinos
expected from integrating the contributions from all AGN cores over
the history of the universe.

Neutrinos can also be produced within AGN jets. Jets are relativistic
outflows of charged particles collimated along the black hole's rotation
axis. The mechanism that powers the jets is not yet well-understood
in detail, but the basic idea is that a varying magnetic field linked
to the rotation of the black hole or the accretion disk generates
an electric field that can accelerate charged particles. In the case
of jets, the shocks needed for first-order Fermi acceleration to extremely
high energies are formed by inhomogeneities in the matter traveling
along the jet. The required photon field is provided by synchrotron
radiation from electrons spiraling around magnetic field
lines. A model of the predicted diffuse neutrino flux from AGN jets
can be found in \citet{Mannheim:1998wp}. 

It is not known whether jets consist mainly of electrons and positrons
or if there is a hadronic component as well. The observed gamma rays
coming from jets could be explained by either leptonic or hadronic
processes. In the leptonic process, gamma rays are produced via low-energy
synchrotron photons gaining energy via inverse Compton scattering
with the accelerated electrons -- this is known as synchrotron-self-Compton
radiation. Alternatively, accelerated hadrons could produce gamma
rays via the decay of neutral pions, as in equation~\eqref{eq:gamma}.
Since neutrinos are only produced in hadronic processes, observation
of a neutrino flux from these sources would help distinguish between
these models. The question of whether hadrons are accelerated to extremely
high energies by AGNs is also an important question for determining
the origin of ultra-high energy cosmic rays. Recent results from the
Auger Observatory \citep{auger_agns} indicate that the arrival directions of the
highest energy cosmic rays are statistically correlated with the locations
of nearby AGNs -- this is a tantalizing hint that AGNs could be the
answer to the long-standing puzzle of ultra-high energy cosmic rays.
A definitive observation of a neutrino flux associated with AGNs would
lend strong support to this picture \citep{2008arXiv0802.0887H}.

\subsection{Other sources}

In addition to AGNs, there are many other possible astrophysical sources
of neutrinos. For example, gamma ray bursts (GRBs) -- the
highest-energy explosions in the universe -- are likely to provide
the conditions needed for diffusive shock acceleration and photo-pion
production as well \citep{1999PhRvD..59b3002W}. There are also a number of more
exotic phenomena that could produce high-energy neutrinos, including
decays of topological defects such as cosmic strings or magnetic monopoles
and self-annihilation of weakly interacting dark matter particles
(WIMPs) \citep{2004IJMPA..19..317S}.

Another source of high-energy neutrinos is linked to the propagation
of ultra-high energy cosmic rays through the universe: protons with
energies above ${\sim}10^{20}$ eV will undergo photo-pion production
as in equation~\eqref{eq:gamma} via collisions with cosmic microwave
background (CMB) photons (see \S\ref{sec:CMB} for more about the
CMB). This is known as the GZK cutoff \citep{GZK1,GZK2}, and recent cosmic ray
experiments \citep{2007JPhG...34..401S} claim to have 
observed the predicted break
in the cosmic ray spectrum due to this effect. Thus there should be
a corresponding neutrino flux from the decay of the pions produced
-- experiments such as ANITA \citep{2006PhRvL..96q1101B} are attempting to detect these
neutrinos.

\section{Neutrino telescopes}

\subsection{How can we detect the astrophysical neutrino flux?}

The flux of high-energy neutrinos from active galactic nuclei is quite
small and the probability of a neutrino interacting with material
in a detector is also quite low. These two basic facts point to the
need for an extremely large collecting area. Modern neutrino {}``telescopes''
are thus actually large naturally-occurring volumes of ice or water
instrumented with photomultiplier tubes (PMTs). Neutrinos that interact
in or near the detector volume produce charged particles that emit
Cerenkov light as they pass through the water or ice; this light is
then recorded by the PMTs.

In order to make a significant observation of astrophysical
neutrino flux, it is expected that we will need a detector with a volume
on the order of $1\,\mathrm{km}^3$ \citep{2002RPPh...65.1025H}.  There are two main efforts underway
aiming towards this goal: IceCube, a detector currently under construction
at the South Pole which uses the Antarctic ice shelf as its detection medium
that is expected to be completed in 2011 \citep{2007arXiv0711.0353T},
and KM3NeT, a detector to be built in the Mediterranean Sea 
\citep{2006NIMPA.567..457K,2007arXiv0711.0563K} with the combined efforts of 
the ANTARES \citep{2004EPJC...33S.971K}, NEMO \citep{2006astro.ph.11105S}, and 
NESTOR \citep{2005ICRC....5...91A}
collaborations. Additionally, the AMANDA
detector \citep{2004NIMPA.524..169A}, a pathfinder for IceCube covering a 
volume of $0.015\,\mathrm{km}^{3}$ of Antarctic ice that was completed 
in 2000, has already produced a number of interesting results \citep{2007arXiv0712.4406B}.

\subsection{Super-K as a neutrino telescope}

The Super-K volume is a large tank of water with a volume of $5\times10^{-5}\,\mathrm{km}^{3}$.
It was designed primarily to detect neutrinos from the sun, supernovae,
and particle accelerators -- its volume is quite small compared to
the neutrino telescopes currently being built whose primary science
target is the high-energy astrophysical neutrino flux. Thus the chances
of actually seeing this signal in Super-K are not very high.

Nonetheless, it is still an interesting exercise to look for it in
the Super-K data: it provides an important consistency check, and
Super-K can complement IceCube and KM3NeT in terms of sky coverage
and livetime. Furthermore, observing a high-energy neutrino event
in Super-K would provide much more detailed directional information
than the more sparsely instrumented kilometer-scale detectors.

To this end, a number of neutrino astronomy studies have been undertaken
by the Super-K collaboration, including searches for neutrino point
sources \citep{2006ApJ...652..198A,2008APh....29...42D}, supernova bursts \citep{2007ApJ...669..519I}, and annihilations of WIMPs
in the Sun, Earth, and galactic center \citep{2008APh....29...42D}. Here we focus on one
such study: chapter~\ref{chap:superk} describes a search for a diffuse
neutrino flux from unresolved sources performed by looking for extremely
high energy events in 1679.6 live days of Super-K data and comparing
the observed flux to the expected atmospheric neutrino background.
We use Super-Kamiokande's highest energy data sample to set an upper
limit on the diffuse flux of neutrinos from astrophysical sources.

\chapter{Search for diffuse astrophysical neutrino flux using ultra-high energy upward-going
muons in Super-Kamiokande I}
\label{chap:superk}
\begin{quote}
This chapter is adapted from the paper ``Search for diffuse astrophysical neutrino flux using ultra-high energy upward-going muons in Super-Kamiokande I'' by Molly E.~C. Swanson et al. (the Super-Kamiokande Collaboration), which was previously published in the \emph{Astrophysical Journal} 652, pp. 206-215
\citep{2006ApJ...652..206S}.
\end{quote}

\section{Introduction}

\label{sec:Introduction}

Neutrinos have great potential as astronomical messengers. The detection
of neutrinos from astronomical sources has led to 
many significant discoveries.  
Two sources of extraterrestrial neutrinos in 
the MeV energy range have been detected so far: the
Sun \citep{2003RvMP...75..985D,2003RvMP...75.1011K} 
and supernova 1987a \citep{1987PhRvL..58.1490H,1987PhRvL..58.1494B}.
The next step for neutrino astronomy is to detect neutrinos in the
GeV-PeV energy range, which will open a new window on the high energy
universe. 
A wide variety of astrophysical phenomena are expected to
produce extremely high energy neutrinos, ranging from active galactic
nuclei (AGNs) and gamma-ray bursts (GRBs; 
\citealt{2002RPPh...65.1025H,1995PhR...258..173G}) 
to more exotic sources such
as dark matter annihilation or decays of topological 
defects \citep{2004IJMPA..19..317S}.

The flux of neutrinos at such high energies is quite small; therefore,
large-scale detectors are required. One effective technique for observing
high-energy neutrinos with an underground detector is to look for
muons produced by $\nu_{\mu}$ or $\bar{\nu}_{\mu}$ interacting in
the surrounding rock. (Throughout this chapter, ``muons'' will refer
to both $\mu^{+}$ and ${\mu^{-}}$.) The muon range in rock increases
with muon energy, which expands the effective interaction volume for
high-energy events. Downward-going neutrino-induced muons cannot be distinguished
from the much larger flux of downward cosmic-ray muons, but since cosmic 
ray muons cannot travel through the entire Earth, upward-going muons are 
almost always neutrino-induced. Thus, upward-going muons provide a suitable 
high-energy neutrino signal.

At muon energies above $1-10{\rm {\; TeV}}$, the upward-going muon
flux due to neutrinos from AGNs is expected to exceed the upward-going muon flux due
to atmospheric neutrinos~\citep{Stecker:1995th,Mannheim:1998wp}.
This cosmic
neutrino flux could be detected either by searching for point sources
of high-energy neutrinos or by detecting a diffuse, isotropic flux
of neutrinos from unresolved astrophysical sources. A diffuse cosmic
neutrino flux would be observed as an excess to the expected atmospheric
neutrino flux at high energies. In this analysis, we focus on searching
for a diffuse flux of upward-going muons due to neutrinos from astrophysical
sources using the highest energy data sample in Super-Kamiokande (Super-K).  

This study
complements other Super-K searches for astrophysical point sources
of high energy neutrinos that use data over a larger
energy range \citep{2006ApJ...652..198A}.
In this chapter we describe a search for evidence of a high energy astrophysical
neutrino flux in Super-K's highest energy upward-going muon sample. 
In \S\ref{sec:Detector} we describe the 
Super-Kamiokande detector, and in \S\ref{sec:Event-Selection} we give the 
details of how we selected candidate events from Super-K's ultra--high-energy sample. 
We evaluate 
our selection process with Monte Carlo 
in \S\ref{sec:High-Energy-Isotropic-MC} 
and calculate the observed upward-going muon flux in \S\ref{sec:Flux-Calculation}. 
Section \ref{sec:Expected-Atmospheric-Background} and \S\ref{sec:Analytical-Estimate}
discuss the background due to the atmospheric neutrino flux. Based on the 
results, we set an upper limit in \S\ref{sec:Upper-Limit} and conclude in 
\S\ref{sec:Conclusions}. Any necessary estimates and approximations have been
 made so that they lead to a conservative result for this upper
limit.

\section{The Super-Kamiokande Detector}

\label{sec:Detector}

\begin{figure*}
\includegraphics[width=1\textwidth]{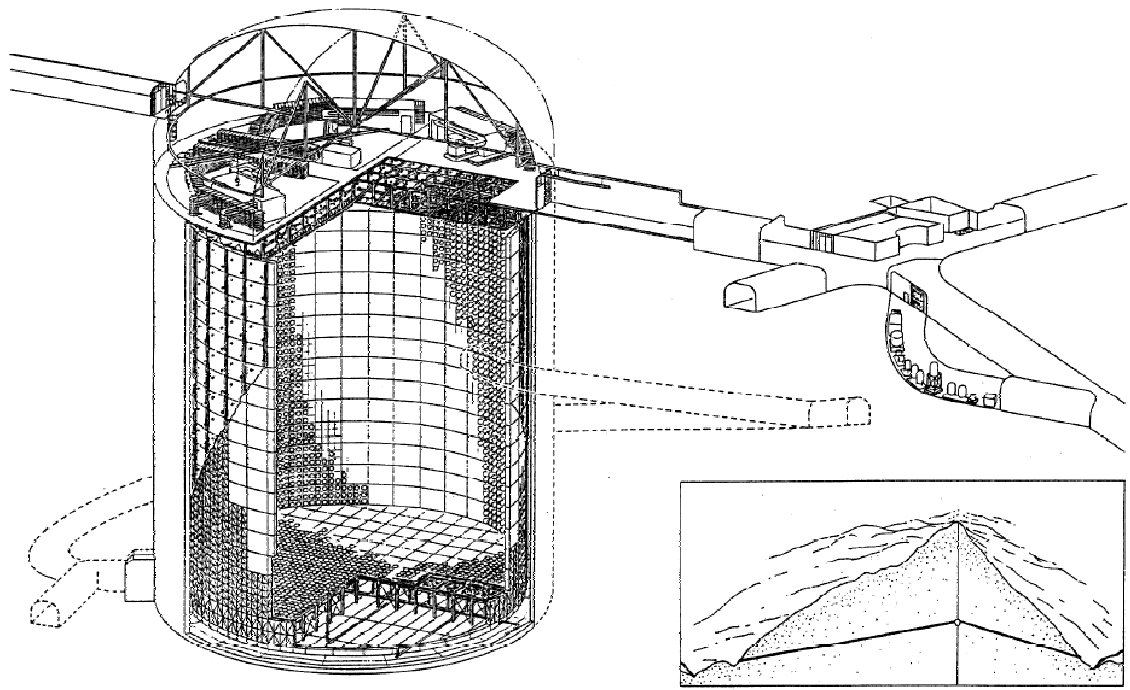}
\caption[Schematic drawing of the Super-K detector.]{
Schematic drawing of the Super-K detector. Inset at bottom right shows the location of the detector within Mount Ikenoyama (cutaway view).
\label{fig:sk_detector}}
\end{figure*}

The Super-K detector,
shown in
Figure~\ref{fig:sk_detector},%
 is a cylindrical 50 kiloton
water Cerenkov detector, located in the Kamioka-Mozumi mine in Japan.
It is 41.4 m tall and 39.3 m in diameter.
The detector was constructed under the peak of Mount Ikenoyama, which
provides an average rock overburden
of 1000~m (2700~m water equivalent). Its geodetic location is at 
36.4{$^\circ$}~north, 137.3{$^\circ$}~east, and altitude 370~m. 

Super-K consists of 
two concentric, optically separated detectors. 
Until 2001 July the inner detector (ID) was 
instrumented with 11,146 inward-facing 50 cm diameter photomultiplier tubes 
(PMTs). The outer detector (OD) is a cylindrical shell of water surrounding
the ID and is instrumented with 1885 outward-facing 20~cm diameter
PMTs. Between the ID and the OD, there is a 50~cm thick shell.
Photons coming from this
region will not be detected by either the OD or the ID, so we refer
to it as the insensitive region. 

More details about the detector can
be found in \citet{2003NIMPA.501..418T}.
The data sample used in this analysis
was taken from 1996 April to 2001 July, corresponding to 1679.6 days
of detector livetime. This data run is referred to as SK-I.

Super-K is primarily designed to detect lower energy neutrinos from
the Sun, the atmosphere, and particle accelerators 
but can potentially
detect the extremely high energy neutrinos expected from astrophysical
sources as well. This chapter focuses on the
events at the highest energy end of Super-K's detection range.

\section{Event Selection}
\label{sec:Event-Selection}
The ultra--high-energy sample in SK-I consists of events that
deposit $\ge1.75\times10^{6}$ photoelectrons (pe) in the ID.
In the low-energy regime, on average about 9 pe are recorded by the ID PMTs
for each MeV of energy deposited in the tank;
the electronics for the ID PMTs saturate at about 300 pe. Thus an event
with $\ge1.75\times10^{6}$ pe in the ID corresponds to a minimum energy
deposition of approximately 200 GeV, but the actual energy deposition
could be much higher, since the saturation effect prevents
all of the produced pe from being recorded.

At high energies, muons have some probability to lose energy through radiative
processes such as bremsstrahlung, resulting in an electromagnetic shower
that deposits large quantities of pe in the detector. For comparison, a muon
that traverses the maximum path length through the ID (50 m) but does not
produce any electromagnetic showers will deposit approximately 11 GeV via 
ionization energy loss, corresponding to  $\sim 10^{5}$ pe deposited in the 
ID. Thus a high-pe
cutoff offers a means of selecting high energy events. 

At the high-pe threshold of $\ge1.75\times10^{6}$ pe, the high level of 
saturation in the ID PMT electronics 
can cause Super-K's precision
muon fitting algorithms to fail. Therefore, these extremely energetic events
are not included in other studies of upward-going muons in SK-I 
\citep{2004PhRvD..70h3523D,1999PhRvL..82.2644F,2005PhRvD..71k2005A}.
In this study we analyzed this ultra--high-energy data sample separately 
using a different fitting method based on information
from the OD.

\subsection{Outer Detector Linear Fit}

\label{sub:Linear-Fit-Method}

\begin{figure*}
\includegraphics[width=.5\textwidth]{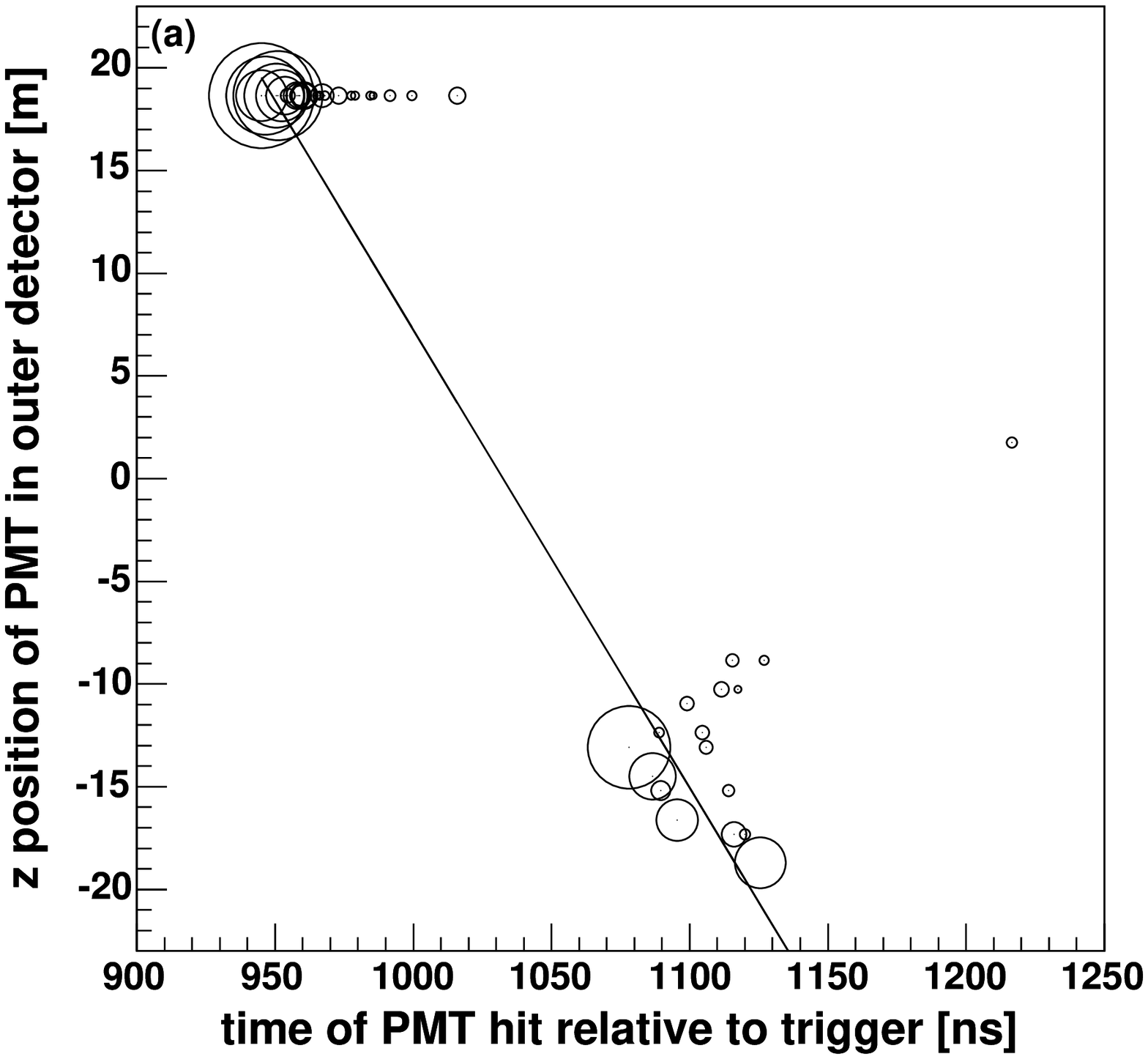}\includegraphics[width=.5\textwidth]{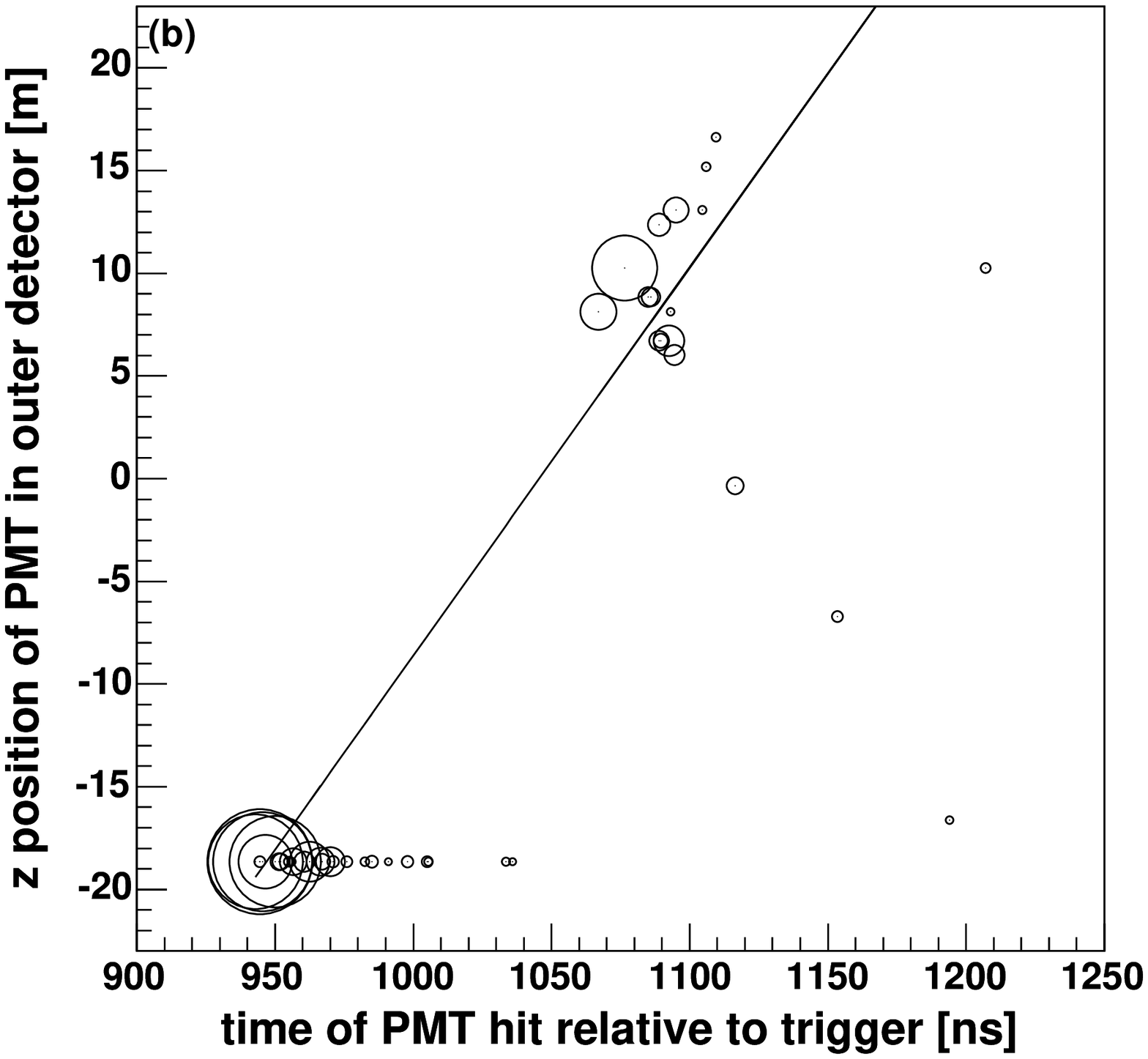}
\caption[OD-based muon trajectory fit applied to example MC muon events.]{
(\emph{a}) OD-based muon trajectory fit applied
to an example MC downward-going muon event. (\emph{b}) OD-based fit applied to an example
MC upward-going muon event. The size of the circle around each point is proportional
to the charge detected in the PMT.\label{fig:fit_examples}}
\end{figure*}

SK-I's ultra--high-energy data sample contains a total
of 52214 events. Most of these are either very energetic downward-going
cosmic-ray muons or multiple muon events where two or more downward-going
muons hit the detector simultaneously. In order to select candidate
upward-going muons from this sample, we applied a simple linear fit to the OD 
data for each event. A linear
fit was done on the $z$-position of each OD PMT versus the time it fired,
weighted by the total charge in the PMT. Example fits of simulated
downward-going and upward-going muon events
are shown in
Figure~\ref{fig:fit_examples}.%
The slope of this fitted line is an estimate of $-\cos{\Theta}$,
where $\Theta$ is the zenith angle of the incoming muon. A positive
fitted slope indicates that the muon is upward-going. A similar linear
fit was done on the $x$- and $y$-positions to determine the full muon trajectory
through the detector.

Since this fitting method is based on the OD (which has a lower
resolution than the ID), it is not as precise as the muon fitting algorithms
used in the lower energy upward-going muon analysis. However, it works even 
when the ID PMT electronics are completely saturated and the precision 
ID-based algorithms fail.

\subsection{Selection Cuts}

\label{sub:Selection-Cuts}

To select candidate upward-going muons, we applied 
the OD-based fit to all 52214
events in the ultra--high-energy data sample. A cut of $\cos{\Theta}\le0.1$
was used to eliminate the bulk of the downward-going single and multiple
muon events. In addition, the fitted trajectory was required to have a path
length of $>7{\rm {\; m}}$ in the ID.

To ensure high-quality fit results, we looked at the number of OD
PMTs hit near the projected entry and exit points. For a true
throughgoing muon, there will be a cluster of hit PMTs around the
entry and exit points. If the fit is accurate,
the projected path should pass through both of these clusters, so
we made an additional cut on the number of OD PMTs hit within $10{\rm {\; m}}$
of the projected OD entry and exit points, $N_{{\rm ODentry}}$
and $N_{{\rm ODexit}}$. We required both $N_{{\rm ODentry}}$
and $N_{{\rm ODexit}}$ to be 10 or greater, indicating significant 
clusters of OD PMTs that were hit near the entry and exit points of 
the fitted trajectory.

Events that do not have $N_{{\rm ODentry}}$ and $N_{{\rm ODexit}}\ge10$ are generally either
stopping muons, partially contained events, or poorly fitted throughgoing muons.
Stopping muons are muons that stop in the detector and only form an OD entry cluster.  
They typically have energies of 1-10 GeV, well below the range of the 
expected astrophysical signal, so it
is appropriate to discard events that look like stopping muons.
Partially-contained events are neutrino interactions that take place inside the detector 
and only
form an OD exit cluster --- they are not part of the upward-going muon flux incident on the 
detector, so we want to discard these as well.
This cut also occasionally eliminates inaccurately fitted
throughgoing muon events, which
reduces the efficiency somewhat but improves the accuracy
of the fit results. 

Another possible type of event that can masquerade as a throughgoing
muon is a partially-contained event with multiple exiting particles.
Such an event will create two (or more) clusters of hit PMTs in the
OD, which could be mistaken as the entry and exit points of a throughgoing
muon. In order to eliminate such events, we looked at the timing between
the OD and ID entry points. If the event is truly a throughgoing muon,
the OD PMTs near the entry point should fire before the ID PMTs. We
determined OD and ID entry clusters via a simple time-based clustering
method and evaluated the mean time of hits within $6{\rm {\; m}}$
of the OD and ID entry points, $t_{{\rm IDentry}}$ and $t_{{\rm ODentry}}$.

In an ideal measurement, $t_{{\rm IDentry}}<t_{{\rm ODentry}}$
would indicate that the perceived entry cluster is actually caused
by an exiting particle. However, the timing determination is complicated
by an effect known as prepulsing.  
Prepulsing occurs when there are so many photons incident on the PMT that 
they are not all converted to photoelectrons at the photocathode - some photons
hit the first dynode instead and are converted to photoelectrons there.
When this happens
in an ID PMT, the time it was hit appears earlier than it actually
was, which will artificially reduce $t_{{\rm {ID\, entry}}}$, making
the ID light appear earlier compared to the OD light.
If this occurs for 
an ultra-high energy throughgoing muon event, it could cause
 $t_{{\rm IDentry}}-t_{{\rm ODentry}}$ to be negative.
To allow for this effect, we used a fairly loose
cut of $-40{\rm {\; ns}}$: if $t_{{\rm IDentry}}-t_{{\rm ODentry}}<-40{\rm {\; ns}}$,
indicating very early ID light, then the event was rejected as a likely
neutrino interaction in the ID.

After these cuts on $\cos{\Theta}$, path length, $N_{{\rm ODentry}}$,
$N_{{\rm ODexit}}$, and $t_{{\rm IDentry}}-t_{{\rm ODentry}}$
were applied to the 52214 events in the sample, 343 candidate events
remained. These remaining events were then evaluated by a visual scan
and a manual direction fit by two independent researchers
to select events with $\cos{\Theta}<0$. 

The visual scan eliminates events that can pass the automatic reduction but are obvious to the trained human eye as noise, i.e., mainly downward-going multiple muon events and ``flashers.''  Multiple muon background events occur when two or more downward-going muons from an atmospheric shower pass through the detector simultaneously.  They have extra energy deposition (and thus are expected to be more common in the ultra--high-energy sample) and typically give poor OD fit results due to multiple OD clusters, but they are easy to identify visually.  Flashers are events caused by malfunctioning PMTs that emit light and create characteristic patterns.  After the multiple muon events and flashers are removed, the manual direction fit separates truly upward-going muons from mis-fitted downward-going muons.

From the 343 candidates, only one event passed the visual scan and manual direction
fit selection as being truly upward-going. The breakdown of the visual
scan and manual fit classifications is shown in Table~\ref{table:visual scan}.
\begin{table}
\begin{center}
\caption{Visual scan of candidate upward-going muon events
\label{table:visual scan}}
\begin{tabular}{cc}
\tableline 
\tableline
Visual scan classification&Number of events\\
\tableline 
Multiple muon events&164\\
PMT ``flashers'' (malfunctioning PMTs)&5\\
Other noise events&2\\
Downgoing muons (manual fit $\cos\Theta\ge0$)&171\\
Upgoing muons (manual fit $\cos\Theta<0$)&1\\
\tableline 
Total&343\\
\tableline
\end{tabular}
\end{center}
\end{table}

This upward-going muon event selected from the $\ge1.75\times10^{6}{\rm {\; pe}}$
sample is the ultra-high energy upward-going muon signal observed by SK-I.  This event
occurred on 2000 May 12 at 12:28:07 UT and deposited 1,804,716 pe in the ID.  
Based on the manual fit results, the path length through the ID was 40 m, and the 
zenith angle was $\cos{\Theta}=-0.63$, corresponding to a direction of origin of 
$({\rm{R.A.}}, {\rm{decl.}})=20^{\rm h}38^{\rm m},-37^{\circ}18\arcmin$.



\section{High-Energy Isotropic Monte Carlo}

\label{sec:High-Energy-Isotropic-MC}

In order to calculate the observed muon flux, we need to determine
the resolution and efficiency of the OD-based fit and other cuts on high-energy
muons, and we need to estimate the probability that a muon of
a given energy will deposit $\ge1.75\times10^{6}{\rm {\; pe}}$ in
the ID.

To determine these quantities, we generated a high-energy isotropic
Monte Carlo (MC) sample. This MC consists of an isotropic flux
of muons in monoenergetic bins impinging on the Super-K detector,
representing a flux of muons from neutrino interactions in the surrounding
rock. The MC simulates the detector response -- i.e., the time hit and
charge deposited for each PMT.

Seven monoenergetic bins were used, with muon energies ranging
from $100{\rm {\; GeV}}$ to $100{\rm {\; TeV}}$, and 10000 events
with a path length in the ID of $>7{\rm {\; m}}$ were generated in
each bin. The simulation was performed using a {\tt GEANT}-based detector
simulation. {\tt GEANT}'s muon propagation has been shown to agree with
theoretical predictions up to muon energies of $100{\rm {\; TeV}}$
\citep{2001NIMPA.459..319B,2003ICRC....3.1673D}.

\subsection{Resolution and Efficiency of Event Selection}

\label{sub:Resolution-and-Efficiency}

\begin{figure*}
\includegraphics[width=.5\textwidth]{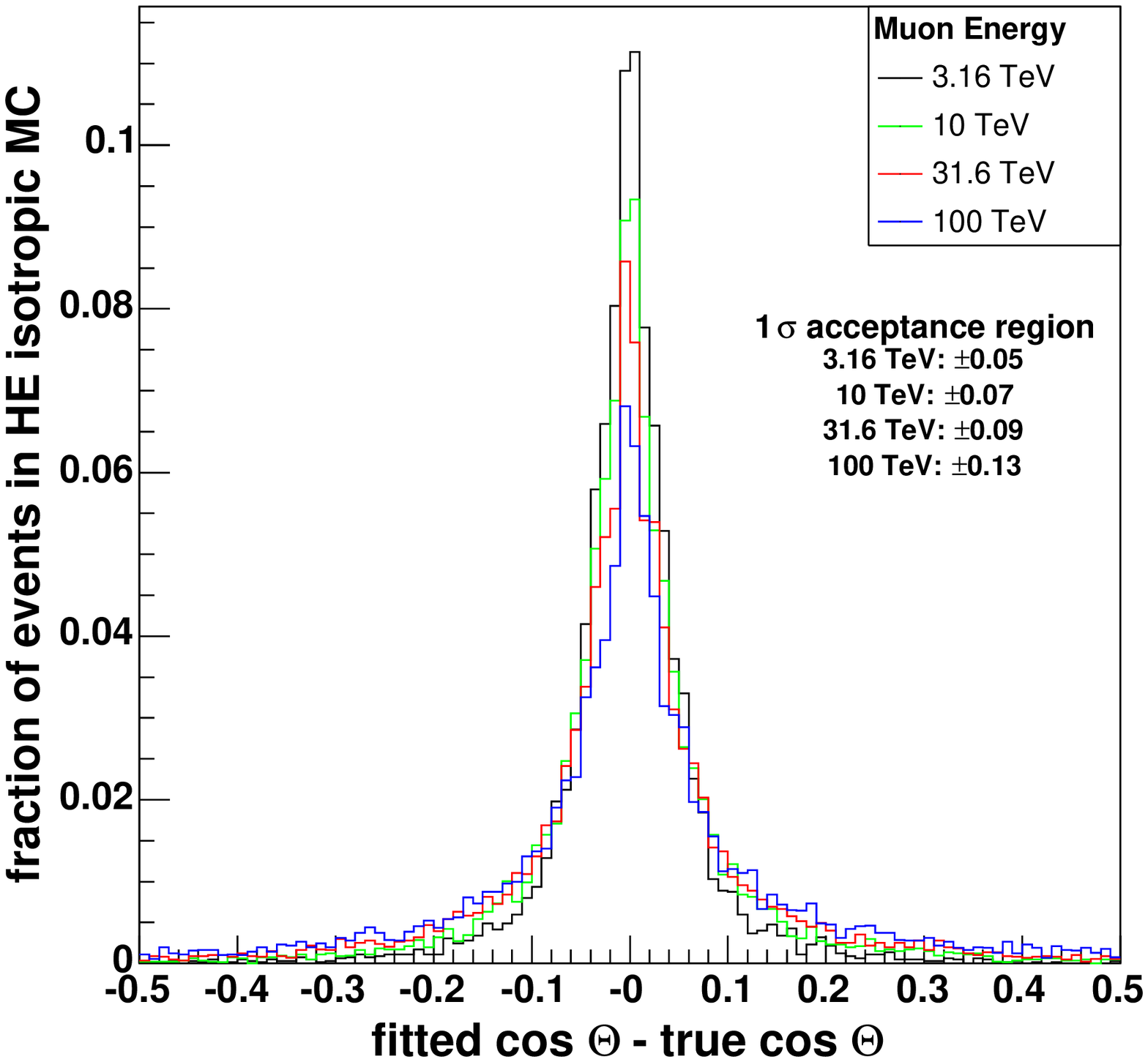}\includegraphics[width=.5\textwidth]{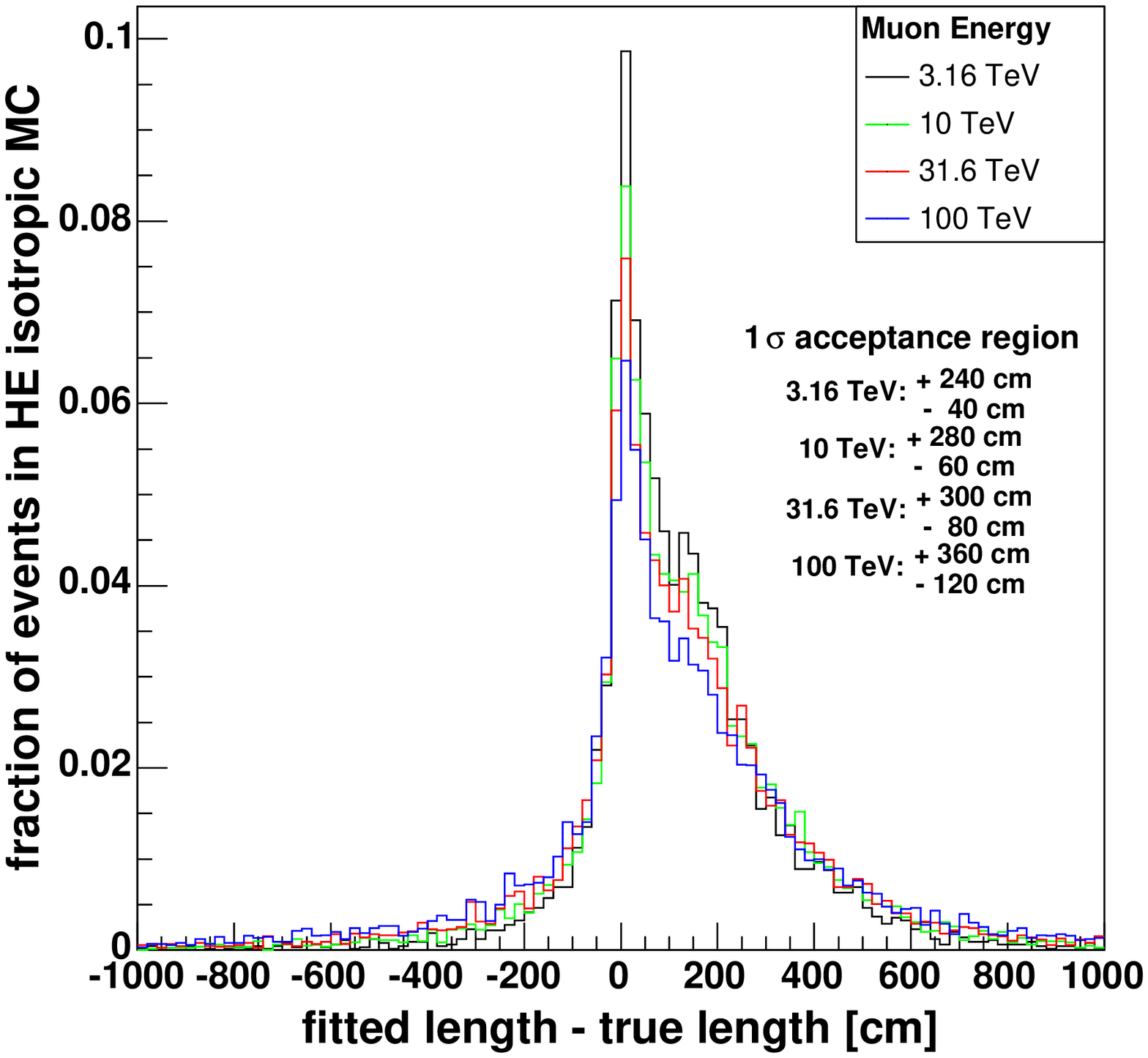}\\
\includegraphics[width=.5\textwidth]{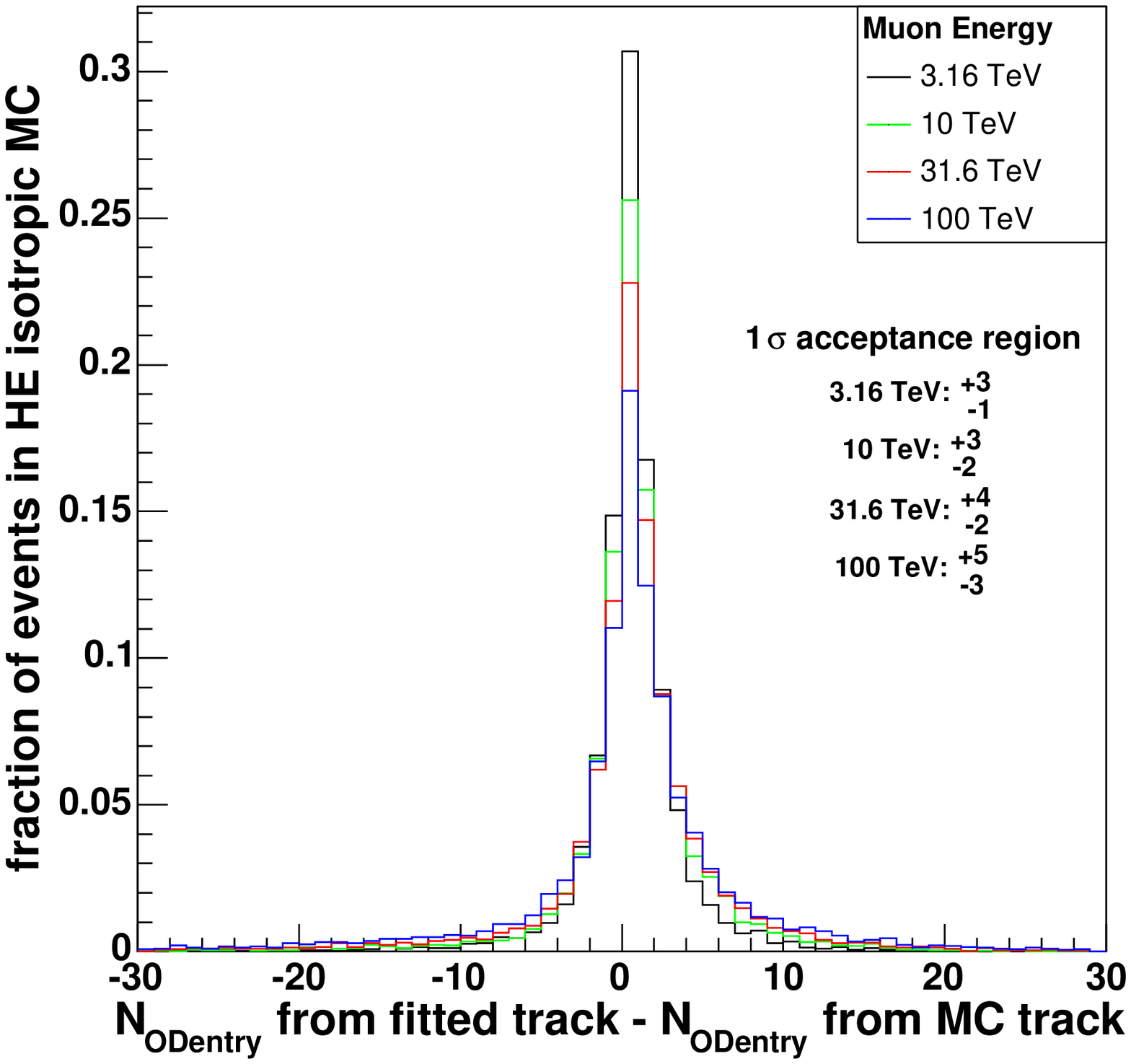}\includegraphics[width=.5\textwidth]{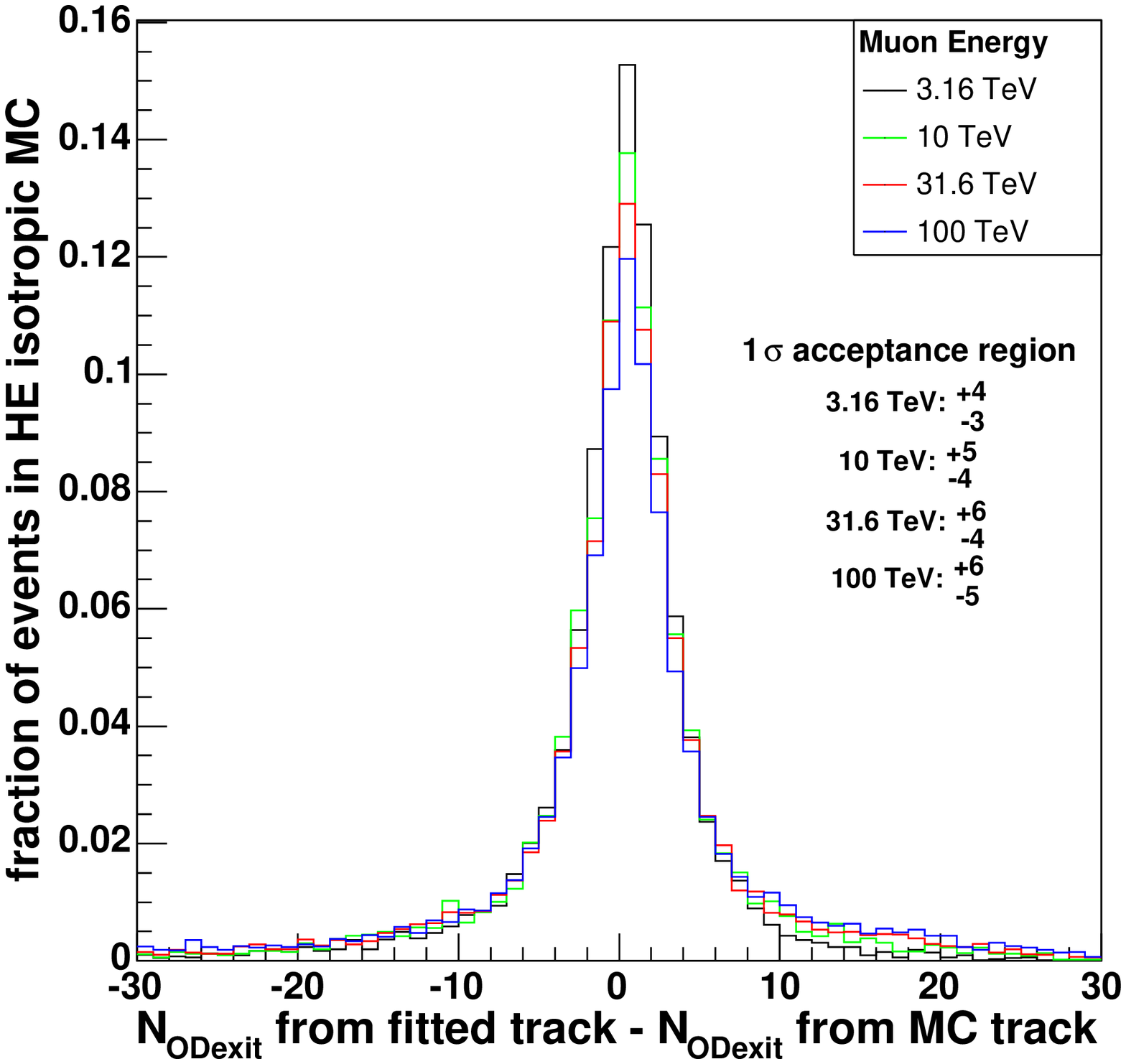}
\caption[Resolution of OD-based fit on determining track parameters for events
from the high energy isotropic MC.]{Resolution of OD-based fit on determining track parameters for events
from the high energy isotropic MC with $>10^{5}{\rm {\; pe}}$
in the inner detector. 
Parameters shown are $\cos\Theta$, path length, and the number of OD PMTs hit 
within 8 meters of the OD entry and exit points.
The $1\,\sigma$ acceptance regions contain
68.27\% of the total area under each histogram in each plot.\label{fig:histograms}}
\end{figure*}

\begin{figure}
\begin{center}%
\includegraphics[width=.5\textwidth]{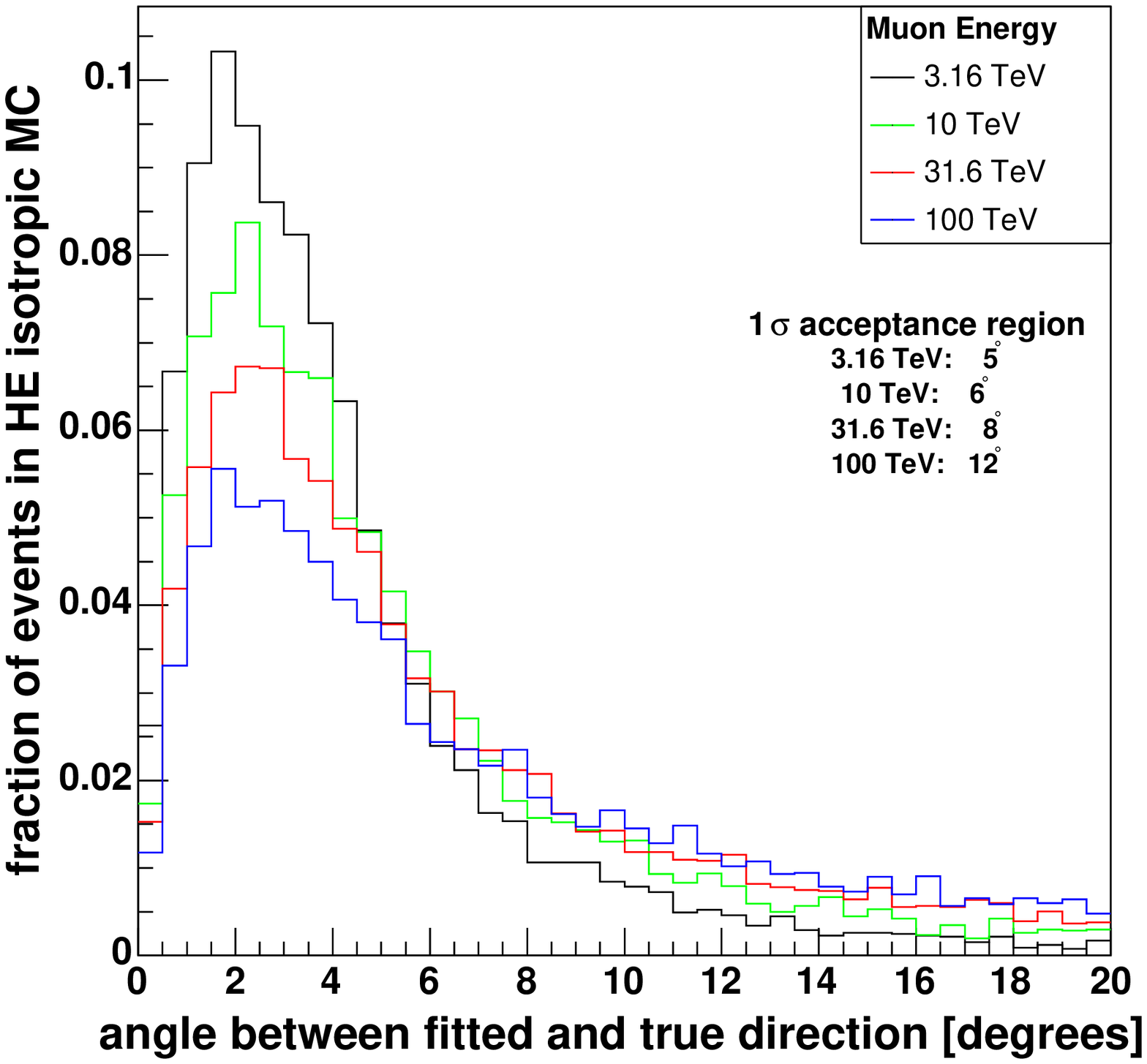}%
\end{center}%
\caption[Angular resolution of OD-based fit for events 
from the high energy isotropic MC.]{Angular resolution of OD-based fit for events 
from the high energy isotropic MC with $>10^{5}{\rm {\; pe}}$
in the ID.\label{fig:angle_rock}}
\end{figure}

The muon trajectory fitting algorithm
 discussed in \S\ref{sub:Linear-Fit-Method}
was applied to the high-energy isotropic MC, and the fitted values
for $\cos\Theta$, path length, number of OD PMTs hit near the entry
and exit points, and direction of trajectory were compared with the
true MC values.
These plots are shown in Fig.~\ref{fig:histograms}.%
A plot 
illustrating the angular resolution
of the fit is shown in Fig.~\ref{fig:angle_rock}.
The resolution of 5--12$^\circ$ is poor compared to the typical $1^\circ$ 
resolution of precision ID fitting algorithms, but those
algorithms do not work well on
these high-energy saturated events. 

The efficiency of the upward-going cuts was estimated by considering all of 
the MC events with true
values of $\cos{\Theta}<0$ and ID path length $>7{\rm {\; m}}$
and then determining the fraction of these events that pass the selection 
cuts described in \S\ref{sub:Selection-Cuts}. This was done using 
isotropic MC events
with $>10^{5}{\rm {\; pe}}$ since there were relatively few events in the MC with 
$\ge1.75\times10^{6}{\rm {\; pe}}$. (Above $10^{5}{\rm {\; pe}}$ the efficiency and 
resolution do not depend strongly on the number of ID pe deposited.) Also, as 
discussed in \S\ref{sub:High-Energy-Fraction},
only the energy bins in the range $3.16-100{\rm {\; TeV}}$ were considered.
The efficiency
was calculated as a function of muon energy and $\cos\Theta$. 

Statistical uncertainties on the efficiency determination were calculated
using the Bayesian method discussed by 
\citet{Conway:2002}.
Systematic
uncertainties due to uncertainties in the fitted values of the zenith
angle, path length, and number of OD PMTs hit near the entry and exit
points were 
calculated using the resolution histograms for $\cos\Theta$,
path length, $N_{{\rm OD\, entry}}$ and $N_{{\rm {OD\, exit}}}$
shown in Fig.~\ref{fig:histograms}.
First, a $1\,\sigma$ acceptance region containing 68.27\% of the
total area was selected for each resolution histogram. Then the cuts
on these parameters were varied by $1\,\sigma$ in either direction
to determine the effect on the efficiency.


Another source of systematic uncertainty on the efficiency is 
prepulsing, which is not included in the MC simulation and must be estimated
separately.
To do this, we compared the results from
the high-energy isotropic MC to a sample of 627 ultra--high-energy 
downward-going muon data events with $\ge1.75\times10^{6}{\rm {\; pe}}$ in the ID
selected by a visual scan. For many of these data events,
the time difference $t_{{\rm IDentry}}-t_{{\rm ODentry}}$ is negative, showing 
evidence of prepulsing not seen in the MC sample. 

\begin{figure}
\begin{center}%
\includegraphics[width=.5\textwidth]{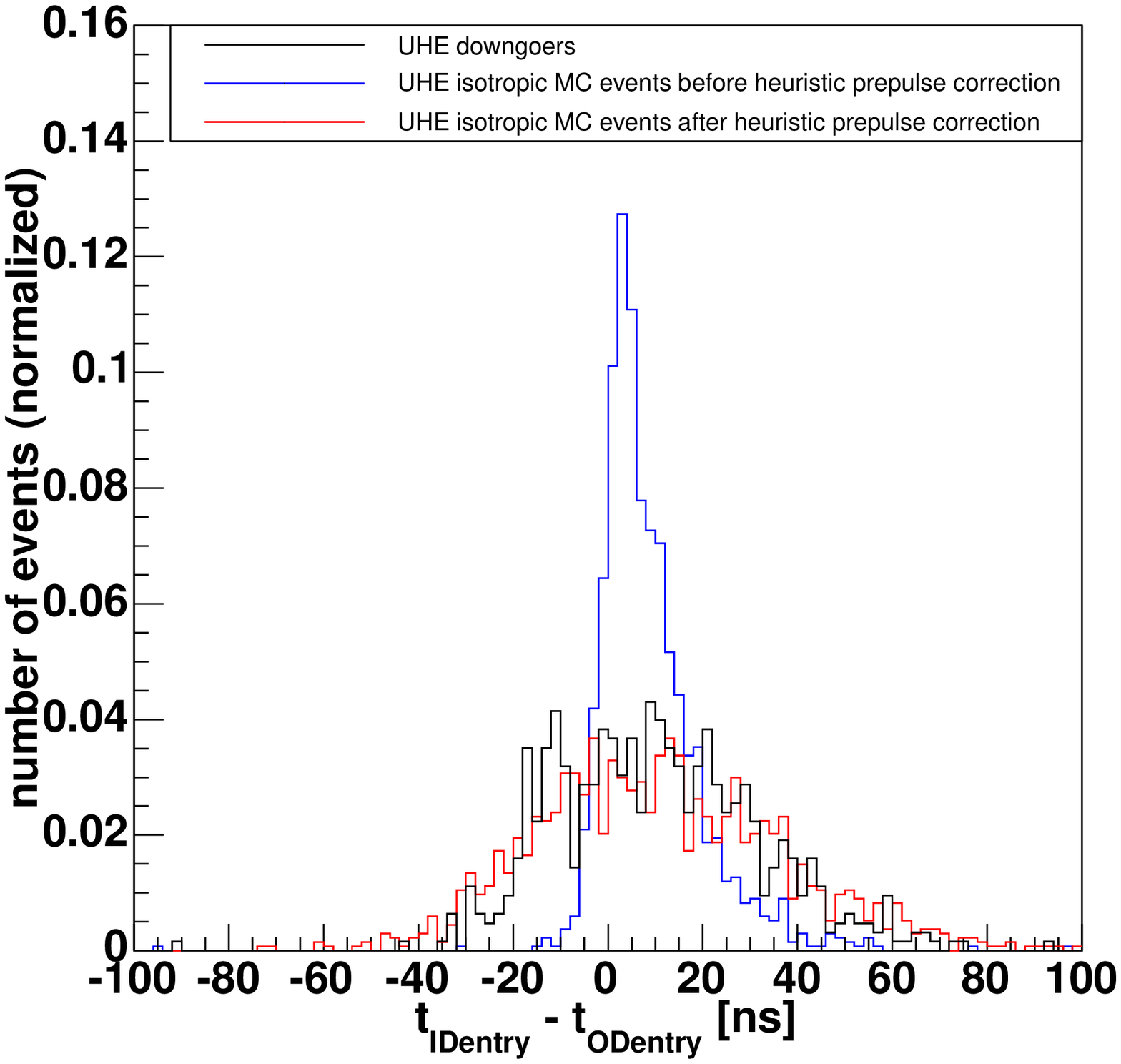}%
\end{center}%
\caption[Distribution of the time difference between OD entry
and ID entry clusters for UHE downgoing muon events from data, compared
to the distribution for the UHE events in the high-energy isotropic
MC.]{\label{fig:odid}Distribution of the time difference between OD entry
and ID entry clusters for UHE downgoing muon events from data, compared
to the distribution for the UHE events in the high-energy isotropic
MC. After applying a heuristic correction to account for prepulsing,
the MC distribution looks similar to the data distribution.}
\end{figure}

Histograms of the time difference $t_{{\rm {ID\, entry}}}-t_{{\rm {OD\, entry}}}$
for the data events and the MC events are shown in Fig.~\ref{fig:odid}.

Note that for the MC events, the histogram is sharply peaked around
a time difference slightly above zero, and has very few events with
negative time differences. In contrast, the histogram for the data
events has a wider spread, and has several events with negative time
differences. The difference between the data and the MC distributions
is expected to be due mainly to prepulsing.

To adjust for this, we derive a heuristic correction for prepulsing
by drawing a random value from the distribution defined by the data
histogram and another random value from the distribution defined by
the MC histogram, and then subtracting the data value from the MC
value. By repeating this process several thousand times, we build
up a distribution of the expected difference between the data and
MC values of $t_{{\rm {ID\, entry}}}-t_{{\rm {OD\, entry}}}$. Then
for each event in the MC sample, a random value is drawn from the
distribution of the expected difference between data and MC, and then
adding it to $t_{{\rm {ID\, entry}}}-t_{{\rm {OD\, entry}}}$ for
the MC event. The result is a rough approximation of what the time
difference might be if the MC simulation included prepulsing.

The histogram of the heuristically-corrected ultra-high energy MC
events is also shown in Fig.~\ref{fig:odid} --- the distribution
looks similar to the data events, as expected. To estimate the systematic
uncertainty on the efficiency calculation due to prepulsing, we
applied this adjustment to the events in the MC sample and found that an 
additional 1\% of the MC events would be cut after accounting for prepulsing, which 
lengthens the negative-side error bars on the efficiency estimates by approximately 0.01. 

Finally, one additional correction must be made: the efficiency has
been estimated \emph{at} different values for $E_{\mu}$, but the
flux calculation is for the flux \emph{above} a threshold energy $E_{\mu}^{\rm min}$.
Since the efficiency decreases with energy, $\varepsilon\left(E_{\mu},\Theta\right)$
is an overestimate for $\varepsilon\left(\ge E_{\mu}^{\rm min},\Theta\right)$.
This does \emph{not} lead to a conservative upper limit for the flux, so
a correction must be made.  This requires knowledge of the energy spectrum
of the expected signal, so we modeled the signal as an isotropic flux
of neutrinos with ${d\Phi_{\nu}\left(E_{\nu}\right)}/{dE_{\nu}}\propto E_{\nu}^{-2}$
(a plausible astrophysical spectrum \citealt{1990cup..book.....G}), used the method of
 \S\ref{sec:Analytical-Estimate} to estimate the muon flux
$\Phi_{\mu}\left(\geq E_{\mu}^{\rm min}\right)$, and used this muon flux 
to extrapolate our efficiency calculations to find 
$\varepsilon\left(\ge E_{\mu}^{\rm min},\Theta\right)$.

The efficiency as a function of energy was estimated using a linear
fit on $\varepsilon\left(E_{\mu},\Theta\right)$ vs.~$\log E_{\mu}$
in each angular bin, using the MC results from the energy
bins in the range $3.16-100{\rm {\; TeV}}$. 
This gives an estimator
$\hat{\varepsilon}\left(E_{\mu},\Theta\right)$ for the efficiency:\begin{equation}
\hat{\varepsilon}\left(E_{\mu},\Theta\right)=A\left(\Theta\right)+B\left(\Theta\right)\log E_{\mu},\label{eq:extrapolation}\end{equation}
where $A$ and $B$ are the fit parameters. 

Then $\varepsilon\left(\ge E_{\mu}^{\rm min},\Theta\right)$
is estimated by weighting the fitted efficiency
$\hat{\varepsilon}\left(E_{\mu},\Theta\right)$ 
by the $E_{\nu}^{-2}$ model muon flux.

\begin{equation}
\hat{\varepsilon}\left(\ge E_{\mu}^{\rm min},\Theta\right)=\frac{\int_{E_{\mu}^{\rm min}}^{\infty}\hat{\varepsilon}\left(E_{\mu},\Theta\right)\frac{d\Phi_{\mu}\left(\geq E_{\mu}\right)}{dE_{\mu}}dE_{\mu}}{\int_{E_{\mu}^{\rm min}}^{\infty}\frac{d\Phi_{\mu}\left(\geq E_{\mu}\right)}{dE_{\mu}}dE_{\mu}}.\label{eq:eff_correction}\end{equation}

This procedure yields small downward adjustments ($0.6-9\%$) to the calculated
efficiency in each angular bin. 

Additionally, we must also account for the efficiency of the visual
scan and manual fit procedure. To do this, we did a visual scan of
the 605 events from the high-energy isotropic MC with
true $\cos{\Theta}<0$, true ID path length $>7{\rm {\; m}}$, and
$\ge1.75\times10^{6}{\rm {\; pe}}$ in the ID and found that 10 of
these events were eliminated in the visual scan. This gives an efficiency
of roughly $98\%$, so we adjust our efficiency results by a factor
of $0.98$. 

\begin{figure}
\begin{center}
\includegraphics[width=.5\textwidth]{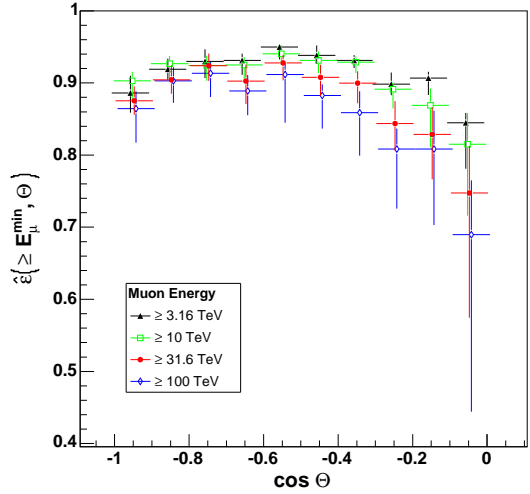}
\end{center}
\caption[Efficiency of our data reduction on 
upward, throughgoing, $\ge 1.75\times10^{6}{\rm {\; pe}}$ muons.]{Efficiency of our data reduction on 
upward, throughgoing, $\ge 1.75\times10^{6}{\rm {\; pe}}$ muons  
as determined by the high-energy isotropic MC. Error
bars include statistical and systematic errors.\label{fig:eff_total}}
\end{figure}

The final results for the efficiency of our cuts on upward, throughgoing, $\ge1.75\times10^{6}{\rm {\; pe}}$ muons (including all corrections
discussed above) are plotted in Figure~\ref{fig:eff_total}.%

\subsection{Ultra--High-Energy Fraction}

\label{sub:High-Energy-Fraction}

In order to set an upper limit on the flux of upward-going
muons from cosmic neutrinos one must make an inference about the energies
of the upward-going muons in the ultra--high-energy sample. Above energies
of $\sim1{\rm {\; TeV}}$, muon energy loss in water is dominated
by radiative processes such as bremsstrahlung,  
so high-energy muons
have some probability of depositing large numbers of photoelectrons
in the Super-K detector and contributing to the ultra--high-energy sample.
Since this energy loss is not continuous, it is not possible to estimate
the muon energy for a single ultra--high-energy upward-going muon event.
Rather, MC is used to make a statistical statement about
the energies of the muons that make up the $\ge1.75\times10^{6}{\rm {\; pe}}$
sample.

The high-energy isotropic MC has been used to determine the
fraction $k\left(E_{\mu}\right)$ of muons with energy $E_{\mu}$
that will deposit $\ge1.75\times10^{6}{\rm {\; pe}}$ in the ID, thus
contributing to the ultra--high-energy sample. Results are shown in
Table~\ref{table:highe_frac}.%
\begin{table}
\begin{center}
\caption{Fraction of high-pe events $k\left(E_{\mu}\right)$
in high energy isotropic MC\label{table:highe_frac}}
\begin{tabular}{ccc}
\tableline 
\tableline
&
number of MC events&
$k\left(E_{\mu}\right)$ \\
$E_{\mu}$&
(out of 10000) with&
(with statistical\\
(TeV)&
$\ge1.75\times10^{6}{\rm {\; pe}}$ in ID&
uncertainties)\\
\tableline
0.1, 0.316, 1&
0&
$0.0000\begin{array}{c}
+0.0001\\
-0.0000\end{array}$\\
$3.16$&
35&
$0.0035\pm0.0006$\\
$10$&
113&
$0.0113\begin{array}{c}
+0.0011\\
-0.0010\end{array}$\\
$31.6$&
293&
$0.0293\pm0.0017$\\
$100$&
879&
$0.0879\begin{array}{c}
+0.0029\\
-0.0028\end{array}$\\
\tableline
\end{tabular}
\end{center}
\end{table}
 Statistical uncertainties were calculated using the Bayesian method
discussed by \citet{Conway:2002}. As can be seen in Table~\ref{table:highe_frac},
the three lowest energy bins --- from $100\;{\rm {GeV}}$ to $1\;{\rm {TeV}}$
--- do not make a significant contribution to the $\ge1.75\times10^{6}{\rm {\; pe}}$
sample. Hence, the rest of this analysis was done using only the four
highest energy bins --- from $3.16$ to $100\;{\rm {TeV}}$.

These results for $k\left(E_{\mu}\right)$ are used in \S\ref{sec:Flux-Calculation}
to calculate the flux $\Phi_{\mu}\left(\geq E_{\mu}^{\rm min}\right)$.
Since $k\left(E_{\mu}\right)$ increases with energy, it is an underestimate
for $k\left(\geq E_{\mu}^{\rm min}\right)$, which leads to a conservative
upper limit for $\Phi_{\mu}\left(\geq E_{\mu}^{\rm min}\right)$. 

Also, since this MC only simulates the muon and not the actual
neutrino interaction, the effect of lower energy debris from deep inelastic
scattering events that make it into the detector has been neglected,
again leading to an underestimate of $k\left(E_{\mu}\right)$. This
effect is expected to be small in the energy range considered here
--- above $E_{\mu}=1\;{\rm {TeV}}$ at the detector entry, 
over 80\% of the neutrino-induced upward-going
muons come from over $200\;{\rm{m}}$ away from the detector, so most of
the debris is absorbed by the surrounding rock, as
determined using the atmospheric neutrino MC discussed in
\S\ref{sec:Expected-Atmospheric-Background}.


\section{Flux Calculation}

\label{sec:Flux-Calculation}

The flux of upward-going muons above a threshold energy $E_{\mu}^{\rm min}$
is given by
\begin{equation}
\Phi_{\mu}\left(\geq E_{\mu}^{\rm min}\right)=\frac{1}{2\pi T k\left(\geq E_{\mu}^{\rm min}\right)}{\displaystyle \sum\limits _{j=1}^{n}\frac{1}{\varepsilon\left(\ge E_{\mu}^{\rm min},\Theta_{j}\right) A\left(\Theta_{j}\right)}},\label{eq:flux}\end{equation}
where $n$ is the total number of upward-going muon events observed and
$\Theta_{j}$ is the zenith angle of the $j$th event. The efficiency
$\varepsilon\left(\ge E_{\mu}^{\rm min},\Theta_{j}\right)$ and the ultra--high-energy
fraction $k\left(\geq E_{\mu}^{\rm min}\right)$ are calculated in \S\ref{sec:High-Energy-Isotropic-MC}.
$T$ is the detector livetime, which is 1679.6 days for SK-I. $A\left(\Theta_{j}\right)$
is the effective area of the Super-K detector perpendicular to the direction
of incidence for tracks with a path length of $>7{\rm {\; m}}$ in
the ID. The average effective area of the detector is $\sim 1200{\rm {\; m}}^{2}$.

Equation~(\ref{eq:flux}) has been applied to the detected upward-going muon
event discussed in \S\ref{sub:Selection-Cuts} to calculate $\Phi_{\mu}\left(\geq E_{\mu}^{\rm min}\right)$
for $E_{\mu}^{\rm min}$ in the range $3.16-100{\rm {\; TeV}}$. Results
are shown in Table~\ref{table:flux_results}.%
\begin{table}
\begin{center}
\caption{The flux of ultra-high energy upward-going
muons as observed by SK-I.\label{table:flux_results} }
\begin{tabular}{cc}
\tableline 
\tableline
$E_{\mu}^{\rm min}$&
$\Phi_{\mu}\left(\geq E_{\mu}^{\rm min}\right)$\\
(TeV)&
(${\rm {cm^{-2}s^{-1}sr^{-1}}}$)\\
\tableline
$3.16$&
$2.64\times10^{-14}\begin{array}{c}
+16.1\%\\
-17.9\%\end{array}$\\
$10$&
$8.23\times10^{-15}\begin{array}{c}
+9.48\%\\
-9.73\%\end{array}$\\
$31.6$&
$3.25\times10^{-15}\begin{array}{c}
+6.67\%\\
-6.40\%\end{array}$\\
$100$&
$1.10\times10^{-15}\begin{array}{c}
+4.96\%\\
-4.09\%\end{array}$\\
\tableline
\end{tabular}
\tablecomments{These fluxes include both atmospheric background and potential 
astrophysical signal at each threshold energy.}
\end{center}
\end{table}
 Systematic uncertainties include a $0.1\%$ uncertainty on the live time
$T$, a $0.3\%$ uncertainty on the effective area $A$, the total
efficiency uncertainties shown in Figure~\ref{fig:eff_total}, and
the statistical uncertainties on $k$ shown in Table~\ref{table:highe_frac}.
This flux includes both the potential signal from astrophysical neutrinos
and a background of atmospheric neutrinos.

\section{Expected Atmospheric Background from Monte Carlo}

\label{sec:Expected-Atmospheric-Background}

When searching for neutrinos from astrophysical sources, the dominant
background is the high-energy tail of the atmospheric neutrino spectrum.
Atmospheric neutrinos are produced by decays of pions and kaons formed
when cosmic rays interact with particles in the atmosphere. 
In order
to set a limit on the flux of cosmic neutrinos, the expected flux
of upward-going muons due to the atmospheric background must be carefully
analyzed. To do this, 
we have used an atmospheric neutrino MC that
is a 100 yr equivalent sample of events
due to the atmospheric neutrino flux.
The neutrino flux in \citet{2004PhRvD..70d3008H} was used up to neutrino
energies of 1 TeV. At 1 TeV, the calculated flux in \citet{1980SvJNP..31..784V}
was rescaled to the Honda et al. flux. Above 1 TeV, the
rescaled flux from Volkova was used up to 100 TeV. Neutrino
interactions were modeled using the GRV94 parton distribution functions
\citep{1995ZPhyC..67..433G}, and muon propagation through the rock and water was modeled 
using {\tt GEANT}. Further details on the atmospheric neutrino MC can
be found in \citet{2005PhRvD..71k2005A}.
No correction is made for neutrino oscillations, because based on the 
oscillation parameters determined in \citet{2005PhRvD..71k2005A}, the neutrino 
oscillation probability is negligible for neutrinos above $1{\rm {\; TeV}}$.

The model flux is weighted to represent 3 years of solar minimum,
1 year of changing activity, and 1 year of solar maximum, which corresponds
to the actual solar activity during the run of SK-I.
This atmospheric MC is
split into two parts: a partially-contained/fully-contained (PC/FC)
sample, which consists of events with neutrino interaction points
inside the ID plus a shell 50 cm thick surrounding the ID (the insensitive
region), and an upward-going muon sample, which consists of
events with neutrino interaction points outside the ID. Note that
these two samples overlap because they both cover the 50 cm insensitive
region.

The OD-based fit was applied to the events in the atmospheric MC, using
the same cuts that were applied to the SK-I data. A total of 11 MC events passed the 
$\ge1.75\times10^{6}{\rm {\; pe}}$, $\cos{\Theta}\le0.1$,
path length $>7{\rm {\; m}}$, $N_{{\rm ODentry}}$ and $N_{{\rm ODexit}}\ge10$,
OD/ID timing, and manual fit cuts. Out of these 11, 2 are from the
PC/FC sample, both with interaction points inside the ID. The remaining
9 events are from the upward-going muon sample: 3 events with interaction points
in the 50 cm insensitive region, 1 event in the water of the OD, and
5 events in the rock surrounding the detector. 

All of these background events are deep inelastic scattering (DIS) events where an interaction between a muon neutrino and a nucleon produces a muon plus a spray of lower energy particles.  The 6 events with interaction points within the detector (ID or OD) have muon energies of $0.1-0.8{\rm {\; TeV}}$, and the 5 events occurring in the rock have muon energies of $1-20{\rm {\; TeV}}$.  This difference in the energy range can be understood as follows: For DIS events occurring a long distance ($>2{\rm {\; m}}$ or so) from the detector, only the muon will reach the detector since the lower energy debris will be absorbed by the surrounding rock, but for nearby events or events occurring in the water of the OD, some of these lower energy particles will enter the detector as well.  This means that nearby events can be included in the $\ge1.75\times10^{6}{\rm {\; pe}}$ sample with lower muon energies than more distant events.    

Since the insensitive region is covered by both the PC/FC and
the upward-going muon MC samples, we divided the 3 events originating from
this region in half, for a total of 1.5 events in the insensitive
region. This gives a total of 9.5 MC events in 100 yr
of simulated live time. Scaling the 100 yr MC to SK-I's
live time of 1679.6 days gives an expected background of $0.44$ events
due to atmospheric neutrinos during the operation of SK-I.

The statistical uncertainty in this background measurement
of the MC events is 
\begin{equation}
\frac{1}{9.5}\sqrt{\left(\sqrt{8}\right)^2+\left(\frac{\sqrt{3}}{2}\right)^2}=31\%.\label{eq:stat}
\end{equation}
There are also significant systematic uncertainties:
the normalization of the atmospheric neutrino flux has a theoretical
uncertainty of $\pm10\%$ at neutrino energies below $10{\rm {\; GeV}}$
\citep{2005PhRvD..71k2005A}. In order to extend this to the energy range of the
expected background, we must also account for the uncertainty of 0.05 in the spectral index 
of the primary cosmic ray spectrum above 100 GeV, which leads to a 0.05 uncertainty
in the spectral index for atmospheric neutrinos above $10{\rm {\; GeV}}$
\citep{2005PhRvD..71k2005A}. 


To determine how much this uncertainty affects our
result for the background, we consider the spectral index $\gamma$
to be known at $E_{\nu}^{\rm pivot}=10{\rm {\; GeV}}$, and we calculate
the uncertainty of the total flux $\Phi_{\nu}$ above a threshold
energy $E_{\nu}^{\rm min}=10.6{\rm {\; TeV}}$, the average neutrino energy
of the 100 year MC events passing our cuts. For a differential
flux of ${d\Phi_{\nu}}/{dE_{\nu}}\propto E_{\nu}^{-\gamma}$,
the uncertainty in $\Phi_{\nu}$ due to the uncertainty in $\gamma$
is given by \begin{equation}
\frac{\delta\Phi_{\nu}}{\Phi_{\nu}}=\left(\frac{1}{-\gamma+1}-\ln\left(\frac{E_{\nu}^{\rm min}}{E_{\nu}^{\rm pivot}}\right)\right)\delta\gamma.\label{eq:spectralindex}\end{equation}
With $\gamma=3.7$, $\delta\gamma=0.05$, $E_{\nu}^{\rm min}=10.6{\rm {\; TeV}}$,
and $E_{\nu}^{\rm pivot}=10{\rm {\; GeV}}$, this formula yields ${\delta\Phi_{\nu}}/{\Phi_{\nu}}=0.37$.
Thus the primary spectral index uncertainty gives us an additional
$\pm37\%$ uncertainty on the atmospheric neutrino flux $\Phi_{\nu}$. 

Finally, the
neutrino cross-section at high energies is thought to be known to
within $10\%$ or less, so we include an additional $10\%$ uncertainty
to account for this. These uncertainties are summarized in Table~\ref{Table:bg_sys}
and lead to a total uncertainty on the background of $50\%$.

\begin{table}
\begin{center}
\caption{Systematic uncertainties in atmospheric
neutrino background\label{Table:bg_sys}}
\begin{tabular}{cc}
\tableline 
\tableline
Source of Uncertainty&
Uncertainty\\
\tableline
Statistical&
31\%\\
Absolute normalization of atm $\nu$ flux&
10\%\\
Primary spectral index&
37\%\\
Neutrino cross-section uncertainty&
10\%\\
\tableline 
Total uncertainty in background flux&
50\%\\
\tableline
\end{tabular}
\end{center}
\end{table}

The atmospheric MC does not include neutrino oscillations, so this
is another potential systematic uncertainty. However, for neutrinos
above $1{\rm {\; TeV}}$, the neutrino oscillation probability is
negligible, so it does not make a significant contribution
for this analysis. 
Other potential errors (uncertainties
in the simulation of the SK detector and varying hadron multiplicities
in different deep inelastic scattering models) were tested and shown
to not make a significant contribution to the systematic uncertainty. 

Another potential background source of high-energy neutrinos not included
in the 100 yr atmospheric MC is the prompt atmospheric
neutrino flux, which arises from decays of short-lived charmed particles
produced when cosmic rays interact with particles in the atmosphere.
This flux is not as well-understood as the conventional atmospheric
neutrino flux from decays of pions and kaons, but it is expected to
have a harder spectrum and therefore is expected to become more important
as we push towards higher energy scales. 

Based on the 100 yr atmospheric MC, we calculate
an expected background for this analysis of $0.44\pm0.22$ events,
compared to the 1 event observed. However, there are three effects
that this MC does not take into account: it does not
include neutrinos over $100{\rm {\; TeV}}$, it does not include the
prompt atmospheric neutrino flux, and it does not account for attenuation
of neutrinos passing through the Earth. We account for these
issues by making corrections based on an analytical calculation discussed
in \S\ref{sec:Analytical-Estimate}.

Also, it is important to note 
that the atmospheric background --- both conventional and prompt
--- comes from a lower energy range than that for which we expect to observe
the possible signal of neutrinos from astrophysical sources. Roughly speaking,
the peak 90\% of the expected upward-going muon events 
come from muons with energies in the range $E_{\mu}=0.02-10{\rm \; TeV}$
for conventional atmospheric neutrinos, and $E_{\mu}=0.2-200{\rm \; TeV}$
for prompt neutrinos. As we learned from the 11 background events
in the 100 yr atmospheric MC, 
muons with energies
 $E_{\mu}<1{\rm \; TeV}$ contribute to the  $1.75\times10^{6}{\rm {\; pe}}$ sample mainly via debris from DIS events very close to the detector rather than from catastrophic energy loss of the muon. 

Since there are many more low-energy events in the atmospheric spectrum, 
they will dominate
even though each one only has a tiny probability of depositing a large
amount of energy in the detector. In contrast, the peak 90\% of the
expected muon events from a harder spectrum --- a hypothetical $E_{\nu}^{-2}$
astrophysical flux --- come from the range $E_{\mu}=7-6000{\rm \; TeV}$.
Thus, even though the atmospheric flux is very small in the energy
range $E_{\mu}=3-100{\rm \; TeV}$ and above where we are setting
our limit, the high-pe tails of the distribution from lower energy
events dominate our background simply because of the much larger flux
of lower energy atmospheric events.

\section{Analytical Estimate of Expected Muon Flux}

\label{sec:Analytical-Estimate}

\subsection{Method for Calculating Muon Flux}

\label{sub:Method-for-Calculating-Muon-Flux}

In order to better understand our observed flux of high-energy upward-going
muons, we developed a method to calculate the expected upward-going muon
event rate due to a predicted flux of neutrinos. We have used this
to plot curves for theoretical muon fluxes in Figure~\ref{fig:limits} and also to adjust
the atmospheric background calculated with MC in \S\ref{sec:Expected-Atmospheric-Background}
by correcting for effects not included in the simulation. 

To convert a model neutrino flux into an expected upward-going muon event
rate, we follow the calculation detailed in \citet{1995PhR...258..173G} and
 \citet{1996APh.....5...81G}.
The flux of muons $\Phi_{\mu}\left(\geq E_{\mu}^{\rm min}\right)$ above
an energy threshold $E_{\mu}^{\rm min}$ is given by

\begin{equation}
\Phi_{\mu}\left(\geq E_{\mu}^{\rm min}\right)=\int_{E_{\mu}^{\rm min}}^{\infty}dE_{\nu}P_{\mu}\left(E_{\nu},\, E_{\mu}^{\rm min}\right)\frac{d\Phi_{\nu}^{\rm av}\left(E_{\nu}\right)}{dE_{\nu}},\label{eq:muflux}\end{equation}
where $P_{\mu}\left(E_{\nu},\, E_{\mu}^{\rm min}\right)$ is the probability
that an incoming neutrino with energy $E_{\nu}$ will produce a muon
with energy above the threshold $E_{\mu}^{\rm min}$ at the detector,
and ${d\Phi_{\nu}^{\rm av}\left(E_{\nu}\right)}/{dE_{\nu}}$ is the
differential neutrino flux averaged over solid angle and reduced by an exponential
factor due to attenuation of the neutrinos as they pass through the Earth. 

given by
\begin{equation}
\frac{d\Phi_{\nu}^{\rm av}\left(E_{\nu}\right)}{dE_{\nu}}=\frac{\int_{-1}^{0}d\cos\left(\Theta\right)\int_{0}^{2\pi}d\phi\, A\left(\Theta\right)e^{\left(-z\left(\Theta\right)\sigma\left(E_{\nu}\right)N_{A}\right)}\frac{d^{2}\Phi_{\nu}\left(E_{\nu},\Theta,\phi\right)}{d\Omega dE_{\nu}}}{2\pi A_{\rm av}}.\label{eq:ang_av}\end{equation}
Here, the integral over solid angle extends over $\cos{\Theta}<0$, $\left[{d^{2}\Phi_{\nu}\left(E_{\nu},\Theta,\phi\right)}\right]/\left[{d\Omega dE_{\nu}}\right]$
is the angle-dependent neutrino flux, $A\left(\Theta\right)$ is the
effective area of the SK detector at a zenith angle $\Theta$ 
, and $A_{\rm av}$ is the average effective area
$\int_{-1}^{0}d\cos\left(\Theta\right)A\left(\Theta\right)$. The
exponential factor accounts for attenuation of neutrinos as they pass
through the Earth, which becomes important at higher energies. 

$N_{A}=6.022\times10^{23}{\rm {\; mol^{-1}}=}6.022\times10^{23}{\rm {\; cm^{-3}\;(water\, equivalent)}}$
is Avogadro's number, $\sigma\left(E_{\nu}\right)$ is the total neutrino-nucleon
cross-section, and $z\left(\Theta\right)$ is the column depth through
the Earth at a zenith angle $\Theta$. The column depth has been calculated in 
\citet{1996APh.....5...81G}
using the Preliminary Earth Model
\citep{earthmodel} --- we adopt this column depth here.

As discussed in \citet{1996APh.....5...81G},
there is an uncertainty in choosing a value for the cross-section
$\sigma\left(E_{\nu}\right)$ to use in the neutrino attenuation 
factor. Neutrinos interacting in the Earth
via charged-current (CC) interactions will disappear from the incoming
neutrino beam. However, neutrinos interacting via neutral current
(NC) interactions will only be scattered and reduced in energy, so
they may still contribute to the overall neutrino flux. Thus, we can
put upper and lower limits on the appropriate value of $\sigma$ to
use in Eq.~\ref{eq:ang_av}: $\sigma_{CC}<\sigma<\sigma_{CC}+\sigma_{NC}$.
Here we calculate the angle-averaged flux using both the upper and
lower limits to estimate the resulting uncertainty, and then average
the resulting fluxes together.

If the model neutrino flux is isotropic, it can be taken out of the
angular integral in equation~(\ref{eq:ang_av}), and we can derive a shadow
factor $S\left(E_{\nu}\right)$ given by

\begin{equation}
S\left(E_{\nu}\right)=\frac{1}{A_{\rm av}}\int_{-1}^{0}d\cos\left(\Theta\right)\, A\left(\Theta\right)e^{\left(-z\left(\Theta\right)\sigma\left(E_{\nu}\right)N_{A}\right)}.\label{eq:shadow}\end{equation}

Thus, for an isotropic neutrino flux, equation~(\ref{eq:muflux}) simplifies
to

\begin{equation}
\Phi_{\mu}\left(\geq E_{\mu}^{\rm min}\right)=\int_{E_{\mu}^{\rm min}}^{\infty}dE_{\nu}P_{\mu}\left(E_{\nu},\, E_{\mu}^{\rm min}\right)S\left(E_{\nu}\right)\frac{d\Phi_{\nu}\left(E_{\nu}\right)}{dE_{\nu}}.\label{eq:iso_muflux}\end{equation}

The probability $P_{\mu}\left(E_{\nu},\, E_{\mu}^{\rm min}\right)$
depends on both the charged-current neutrino cross-section and the
energy lost by the resulting muon as it propagates through the
Earth. It is given by
%
%
\begin{equation}
P_{\mu}\left(E_{\nu},\, E_{\mu}^{\rm min}\right)=N_{A}\int_{0}^{E_{\nu}}dE_{\mu}\frac{d\sigma_{\rm CC}}{dE_{\mu}}\left(E_{\mu},\, E_{\nu}\right)R\left(E_{\mu},\, E_{\mu}^{\rm min}\right),\label{eq:prob}\end{equation}
where ${d\sigma_{\rm CC}}/{dE_{\mu}}$
is the differential (with respect to muon energy) charged-current
neutrino cross section and $R\left(E_{\mu},\, E_{\mu}^{\rm min}\right)$
is the average range that a muon with energy $E_{\mu}$ will travel
before its energy is reduced to $E_{\mu}^{\rm min}$. 

In our calculation,
we calculate the cross-section using the GRV94 parton distribution
functions \citep{1995ZPhyC..67..433G} applied with the code used in \citet{1996APh.....5...81G} provided 
by M.~Reno (2005, private communication),
and we use an effective muon range calculated
using MC methods \citep{1991PhRvD..44.3543L}.



We integrated equation~(\ref{eq:muflux}) 
numerically using various predicted models of the neutrino
flux ${d\Phi_{\nu}}/{dE_{\nu}}$ to calculate theoretical 
$\Phi_{\mu}\left(\geq E_{\mu}^{\rm min}\right)$ curves.  These are shown in 
Figure~\ref{fig:limits} for comparison to our experimental limits. 

%

\subsection{Analytical Estimate of Expected Background}

\label{sub:Analytical-Background}

Of particular interest here are the model fluxes of the background
due to atmospheric neutrinos, both conventional and prompt. We
use the method discussed in \S\ref{sub:Method-for-Calculating-Muon-Flux} to correct 
for the omissions in the 100 yr atmospheric MC
discussed in \S\ref{sec:Expected-Atmospheric-Background} by accounting for
neutrinos over $100{\rm {\; TeV}}$, attenuation of neutrinos in the
Earth, and the flux of prompt neutrinos.

To estimate the number of expected background events with our analytical
calculation, we need to factor in the number of muons at each energy
that will make it above the $\ge1.75\times10^{6}{\rm {\; pe}}$ cutoff.
The expected number of events $N$ seen by Super-K in livetime $T$
is given by

\begin{equation}
N=2\pi TA_{\rm av}\int_{0}^{\infty}dE_{\mu}^{\rm min}\frac{d\Phi_{\mu}\left(\geq E_{\mu}^{\rm min}\right)}{dE_{\mu}^{\rm min}}k\left(E_{\mu}^{\rm min}\right),\label{eq:sk_background}\end{equation}
where ${d\Phi_{\mu}\left(\geq E_{\mu}^{\rm min}\right)}/{dE_{\mu}^{\rm min}}$
is the derivative of the curve calculated by the method in \S\ref{sub:Method-for-Calculating-Muon-Flux}
and $k\left(E_{\mu}^{\rm min}\right)$ is the fraction of muons with energy
above $E_{\mu}^{\rm min}$ that will deposit $\ge1.75\times10^{6}{\rm {\; pe}}$
in the ID. 
We estimated $k\left(E_{\mu}^{\rm min}\right)$ using the results of the
high-energy isotropic MC discussed in \S\ref{sub:High-Energy-Fraction}.
The four non-zero values of $k\left(E_{\mu}\right)$ shown in Table~\ref{table:highe_frac}
are used to find an estimator $\hat{k}_{{\rm PL}}$ by fitting with
a simple power law: $\log\left[\hat{k}_{{\rm PL}}\left(E_{\mu}\right)\right]=A+B\log\left[E_{\mu}\right]$
. Additionally, we impose a physically motivated high-energy cutoff:
since $k$ represents a probability, it should never exceed 1. Thus
our estimator for $k\left(E_{\mu}^{min}\right)$ is given by

\begin{equation}
\hat{k}\left(E_{\mu}^{min}\right)=\max\left(\hat{k}_{{\rm PL}}\left(E_{\mu}^{min}\right),\,1\right)\label{eq:kfit}\end{equation}

\begin{figure}
\begin{center}%
\includegraphics[width=.5\textwidth]{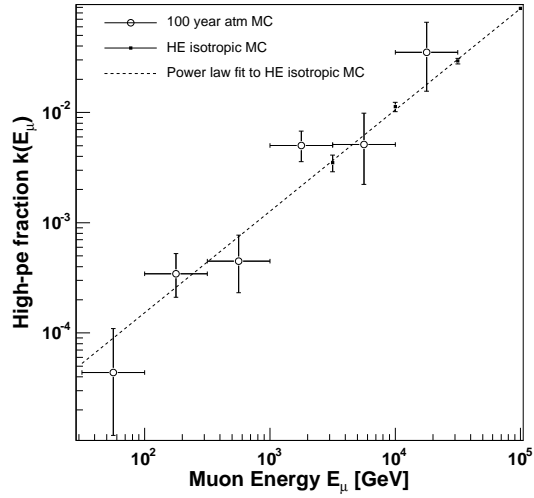}%
\end{center}%
\caption[Fraction of events $k\left(E_{\mu}\right)$
that deposit $\ge1.75\times10^{6}{\rm {\; pe}}$ in the ID
for the 100 yr atmospheric MC, the high-energy isotropic
MC, and a power-law fit to the results from the latter.]{Fraction of events $k\left(E_{\mu}\right)$
that deposit $\ge1.75\times10^{6}{\rm {\; pe}}$ in the ID
for the 100 yr atmospheric MC (\emph{open circles}), the high-energy isotropic
MC (\emph{filled squares}), and a power-law fit to the results from high-energy isotropic
MC. The power law is a reasonable approximation for the dominant
energy range of the atmospheric flux ($E_{\mu}=100{\rm \; GeV}-30{\rm \; TeV}$).\label{fig:kfit}}
\end{figure}

The energy at which $\hat{k}_{{\rm PL}}$ crosses 1 is approximately
$1.5\times10^{6}{\rm \; GeV}$, well above the energy range that produces
most of the flux. 
%
%
Figure~\ref{fig:kfit} compares this power-law fit to results 
from 100 yr atmospheric MC, illustrating that the power law is reasonable
in the energy range we are considering.%

We analytically estimated the expected
number of events in the 100 yr atmospheric MC by
starting with the same input neutrino flux,
applying equation~(\ref{eq:muflux}) \emph{without}
the exponential neutrino attenuation factor and integrating up to a maximum 
neutrino energy of $E_{\nu}^{\rm max}=100{\rm \; TeV}$.

As a cross-check on the analytical calculation, we compared the resulting angle-averaged 
muon flux with the data from the 100 year atmospheric MC. %
\begin{figure}
\begin{center}%
\includegraphics[width=.5\textwidth]{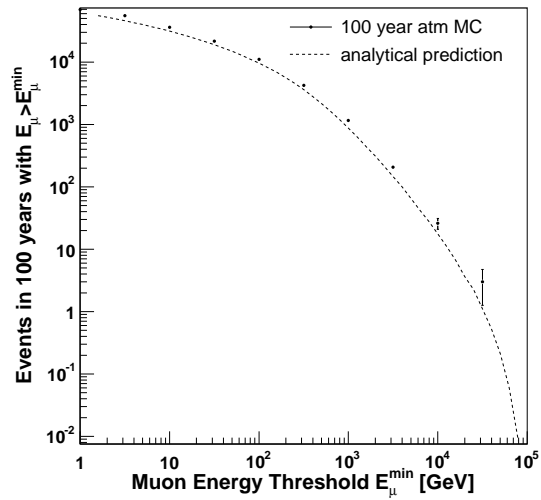}%
\end{center}
\caption[The number of upward-going muon events above energy
threshold $E_{\mu}^{\rm min}$ induced by atmospheric neutrinos, from our analytical calculatoin and from the 100 year atmospheric MC.]{The number of upward-going muon events above energy
threshold $E_{\mu}^{\rm min}$ induced by atmospheric neutrinos. Predictions
from our analytical calculation are compared with the results from
the 100 year atmospheric MC.\label{fig:analytical}}
\end{figure}
Figure~\ref{fig:analytical} shows that the analytical calculation predicts the shape
of the flux curve accurately, but has a slightly lower normalization.
Since we only use the analytical calculation to derive relative scalings,
we consider it to be in good agreement with the MC.

We then used this muon flux in equation~(\ref{eq:sk_background}) to find
the expected number of background events in 100 yr and obtained
 $N_{100}=7.39$ expected events. (The 100 subscript denotes 100 yr of exposure.)
This matches our results from the 100 yr atmospheric MC to within 
statistical uncertainties.

To estimate the effects
of neutrinos above 100 TeV, we repeated the above calcuation 
without the $E_{\nu}^{\rm max}=100{\rm \; TeV}$ cutoff 
and obtained $N_{100}=7.47{\rm \; events}$, 
an increase of $1.1\%$.
Including the neutrino attenuation factor as well, we obtained
$N=7.32{\rm \; events}$ for $\sigma=\sigma_{CC}$ and $N=7.27{\rm \; events}$
for$\sigma=\sigma_{CC}+\sigma_{NC}$, corresponding to an average
of 
$N_{100}=7.29\pm0.02{\rm \; events}$, 
a decrease of $2.4\%\pm0.3\%$. 


We also used this calculation to correct for the prompt atmospheric neutrino 
flux.  To account for the theoretical uncertainty in the prompt flux due
to differences between various flux models, we defined a 
high model and a low model for the prompt flux that bracket the models shown 
in Figure~1 of \citet{2003PhRvD..67a7301G} that are not ruled out by 
experimental limits.
The models used are discussed in more
detail in \citet{1996APh.....5..309T}, \citet{1993APh.....1..297Z}, \citet{2002NuPhS.110..531R}, \citet{1989NCimC..12...41B}, \citet{1999PhRvD..59c4020P}, and \citet{2000PhRvD..61e6011G, 2000PhRvD..61c6005G}.
Our analytical calculation 
gives $N_{100}=0.033{\rm \; events}$  
for the low model and $N_{100}=0.94{\rm \; events}$
for the high model, corresponding to a prompt flux that is $6.7\%\pm6.2\%$ 
of the flux of conventional atmospheric neutrinos.

Based on these results, we correct the background estimate made using 
the 100 yr atmospheric MC in
\S\ref{sec:Expected-Atmospheric-Background} by 
applying these relative scalings to the result of $0.44{\rm \; events}$
in the SK-I livetime from \S\ref{sec:Expected-Atmospheric-Background}.
This gives us a final result for the background of 
$0.46\pm0.23{\rm \; events}$ for the SK-I exposure. 


\section{Upper Limit for Muon Flux from Cosmic Neutrinos}

\label{sec:Upper-Limit}

Using the observed ultra--high-energy upward-going muon signal of 1 event
and the expected atmospheric neutrino background of $0.46\pm0.23$
events, we have calculated $90\%$ confidence upper limits for the
upward-going muon flux in the $3.16-100{\rm {\; TeV}}$ range due to neutrinos
from astrophysical sources (or any other non-atmospheric sources). 

This was done using the method of \citet{1998PhRvD..57.3873F}, with the
systematic uncertainties incorporated using the method of \citet{1992NIMPA.320..331C}, as implemented by \citet{2003PhRvD..67a2002C}
and improved by \citet{2003PhRvD..67k8101H}. This method incorporates both uncertainties
in the background flux and uncertainties in the flux factor $f$ relating
the observed number of events $n$ to the observed flux $\Phi$: $\Phi=f\, n$.
The uncertainty in $f$ includes systematic errors in the livetime,
effective area, efficiency, and ultra--high-energy fraction. For the confidence
interval calculation, the largest percent error for each energy bin
from Table~\ref{table:flux_results} was used as the percent uncertainty
in $f$. The uncertainties in both the background and the flux factor
were assumed to have a Gaussian distribution.

\begin{figure}
\includegraphics[width=1\textwidth]{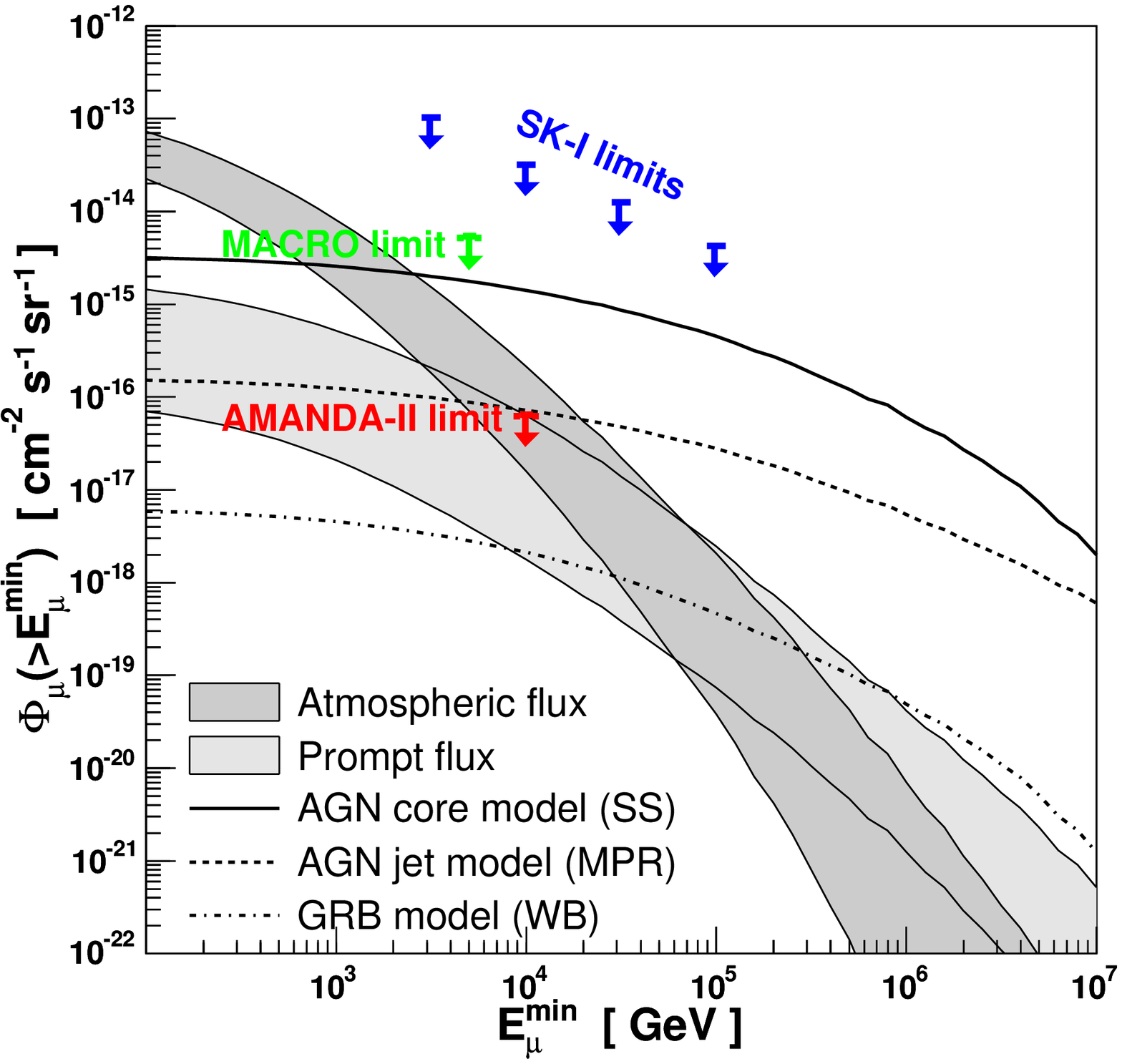}
\caption[Upper limits from this analysis on muon $\left(\mu^{+}+\mu^{-}\right)$
flux above energy threshold $E_{\mu}^{\rm min}$, compared to various
model fluxes.]{Upper limits from this analysis on muon $\left(\mu^{+}+\mu^{-}\right)$
flux above energy threshold $E_{\mu}^{\rm min}$, compared to various
model fluxes. Models shown for muon flux due to astrophysical neutrinos
are AGN models from SS \citep{Stecker:1995th} and
MPR \citep{Mannheim:1998wp}, and a GRB model from WB \citep{1999PhRvD..59b3002W}.
Also shown is the atmospheric background, as modeled by
\citet{2004PhRvD..70d3008H} below $E_{\nu}=1{\;{\rm TeV}}$ and by \citet{1980SvJNP..31..784V}
rescaled to match the Honda model above $E_{\nu}=1{\;{\rm TeV}}$.
The upper edge of the atmospheric band represents the horizontal flux,
and the lower edge represents the vertical flux. The background due
to muons from prompt atmospheric neutrinos (assumed to be isotropic)
is shown for a range of possible models
as summarized in \citet{2003PhRvD..67a7301G}. 
Finally, 
we also show limits on a hypothetical $E_{\nu}^{-2}$ isotropic neutrino flux 
set by MACRO \citep{2003APh....19....1M}
and AMANDA-II \citep{2007PhRvD..76d2008A}. The models of the neutrino flux have
been converted to a muon flux with equation~(\ref{eq:muflux}) using the
GRV94 parton distributions \citep{1995ZPhyC..67..433G} and the
effective muon range from \citet{1991PhRvD..44.3543L}.\label{fig:limits}}
\end{figure}

The final results are shown in Table~\ref{Table:flux limits} and
plotted in Figure~\ref{fig:limits}, along with models of various possible
signals from AGNs \citep{Stecker:1995th,Mannheim:1998wp} and GRBs \citep{1999PhRvD..59b3002W},
as well as the backgrounds due to atmospheric neutrinos 
\citep{2004PhRvD..70d3008H,1980SvJNP..31..784V} and prompt
neutrinos \citep{2003PhRvD..67a7301G}.
%
\begin{table}
\begin{center}
\caption{Confidence intervals for the upward-going muon
flux due to neutrinos from astrophysical sources.\label{Table:flux limits}}
\begin{tabular}{cc}
\tableline 
\tableline
$E_{\mu}^{\rm min}$&
90\% C.L. range\\
(TeV)&
(${\rm {\; cm^{-2}s^{-1}sr^{-1}}}$)\\

\tableline 
$3.16$&
$0-1.03\times10^{-13}$\\
$10$&
$0-3.19\times10^{-14}$\\
$31.6$&
$0-1.26\times10^{-14}$\\
$100$&
$0-4.28\times10^{-15}$\\
\tableline
\end{tabular}
\end{center}
\end{table}

The upper limits calculated here are consistent with the models of
astrophysical signals. Also shown are limits on a hypothetical $E_{\nu}^{-2}$
isotropic neutrino flux set
by MACRO \citep{2003APh....19....1M} and AMANDA-II \citep{2007PhRvD..76d2008A}.
The model neutrino fluxes were converted into muon fluxes using equation~(\ref{eq:muflux})
as discussed in \S\ref{sec:Analytical-Estimate}. 

%
To facilitate easier comparison with other experiments, we also
convert our limits on the muon flux into approximate limits on the
neutrino flux. In order to do this, we assume a model neutrino
flux that is isotropic and proportional to $E_{\nu}^{-2}$. To get
an approximate neutrino limit, we find normalization factors for an
$E_{\nu}^{-2}$ muon flux curve such that the curve passes through
each of our four limit points in Figure~\ref{fig:limits}, and we use
these factors to find the implied limits on $E_{\nu}^{-2}$ flux. 

In order to determine the approximate neutrino energy range in which
these limits are valid, we use equation~(\ref{eq:muflux}) to determine
the neutrino energy range that produces the bulk of the muon signal
for a given value of the muon energy threshold $E_{\mu}^{\rm min}$. We
define the energy range as the range that (1) produces 90\% of the
muon flux above $E_{\mu}^{\rm min}$ and (2) has a higher value of the
integrand of equation~(\ref{eq:muflux}) within the range than anywhere 
outside the range. This
definition is based on the definition of highest posterior density
intervals as described by \citet{Conway:2002}. 

Explicitly, the
energy range is $\left(E_{a},\, E_{b}\right)$, where\begin{equation}
.9=\frac{\int_{E_{a}}^{E_{b}}dE_{\nu}I\left(E_{\nu}\right)}{\int_{E_{\mu}^{\rm min}}^{\infty}dE_{\nu}I\left(E_{\nu}\right)},\label{eq:range_integral}\end{equation}%
with the integrand defined as\begin{equation}
I\left(E_{\nu}\right)=P\left(E_{\nu},\, E_{\mu}^{\rm min}\right)S\left(E_{\nu}\right)\frac{d\Phi_{\nu}\left(E_{\nu}\right)}{dE_{\nu}}\label{eq:integrand}\end{equation}%
as in equation~(\ref{eq:iso_muflux}), with\begin{equation}
\frac{d\Phi_{\nu}\left(E_{\nu}\right)}{dE_{\nu}}\propto E_{\nu}^{-2}.\label{eq:E-2 flux}\end{equation}%
Since this requirement alone does not define a unique range, we also
require that the value of $I\left(E_{\nu}\right)$ at any $E_{\nu}\in\left(E_{a},\, E_{b}\right)$
be greater than the value of $I\left(E_{\nu}\right)$ at any $E_{\nu}\notin\left(E_{a},\, E_{b}\right)$.

\begin{figure}
\includegraphics[width=1\textwidth]{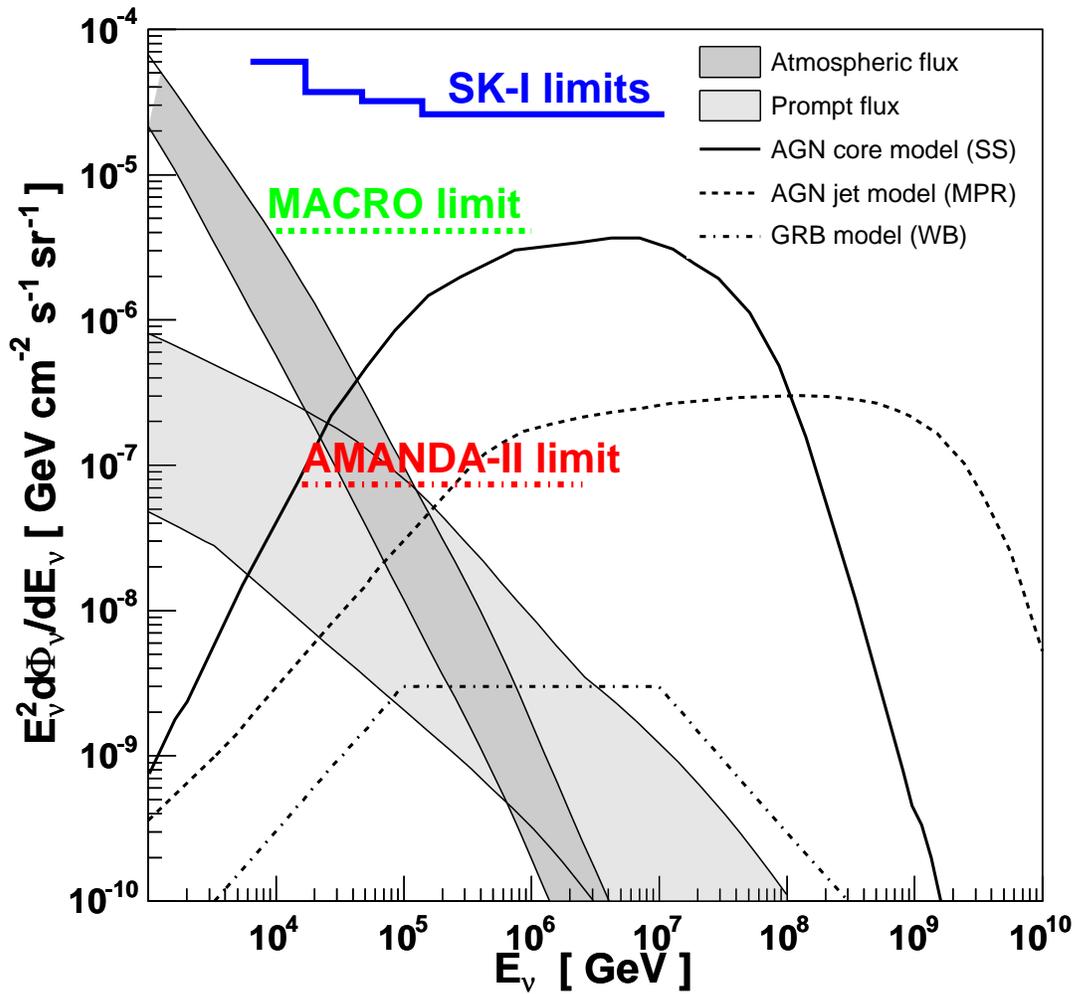}
\caption[Approximate upper
limits from SK-I on astrophysical neutrinos $\left(\nu_{\mu}+\bar{\nu}_{\mu}\right)$.]{\label{fig:neutrino-upper-limits}Approximate upper
limits from SK-I on astrophysical neutrinos $\left(\nu_{\mu}+\bar{\nu}_{\mu}\right)$.
Models shown are the same as in Figure~\ref{fig:limits}. Note that
converting from a muon flux limit to a neutrino flux limit requires
additional assumptions - our limits on the muon flux are shown in
Figure~\ref{fig:limits}. }
\end{figure}

The results of these approximations for neutrino limits and energy
ranges are shown in Table~\ref{table:neutrino-upper-limits}%
\begin{table}
\begin{center}
\caption{Approximate upper limits from SK-I on astrophysical neutrinos 
$\left(\nu_{\mu}+\bar{\nu}_{\mu}\right)$.\label{table:neutrino-upper-limits}}
\begin{tabular}{ccc}
\tableline
\tableline 
$E_{\mu}^{\rm min}$&
90\% C.L. upper limit&
Neutrino energy range\\
(TeV)&
(${\rm {GeV\, cm^{-2}s^{-1}sr^{-1}}}$)&
(GeV)\\
\tableline 
$3.16$&
$6.0\times10^{-5}$&
$6.3\times10^{3}-1.4\times10^{6}$\\
$10$&
$3.7\times10^{-5}$&
$1.7\times10^{4}-2.4\times10^{6}$\\
$31.6$&
$3.2\times10^{-5}$&
$4.7\times10^{4}-5.1\times10^{6}$\\
$100$&
$2.6\times10^{-5}$&
$1.4\times10^{5}-1.1\times10^{7}$\\
\tableline
\end{tabular}
\tablecomments{Upper limits on $E_{\nu}^{2}\left({d\Phi_{\nu}}/{dE_{\nu}}\right)$.
Note that converting from a muon flux limit to a neutrino flux limit
requires additional assumptions --- our limits on the muon flux are
shown in Table~\ref{Table:flux limits}.}
\end{center}
\end{table}
 and are also plotted in Figure~\ref{fig:neutrino-upper-limits}, %
with the same models and experimental limits as shown in Figure~\ref{fig:limits}.
To draw our limits on this plot, we chose to draw only the most sensitive
limit in each energy range.

Note that these results for the limits on the neutrino flux are only
approximations made to facilitate comparison to other experiments
--- our primary results are the limits on the muon flux shown in Table~\ref{Table:flux limits}
and Figure~\ref{fig:limits}.

\section{Conclusions}

\label{sec:Conclusions}

In conclusion, we have developed a method for analyzing Super-K's highest energy data to search for evidence of high-energy neutrino flux from astrophysical sources.  We have done a thorough study of the efficiency and the expected backgrounds from this method and applied our method to the SK-I data sample. Our study of the highest energy events in SK-I does not show evidence of a high-energy cosmic neutrino signal.

We have set upper limits on the muon flux due to cosmic neutrino sources.
These limits are consistent with the results of other experiments
\citep{2003APh....19....1M,2007PhRvD..76d2008A}. 
It is possible that an astrophysical neutrino signal could
be within the grasp of the next generation of neutrino detectors such
as IceCube \citep{2007arXiv0711.0353T} and KM3NeT \citep{2006NIMPA.567..457K,2007arXiv0711.0563K}.

\cleardoublepage
\part{Particle Physics in the Sky}

\chapter{Doing cosmology with galaxies}

\label{chap:cosmology}

\section{Cosmology Basics}

\subsection{What is cosmology?}

Cosmology is the study of our universe as a whole, and thus focuses
on the largest scales accessible to science. It strives to answer
a number of {}``big questions'': What is our universe made of? How
did it begin? How did the matter assemble into the structures we see
today? What is its ultimate fate? In order to address these questions,
cosmologists use a wide variety of astronomical observations and draw
on theory from across essentially all fields of physics, ranging from
general relativity to quantum field theory.

One remarkable feature of the study of cosmology is that it provides
insights into particle physics -- i.e., the study of the smallest
scales in physics -- that could never be observed in a terrestrial
laboratory. The tremendous energy in the hot, early universe shortly
after the Big Bang far exceeds the energies that could ever be produced
by the most sophisticated particle accelerators on Earth. Now that
cosmology is becoming a precision science, particle physicists are
turning to cosmology as a tool, using the study of the largest scales in
physics to learn more about the smallest.

\subsection{The standard cosmological model}

\label{sub:Standard-model}

Over the past few decades, cosmology has evolved from a highly speculative
field dubbed {}``A search for two numbers'' by \citet{1970PhT....23b..34S}
into a full-fledged observational science, with vast quantities of
data supporting a detailed theoretical model that is now well-established
enough to be known as the {}``standard cosmological model''. We
are commonly said to be living in the age of precision cosmology.
However, many aspects of the standard cosmological model are quite
surprising, and have led to more questions than answers.

The basic premises of the standard cosmological model are as follows:

\begin{itemize}
\item Our universe is expanding -- the gravitationally bound structures
in our universe (e.g. clusters of galaxies) are all moving away from
each other.
\item Our universe used to be much hotter, denser, and smoother than it
is today - the early universe was a hot soup of quarks and elementary
particles.
\item The large-scale structures in our universe grew through gravitational
instabilities seeded by quantum fluctuations in the early universe.
\item These quantum fluctuations were stretched to macroscopic size during a phase
of rapidly accelerating expansion of the early universe called \emph{inflation}.
\item Ordinary matter (protons and neutrons, a.k.a. baryons) only make up
4\% of the mass density in our universe. 
\item About 21\% is made of \emph{dark matter}, a mysterious substance
that is gravitationally attractive but does not interact with light.
\item About 75\% is made of \emph{dark energy}, an even more mysterious
substance that effectively gives rise to a repulsive gravitational
force and is causing the current expansion of our universe to accelerate.
\end{itemize}
This basic picture is consistent with a wide variety of different
types of measurements, ranging from the tiny fluctuations in the microwave
radiation produced by the early universe to the rates at which distant
supernovae are moving away from us to the clustering patterns observed
in the distribution of galaxies today. The agreement between all of
these different measurements is quite remarkable, and has forced scientists
to take seriously these seemingly preposterous concepts of dark matter,
dark energy, and inflation. 

Today the study of cosmology is focused on gaining a deeper understanding
of the physics behind these concepts. What triggered inflation and
how did it stop? What is the nature of dark matter? How much of the
dark matter might be neutrinos? What sort of substance could
have such bizarre properties as dark energy? Is dark energy a substance
at all, or is the observed acceleration really due to a breakdown
of general relativity at cosmological distance scales? These are the
types of questions currently being investigated.

\subsection{Zeroth order cosmology}

\label{sub:Zeroth-order}

As a first approximation, our universe can be treated as being homogenous
and isotropic -- that is, the matter in our universe is distributed
uniformly and it looks the same in every direction. Looking around
our neighborhood might lead us to believe that this is not a very
good assumption: matter is packed together much more densely within
the Earth than a thousand kilometers above its surface, for example. 

However, cosmologists think on much larger scales: if we average over
scales of several hundred $\mathrm{Mpc}$ (i.e. about a billion light
years -- that's 100 trillion times the distance from the Earth to
the Sun), we will smooth out structures as small as a solar system,
a galaxy, or even a galaxy cluster, and the matter in our universe
begins to look quite uniform. Historically, the homogeneity and isotropy
of our universe on large scales has been a fundamental assumption
of cosmology, but in recent years our surveys of our universe have
gotten large enough to confirm this directly (see, e.g., \citealt{2005MNRAS.364..601Y}).

\begin{figure}
\begin{centering}\includegraphics[width=0.75\textwidth]{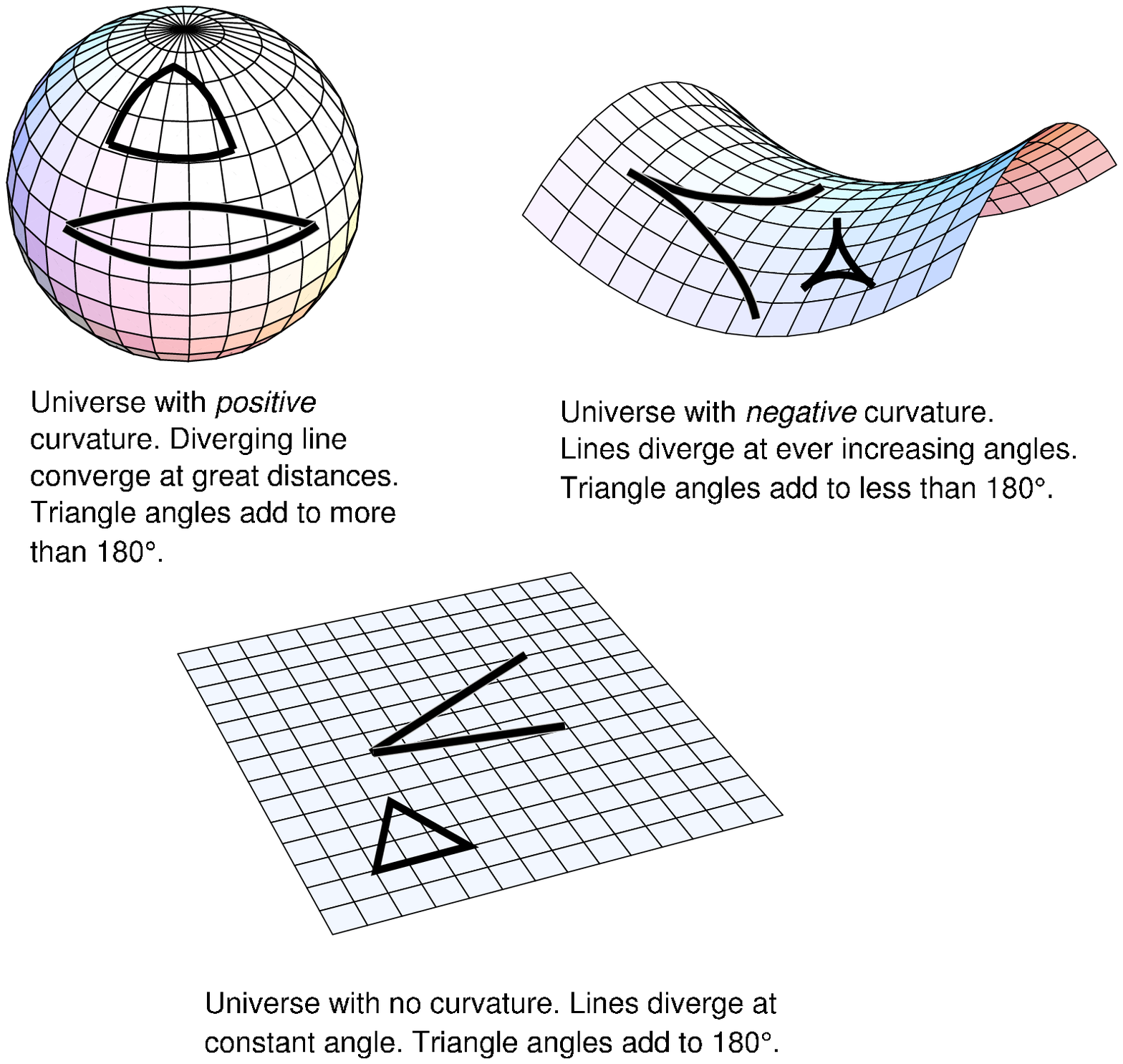}\par\end{centering}

\caption[The curvature of space illustrated for two-dimensional
universes.]{\label{fig:curvature}The curvature of space illustrated for two-dimensional
universes. Figure copyright Nick Strobel, \protect\url{http://www.astronomynotes.com}.
Used with permission.}
\end{figure}

The physics of a homogenous and isotropic universe can be described
in Einstein's theory of general relativity (or more generally, any
metric theory of gravity) using the Friedmann-Robertson-Walker (FRW)
spacetime metric:\begin{equation}
ds^{2}=-c^{2}dt^{2}+a\left(t\right)\left[\frac{dr^{2}}{1-Kr^{2}}+r^{2}\left(d\theta^{2}+\sin^{2}\theta d\phi^{2}\right)\right].\label{eq:FRW}\end{equation}
This resembles the more familiar Minkowski metric used in special
relativity, given by

\begin{equation}
ds^{2}=-c^{2}dt^{2}+dr^{2}+r^{2}\left(d\theta^{2}+\sin^{2}\theta d\phi^{2}\right)\label{eq:flat}\end{equation}
with two additional pieces: a constant factor $K$ describing the
global curvature of space, and an arbitrary function of time $a\left(t\right)$.
The curvature of space is defined by the behavior of parallel lines:
in flat space ($K=0$), parallel lines always remain parallel, in
a space with negative curvature ($K<0$ ), they diverge, and in a
space with positive curvature ($K>0$), they converge. An illustration
of this is shown for two-dimensional spaces in Fig.~\ref{fig:curvature};
our universe is a three-dimensional analog of this. %
The function $a\left(t\right)$ is known as the scale factor, which
allows space itself to expand or contract as our universe evolves.
Here we normalize $a\left(t\right)$ such that $a\left(t_{0}\right)=1$
at the present time $t_{0}$. Doing cosmology at zeroth order fundamentally
boils down to determining $K$ and $a\left(t\right)$. 

By applying the Einstein field equations 
for a homogeneous, isotropic universe to the FRW metric 
(see, e.g.,
\citealt{2003moco.book.....D}), $K$ and $a\left(t\right)$ can be
linked to to the composition of our universe via the Friedmann equation:\begin{equation}
H^{2}\equiv\left(\frac{\dot{a}}{a}\right)^{2}=\frac{8\pi G}{3}\rho\left(t\right)-\frac{Kc^{2}}{a^{2}}.\label{eq:friedmann}\end{equation}

We will discuss each term in this equation in turn. The factor $\dot{a}/a$
is the expansion rate of our universe. It is
called the Hubble parameter and denoted by $H$. The value of $H$
today is known as the Hubble constant $H_{0}\equiv\dot{a}\left(t_{0}\right)/a\left(t_{0}\right)$.
This is the parameter that Edwin Hubble measured in his initial observations
of the expansion of our universe \citep{1931ApJ....74...43H}. He
observed that the further away a galaxy was, the faster it appeared
to be moving away from us:\begin{equation}
v=H_{0}d\label{eq:hubble}\end{equation}
where $v$ is a galaxy's recession velocity and $d$ is its distance
from us. Hubble's initial measurements gave $H_{0}\sim550\,\mathrm{km\ s^{-1}Mpc^{-1}}$;
the modern accepted value is $H_{0}=73\,\mathrm{km\ s^{-1}Mpc^{-1}}$.
It is often parameterized using the Hubble parameter $h$, where $h\equiv H_{0}/\left[100\,\mathrm{km\ s^{-1}Mpc^{-1}}\right]$.
Cosmological distances are conventionally written in units of $h^{-1}\mathrm{Mpc}$
-- we use this convention in chapter~\ref{chap:bias}.

The recession velocity $v$ is inferred from the wavelength shift
of the galaxy's spectral lines due to the Doppler effect:\begin{equation}
\frac{v}{c}\approx\frac{\lambda_{\mathrm{obs}}-\lambda_{\mathrm{rest}}}{\lambda_{\mathrm{rest}}}\equiv z.\label{eq:redshift}\end{equation}
where $c$ is the speed of light, $\lambda_{\mathrm{rest}}$ is the
wavelength of the spectral line in the rest frame of the emitter,
$\lambda_{\mathrm{obs}}$ is the wavelength observed at Earth, and
the quantity $z$ is called the {}``redshift'' since objects moving
away from us appear redder. The approximation holds when $v\ll c$
and the galaxy is close enough that it is sensible to interpret the
redshift as a recession velocity. However, it is more accurate to
think of the redshift as being caused by the expansion of space as
the light is propagating through our universe. The wavelength of the
light will get stretched along with space, and thus the wavelength
change can be directly related to the scale factor $a$: \begin{equation}
\frac{\lambda_{\mathrm{obs}}}{\lambda_{\mathrm{rest}}}=\frac{a\left(t_{e}\right)}{a\left(t_{0}\right)}\label{eq:scalefactor_wavelength}\end{equation}
where $t_{e}$ is the time the light was emitted from the galaxy.
Light from galaxies that are further away will take longer to reach
us, so it will have a longer amount of time in which to get stretched
and thus appears more redshifted. Using equations~\eqref{eq:redshift}~and~\eqref{eq:scalefactor_wavelength}
along with the normalization convention $a\left(t_{0}\right)=1$,
we can relate $z$ to $a$:\begin{equation}
a=\frac{1}{1+z}.\label{eq:az}\end{equation}

Moving on to the first term on the left-hand side, we have the the
matter-energy density of our universe $\rho\left(t\right)$. The time
evolution of this quantity depends on the type of matter. For the
case of a perfect isotropic fluid with density $\rho$ and pressure
$P$, the Einstein field equations and the FRW metric tell us that
(see, e.g., \citealt{2003moco.book.....D}) \begin{equation}
\dot{\rho}+3\frac{\dot{a}}{a}\left(\rho+\frac{P}{c^{2}}\right)=0,\label{eq:fluid}\end{equation}
Determining the evolution of $\rho$ requires a relation between $P$
and $\rho$, known as the equation of state. Here we will consider
three different types of contributions to $\rho$:

\begin{itemize}
\item Cold, pressureless matter $\rho_{m}$ (includes baryons and dark matter):
$P=0$, so equation~\eqref{eq:fluid} yields $\rho_{m}\propto a^{-3}$.
Physically this means that the number of particles stays constant
while the volume expands as $\mathrm{length}^{3}$.
\item Radiation or relativistic matter $\rho_{r}$ (includes photons and
neutrinos): $P=\rho/3$, so equation~\eqref{eq:fluid} yields $\rho_{r}\propto a^{-4}$.
Again the number of particles stays constant and the volume expands
as $\mathrm{length}^{3}$, but we gain an additional factor of $a^{-1}$
since the energy of each particle is $\hbar\omega\propto\lambda^{-1}$,
so the energy per particle decreases as $a^{-1}$ as the wavelength
gets redshifted. 
\item Unknown equation of state, i.e. dark energy $\rho_{\Lambda}$: $P=w\rho$,
so equation~\eqref{eq:fluid} yields $\rho_{\Lambda}\propto a^{-3(1+w)}$.
Current observations of dark energy are consistent with a cosmological
constant, i.e. an energy density that remains constant as our universe
expands. This corresponds to $\rho_{\Lambda}\propto a^{0}$, so $w=-1$. In the above
examples, pressureless matter has $w=0$ and radiation has $w=1/3$.
More general parameterizations of the dark energy equation of state
also allow $w$ to vary in time.
\end{itemize}
Finally, the second term on the right-hand side of equation~\eqref{eq:friedmann},
which includes the parameter $K$ defining the overall curvature of
space in our universe. One important insight to be gained from this
equation is that the global geometry of our universe is directly linked
to its matter content. There is a special value of the matter density
that corresponds to flat space (i.e. $K=0$) that is known as the
{}``critical density'' $\rho_{\mathrm{cr}}$. By plugging $K=0$
into equation~\eqref{eq:friedmann} we find that the present value
of the critical density is given by\begin{equation}
\rho_{\mathrm{cr0}}=\frac{3H_{0}^{2}}{8\pi G}.\label{eq:rhocrit}\end{equation}
The Friedmann equation takes on a particularly elegant form if we
express all densities as fractions of the critical density. We define\begin{equation}
\Omega_{m}\equiv\frac{\rho_{m0}}{\rho_{\mathrm{cr}0}},\Omega_{r}\equiv\frac{\rho_{r0}}{\rho_{\mathrm{cr}0}},\Omega_{\Lambda}\equiv\frac{\rho_{\Lambda0}}{\rho_{\mathrm{cr}0}},\label{eq:Omegas}\end{equation}
where again the 0 subscript denotes the present-day value. Now the
Friedmann equation evaluated at the present is\[
1=\Omega_{m}+\Omega_{r}+\Omega_{\Lambda}-\frac{Kc^{2}}{H_{0}^{2}}.\]
 Using this we can define an effective $\Omega$ for the curvature:\begin{equation}
\Omega_{K}\equiv1-\Omega_{m}-\Omega_{r}-\Omega_{\Lambda}=-\frac{Kc^{2}}{H_{0}^{2}}.\label{eq:omegak}\end{equation}
Thus the curvature is determined by the total matter density: if $\Omega_{m}+\Omega_{r}+\Omega_{\Lambda}=1$,
i.e. the total density of matter is exactly equal to the critical
density, then $\Omega_{K}=0$ and our universe is flat. If the total
density is less than the critical density, then $K<0$ and our universe
has negative curvature. If the total density is greater than the critical
density, then $K>0$ and our universe has positive. 

To write our elegant Friedmann equation, we use the definitions in
equations \eqref{eq:Omegas} and \eqref{eq:omegak} and the scalings
with $a$ of different types of matter density determined by equation~\eqref{eq:fluid}
to re-express equation~\eqref{eq:friedmann} as\begin{equation}
\left(\frac{\dot{a}}{a}\right)^{2}=H_{0}^{2}\left(\Omega_{r}a^{-4}+\Omega_{m}a^{-3}+\Omega_{K}a^{-2}+\Omega_{\Lambda}a^{-3(1+w)}\right)\label{eq:prettyfriedmann}\end{equation}
This relatively simple differential equation for the scale factor
$a\left(t\right)$ is the cornerstone of zeroth order cosmology. Given
values for the parameters $H_{0}$, $\Omega_{m}$, $\Omega_{r}$,
$\Omega_{\Lambda}$, and $w$, one can completely describe the evolution
of a homogenous universe containing matter, radiation, and dark energy.
(Or to be more precise, a dark-energy-like component with an equation
of state that can be parameterized by a constant $w$.)

\subsection{First order cosmology}

\label{sub:First-order}

The next step in cosmology is to study the evolution of small deviations
from homogeneity. The early universe was extremely smooth, with tiny
perturbations in the density on the order of one part in $10^{5}$.
As our universe evolved, these small fluctuations grew under the influence
of gravity into the complex cosmic web we see today. The behavior
of these fluctuations is properly understood by perturbing the Einstein
equations of general relativity around the homogenous solution described
in \S\ref{sub:Zeroth-order} -- the details of this can be found
in, e.g., \citet{2003moco.book.....D}. Here we go through a Newtonian
approximation that elucidates much of the key physics governing cosmological
perturbations, and then discuss the statistical measures used to analyze
them.

\subsubsection{Linear growth of structure}

The equations governing the behavior of a non-relativistic, self-gravitating
fluid are the continuity equation, Euler's equation, and the Poisson
equation, respectively given by\begin{eqnarray}
\frac{\partial\rho}{\partial t}+{\bf \nabla}_{{\bf r}}\cdot\left(\rho{\bf v}\right) & = & 0\label{eq:continuity}\\
\frac{\partial{\bf v}}{\partial t}+\left({\bf v}\cdot{\bf \nabla}_{{\bf r}}\right){\bf v} & = & -\frac{{\bf {\bf \nabla}_{{\bf r}}}P}{\rho}-{\bf \nabla}_{{\bf r}}\Phi\label{eq:euler}\\
\nabla_{{\bf r}}^{2}\Phi & = & 4\pi G\rho\label{eq:poisson}\end{eqnarray}
where $\rho\left({\bf r},t\right)$ is the density of the fluid, $P\left({\bf r},t\right)$
is the pressure, ${\bf v}\left({\bf r},t\right)$ is the velocity
of a fluid element, $\Phi\left({\bf r},t\right)$ is the gravitational
potential., and ${\bf \nabla}_{{\bf r}}$ denotes the gradient with
respect to ${\bf r}$. The spirit of first order cosmology is to start
with a homogenous universe and then consider small perturbations to
$\rho$, ${\bf v}$, $\Phi,$and $P$. We will use\begin{equation}
\left\{ \begin{array}{l}
\rho_{0}\left({\bf r},t\right)=\frac{\rho_{C}}{a\left(t\right)^{3}}\\
{\bf v}_{0}\left({\bf r},t\right)=\frac{\dot{a}\left(t\right)}{a\left(t\right)}{\bf r}\\
\Phi_{0}\left({\bf r},t\right)=\frac{2\pi G}{3}\rho_{0}\left({\bf r},t\right)r^{2}\\
P_{0}\left({\bf r},t\right)=0\end{array}\right.\label{eq:homogenous}\end{equation}
for our homogenous solution, which satisfies equations \eqref{eq:continuity},
\eqref{eq:euler}, and \eqref{eq:poisson} if $a\left(t\right)$ satisfies
our zeroth order cosmology equations \eqref{eq:friedmann} and \eqref{eq:fluid}.
This corresponds to a universe of pressureless matter (such as dark
matter) undergoing Hubble expansion ${\bf v}_{0}=H{\bf r}$.

Now we solve for the behavior of small perturbations: 

\begin{equation}
\left\{ \begin{array}{l}
\rho\left({\bf r},t\right)=\rho_{0}\left({\bf r},t\right)+\rho_{1}\left({\bf r},t\right)\\
{\bf v}\left({\bf r},t\right)={\bf v}_{0}\left({\bf r},t\right)+{\bf v}_{1}\left({\bf r},t\right)\\
\Phi\left({\bf r},t\right)=\Phi_{0}\left({\bf r},t\right)+\Phi_{1}\left({\bf r},t\right)\end{array}\right.\label{eq:perturbations}\end{equation}
where quantities with 1 subscripts are small enough that we can ignore
all but the linear terms. Inserting these into the continuity, Euler,
and Poisson equations, subtracting the homogenous solution, and dropping
higher order terms of the perturbations gives

\begin{eqnarray}
\frac{\partial\rho_{1}}{\partial t}+\left({\bf v}_{0}\cdot{\bf \nabla}_{{\bf r}}\right)\rho+\rho_{0}{\bf \nabla}_{{\bf r}}\cdot{\bf v}_{1}+3\frac{\dot{a}}{a}\rho_{1} & = & 0\label{eq:continuity1}\\
\frac{\partial{\bf v}_{1}}{\partial t}+\left({\bf v}_{0}\cdot{\bf \nabla}_{{\bf r}}\right){\bf v}_{1}+\frac{\dot{a}}{a}{\bf v}_{1} & = & -{\bf \nabla}_{{\bf r}}\Phi_{1}\label{eq:euler1}\\
\nabla_{{\bf r}}^{2}\Phi_{1} & = & 4\pi G\rho_{1}.\label{eq:poisson1}\end{eqnarray}

Cosmology is often best understood in \emph{comoving} coordinates
defined by ${\bf x}\equiv{\bf r}/a\left(t\right)$. Comoving coordinates
are defined such that they remain fixed for observers moving along
with the Hubble expansion in a homogenous universe. The comoving peculiar
velocity is given by ${\bf u}\equiv{\bf v}_{1}/a\left(t\right)$,
and we define the density contrast as $\delta\equiv\rho_{1}/\rho_{0}$.
Re-writing the above equations in comoving coordinates gives

\begin{eqnarray}
\dot{\delta} & = & -{\bf \nabla}\cdot{\bf u}\label{eq:continuity2}\\
\dot{{\bf u}}+\frac{\dot{a}}{a}{\bf u} & = & -\frac{1}{a^{2}}{\bf \nabla}\Phi_{1}\label{eq:euler2}\\
\nabla^{2}\Phi_{1} & = & 4\pi Ga^{2}\rho_{0}\delta,\label{eq:poisson2}\end{eqnarray}
where ${\bf \nabla}=a\left(t\right){\bf \nabla}_{{\bf r}}$ is the
gradient with respect to the comoving coordinates ${\bf x}$, dots
represent $d/dt\equiv\partial/\partial t+{\bf v}_{0}\cdot{\bf \nabla}_{{\bf r}}$,
the time derivative for comoving observers, and $c_{s}\equiv\left(\partial P/\partial\rho\right)^{1/2}$
is the sound speed.

By taking the time derivative of equation~\eqref{eq:continuity2}
and the divergence of equation~\eqref{eq:euler2} we can derive a
second order differential equation for $\delta$:\begin{equation}
\ddot{\delta}+2\frac{\dot{a}}{a}\dot{\delta}-4\pi G\rho_{0}\delta=0.\label{eq:lineargrowth}\end{equation}
This is the basic equation governing the growth of small density perturbations
in a universe of pressureless matter, which is quite applicable to
our universe since the bulk of large-scale structure formation in
our universe occurs while dark matter is the dominant component.

Equation~\eqref{eq:lineargrowth} has two linearly independent solutions:

\begin{equation}
\begin{array}{ll}
\delta_{G}\left(t\right)=\frac{\dot{a}}{a}\int_{0}^{a}\frac{1}{\dot{a}^{3}}da & \mathrm{\;(growing\, mode)}\\
\delta_{d}\left(t\right)=\frac{\dot{a}}{a} & \;\mathrm{(decaying\, mode)}\end{array}\label{eq:linear_soln}\end{equation}
where the growing mode increases with time and the decaying mode decreases.
The general solution is a linear combination:\begin{equation}
\delta\left({\bf x},t\right)=A\left({\bf x}\right)\delta_{G}\left(t\right)+B\left({\bf x}\right)\delta_{d}\left(t\right),\label{eq:linear_comb}\end{equation}
where $A$ and $B$ are arbitrary functions of ${\bf x}.$ The decaying
mode will eventually become negligible as the universe expands, and
is thus not relevant to structure formation, so we focus on the growing
mode. We can then write the density contrast as

\begin{equation}
\delta\left({\bf x},t\right)=\delta\left({\bf x},t_{i}\right)\frac{\delta_{G}\left(t\right)}{\delta_{G}\left(t_{i}\right)},\label{eq:growing_soln}\end{equation}
where the perturbations are specified at some initial time $t_{i}$.
Thus an initial perturbation field $\delta\left({\bf x},t_{i}\right)$
will retain its shape but increase in amplitude until $\delta\sim1$
and our linear approximation no longer holds. This is the basic picture
of how small perturbations start growing into large structures through
gravitational instability. For a more detailed, properly relativistic
treatment, see \citet{1993sfu..book.....P,2003moco.book.....D}. 

\[
\]

\subsubsection{Statistics of random fields}
\label{sub:randomfields}
A general theory of how structure formed in our universe could never
hope to predict the value of the density contrast $\delta$ at a specific
point ${\bf x}$. Instead the goal is to predict its statistical properties,
which are typically described in the language of random fields. A
random field is an infinite dimensional random variable that has some
probability to take on a given value at every point in space. The
function $\delta\left({\bf x},t_{i}\right)$ in equation~\eqref{eq:growing_soln},
for example, is a realization of a random field.

The statistical properties of a random field $\delta$ are specified
by a series of joint probability distributions $P_{n}\left(\delta_{1},\delta_{2},\dots\delta_{n}\right)$
for its values $\delta_{i}\equiv\delta\left({\bf x}_{i}\right)$ at
$n$ points ${\bf x}_{1},{\bf x}_{2},\dots{\bf x}_{n}$ for $n=1,2,3,\dots$.
The most commonly used random field in cosmology is a zero-mean Gaussian
random field, where the probability distribution is a multivariate
Gaussian:

\begin{equation}
P_{n}\left(\delta_{1},\delta_{2},\dots\delta_{n}\right)d\delta_{1}d\delta_{2}\dots d\delta_{n}=\frac{1}{\sqrt{\left(2\pi\right)^{n}\det\left(\mathbfss{M}\right)}}\exp\left({\bf \delta}^{T}\mathbfss{M}^{-1}{\bf \delta}\right)d\delta_{1}d\delta_{2}\dots d\delta_{n},\label{eq:gaussian}\end{equation}
where ${\bf \delta}$ is a column vector of the $\delta_{i}$'s and
$\mathbfss{M}$ is a covariance matrix that defines the distribution.
The expectation value of any function $f$ of the $\delta_{i}$'s
is defined as\begin{equation}
\left\langle f\left(\delta_{1},\delta_{2},\dots\delta_{n}\right)\right\rangle \equiv\int f\left(\delta_{1},\delta_{2},\dots\delta_{n}\right)P_{n}\left(\delta_{1},\delta_{2},\dots\delta_{n}\right)d\delta_{1}d\delta_{2}\dots d\delta_{n}.\label{eq:expectation}\end{equation}

In typical cosmological applications, we try to infer the statistical
properties of a random field, e.g. the density contrast, from observing
the one realization of it provided by our universe. We do not have
the luxury of observing multiple universes drawn from the ensemble
defined by our model probability distribution, so we cannot really
measure an ensemble average like equation~\eqref{eq:gaussian} specifies.
To connect our statistical model with what we can actually observe,
we have to make an assumption that the random fields are \emph{ergodic},
which means that averaging over a sufficiently large volume of space
is equivalent to averaging over a statistical ensemble, so we can
use

\begin{equation}
\left\langle f\left(\delta_{1},\delta_{2},\dots\delta_{n}\right)\right\rangle =\lim_{V_{i}'s\rightarrow\infty}\int_{V_{1}}\int_{V_{2}}\dots\int_{V_{n}}f\left(\delta\left({\bf x}_{1}\right),\delta\left({\bf x}_{2}\right),\dots\delta\left({\bf x}_{n}\right)\right)d^{3}x_{1}d^{3}x_{2}\dots d^{3}x_{n}\label{eq:ergodic}\end{equation}
for a practical means of defining expectation values.

The most commonly used statistical measures of random fields are moments
of the probability distribution, called $n$-point functions. The
1-point function, i.e. the mean, is given by $\left\langle \delta_{i}\right\rangle $,
the 2-point function is given by $\left\langle \delta_{i}\delta_{j}\right\rangle $,
and so forth. For example, a zero-mean Gaussian random field with
$P_{n}$ given by equation~\eqref{eq:gaussian} has $\left\langle \delta_{i}\right\rangle =0$
(hence the {}``zero-mean'') , $\left\langle \delta_{i}\delta_{j}\right\rangle =\mathbfss{M}_{ij}$,
and $\left\langle \delta_{1}\delta_{2}\dots\delta_{n}\right\rangle =0$
for $n>2$. 

The 2-point function is the most basic statistical measure of inhomogeneities
in a random field and thus plays an important role in first-order
cosmology. It is known as the correlation function $\xi\left({\bf x}_{i},{\bf x}_{j}\right)\equiv\left\langle \delta_{i}\delta_{j}\right\rangle $,
and for a universe that is statistically homogenous and isotropic,
the the correlation function will only depend on the distance between
the two points ${\bf x}_{i}$ and ${\bf x}_{j}$ and not on their
absolute position or orientation: $\xi\left({\bf x}_{i},{\bf x}_{j}\right)=\xi\left(\left|{\bf x}_{i}-{\bf x}_{j}\right|\right)$.

The 2-point function has some particularly convenient properties if
we transform into Fourier space. Defining the Fourier transform as

\begin{equation}
\hat{\delta}\left({\bf k}\right)\equiv\int e^{i{\bf k}\cdot{\bf x}}\delta\left({\bf x}\right)d^{3}x,\label{eq:fourier}\end{equation}
the Fourier space version of the 2-point function is\begin{equation}
\left\langle \hat{\delta}^{*}\left({\bf k}_{i}\right)\hat{\delta}\left({\bf k}_{j}\right)\right\rangle =\int e^{i{\bf k}_{j}\cdot{\bf x}_{j}-i{\bf k}_{i}\cdot{\bf x}_{i}}\xi\left({\bf x}_{i},{\bf x}_{j}\right)d^{3}x_{i}d^{3}x_{j},\label{eq:fourier2pt}\end{equation}
and if the correlation function depends only on the separation, i.e.
$\xi\left({\bf x}_{i},{\bf x}_{j}\right)=\xi\left({\bf x}_{i}-{\bf x}_{j}\right)$,
then equation~\eqref{eq:fourier2pt} reduces to \begin{equation}
\left\langle \hat{\delta}^{*}\left({\bf k}_{i}\right)\hat{\delta}\left({\bf k}_{j}\right)\right\rangle =\left(2\pi\right)^{3}\delta_{D}\left({\bf k}_{i}-{\bf k}_{j}\right)P\left({\bf k}_{i}\right),\label{eq:Pkdef}\end{equation}
 where $\delta_{D}$ is the Dirac delta function and $P\left({\bf k}\right)$,
called the \emph{power spectrum}, is the Fourier transform of the
correlation function:\begin{equation}
P\left({\bf k}\right)\equiv\int e^{i{\bf k}\cdot{\bf x}}\xi\left({\bf x}\right)d^{3}x.\label{eq:fourierPk}\end{equation}
Furthermore, if the correlation function depends only on the magnitude
of the separation, i.e. $\xi\left({\bf x}_{i},{\bf x}_{j}\right)=\xi\left(\left|{\bf x}_{i}-{\bf x}_{j}\right|\right)$,
then the power spectrum will depend only on the magnitude $k\equiv\left|{\bf k}\right|$
of the wavevector ${\bf k}$: $P\left({\bf k}\right)=P\left(k\right)$.

Equation~\eqref{eq:Pkdef} illustrates why theorists prefer to work
with the power spectrum rather than the correlation function: the
Dirac delta function indicates that for a random field that is homogenous
in real space, its Fourier modes will be \emph{uncorrelated} for different
values of ${\bf k}$. In other words, its covariance matrix is diagonal:
if $\delta\left({\bf x}\right)$ is a homogenous, zero-mean Gaussian
random field described by the probability distribution in equation~\eqref{eq:gaussian},
then $\hat{\delta}\left({\bf k}\right)$ is a Gaussian random field
as well, with the additional feature that the matrix $\mathbfss{M}$
in equation~\eqref{eq:gaussian} is a diagonal matrix. This makes
the power spectrum much more convenient for many theoretical calculations.
For more details see \citet{1998ASSL..231..185H} for an excellent
discussion of correlation functions and power spectra.

The primary goal of first order cosmology is to predict the power
spectrum $P\left(k,t\right)$ of the density contrast $\delta\left({\bf x},t\right)$
of matter in our universe as a function of cosmic time. This can be
modeled as\[
P\left(k,t\right)=\left(\frac{\delta_{G}\left(t\right)}{\delta_{G}\left(t_{i}\right)}\right)^{2}T^{2}\left(k\right)P\left(k,t_{i}\right),\]
where $\delta_{g}\left(t\right)$ is (an appropriately relativistic
version of) the growing mode in equation~\eqref{eq:linear_soln}
and $T\left(k\right)$, known as the transfer function, accounts for
a bunch of physics we have ignored in the simple treatment here --
interactions between matter and radiation, the behavior of modes with
wavelengths larger than the horizon (the distance a photon can travel
through the universe since time $t=0$) and so forth. See \citet{1993sfu..book.....P}
for a detailed discussion of the transfer function.

The function $P\left(k,t_{i}\right)$ is the spectrum of primordial
fluctuations that seeds the growth of structure. Theories of inflation
predict that this initial spectrum was produced by quantum fluctuations
that were stretched to macroscopic size by a period of exponential
expansion in the very early universe. This process typically predicts
that the initial fluctuations can be described by a Gaussian random
field with a power-law power spectrum:

\[
P\left(k,t_{i}\right)=A_{s}k^{n_{s}},\]
where $A_{s}$ and $n_{s}$ are the basic first order cosmological
parameters that are included along with the zeroth order parameters
$H_{0}$, $w$, and various $ $$\Omega$'s in parameter estimation
analyses such as the one discussed in \S\ref{sec:cosmological-parameters}.
Most models of inflation predict that $n_{s}$ to be slightly less
than 1.

Theoretical predictions for $P\left(k,t\right)$ for matter in our
universe given a set of cosmological parameters can be computed exactly
in the linear regime and is typically done using the handy computer
program CMBFAST \citep{1996ApJ...469..437S}. With the first order
approximation that $\delta\ll1$, each Fourier mode evolves independently,
so this makes the theoretical calculations much easier in Fourier
space than in real space. However, there are two issues with comparing
this predicted power spectrum with what we observe: first, for perturbations
with $\delta\sim1$ the linear theory no longer applies. Secondly,
our observations rely on using galaxies to trace the overall matter
distribution. The issue of nonlinear evolution is discussed in the
next section, and the issue of galaxies as tracers is explored in
depth in chapter~\ref{chap:bias}. Here we merely note the usual
assumption made when measuring the power spectrum from galaxy surveys:
that the power spectrum of the galaxies has the same shape as that
of the overall matter, but scaled by a normalization factor:\begin{equation}
P_{\mathrm{galaxy}}\left(k\right)=b^{2}P_{\mathrm{matter}}\left(k\right),\label{eq:biaspk}\end{equation}
where the parameter $b$ is called the bias. On scales larger than
the size of the largest gravitationally bound structures in our universe
($k\simlt0.09h/\mathrm{Mpc}$, or $\lambda\equiv2\pi/k\simgt70h^{-1}\mathrm{Mpc}$),
the simplifying assumptions of linear evolution and simple galaxy
bias have been found to be quite reasonable \citep{2006PhRvD..74l3507T}.

\subsection{Higher order cosmology}

\label{sub:Higher-order}

The zeroth and first order cosmology discussed in the previous sections
can take us a long way, but a full theory of structure formation in
our universe requires a nonlinear treatment of differential equations
such as equations\eqref{eq:continuity}, \eqref{eq:euler}, and \eqref{eq:poisson}
that can be applied to systems like galaxies and galaxy clusters that
have $\delta\gg1$. Unfortunately, this turns out to be quite challenging
-- the nonlinear versions of the equations couple together the evolution
of different Fourier modes, so most of the techniques used in first
order cosmology cannot be applied and exact solutions do not exist. 

This difficult and unsolved problem has been tackled from a number
of different angles: analytical models that apply under various simplified
conditions, numerical simulations that attempt to include all of the
relevant physics, empirical fitting formulas tuned to match simulations
or observations, and various hybrid techniques that endeavor to combine
analytical and numerical results to build up a complete picture. Numerical
simulations are arguably the only way to obtain a complete solution,
and much work has been done on both $N$-body simulations that model
purely gravitational systems (e.g. \citealt{2005Natur.435..629S})
and hydrodynamical simulations that include gas physics as well (see
\citealt{2008SSRv..tmp...26D} for a recent review). However, such
simulations are always subject to a variety of approximations and
the results can be challenging to interpret \citep{2004ApJ...614..533T,2008arXiv0804.0294R},
so analytical and semi-analytical methods continue to play an important
role. Here we outline perhaps the most widely-used analytical model
for nonlinear structure evolution and discuss some of its extensions,
and then we give an overview of a hybrid approach known as the halo
model that has had much success in recent years.

\subsubsection{Spherical collapse and Press-Schechter theory}

The analytical model we will discuss here starts with a simple approximation
for the collapse of a spherical overdensity. Consider a universe that
contains only matter and has a density equal to the critical density
$\rho_{\mathrm{cr}}$ such that $\Omega_{m}=1$, except in a spherical
region of comoving radius $R_{0}$ enclosing mass $M=4\pi\Omega_{m}^{\prime}\rho_{\mathrm{cr0}}R_{0}^{3}/3$
in which the density is slightly higher than the critical density,
with $\Omega_{m}^{\prime}\simgt1$. Equation~\eqref{eq:prettyfriedmann}
can be solved for the $\Omega_{m}=1$ background universe to give
\begin{equation}
a_{b}\left(t\right)=\left(\frac{3}{2}H_{0}t\right)^{2/3},\;\rho_{b}\left(t\right)=\frac{1}{6\pi Gt},\label{eq:flatsoln}\end{equation}
where the $b$ subscript denotes the background universe. The perturbed
region also evolves according to Equation~\eqref{eq:prettyfriedmann},
but with $\Omega_{m}^{\prime}>1$. A parametric solution is given
by\begin{equation}
\left\{ \begin{array}{l}
a_{p}\left(\eta\right)=\frac{1}{2}\frac{\Omega_{m}^{\prime}}{\Omega_{m}^{\prime}-1}\left(1-\cos\eta\right)\\
t\left(\eta\right)=\frac{1}{2H_{0}}\frac{\Omega_{m}^{\prime}}{\left(\Omega_{m}^{\prime}-1\right)^{3/2}}\left(\eta-\sin\eta\right)\end{array}\right.\label{eq:cycloid}\end{equation}
where the $p$ subscript denotes the perturbation. The perturbation
will expand until $\eta=\pi$, and then start to contract until $\eta=2\pi$,
at which point it will be fully collapsed. 

In reality, the density will not become infinite -- if the matter
is baryonic, the collapse will eventually be stopped by gas pressure,
and if it is dark matter, the particles will whiz past each other
and get slowed down by gravitational forces. Either way, the system
is roughly expected to reach an equilibrium state where its kinetic
and potential energy should satisfy the virial theorem: $U+2K=0.$
The virialized radius can be found by considering the energy at the
turnaround point $\eta=\pi$$ $ is purely potential. Since the gravitational
potential energy of a sphere is proportional to $R^{-1}$, energy
conservation and the virial theorem imply that the virialized radius
is half the radius at turnaround, so $ $the collapsed density will
be \begin{eqnarray}
\rho_{\mathrm{collapsed}} & = & 8\rho_{p}\left(\eta=\pi\right)\nonumber \\
 & = & 18\pi^{2}\rho_{b}\left(\eta=2\pi\right).\label{eq:rhocollapsed}\end{eqnarray}
After the collapse, the virialized object will retain this physical
density as the background universe continues to expand.

The density contrast between the perturbation and the background universe
can be found from equations \eqref{eq:flatsoln} and \eqref{eq:cycloid}
to be\begin{eqnarray}
\delta\equiv\frac{\rho_{p}-\rho_{b}}{\rho_{b}} & = & \frac{9}{2}\frac{\left(\eta-\sin\eta\right)^{2}}{\left(1-\cos\eta\right)^{3}}\nonumber \\
 & \approx & \frac{3}{20}\eta^{2}\;\mathrm{for\,}\eta\ll1.\label{eq:deltasphere}\end{eqnarray}
How does this compare to the first order evolution of $\delta$ derived
in \S\ref{sub:First-order}? Equation~\ref{eq:linear_soln} for
a flat, matter-dominated universe gives $\delta_{g}\propto a$, so
the linear theory prediction gives $\delta_{\mathrm{lin}}\propto t^{2/3}$.
Choosing the proportionality constant to match equation~\ref{eq:deltasphere}
for early times, we have\begin{equation}
\delta_{\mathrm{lin}}=\frac{3}{20}\left[6\left(\eta-\sin\eta\right)\right]^{2/3}.\label{eq:deltalin}\end{equation}
Equations \eqref{eq:deltasphere} and \eqref{eq:deltalin} give us
an important tool: a means of connecting an exact solution with linear
theory predictions. We have done these calculations for a the simple
model of a $\Omega_{m}=1$ universe, but the same basic ideas apply
for other cases -- see \citet{1993sfu..book.....P} for more details.
At the time of the collapse, we have $\delta_{\mathrm{lin}}\left(\eta=2\pi\right)=3\left(12\pi\right)^{2/3}/20\approx1.686$.
Based on this, we can construct a recipe for modeling nonlinear evolution:
evolve $\delta$ using linear perturbation theory until it reaches
a critical value of $\delta_{c}=1.686$ in a given region, and then
substitute in a spherically-collapsed, virialized object at that location.

\citet{1974ApJ...187..425P} applied this idea to describe structure
formation in our universe and produced a canonical formula for the
mass function $n\left(M\right)$ giving the comoving number density
of objects of mass $M$ based on the following argument: consider
a universe with mean comoving mass density $\bar{\rho}$ that has
a matter distribution described by the linearly-evolved power spectrum
$P_{\mathrm{lin}}\left(k\right)$. To determine the properties of
the population of mass $M$ objects, smooth the linearly-evolved density
field with a spherical top-hat filter $w\left(r,R\right)\equiv3/(4\pi R^{3})\Theta(R-r)$
with radius $R\equiv\left(3M/\left(4\pi\bar{\rho}\right)\right)^{1/3}$.
The smoothed density field is \begin{equation}
\delta_{M}\left({\bf x}\right)=\int w\left(\left|{\bf x}-{\bf x}^{\prime}\right|,R\right)\delta\left({\bf x}^{\prime}\right)d^{3}{\bf x}^{\prime},\label{eq:deltasmoothed}\end{equation}
and the variance of the smoothed field is

\begin{equation}
\left\langle \delta_{M}\left({\bf x}\right)^{2}\right\rangle \equiv\sigma^{2}\left(M\right)=\int\frac{dk}{k}\frac{k^{3}P_{\mathrm{lin}}\left(k\right)}{2\pi^{2}}\left|\hat{w}\left(k,R\right)\right|^{2},\label{eq:sigmadef}\end{equation}
where $\hat{w}\left(k,R\right)$ is the Fourier transform of the top
hat filter: \begin{equation}
\hat{w}\left(k,R\right)=\frac{3}{\left(kR\right)^{3}}\left[\sin\left(kR\right)-kR\cos\left(kR\right)\right].\label{eq:tophatft}\end{equation}
Now consider any region that has $\delta_{M}\left({\bf x}\right)>\delta_{c}$
to be a virialized object that has mass $\ge M$. By repeating this
for all smoothing scales, we can build up a picture of the mass distribution
of objects in our universe. 

Assuming that the density contrast $\delta$ can be modeled as a Gaussian
random field, the fraction of the mass in our universe contained in
virialized objects with masses $\ge M$ is given in the Press-Schechter
model by \begin{eqnarray}
F\left(>M\right) & = & \frac{1}{\sqrt{2\pi}\sigma\left(M\right)}\int_{\delta_{c}}^{\infty}\exp\left(-\frac{\delta_{M}^{2}}{2\sigma^{2}\left(M\right)}\right)d\delta_{M}\nonumber \\
 & = & \frac{1}{2}\mathrm{erfc}\left(\sqrt{\frac{\nu}{2}}\right),\label{eq:PS_gaussian}\end{eqnarray}
where $\nu\equiv\delta_{c}^{2}/\sigma^{2}\left(M\right)$. This formula
has a curious feature: it says that the maximum fraction of mass that
can be contained in virialized objects is $1/2$, not 1 -- this is
because we have ignored underdense regions that have $\delta<0$.
However, mass in underdense regions is likely to accrete onto the
virialized objects eventually. To account for this omission, Press
and Schechter simply multiply equation~\eqref{eq:PS_gaussian} by
a {}``fudge factor'' of 2 and find they get a sensible final answer. 

We can find the mass function by differentiating equation~\eqref{eq:PS_gaussian}
and multiplying by the extra factor of 2 and by the number density
$\bar{\rho}/M$ that would be expected if \emph{all} of the mass in
a universe with average mass density $\bar{\rho}$ were in mass $M$
objects: \begin{eqnarray}
n\left(M\right)dM & = & -2\frac{\bar{\rho}}{M}\frac{dF}{dM}dM\nonumber \\
 & = & \frac{\bar{\rho}}{M^{2}}\sqrt{\frac{\nu}{2\pi}}\exp\left(-\frac{\nu}{2}\right)\left(\frac{d\left(\ln\nu\right)}{d\left(\ln M\right)}\right)dM\label{eq:PS}\end{eqnarray}
This is the famous Press-Schechter mass function. Although it is based
on a rather heuristic argument, it has proven to be a remarkably valuable
tool -- it agrees well with numerical simulations (e.g. \citealt{1993MNRAS.262.1023W})
and has provided a launching point for further theoretical work, particularly
the extended Press-Schechter theory based on the excursion set formalism
(see \citet{2007IJMPD..16..763Z} for a recent review). Early applications
of the excursion set formalism \citep{1990MNRAS.243..133P,1991ApJ...379..440B}
provided a more formal derivation of equation~\eqref{eq:PS} that
explains the factor of 2 and resolves other issues as well.

An updated version of the mass function that provides a better fit
to the latest numerical simulations \citep{2005Natur.435..629S} is
given by \citet{1999MNRAS.308..119S}:\begin{equation}
n\left(M\right)dM=\frac{A\left(p\right)\bar{\rho}}{m^{2}}\sqrt{\frac{q\nu}{2\pi}}\left(1+\frac{1}{\left(q\nu\right)^{p}}\right)\exp\left(-\frac{q\nu}{2}\right)\left(\frac{d\left(\ln\nu\right)}{d\left(\ln M\right)}\right)dM,\label{eq:ST}\end{equation}
with $q\approx0.707$, $p\approx0.3$, and $A\left(p\right)=\left(1+2^{-p}\Gamma\left(1/2-p\right)/\sqrt{\pi}\right)^{-1}\approx0.3222$.
Note that if $p=0$ and $q=1$ then This mass function can be derived
using excursion set theory \citep{2001MNRAS.323....1S} by considering
ellipsoidal rather than spherical collapse.

The extended Press-Schechter theory also provides a way to make theoretical
predictions for the clustering of halos and their merger histories
-- see \citet{2002sgd..book.....M} for a good summary. For example,
\citet{1991ApJ...379..440B} showed that the fraction of mass of a
large object of mass $M_{0}$ with present linear density contrast
$\delta_{0}$ and variance $\sigma_{0}$ that is made up of merged-in
smaller objects of mass $M_{1}$, present linear density contrast
$\delta_{1}$ and variance $\sigma_{1}$ is given by

\begin{equation}
f\left(\sigma_{1},\delta_{1}|\sigma_{0},\delta_{0}\right)\frac{d\sigma_{1}^{2}}{dM_{1}}dM_{1}=\frac{1}{\sqrt{2\pi}}\frac{\delta_{1}-\delta_{0}}{\left(\sigma_{1}^{2}-\sigma_{0}^{2}\right)^{3/2}}\exp\left[-\frac{\left(\delta_{1}-\delta_{0}\right)^{2}}{2\left(\sigma_{1}^{2}-\sigma_{0}^{2}\right)}\right]\frac{d\sigma_{1}^{2}}{dM_{1}}dM_{1}\label{eq:mergers}\end{equation}

\citet{1996MNRAS.282..347M} use equation~\eqref{eq:mergers} to
calculate the bias factor $b_{M}$ for halos of mass $M$ using the
extended Press-Schechter theory, where $b_{M}$ is defined such that
the power spectrum $P_{M}\left(k\right)$ of virialized objects of
mass $M$ is $b_{M}^{2}P_{\mathrm{lin}}\left(k\right)$, with $P_{\mathrm{lin}}\left(k\right)$
being the linear theory matter power spectrum. They find that\begin{equation}
b_{M}=1+\frac{\nu-1}{\delta_{c}}.\label{eq:PSbias}\end{equation}
The generalized version of this equation using the Sheth-Tormen mass
function (equation~\eqref{eq:ST}) was derived by \citet{1999MNRAS.308..119S}:

\begin{equation}
b_{M}=1+\frac{q\nu-1}{\delta_{{\rm c}}}+\frac{2p/\delta_{{\rm c}}}{1+\left(q\nu\right)^{p}}.\label{eq:STbias}\end{equation}
These and other results from extended Press-Schechter theory are key
ingredients in assembling an analytical description of nonlinear structure
formation.

\subsubsection{Halo model}

\label{sub:Halo-model}

One particular model that has proven useful for describing both nonlinear
clustering and galaxy bias is the known as the halo model, first formalized
by \citet{2000MNRAS.318..203S,2001ApJ...546...20S}. (See \citealt{2002PhR...372....1C}
for a comprehensive review.) The basic idea is this: at late times,
the matter distribution in our universe can be modeled using only
virialized objects made of dark matter that are called \emph{halos},
and the galaxy distribution can be modeled by an appropriate scheme
for placing galaxies in these dark matter halos. The key assumption
of the model is that the average properties of a halo -- its density
profile and its galaxy content -- only depend on the halo mass.

Using this assumption, we can construct an analytical form for the
nonlinear power spectrum by assembling components from first order
cosmology, the extended Press-Schechter theory discussed above, and
$N$-body simulation results. The halo model power spectrum is$ $

\begin{equation}
P\left(k\right)=P^{1h}\left(k\right)+P^{2h}\left(k\right),\label{eq:halopk}\end{equation}
where $P^{1h}\left(k\right)$ is the 1-halo term arising from correlations
between mass within the same halo and $P^{1h}\left(k\right)$ is the
2-halo term arising from correlations between mass in different halos. 

To calculate these terms we need a description of how mass is distributed
within a halo. This is usually modeled by an NFW profile \citep{1996ApJ...462..563N},
which has been found to be an excellent fit to dark matter halos that
form in $N$-body simulations. The normalized profile $u\left(r|M\right)\equiv\rho\left(r|M\right)/M$
is given by

\begin{equation}
u\left(r|M\right)=\frac{fc^{3}}{4\pi r_{{\rm vir}}}\frac{1}{\left(\frac{cr}{r_{{\rm vir}}}\right)\left(1+\frac{cr}{r_{{\rm vir}}}\right)^{2}}\label{eq:NFW}\end{equation}
where $f=1/\left[\ln\left(1+c\right)-c/\left(1+c\right)\right]$ and
the concentration parameter $c$ is a function of the mass, typically
approximated as \citep{2001MNRAS.321..559B}\begin{equation}
\bar{c}\left(M,z\right)=\frac{9}{1+z}\left(\frac{M}{M_{*}\left(z\right)}\right)^{-0.13}\label{eq:concentration}\end{equation}
with the characteristic mass scale $M_{*}$ is defined as the point
where $\nu=1$: $\delta_{{\rm c}}^{2}\equiv\sigma^{2}\left(M_{*}\right)$.
The virial radius $r_{{\rm vir}}$ is defined by the mass: $M=\Delta\bar{\rho}4\pi r_{{\rm vir}}^{3}/3$,
where $\Delta$ is the typical overdensity of a virialized structure
compared to the background density -- we found in equation~\eqref{eq:rhocollapsed}
that $\Delta=18\pi^{2}\approx178$ for a $\Omega_{m}=1$. Equation~\eqref{eq:NFW}
is normalized such that the total mass within the virial radius is
equal to $M$.

The 1- and 2-halo terms are then given by

\begin{equation}
P^{1h}\left(k\right)=\frac{1}{\bar{\rho}^{2}}\int dM\, n\left(M\right)\left|\hat{u}\left(k|M\right)\right|^{2}\label{eq:1halo}\end{equation}

\begin{equation}
P^{2h}\left(k\right)=\frac{1}{\bar{\rho}^{2}}\int dM_{1}\, n\left(M_{1}\right)\hat{u}\left(k|M_{1}\right)\int dM_{2}\, n\left(M_{2}\right)\hat{u}\left(k|M_{2}\right)P^{hh}\left(k|M_{1},M_{2}\right),\label{eq:2halo}\end{equation}
where $\hat{u}\left(k|M\right)$ is the Fourier transform of the NFW
profile in equation~\eqref{eq:NFW} and $n\left(M\right)$ is the
mass function given by equation~\eqref{eq:ST}. $P^{hh}\left(k|M_{1},M_{2}\right)$
is the halo-halo power spectrum, most simply modeled by\begin{equation}
P^{hh}\left(k|M_{1},M_{2}\right)=b_{M_{1}}b_{M_{2}}P_{{\rm lin}}\left(k\right),\label{eq:halohalo}\end{equation}
where $b_{M}$ is bias for halos of mass $M$ given in equation~\eqref{eq:STbias},
although more sophisticated models have been used in recent work \citep{2005ApJ...630....1Z,2005ApJ...631...41T}.

In addition to modeling the dark matter distribution, the halo model
is also a useful tool for modeling galaxy bias. The galaxy distribution
then be modeled by first determining the halo distribution and then
populating the halos with galaxies. The second step requires a model
for $P\left(N|M,L,C,\dots\right)$$ $, the probability that a halo
of mass $M$ will host $N$ galaxies with luminosity $L$, color $C$,
and any other galaxy properties the modelers wish to include. Multiple
parameterizations for this probability distribution have been developed
\citep{2000MNRAS.318.1144P,2002ApJ...575..587B,2005PhRvD..71f3001S,2003MNRAS.339.1057Y,2003MNRAS.340..771V}.
The algorithm used for placing the galaxies in halos encodes the physics
of galaxy formation and can be used to compare semi-analytical galaxy
formation models \citep{2006RPPh...69.3101B} with observational results
such as those discussed in chapter~\eqref{chap:bias}. This procedure
yields a model of galaxy bias that is more closely linked with the
actual physical processes than the simple assumption of equation~\eqref{eq:biaspk}. 

The halo model has been successfully used to model several different
types of observations, such as galaxy correlation functions \citep{2005ApJ...630....1Z},
void statistics \citep{2007arXiv0707.3445T}, and counts of galaxies
in groups \citep{2008ApJ...676..248Y}. However, as we will see in
\S\ref{sec:cosmological-parameters} it has yet to be incorporated
into full cosmological parameter analyses -- this could potentially
be a promising direction for future work.

\subsection{What does this teach us about particle physics?}

What can we hope to learn about physics on at the sub-atomic level
by studying our universe on such a grand scale? Quite a lot, as it
turns out! One basic tenet of particle physics is that higher energies
can probe shorter length scales. In order to examine extremely small
length scales -- the quarks within a proton, for example -- particles
are collided with each other at extremely high energies. The current
generation of particle accelerators can achieve energies of 2 TeV,
and the Large Hadron Collider, set to turn on in 2008, will push this
up to 14 TeV \citep{2008arXiv0801.1334F}. However, the extremely
hot, dense conditions in the first fraction of a second after the
Big Bang provide a natural particle accelerator -- a split-second
after the Big Bang, the average thermal energy of particles could
have been as high as $10^{13}$ TeV.
Thus by studying cosmology, particle physicists can learn about
phenomena that could not be created in a terrestrial experiment.

Of course, the problem with this plan is that after 14 billion years,
much of the action has faded away -- unstable particles created early
on have long since decayed and the present-day density of any remaining
stable particles has been so diluted by the expansion that they are
no longer interacting with each other. Particle physicists who endeavor
to use our early universe as a laboratory must find ways to predict
detectable signatures left behind by the processes they hope to investigate.
Particle physicists also work to develop models that can explain the
bizarre features of the standard cosmological model described in \S\ref{sub:Standard-model}.
This is especially interesting because the ideas of dark matter and
dark energy cannot be explained within the framework of the standard
model of particle physics and thus hint at physics beyond our current
understanding.

From cosmological and other astronomical observations, we know that
some form of non-baryonic dark matter that does not couple strongly
to photons makes up most of the mass in our universe. Theorized particles
that have properties similar to dark matter are actually well-motivated
in a particle physics context. One class of proposed extensions to
the standard model of particle physics, known as supersymmetry, postulate
that every particle has a supersymmetric partner particle whose spin
differs by $1/2$ \citep{2008arXiv0801.1928P}. The supersymmetric
partners of known particles are expected to have such high masses
that we would not have been produced by the current generation of
particle accelerators. Supersymmetry was proposed to explain a number
of seemingly {}``unnatural'' features of the standard model, such
as why the masses of standard model particles are so much smaller
than the natural energy scale of the theory. It also provides a candidate
particle for dark matter: the lightest supersymmetric particle. At
the extremely high temperatures and densities present in the early
universe, an abundance of supersymmetric particles might have been
created. Most of these will be unstable and decay quickly into lighter
supersymmetric particles, but the lightest supersymmetric particle
is predicted to be stable. If this is the case, it would be present
in our universe today and act like dark matter -- that is, like a
weakly interacting massive particle, a.k.a. a WIMP. This is currently
regarded as the most natural explanation for dark matter, but there
are a number of other particle physics theories that can generate
particles with dark-matter-like properties as well, such as axions,
solitions, or WIMPZILLAS \citep{2004astro.ph..3064G}. 

One might ask why we need to resort to such exotic-sounding models
to explain dark matter -- after all, we know that neutrinos must have
some mass, as discussed in \S\ref{chap:nuastro}, so could the dark
matter simply be neutrinos? The answer comes from first order cosmology,
considering the nature of the perturbations in the density field.
The three types of neutrinos in the standard model, while we know
they must have \emph{some} mass, are still light enough that their
average velocity is quite high at the time when structure is forming, so they are
classified as \emph{hot} dark matter. 
The main effect of hot dark matter is to wash out the formation of 
structure on small scales -- heuristically, neutrinos will not collapse
into a structure if their average velocity exceeds the escape velocity of 
the a given overdense clump.  For small (i.e., galaxy-sized) clumps, the escape velocity
is small enough that we would not expect neutrinos to collapse into
structures of this size.
See \citet{2006PhR...429..307L} for
a detailed explanation of this effect. 

Such a picture is not consistent
with the structure we see today, which implies that the dark matter
must be \emph{cold} -- that is, massive enough to have very low average velocity
at the time structure starts to form. In fact, measurements of the
power spectrum can be used to set a strong upper limit on the sum
of the masses of the three types of neutrinos -- we discuss an example
of this in \S\ref{sec:cosmological-parameters}. In fact, cosmology
currently place stronger upper limits on the neutrino mass than current
particle physics experiments \citep{2006PhST..127..105E,2008arXiv0803.1585H},
and the next generation of cosmological measurements may be sensitive
enough to detect masses as small as the mass differences seen by Super-K. 

Another cosmological puzzle particle physicists have sought to tackle
is that of dark energy. What kind of substance could lead to the bizarre
repulsive gravitational effects that are driving the current acceleration
of our universe's expansion? The most basic possibility from the point
of view of quantum field theory is that there is an inherent energy
density associated with empty space called \emph{vacuum energy}. The
vacuum energy density would remain constant as the universe expands,
which is consistent with the observed properties of dark energy. Theoretical
predictions of what the value of the vacuum energy density should
be are rather problematic -- a basic quantum field theory calculation
gives a divergence, predicting infinite vacuum energy density. Imposing
a natural cutoff to this divergence gives an estimate of ${\sim}10^{110}\,\mathrm{erg/cm^{3}}$
for the vacuum energy density. Prior to the discovery of dark energy,
particle theorists devoted much effort to searching for natural ways
this vacuum energy could cancel out to be exactly zero. Finding that
there is indeed something that looks like vacuum energy in our universe
made the problem even more difficult for the theorists, since the
cosmological observations indicate that the density of dark energy
is ${\sim}10^{-10}\,\mathrm{erg/cm^{3}}$, 120 orders of magnitude
less than the estimated value! A theory where the vacuum energy is
almost entirely canceled out except for a tiny fraction would be much
more bizarre than perfect cancellation. Supersymmetry provides some
help here, but not enough -- it can reduce the discrepancy to {}``just''
60 orders of magnitude. Dark energy has proven to be a fascinating
challenge from a particle theory perspective (see \citealt{2008arXiv0803.0982F}
for a review of possible models) and is particularly interesting from
the point of view of connecting particle physics and astronomy because
as far as we know, the \emph{only} way to measure its properties is
through cosmological observations (see \citealt{2006astro.ph..9591A}
for more details).

\section{Tools of cosmology}

\subsection{Combining cosmological probes}

The great strength of modern observational cosmology is that such
a wide variety of different experimental measurements point to the
same theoretical model. Some measurements, such as using supernovae
as standard candles to trace the expansion history of our universe,
target the zeroth order cosmology discussed in \S\ref{sub:Zeroth-order}
-- that is, they are effectively probing the function $a\left(t\right)$.
Other measurements, such as studying the tiny fluctuations in the
cosmic microwave backround or the clustering patterns in the distribution
of galaxies, are fundamentally probes of the first order cosmology
discussed in \S\ref{sub:First-order}: they measure variations on
the power spectrum $P\left(k\right)$. Still others, such as measuring
the abundances of elements formed by fusion in the hot, dense plasma
of the early universe, target one particular cosmological parameter
-- in this case, the ratio of baryon density to photon density.

Combining all of these complementary probes is an extremely powerful
way to study our universe. First of all, since different probes are
sensitive to different combinations of cosmological parameters, combining
two or more different measurements allows us to break the degeneracies
between the parameters we want to measure. Furthermore, because each
technique is subject to different systematic effects, we can use them
as consistency checks. If two experiments disagree, it would be a
sign that either there is some systematic effect that has not been
accounted for or that our theoretical model is flawed. The fact that
there is broad agreement over so many different experiments lends
much stronger support to the standard cosmological model than any
single experiment could ever do. In the following sections we describe two key probes -- the cosmic microwave background and galaxy surveys -- in some detail, and give a brief overview of other complementary techniques.

\subsection{Galaxy surveys}

\label{sub:Galaxy-surveys}The work in chapters \ref{chap:mangle}
and \ref{chap:bias} focuses on issues with using such surveys as
cosmological tools, so we describe this particular cosmological probe
in some detail here. The basic idea behind a galaxy survey is to map
out the locations of galaxies over a large volume and use them as
point tracers of the overall matter distribution in our universe.
This provides a way to observe the statistical properties of the large-scale
structure predicted from the first order and higher order cosmological
models discussed in \S\ref{sub:First-order} and \S\ref{sub:Higher-order}.

Galaxy surveys come in two basic flavors: angular surveys and redshift
surveys. Angular surveys indentify galaxies over a large area of the
sky, and thus only see the two-dimensional projection of the galaxy
distribution. Redshift surveys, on the other hand, also measure spectra
to obtain the redshift of each galaxy and thus its distance from us
according to Hubble's law (equation~\eqref{eq:hubble}), which produces
a three-dimensional map of galaxy locations. Recently a hybrid approach
between these two types of surveys has become popular: photometric
redshift surveys, whereby the redshift is estimated based on the photometric
information from multiple wavelength bands of an angular survey. This
provides rough measure of the distance that eliminates the need for
time-consuming spectroscopy and the accuracy of the redshift estimation
techniques is continuing to improve \citep{2007MNRAS.375...68C,2008ApJ...674..768O,2007arXiv0711.1059B}.

Furthermore, surveys can either be wide and shallow -- covering a
large area of the sky but only extending a modest distance from us,
or narrow and deep -- focusing on a small patch of sky for an extended
time to find very faint and therefore extremely distant objects. Typical
surveys strike a balance between these two choices that is optimal
for a particular set of science goals.

\subsubsection{Early observations}

Astronomers have been interested in studying the distribution of galaxies
since the time of Hubble (see, e.g., \citealt{1934CMWCI.485....1H,1938ApJ....88..344C}).
The early angular catalogs studied were the Palomar \citep{1959ASPL....8..121A}
and the Lick \citep{1974SouSt..25..107D} sky surveys, and redshifts
of over 800 galaxies from various spectroscopic observations were
collected in \citet{1956AJ.....61...97H}. These early studies showed
evidence for clumpiness in the overall distribution, although its
exact nature was difficult to discern (see, e.g. \citealt{1952PASP...64..247Z,1958AJ.....63..253D,1961AJ.....66..607A}).
These observations inspired the theoretical work of \citet{1952ApJ...116..144N}
which formed the seeds of the present-day halo model discussed in
\S\ref{sub:Higher-order} as well as the first effort to compare
observations with a simulated galaxy distribution \citep{1954ApJ...119...91S}.

\begin{figure}
\begin{centering}\includegraphics[width=0.50\textwidth, angle=-90]{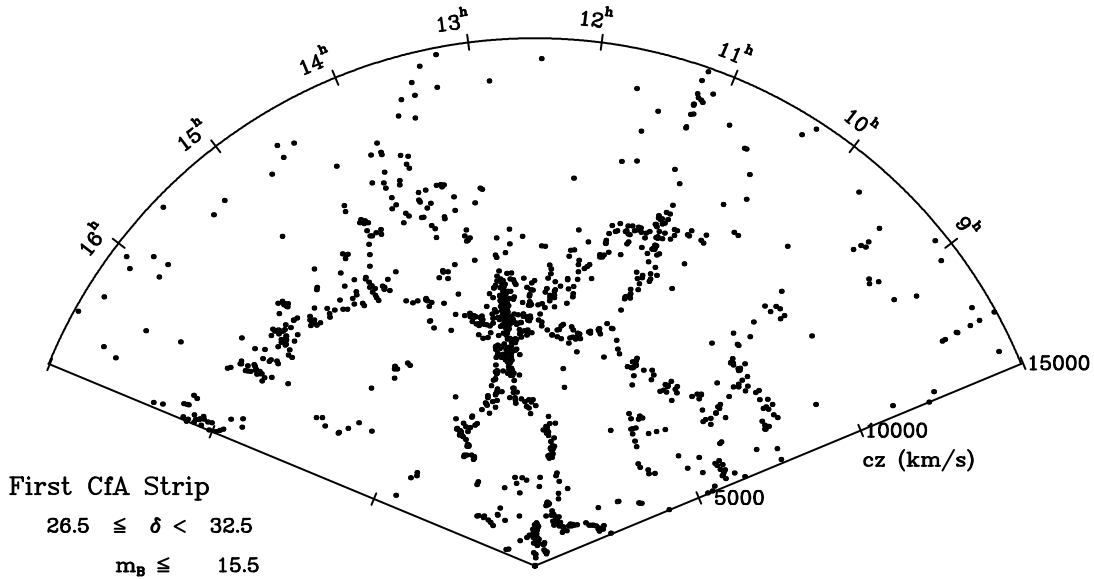}\par\end{centering}

\caption[A slice through the CfA redshift survey.]{\label{fig:cfa}A slice through the CfA redshift survey \citep{1983ApJS...52...89H}
done at the Smithsonian Astrophysical Observatory. Figure from \protect\url{http://cfa-www.harvard.edu/~huchra/zcat/}. }
\end{figure}

The first true redshift survey was the Center for Astrophysics Redshift
Survey (CfA; \citealt{1983ApJS...52...89H}), which started in 1978
-- prior to this only $\sim2000$ galaxies had measured redshifts
\citep{1999elss.conf...87D}. This survey measured redshifts of the
2401 galaxies with an apparent $Z$-band magnitude brighter than $14.5\, m_{Z}$.
A slice through the CfA survey is shown in Fig.~\ref{fig:cfa}
This survey revealed that galaxies are distributed in a complex web
of filaments, sheets, and voids.

Over the following decade a number of other surveys were undertaken,
including the Southern Sky Redshift Survey (SSRS1; \citealt{1991ApJS...75..935D}),
surveys done by the Infrared Astronomy Satellite (IRAS; \citealt{1992ApJS...83...29S}),
the Automatic Plate Measurement angular survey (APM; Maddox et al.~1990a,b, 1996),
\nocite{1990MNRAS.243..692M,1990MNRAS.246..433M,1996MNRAS.283.1227M}
and the Las Campanas Redshift Survey (LCRS; \citealt{1996ApJ...470..172S}).
\citet{1994MNRAS.267.1020P} compiled power spectrum measurements
from this generation of surveys and fit them with several cosmological
models -- they found that the data favored models with low matter
density: $ $$\Omega_{M}<1$, in agreement with present day standard
model. Further details on the history of this field can be found in
\citet{1999elss.conf...87D,2000cucg.confE...1B}.

\subsubsection{Modern surveys}

The current state of the art in galaxy survey cosmology is exemplified
by two surveys: the Two Degree Field Galaxy Redshift Survey (2dFGRS;
\citealt{2001MNRAS.328.1039C,2003astro.ph..6581C}) and the Sloan
Digital Sky Survey (SDSS; \citealt{2000AJ....120.1579Y}).

\begin{figure}
\begin{centering}\includegraphics[width=1\textwidth]{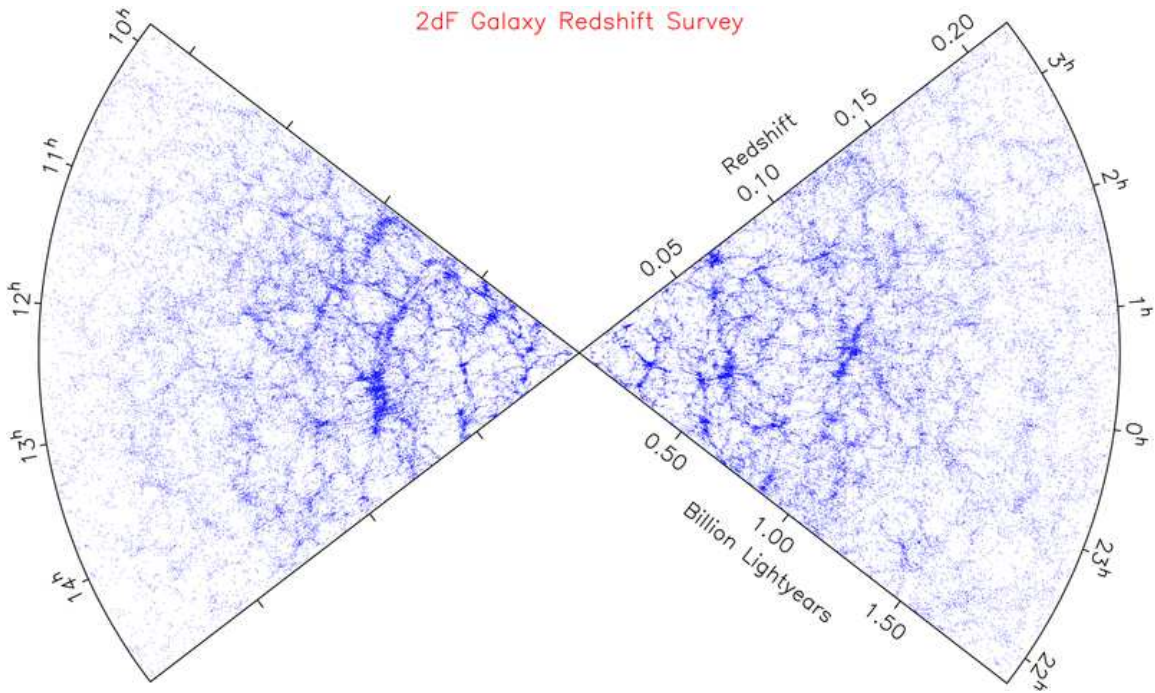}\par\end{centering}

\caption[A slice through 2dFGRS.]{\label{fig:2df}A slice through 2dFGRS \citep{2001MNRAS.328.1039C,2003astro.ph..6581C}.
Figure from \protect\url{http://www.mso.anu.edu.au/2dFGRS}. }
\end{figure}

2dFGRS is a redshift survey that measured the redshifts of galaxies
observed by the APM angular survey. Target galaxies with an apparent
blue-band magnitude brighter than $b_{J}<19.5$ were selected from
APM's $5^{\circ}\times5^{\circ}$ photographic plates and spectra
were taken using the Two Degree Field system at the Anglo-Australian
Telescope in New South Wales, Australia. This is a 4 meter telescope
with a $2^{\circ}$ field of view and a set of spectrographs which
accept 400 optical fibers such that spectra of 400 objects in the
field of view can be taken simultaneously. The redshift survey was
completed in 2002 and took 272 nights over 5 years. It measured redshifts
for 221,414 galaxies with a mean redshift of $z=0.11$ and covered
an area of about 1500 square degrees. A slice through the galaxy distribution
measured by 2dFGRS is shown in Fig.~\ref{fig:2df}.

SDSS, our source of data for the work in chapter~\ref{chap:bias},
is the largest galaxy redshift survey to date, and is continuing to
gather more data. It is being conducted using a dedicated 2.5 meter
telescope at the Apache Point Observatory in New Mexico that has a
$3^{\circ}$ field of view and $\sim1.4^{\prime\prime}$ \citep{2006AJ....131.2332G}.
The final SDSS data set will cover nearly $1/4$ of the sky and have
spectra of nearly 1 million galaxies and about 105,000 quasars (a.k.a.
AGNs; see \S\ref{sec:AGNs}). The most recent data release, DR6,
covers $17\%$ of the sky and has 790,860 galaxies and 103,647 quasars.

The camera used for the photometric survey is a $5\times6$ array
of CCDs arranged such that each of the 5 rows observes through a different
filter, ranging from near-ulraviolet to near-infrared wavelengths
\citep{1996AJ....111.1748F,1998AJ....116.3040G}. The survey method
used is known as drift-scanning: as the sky moves overhead due to
the rotation of the Earth, points on the sky pass through each filter
in succession. This produces a long strip of 6 scanlines across the
sky with photometric data from all 5 filters simultaneously. It takes
over 8 hours to scan a strip $130^{\circ}$ long.

Based on this photometric information, targets are selected for the
spectroscopic survey to obtain redshifts. There are several types
of targets selected, including {}``main'' galaxies, the luminous
red galaxies (LRGs), quasars, and stars within our galaxy. The main
galaxy sample includes all galaxies with an $r$-band apparent magnitude
brighter than $r<17.77$ -- a typical galaxy at a redshift of $z=0.1$
has $r=17.5$. The brightest galaxies in the main sample can be seen
out to a redshift of $z\sim0.3$. This is the sample used in chapter~\ref{chap:bias}.
The LRG sample \citep{2001AJ....122.2267E} is a sparser sample extending
to higher redshifts that contains only galaxies of a particular type:
very luminous early-type elliptical galaxies. The LRGs is roughly
volume-limited and extends to a redshift of $z\sim0.5$. The properties
of the LRG sample make it especially useful for certain types of cosmological
measurements, such as the study discussed in \S\ref{sec:cosmological-parameters}.

\begin{figure}
\begin{centering}\includegraphics[bb=100bp 260bp 1160bp 1000bp,clip,width=1\textwidth]{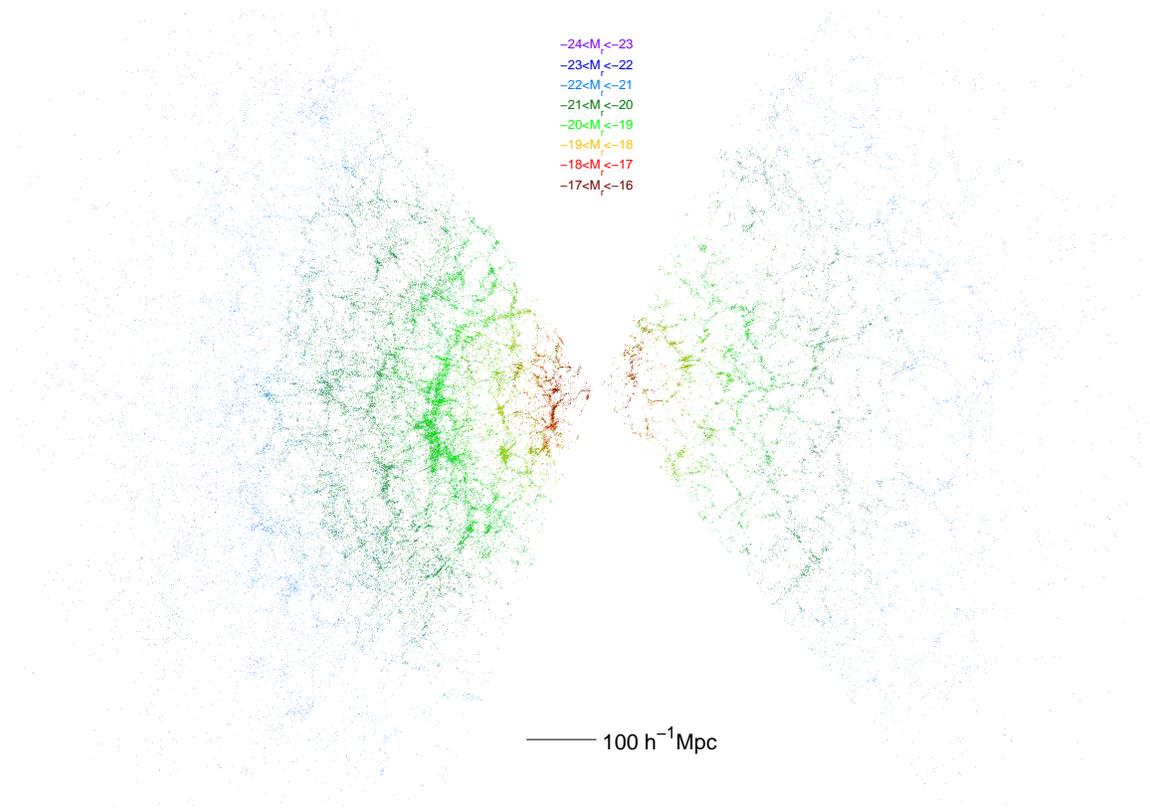}\par\end{centering}

\caption[A slice through SDSS.]{\label{fig:sdss}A slice through SDSS \citep{2000AJ....120.1579Y}.
The galaxies are colored according to their absolute $r$-band magnitude
$M_{r}$, using the same luminosity bins as in chapter~\ref{chap:bias}. }
\end{figure}

SDSS spectroscopy is done using a pair of integral field spectrographs
that can take up to 640 spectra simultaneously \citep{2004SPIE.5492.1411U}.
This is done using $3^{\circ}$ diameter aluminum plates drilled with
holes at the locations of the targets and plugged with optical fiber
cables which are fed into the spectrographs. Typically six to nine
spectroscopic fields can be observed on a given night. Once a spectrum
has been measured for a galaxy, its redshift can be determined by
the positions of known spectral lines, and thus we know its three-dimensional
location within the survey volume. A slice of the SDSS galaxy distribution
is shown in Fig.~\ref{fig:sdss}. 

Both 2dFGRS and SDSS have been extraordinarily valuable resources
for the general scientific community, in large part because the data
is made publicly available. The unprecedented volume mapped by these
two surveys have revolutionized studies of large-scale structure in
our universe in many ways. One worth noting here is that the galaxy
samples are now large enough that they can be reasonably divided up
into subsets for more detailed analysis, as is done in chapter~\ref{chap:bias}.

Other current surveys such as the 2-Micron All Sky Survey (2MASS;
\citealt{2005ASPC..329..135H}) and the second phase of the Deep Extragalactic
Evolutionary Probe (DEEP2; \citealt{2003SPIE.4834..161D}) the have
led to quite interesting cosmological results as well -- we have focused
on 2dFGRS and SDSS here because their balance on the wide and shallow
versus narrow and deep spectrum is ideal for large-scale structure
studies. However, the full-sky coverage of 2MASS provides of our nearby
universe has been quite fruitful for matching local velocity flows
to the mass density \citep{2007ApJ...655..790C} and the high redshift
data from DEEP2 allows us to measure the evolution of galaxy properties
and clustering as well \citep{2008ApJ...672..153C}.

There are a number of upcoming surveys planned that will take this
field another giant step forward. Starting in the next year or so,
the Baryon Oscillation Spectroscopic Survey (BOSS;\citealt{boss})
will use the SDSS telescope with updated technology to target the
baryon oscillation signal described in \S\ref{sub:Other-probes}
and the Dark Energy Survey (DES; \citealt{2005astro.ph.10346T}) will
perform a photometric redshift survey that will contain about 300
million galaxies. Further in the future are the Panoramic Survey Telescope
and Rapid Response System (Pan-STARRS; \citealt{2004SPIE.5489...11K})
and the Large Synoptic Survey Telescope (LSST; \citealt{2002SPIE.4836...10T,2004AAS...20510802S}),
which will operate with unprecedented speed, surveying the entire
available sky several times each month, and the Wide-Field Multi-Object
Spectrograph (WFMOS; \citealt{2006PhRvD..74f3525Y}), which will be
able to take spectra of 20,000 galaxies every night.

On the theoretical side, it is important to be prepared for these
upcoming surveys by understanding systematic effects such as nonlinear
structure growth and galaxy bias so that we can extract as much cosmological
information from them as possible. Galaxy surveys will continue to
be an important probe as cosmology moves into a new era.

\subsection{Cosmic microwave background}

\label{sec:CMB}Galaxy surveys map out the present-day or relatively
low redshift universe, so they are particularly well-complemented
by probes of our universe's early stages. Perhaps the richest source
of information we have about the early universe comes from the leftover
radiation from our universe's hot and dense early phase, known as
the cosmic microwave background (CMB). Measurements of the CMB are
complementary to galaxy surveys that aim to map out our universe today
because the CMB is essentially a snapshot of the early universe. Combining
information from galaxy surveys and the CMB provides a way to break
degeneracies between parameters and tighten the constraints on our
theoretical model.

The basic physics behind this early universe snapshot is as follows:
the early universe was filled with a plasma of protons and electrons
coupled to photons in thermal equilibrium. The electrons were tightly
coupled to the photons through Thomson scattering, and the protons
are tightly coupled to the electrons through electromagnetic attraction,
so these three components behave as a single fluid. 

The dark matter, however, does not interact with baryons or light,
so at this point it has already started to clump together under the
influence of gravity around the seed fluctuations generated by inflation.
The mass of the protons drags the baryon-photon fluid into the gravitational
potential wells formed by the clumping dark matter, but the radiation
pressure from the photons resists this infall. This process generates
oscillations which propagate as sound waves through the baryon-photon
plasma.

As our universe continues to expand, it also cools, and when our universe
is about 400,000 years old (0.003\% of its current age), it became
cool enough for the protons and electrons to combine into neutral
hydrogen. At this point, the baryons and photons decouple from each
other. The baryons are now free to collapse into the potential wells
formed by the dark matter, and begin forming into stars and galaxies.
The photons, on the other hand, free-stream through our universe and
undergo (nearly) no further interactions from the time of decoupling
until the present.

Today these CMB photons permeate our universe, and they have been
redshifted to microwave wavelengths as our universe has expanded.
Because they have traveled largely unimpeded since they decoupled,
they provide a snapshot of the early universe at the moment of decoupling.
This leftover glow from the hot and dense early universe has been
observed to be essentially a perfect blackbody with a uniform temperature
of 2.725K in all directions on the sky \citep{1994ApJ...420..439M}.
This fact alone is extremely solid evidence for the basic picture
of the standard cosmological model described in \S\ref{sub:Standard-model}. 

\begin{figure}
\includegraphics[width=1\textwidth]{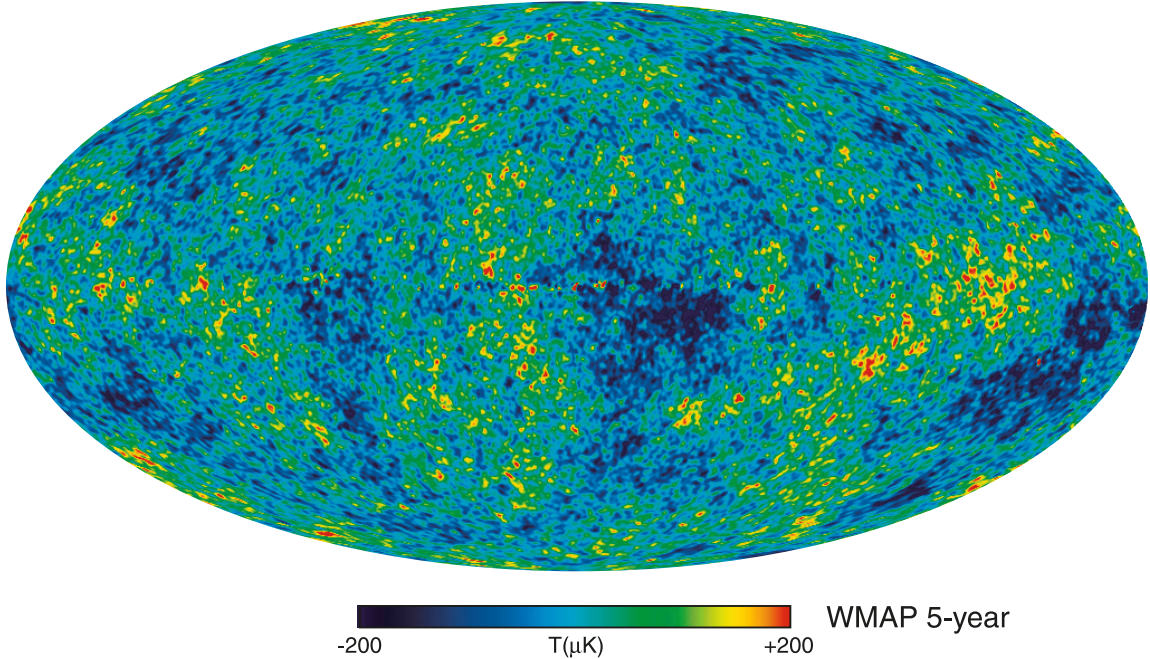}

\caption[Temperature anisotropies in the CMB from WMAP5.]{\label{fig:wmap}Temperature anisotropies in the CMB measured by
the NASA/WMAP Science Team, shown in Hammer-Aitoff projection in Galactic
coordinates. Reproduced from \citet{2008arXiv0803.0732H}; used with
permission. }
\end{figure}

However, the true power of CMB observations lies in tiny deviations
from this uniformity: the CMB anisotropies. The sound waves propagating
through the plasma of the early universe create slight inhomogeneities
in the density, which we can see as tiny variations in the CMB temperature
at different points on the sky. Figure~\ref{fig:wmap} shows the
latest measurement of these anisotropies from the 5-year data release
of the Wilkinson Microwave Anisotropy Probe(WMAP5; \citealt{2008arXiv0803.0732H}).
Since the fluctuations at the time of decoupling are very small (1
part in $10^{5}$), the first order cosmology described in \S\ref{sub:First-order}
can provide extremely accurate predictions for the statistical properties
of these anisotropies, namely their power spectrum.

\begin{figure}
\begin{centering}\includegraphics[width=0.5\textwidth]{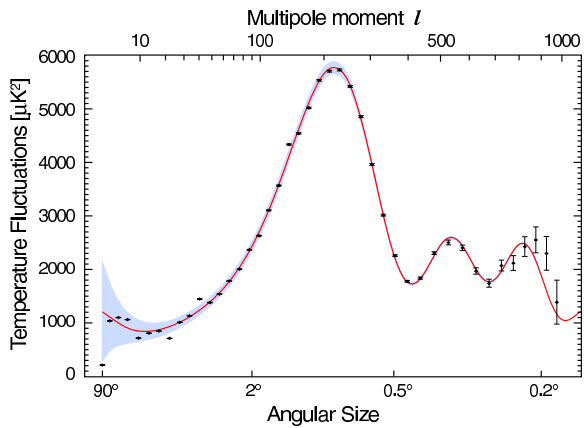}\par\end{centering}

\caption[CMB angular power spectrum from WMAP5.]{\label{fig:wmap_power}CMB angular power spectrum measured by the
NASA/WMAP Science Team from the 5 year data release. This plot shows
the average magnitude of the temperature deviations vs. their angular
size on the sky. Reproduced from \citet{2008arXiv0803.0732H}; used
with permission. }
\end{figure}

To predict the power spectrum, we need only two basic physical assumptions:
that we can apply general relativity to cosmological length scales
and that our understanding of plasma physics is accurate under the
conditions at the time of decoupling (a density of about 300 particles
per cubic centimeter and temperature of about 3000K, within
the range well-studied by solar physics). Figure~\ref{fig:wmap_power}
shows the prediction for the CMB power spectrum from the standard
cosmological model to the power spectrum measured by WMAP5. The theoretical
curve shown has only six free parameters defining its shape, and matches
the data astonishingly well. 

The six parameters describing the {}``vanilla'' standard cosmological
model are the baryon and cold dark matter fractions scaled by $h^{2}$,
i.e., $\Omega_{b}h^{2}$ and $\Omega_{c}h^{2}$, the dark energy fraction
$\Omega_{\Lambda}$, the power law parameters for the power spectrum
of initial fluctuations $A_{s}$ and $n_{s}$, and a nuisance parameter
corresponding to the optical depth $\tau$ looking back through our
universe to the point when the hydrogen atoms were reionized at a
redshift of $z\sim11$. In this simplest possible model that can match
the data, neutrinos are assumed to be massless, dark energy is assumed
to be a cosmological constant, our universe is assumed to be flat,
and a handful of other assumptions are made as well. The measured
shape of the CMB power spectrum provides tight constraints on the
values of these parameters and also can be used to place limits on
deviations from the vanilla model assumptions, e.g., massive neutrinos
or dark energy with $w\neq-1$. Furthermore, measuring the polarization
of the CMB provides another complementary source of information. (See
\citealt{2005ASPC..343..485D} for more details.)

\subsection{Other probes}

\label{sub:Other-probes}

There are several other important cosmological probes that deserve
mention here for completeness, although we will not focus on them
in depth. 

Perhaps the most important for the discovery of dark energy is the
use of type Ia supernovae to trace the expansion of the universe.
These stellar explosions are the result of a white dwarf star in a
binary system accreting mass from its companion. Eventually the white
dwarf will reach a critical mass, known as the Chandrasekhar limit,
where that electron degeneracy pressure can no longer support the
stellar material. This triggers an extremely luminous explosion that
can be seen at cosmological distances. Because the physics of these
systems dictates that the explosion occurs at a particular mass, the
resulting supernovae have sufficiently similar properties that they
can be used as standard (or standardizable) candles: since we know
their absolute brightness, measuring their apparent brightness tells
us the distance. Thus by also measuring their redshift we can do zeroth
order cosmology and effectively measure the scale factor $a\left(t\right)$
as a function of cosmic time. Cosmologists set out to do this expecting
to measure the deceleration of the expansion rate due to gravity,
but instead found that the expansion was accelerating \citep{1998AJ....116.1009R,1999ApJ...517..565P}.
This discovery provided the first evidence for dark energy and has
induced a paradigm shift in cosmological thinking over the past decade.

Another key piece of evidence for the standard cosmological model
comes from Big Bang nucleosynthesis studies. Our universe's hot dense
early phase reached temperatures high enough to induce nuclear fusion
of light elements, including deuterium, helium, and lithium. The fact
that our universe is roughly 25\% helium-4 by mass, too much to have
been fused within stars, is by itself evidence that our universe was,
in fact, once hot and dense. Furthermore, careful measurements of
the abundances of deuterium, lithium, and helium-3 yield an estimate
of the baryonic matter density that is both independent of the CMB
results and in good agreement with them \citep{1996Natur.381..207T,2007ARNPS..57..463S}. 

Another class of cosmological probes focuses on tracing the present
day (or low-redshift) matter distribution by means other than the
galaxy redshift surveys discussed in \S\ref{sub:Galaxy-surveys}.
These include galaxy cluster surveys, gravitational lensing, and Lyman
alpha absorption by cosmic gas clouds. 

Galaxy cluster surveys aim to detect galaxy clusters -- the largest
gravitationally bound and equilibrated clumps of matter in our universe
-- via the X-ray emission of their hot gas \citep{2005astro.ph..7013H}
their effect on the CMB \citep{2007NJPh....9..441H}, or other means,
and measure their abundance as a function of mass. This provides a
sensitive probe of nonlinear structure formation by is complicated
by the difficulty of linking the cluster mass to its observable properties. 

Gravitational lensing takes advantage of the general relativistic
effect that mass bends light and aims to map out the mass distribution
directly by studying the distortion of the light from distant objects.
This has the advantage of dodging the issue of galaxy bias that plagues
the interpretation of galaxy surveys, but is subject to other challenging
systematic effects \citep{2004PhRvD..70l3515B,2005MNRAS.361.1287M}. 

Measurements of Lyman alpha absorption study the light from distant
AGNs that has passed through clumps of hydrogen along the way. Hydrogen
at different redshifts will absorb light at the wavelength of the
Lyman alpha spectral line of atomic hydrogen, creating a series of
absorption lines known as the Lyman alpha forest which provides a
one-dimensional map of the gas density along the line of sight. This
probe of large-scale structure is sensitive to a somewhat higher redshift
than current galaxy surveys, but it is constrained by the geometry
of requiring an AGN backlight which makes it difficult to build up
a true three-dimensional picture \citep{2003AIPC..666..157W}.

Finally, we mention two additional probes that are likely to be important
in future cosmological pursuits: 21cm studies of the epoch of reionization
and measuring baryon oscillations in the galaxy distribution. The
epoch of reionization probes the time period partway between the time
when the CMB was emitted and the time at which we observe the most
distant astronomical objects. This era is known as the cosmological
Dark Ages due to the lack of information we have about it. However,
we do know that our universe went through a phase of reionization
when the first stars and galaxies began to shine, since most of the
hydrogen is ionized today. There are several experiments currently
in the construction or planning phases aiming to probe this era by
detecting the 21cm emission line of neutral hydrogen redshifted to
low radio frequencies \citep{2006SPIE.6267E..76T,2007ApJ...661....1B,2007HiA....14..386F}.
These experiments will open a new window on our universe for cosmological
studies.

We conclude our whirlwind tour of other cosmological probes by discussing
baryon oscillations. The physics behind this probe is as follows:
The overall matter distribution is determined primarily by dark matter,
but the sound wave patterns of the baryon-photon plasma observed in
the CMB leave an imprint of tiny wiggles in the matter power spectrum
as well. These wiggles, called baryon oscillations, encode a characteristic
length scale: the distance a sound wave traveled from the time of
the Big Bang until decoupling. This length scale can be used as a
standard ruler: since we know the physical size, measuring its apparent
angular size on the sky tells us its distance from us. By measuring
this length scale as a function of redshift using galaxy surveys,
we can map out the expansion history of the universe and use it as
a probe of zeroth order cosmology, which will hopefully provide insights
about the nature of dark energy. Current measurements of the baryon
oscillations at two redshifts \citep{2005ApJ...633..560E,2007MNRAS.381.1053P}
have already proved to be a quite valuable constraint on cosmological
parameters \citep{2008arXiv0803.0547K}. However, the effects of nonlinear
matter clustering and galaxy bias discussed in \S\ref{sub:Higher-order}
complicate efforts to measure this standard ruler to high precision
\citep{2007PhRvD..75f3512S,2008PhRvD..77d3525S}. Detailed studies
of galaxy bias such as the one presented in chapter~\ref{chap:bias}
can therefore help us more fully utilize this promising probe in the
future.

\section{Example: combining WMAP and SDSS}

\label{sec:cosmological-parameters} Combining CMB and galaxy survey
data is a particularly powerful strategy: the early-universe snapshot
of the CMB and the present-universe snapshot of the galaxy distribution
provide complementary information that allows us to precisely constrain
the parameters of the cosmological model describing the evolution
from one state to the other. In this section we discuss one example
of this: the combination of CMB data from the WMAP 3 year data release
(WMAP3; \citealt{2007ApJS..170..377S}) and the SDSS LRG sample from
Data Release 4 (DR4; \citealt{2006ApJS..162...38A}) analyzed in \citet{2006PhRvD..74l3507T}.
The goals of this analysis are twofold: first, measure the matter
power spectrum $P\left(k\right)$ on scales large enough to avoid
unfounded assumptions regarding nonlinear clustering and galaxy bias,
and second, combine this $P\left(k\right)$ measurement with the CMB
power spectrum measured by WMAP3 to constrain the parameters of the
standard cosmological model.

\subsection{Power spectrum estimation}

The LRG galaxy sample has a number of features that make it ideal
for measuring the galaxy power spectrum. First of all, its effective
volume $V_{\mathrm{eff}}\left(k\right)$ (as defined by \citealt{1997PhRvL..79.3806T})
is larger by a factor of six than the SDSS main galaxy sample and
over ten times larger than that of the 2dFGRS. The error bars on the
power spectrum scale roughly as $\Delta P\left(k\right)\propto V_{\mathrm{eff}}\left(k\right)^{-1/2}$,
so the SDSS LRGs yield the smallest error bars. Additionally, since
the LRGs are selected to be galaxies of a similar physical type, the
effects of luminosity- and color-dependent bias explored in chapter~\ref{chap:bias}
should not have a significant effect.

\begin{figure*}
\vskip-4.7cm\includegraphics[width=1\textwidth]{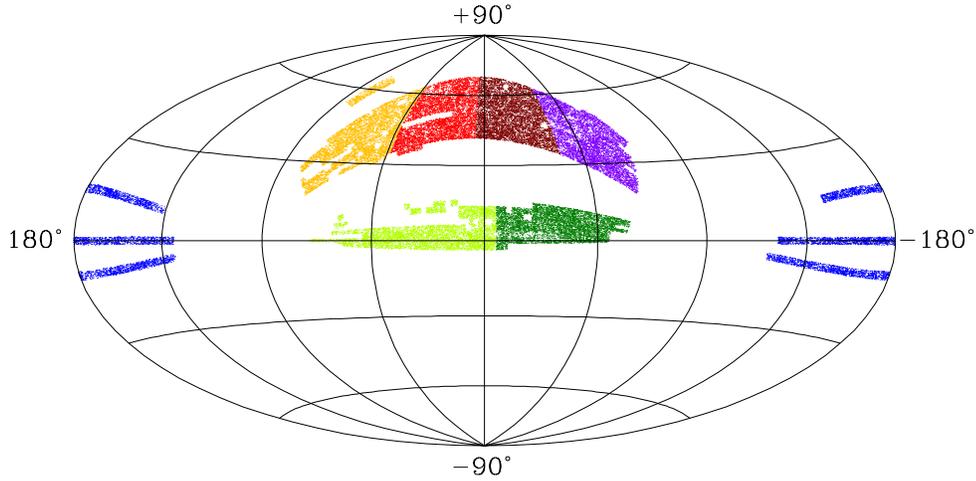}\vskip-5cm

\caption[The angular distribution of the SDSS DR4 luminous
red galaxies showing the seven angular subsamples analyzed.]
{\label{fig:skyfracs}The angular distribution of the SDSS DR4 luminous
red galaxies shown in Hammer-Aitoff projection in celestial coordinates,
with the seven colors/greys indicating the seven angular subsamples
analyzed in \citet{2006PhRvD..74l3507T} constructed with the \noun{mangle}
software discussed in chapter~\ref{chap:mangle}.}
\end{figure*}

The angular distribution of the LRGs from DR4 is shown in Fig.~\ref{fig:skyfracs}.
The regions of the sky containing the galaxies illustrate the area
that has been covered by the SDSS spectroscopic survey at the time
of DR4 -- the task of managing such angular coverage masks is discussed
in detail in chapter~\ref{chap:mangle}. Here we just note that the
updated \noun{mangle} software was used to create the masks for the
angular subsamples shown in different colors in Fig.~\ref{fig:skyfracs}.
The method used to calculate the power spectrum is the Pseudo Kahrunen-Lo\`{e}ve
(PKL) eigenmode method described in \citet{2004ApJ...606..702T},
which, among other advantages, produces uncorrelated error bars on
the power spectrum. In order to speed up the computation time, the
PKL method was applied separately on 21 subvolumes (the 7 angular
subsets shown in Fig.~\ref{fig:skyfracs} $\times$ 3 radial subsets)
and then combined with minimum variance weighting. Additionally, a
global PKL analysis of the full data set was done to specifically
target length scales larger than the individual subvolumes.

\begin{figure}
\begin{centering}\includegraphics[width=0.5\textwidth]{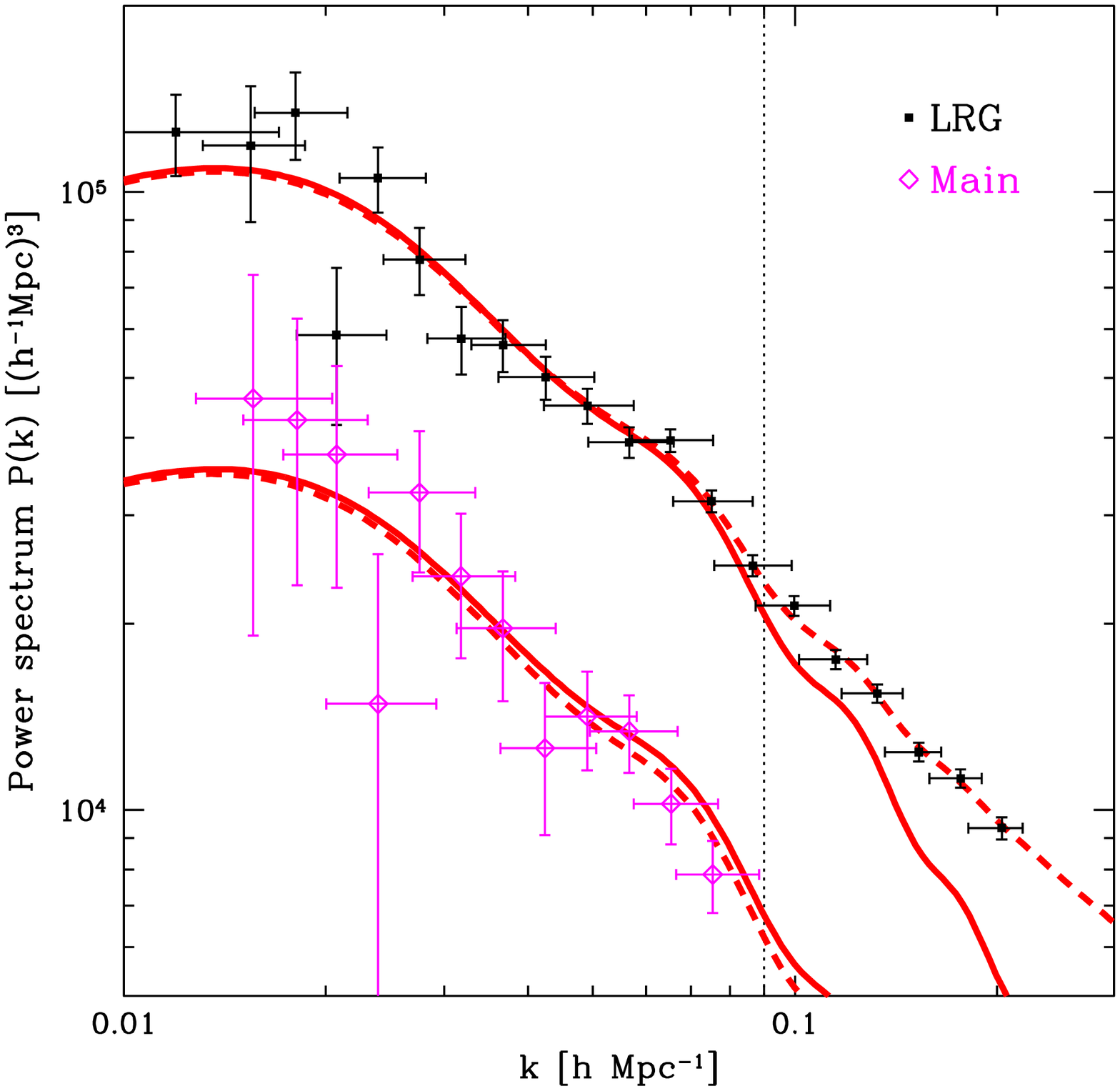}\par\end{centering}

\caption[Measured power spectra for the full LRG and main galaxy
samples.]{\label{fig:pk}Measured power spectra for the full LRG and main galaxy
samples, reproduced from \citet{2006PhRvD..74l3507T}. The solid curves
correspond to the linear theory cosmological model fits to WMAP3 alone
from Table 5 of \citet{2007ApJS..170..377S}, normalized to galaxy
bias $b=1.9$ (top) and $b=1.1$ (bottom) relative to the $z=0$ matter
power. The dashed curves include the nonlinear correction of \citet{2005MNRAS.362..505C}
for $A=1.4$, with $Q_{\mathrm{nl}}=30$ for the LRGs and $Q_{\mathrm{nl}}=4.6$
for the main galaxies; see equation \eqref{eq:coleQ}. The onset of
nonlinear corrections is clearly visible for $k\simgt0.09h$/Mpc (vertical
line).}
\end{figure}

The measured power spectra for both the LRGs and the main sample galaxies
are shown in Fig.~\ref{fig:pk}. Theoretical predictions are also
shown, both for linear theory (a.k.a. first order cosmology; see \S\ref{sub:First-order})
and nonlinear theory (a.k.a. higher order cosmology; see \S\ref{sub:Higher-order}).
The particular flavor of nonlinear modeling used here is based on
\citet{1999ApJ...511....5E,2005MNRAS.362..505C,2007ApJ...664..660E}.
The measured galaxy power spectrum is modeled as\begin{equation}
P_{g}\left(k\right)=P_{\mathrm{dewiggled}}\left(k\right)b^{2}\frac{1+Q_{\mathrm{nl}}k^{2}}{1+1.4k}.\label{eq:coleQ}\end{equation}
$P_{\mathrm{dewiggled}}\left(k\right)$ models the damping of the
baryon oscillations on small scales and is given by \citep{2007ApJ...664..660E}\begin{equation}
P_{\mathrm{dewiggled}}\left(k\right)=W\left(k\right)P\left(k\right)+\left[1-W\left(k\right)\right]P_{\mathrm{nowiggle}}\left(k\right),\label{eq:dewiggled}\end{equation}
where $P\left(k\right)$ is the linear theory power spectrum calculated using CMBFAST \citep{1996ApJ...469..437S}, 
$P_{\mathrm{nowiggle}}\left(k\right)$
is given by the baryon-free fitting formula in \citet{1999ApJ...511....5E},
and $W\left(k\right)\equiv e^{-\left(k/k_{*}\right)^{2}/2}$ is a
weighted averaging factor combining the two. This retains the wiggles
on large scales and fades them out beginning around $k=k_{*}$. The
wiggle suppression scale $k_{*}$ is defined by $1/\sigma$, where
$\sigma\equiv\sigma_{\perp}^{2/3}\sigma_{\parallel}^{1/3}\left(A_{s}/0.6841\right)^{1/2}$,
$\sigma_{\perp}$ and $\sigma_{\parallel}$ are given by equations
(12) and (13) in \citet{2007ApJ...664..660E}, and $A_{s}$ is the
power law normalization for the primordial spectrum. The parameters
$b$ and $Q_{\mathrm{nl}}$ in equation~\eqref{eq:coleQ} parameterize
bias and nonlinear evolution respectively: $b$ is the usual linear
bias factor defined in \S\ref{sub:First-order}, and increasing values
of $Q_{\mathrm{nl}}$ indicate stronger nonlinear evolution. $A_{s}$ is fixed by the
CMB observations -- $b^{2}$ characterizes the difference in normalization between what
one predicts for the matter power spectrum given the CMB data and what is observed for the 
galaxy power spectrum. 

The theoretical curves for linear theory shown as solid lines in Fig.~\ref{fig:pk}
are calculated using the best-fit cosmological parameters from the
WMAP3 CMB data alone \citep{2007ApJS..170..377S} plus a fit to the
data for the normalization $b^{2}$, giving $b=1.9$ for the LRGs
and $b=1.1$ for the main galaxies relative to the predicted matter
power spectrum. The nonlinear models shown as dashed lines again use
the WMAP3 cosmological parameters, plus a fit for $b$ and $Q_{\mathrm{nl}}$,
yielding $Q_{\mathrm{nl}}=4.6$ for the main galaxies and $Q_{\mathrm{nl}}=30$
for the LRGs. The models agree quite well with the data, which is
a good indicator that the nonlinear modeling prescription outlined
above captures the important effects, but also indicate that nonlinear
evolution is quite strong for the LRGs and becomes important at larger
scales. \citet{2006PhRvD..74l3507T} perform several tests to ensure
that their cosmological results are not sensitive to the details of
this largely empirically-based nonlinear modeling, but it is important
to note that if we want to improve the precision with which future
galaxy surveys can do cosmology, it will be essential to understand
bias and nonlinear evolution on a deeper, more physics-based level
-- such work is currently underway \citep{2007PhRvD..75f3512S,2007PASJ...59...93N,2007arXiv0707.3445T,2007ApJ...659....1Z,2008ApJ...674..617T,2008PhRvD..77b3533C,2008PhRvD..77d3525S}.

\subsection{Cosmological parameter estimation}

With the galaxy power spectrum in hand, the next goal is to combine
it with the WMAP3 CMB power spectrum to constrain cosmological parameters.
For simplicity, only the LRG power spectrum on scales larger than
$k\lesssim0.2h$/Mpc are used here. 

The cosmological model is parameterized in terms of 12 standard parameters,
plus the two nuisance parameters $b$ and $\Qnl$ from equation~\eqref{eq:coleQ}:
\begin{equation}
{\bf p}\equiv(\Ot,\Ol,\ob,\ocdm,\on,w,\As,r,\ns,\nt,\al,\tau,b,\Qnl).\label{eq:params}\end{equation}
 Table~\ref{tab:lrg_params} defines these 14 parameters and another
45 that can be derived from them; in essence, $(\Ot,\Ol,\ob,\ocdm,\on,w)$
define the cosmic matter budget, $(\As,\ns,\alpha,r,\nt)$ specify
the seed fluctuations and $(\tau,b,\Qnl)$ are nuisance parameters.
The {}``vanilla'' model, i.e., the minimal model parametrized by
$(\Ol,\ob,\ocdm,\As,\ns,\tau,b,\Qnl)$, setting $\on=\al=r=\nt=0$,
$\Ot=1$ and $w=-1$, is the smallest subset of our parameters that
provides a good fit to the WMAP plus LRG data -- this is the same
vanilla model discussed in \S\ref{sec:CMB} with the addition of
$b$ and $\Qnl$. Constraints on these parameters from the WMAP plus
LRG data were computed using the Monte Carlo Markov Chain (MCMC) approach
\citep{1953JChPh..21.1087M,Hastings,Gilks96,2001CQGra..18.2677C,2002PhRvD..66j3511L,2003astro.ph..7219S,2003ApJS..148..195V}
as implemented in \citet{2004PhRvD..69j3501T}. The best-fit parameter
values and their uncertainties are shown in Table~\ref{tab:lrg_params}
-- more details about the assumptions that have gone into these parameter
estimations as well as tests of how robust these constraints are to
these assumptions can be found in \citet{2006PhRvD..74l3507T}.

\begin{table*}
\hspace{-0.29in}
\begin{minipage}[b][5.1in]{6.5in}
\caption[Cosmological parameters measured from WMAP and SDSS LRG data.]{\label{tab:lrg_params}Cosmological parameters measured from WMAP and SDSS LRG data, reproduced from \citet{2006PhRvD..74l3507T}. References cited in the table are 
C05 \citep{2005MNRAS.362..505C}, 
KT \citep{1990eaun.book.....K}, 
H05 \citep{2005ASPC..339..215H}, 
T94 \citep{1994ApJ...420..484T},
T98 \citep{1998ApJ...499..526T},
T05 \citep{2005PhRvD..71j3523T}, and
H01 \citep{2001ApJ...549..669H}.
}
{\fontsize{5}{6}
\selectfont
\begin{center}
\begin{tabular*}{6.5in}{@{\extracolsep{\fill}}|l|l|l|l|}
\hline
Parameter &Value &Meaning &Definition\\
\hline
\multicolumn{3}{|l|}{Matter budget parameters:}&\\
\cline{1-4}
$        \Omega_{\rm{tot}}$      &$ 1.003^{+ 0.010}_{- 0.009}$                                          &Total density/critical density 	&$\Ot=\Om+\Ol=1-\Ok$\\
$           \Omega_\Lambda$      &$ 0.761^{+ 0.017}_{- 0.018}$                                          &Dark energy density parameter  	&$\Ol\approx h^{-2}\rho_\Lambda(1.88\times 10^{-26}$kg$/$m$^3)$\\
$                 \omega_b$      &$ 0.0222^{+ 0.0007}_{- 0.0007}$                                       &Baryon density		 	&$\ob=\Ob h^2 \approx \rho_b/(1.88\times 10^{-26}$kg$/$m$^3)$\\
$                 \omega_c$      &$ 0.1050^{+ 0.0041}_{- 0.0040}$                                       &Cold dark matter density 	 	&$\ocdm=\Oc h^2 \approx \rho_c/(1.88\times 10^{-26}$kg$/$m$^3)$\\
$               \omega_\nu$      &$< 0.010 \>(95\percent)$                                              &Massive neutrino density		&$\on=\On h^2 \approx \rho_\nu/(1.88\times 10^{-26}$kg$/$m$^3)$\\
$                        w$      &$-0.941^{+ 0.087}_{- 0.101}$                                          &Dark energy equation of state		&$p_\Lambda/\rho_\Lambda$ (approximated as constant)\\
\hline
\multicolumn{3}{|l|}{Seed fluctuation parameters:}&\\
\cline{1-4}
$                      A_s$      &$ 0.690^{+ 0.045}_{- 0.044}$                                          &Scalar fluctuation amplitude		&Primordial scalar power at $k=0.05$/Mpc\\
$                        r$      &$< 0.30 \>(95\percent)$                                               &Tensor-to-scalar ratio		&Tensor-to-scalar power ratio at $k=0.05$/Mpc\\
$                      n_s$      &$ 0.953^{+ 0.016}_{- 0.016}$                                          &Scalar spectral index			&Primordial spectral index at $k=0.05$/Mpc\\
$                    n_t+1$      &$ 0.9861^{+ 0.0096}_{- 0.0142}$                                       &Tensor spectral index			&$\nt=-r/8$ assumed\\
$                   \alpha$      &$-0.040^{+ 0.027}_{- 0.027}$                                          &Running of spectral index		&$\alpha=d\ns/d\ln k$ (approximated as constant)\\
\hline
\multicolumn{3}{|l|}{Nuisance parameters:}&\\
\cline{1-4}
$                     \tau$      &$ 0.087^{+ 0.028}_{- 0.030}$                                          &Reionization optical depth		&\\
$                        b$      &$ 1.896^{+ 0.074}_{- 0.069}$                                          &Galaxy bias factor			&$b=[P_{\rm gal}(k)/P(k)]^{1/2}$ on large scales, where $P(k)$ refers to today.\\
$              Q_{\rm{nl}}$      &$30.3^{+ 4.4}_{- 4.1}$                                                &Nonlinear correction parameter [C05] &$\Pg(k)=P_{\rm dewiggled}(k)b^2(1+Q_{\rm nl}k^2)/(1+1.7k)$\\
\hline\hline
\multicolumn{3}{|l|}{Other popular parameters (determined by those above):}&\\
\cline{1-4}
$                        h$      &$ 0.730^{+ 0.019}_{- 0.019}$                                          &Hubble parameter			&$h = \sqrt{(\ob+\ocdm+\on)/(\Ot-\Ol)}$\\
$                 \Omega_m$      &$ 0.239^{+ 0.018}_{- 0.017}$                                          &Matter density/critical density	&$\Om=\Ot-\Ol$\\
$                 \Omega_b$      &$ 0.0416^{+ 0.0019}_{- 0.0018}$                                       &Baryon density/critical density 	&$\Ob=\ob/h^2$\\
$                 \Omega_c$      &$ 0.197^{+ 0.016}_{- 0.015}$                                          &CDM density/critical density 		&$\Oc=\ocdm/h^2$\\
$               \Omega_\nu$      &$< 0.024 \>(95\percent)$                                              &Neutrino density/critical density 	&$\On=\on/h^2$\\
$                 \Omega_k$      &$-0.0030^{+ 0.0095}_{- 0.0102}$                                       &Spatial curvature			&$\Ok=1-\Ot$\\
$                 \omega_m$      &$ 0.1272^{+ 0.0044}_{- 0.0043}$                                       &Matter density			&$\om=\ob+\ocdm+\on = \Om h^2$\\
$                    f_\nu$      &$< 0.090 \>(95\percent)$                                              &Dark matter neutrino fraction		&$\fn=\rho_\nu/\rho_d$\\
$                      A_t$      &$< 0.21 \>(95\percent)$                                               &Tensor fluctuation amplitude		&$\At=r\As$\\
$                    M_\nu$      &$< 0.94 \>(95\percent)$ eV                                            &Sum of neutrino masses 		&$\Mnu\approx(94.4\>{\rm eV})\times\on$~~~[KT]\\
$                 A_{.002}$      &$ 0.801^{+ 0.042}_{- 0.043}$                                          &WMAP3 normalization parameter		&$\As$ scaled to $k=0.002$/Mpc: $A_{.002} = 25^{1-\ns}\As$ if $\al=0$\\
$                 r_{.002}$      &$< 0.33 \>(95\percent)$                                               &Tensor-to-scalar ratio (WMAP3)	&Tensor-to-scalar power ratio at $k=0.002$/Mpc\\
$                 \sigma_8$      &$ 0.756^{+ 0.035}_{- 0.035}$                                          &Density fluctuation amplitude		&$\sigma_8=\{4\pi\int_0^\infty [{3\over x^3}(\sin x-x\cos x)]^2 P(k) {k^2 dk\over(2\pi)^3}\}^{1/2}$,\\
                                 &                                                                      &            		                &$x\equiv k\times 8h^{-1}$Mpc\\
$   \sigma_8\Omega_m^{0.6}$      &$ 0.320^{+ 0.024}_{- 0.023}$                                          &Velocity fluctuation amplitude&\\
\hline
\multicolumn{3}{|l|}{Cosmic history parameters:}&\\
\cline{1-4}
$              z_{\rm{eq}}$      &$3057^{+105}_{-102}$                                                  &Matter-radiation Equality redshift	&$z_{\rm eq}\approx 24074\om - 1$\\
$             z_{\rm{rec}}$      &$1090.25^{+ 0.93}_{- 0.91}$                                           &Recombination redshift		&$z_{\rm rec}(\om,\ob)$ given by eq.~(18) of [H05]\\
$             z_{\rm{ion}}$      &$11.1^{+ 2.2}_{- 2.7}$                                                &Reionization redshift (abrupt)	&$\zion\approx 92 \left(\frac{0.03h\tau}{\ob}\right)^{2/3}\Om^{1/3}$ (assuming abrupt reionization [T94])\\
$             z_{\rm{acc}}$      &$ 0.855^{+ 0.059}_{- 0.059}$                                          &Acceleration redshift			&$\zacc=[(-3w-1)\Ol/\Om]^{-1/3w}-1$ if $w<-1/3$\\
$              t_{\rm{eq}}$      &$ 0.0634^{+ 0.0045}_{- 0.0041}$ Myr                                   &Matter-radiation Equality time 	&$\teq \approx$($9.778$ Gyr)$\times h^{-1}\int_{\zeq}^\infty [H_0/H(z)(1+z)]dz$~~~[KT]\\
$             t_{\rm{rec}}$      &$ 0.3856^{+ 0.0040}_{- 0.0040}$ Myr                                   &Recombination time 			&$\trec\approx$($9.778$ Gyr)$\times h^{-1}\int_{\zrec}^\infty [H_0/H(z)(1+z)]dz$~~~[KT]\\
$             t_{\rm{ion}}$      &$ 0.43^{+ 0.20}_{- 0.10}$ Gyr                                         &Reionization time			&$\tion\approx$($9.778$ Gyr)$\times h^{-1}\int_{\zion}^\infty [H_0/H(z)(1+z)]dz$~~~[KT]\\
$             t_{\rm{acc}}$      &$ 6.74^{+ 0.25}_{- 0.24}$ Gyr                                         &Acceleration time 			&$\tacc\approx$($9.778$ Gyr)$\times h^{-1}\int_{\zacc}^\infty [H_0/H(z)(1+z)]dz$~~~[KT]\\
$             t_{\rm{now}}$      &$13.76^{+ 0.15}_{- 0.15}$ Gyr                                         &Age of Universe now			&$\tnow\approx$($9.778$ Gyr)$\times h^{-1}\int_0^\infty [H_0/H(z)(1+z)]dz$~~~[KT]\\
\hline
\multicolumn{3}{|l|}{Fundamental parameters (independent of observing epoch):}&\\
\cline{1-4}
$                        Q$      &$ 1.945^{+ 0.051}_{- 0.053}$$\times10^{-5}$                          &Primordial fluctuation amplitude	&$Q=\delta_h\approx A_{.002}^{1/2}\times 59.2384\mu$K$/T_{\rm CMB}$\\
$                   \kappa$      &$ 1.3^{+ 3.7}_{- 4.3}$$\times10^{-61}$                               &Dimensionless spatial curvature [T98]&$\kappa=(\hbar c/k_B T_{\rm CMB} a)^2 k$\\ 
$             \rho_\Lambda$      &$ 1.48^{+ 0.11}_{- 0.11}$$\times10^{-123}\rho_{\rm{Pl}}$             &Dark energy density			&$\rho_\Lambda\approx h^2\Ol\times(1.88\times 10^{-26}$kg$/$m$^3)$\\
$         \rho_{\rm{halo}}$      &$ 6.6^{+ 1.2}_{- 1.0}$$\times10^{-123}\rho_{\rm{Pl}}$                &Halo formation density		&$\rhohalo=18\pi^2 Q^3\xi^4$\\
$                      \xi$      &$ 3.26^{+ 0.11}_{- 0.11}$ eV                                          &Matter mass per photon 		&$\xi =\rhom/\ng$\\
$                    \xi_b$      &$ 0.569^{+ 0.018}_{- 0.018}$ eV                                       &Baryon mass per photon 		&$\xib=\rhob/\ng$\\
$                    \xi_c$      &$ 2.69^{+ 0.11}_{- 0.10}$ eV                                          &CDM mass per photon 			&$\xic=\rhoc/\ng$\\
$                  \xi_\nu$      &$< 0.26 \>(95\percent)$ eV                                            &Neutrino mass per photon 		&$\xin=\rhon/\ng$\\
$                     \eta$      &$ 6.06^{+ 0.20}_{- 0.19}$$\times10^{-10}$                            &Baryon/photon ratio			&$\eta=n_b/n_g=\xib/m_p$\\
$                A_\Lambda$      &$2077^{+135}_{-125}$                                                  &Expansion during matter domination	&$(1+\zeq)(\Om/\Ol)^{1/3}$ [T05]\\  
$      \sigma^*_{\rm{gal}}$      &$ 0.561^{+ 0.024}_{- 0.023}$$\times10^{-3}$                          &Seed amplitude on galaxy scale	&Like $\sigma_8$ but on galactic ($M=10^{12} M_\odot$) scale early on\\
\hline
\multicolumn{3}{|l|}{CMB phenomenology parameters:}&\\
\cline{1-4}
$            A_{\rm{peak}}$      &$ 0.579^{+ 0.013}_{- 0.013}$                                          &Amplitude on CMB peak scales		&$\Ap=\As e^{-2\tau}$\\
$           A_{\rm{pivot}}$      &$ 0.595^{+ 0.012}_{- 0.011}$                                          &Amplitude at pivot point		&$A_{\rm{peak}}$ scaled to $k=\frac{0.028}{\mathrm{Mpc}}$: $A_{\rm pivot}= 0.56^{\ns-1}A_{\rm{peak}}$ if $\al=0$\\
$                      H_1$      &$ 4.88^{+ 0.37}_{- 0.34}$                                             &1st CMB peak ratio			&$H_1(\Ot,\Ol,\ob,\om,w,\ns,\tau)$ given by [H01]\\
$                      H_2$      &$ 0.4543^{+ 0.0051}_{- 0.0051}$                                       &2nd to 1st CMB peak ratio		&$H_2 = \frac{(0.925\om^{0.18} 2.4^{\ns-1})}{[1+(\ob/0.0164)^{12\om^{0.52}})]^{0.2}}$~~~[H01]\\
$                      H_3$      &$ 0.4226^{+ 0.0088}_{- 0.0086}$                                       &3rd to 1st CMB peak ratio		&$H_3 = \frac{2.17 [1+(\ob/0.044)^2]^{-1} \om^{0.59} 3.6^{\ns-1}}{1 + 1.63(1-\ob/0.071)\om}$\\
$        d_A(z_{\rm{rec}})$      &$14.30^{+ 0.17}_{- 0.17}$ Gpc                                         &Comoving ang. diam. dist. to CMB &$d_A(\zrec)={c\over H_0}\sinh\left[\Ok^{1/2}\int_0^\zrec [H_0/H(z)]dz\right]/\Ok^{1/2}$~~~[KT]\\
$        r_s(z_{\rm{rec}})$      &$ 0.1486^{+ 0.0014}_{- 0.0014}$ Gpc                                   &Comoving sound horizon scale		&$\rsound(\om,\ob)$ given by eq.~(22) of [H05]\\
$            r_{\rm{damp}}$      &$ 0.0672^{+ 0.0009}_{- 0.0008}$ Gpc                                   &Comoving acoustic damping scale	&$\rdamp(\om,\ob)$ given by eq.~(26) of [H05]\\
$                 \Theta_s$      &$ 0.5918^{+ 0.0020}_{- 0.0020}$                                       &CMB acoustic angular scale fit ($^{\circ}$)&$\Th(\Ot,\Ol,w,\ob,\om)$ given by \protect[H01]\\
$                   \ell_A$      &$302.2^{+ 1.0}_{- 1.0}$                                               &CMB acoustic angular scale		&$\l_A=\pi d_A(\zrec)/\rsound(\zrec)$\\ 
\hline
\end{tabular*}
\end{center}     
}
\end{minipage}

\end{table*}

The parameter constraints provided by any given experiment define
an allowed region within the $N$-dimensional parameter space (where
$N$ is the number of parameters) that can be quite complicated and
may have degeneracies between some parameters -- for example, CMB
data constrains $\omega_{m}\equiv\Omega_{m}h^{2}$ quite tightly but
gives only weak constraints on $\Omega_{m}$ and $h$ individually.
A common way to illustrate this is to choose two parameters and plot
the constraints in a two-dimensional plane. Such plots illustrate
how combining different data sets breaks various parameter degeneracies,
as discussed in depth in \citet{2006PhRvD..74l3507T}. 

\begin{figure}
\begin{centering}\includegraphics[width=0.5\textwidth]{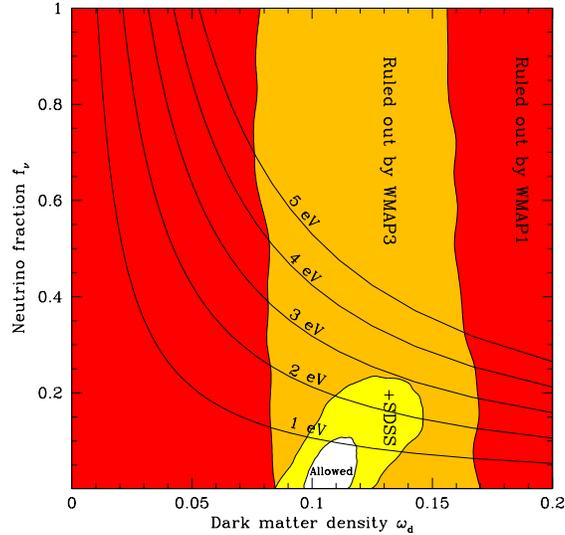}\par\end{centering}

\caption[95\% constraints in the $(\od,\fn)$ plane.]{\label{fig:odfn_contours} 95\% constraints in the $(\od,\fn)$ plane,
reproduced from \citet{2006PhRvD..74l3507T}. The large shaded regions
are ruled out by WMAP1 (red/dark grey) and WMAP3 (orange/grey) when
neutrino mass is added to the 6 vanilla parameters. The yellow/light
grey region is ruled out when adding SDSS LRG information. The five
curves correspond to $M_{\nu}$, the sum of the neutrino masses, equaling
1, 2, 3, 4 and 5 eV respectively -- barring sterile neutrinos, no
neutrino can have a mass exceeding $\sim M_{\nu}/3\approx0.3$ eV
(95\%).}
\end{figure}

As an example here, we show just one of these plots in Fig.~\ref{fig:odfn_contours}
that is particularly relevant for particle physics: the constraints
on the dark matter fraction $\omega_{d}$ and the fraction $f_{\nu}$
of the dark matter that is composed of neutrinos. This plot shows
that the first year WMAP data (WMAP1; \citealt{2003ApJS..148..175S})
only constrains $\omega_{d}$ without saying anything about neutrinos,
and the WMAP3 data indicates that at most about 20\% of the dark matter
mass could be neutrinos. Adding the SDSS LRG data tightens the constraints
still further, allowing only about 10\% of the dark matter to be neutrinos.
Assuming that there are no sterile neutrinos, this implies a constraint
that the mass of any neutrino must be less than $\sim0.3\mathrm{\, eV}$.
This is comparable to the sensitivity of the KATRIN particle physics
experiment currently being conducted to measure the mass of the electron
neutrino via tritium beta decay \citep{2003NuPhA.719..153L}. Thus
by looking at distant galaxies and the radiation that fills our universe,
we can do some extremely interesting and relevant particle physics.

\section{Issues with using galaxies as cosmic tracers}

Measuring the distribution of galaxy clustering has proved to be an
extremely powerful tool for measuring cosmological parameters. However,
there are still issues, both practical and theoretical, that need
to be addressed in order to optimally extract cosmological information
from the next generation of galaxy surveys. The following two chapters
focus on two of these issues: chapter~\ref{chap:mangle} addresses
the technical matter of managing angular masks defining the survey
area on the sky, and chapter~\ref{chap:bias} addresses the more
fundamental question of understanding how closely galaxies trace the
underlying matter distribution.

\subsection{Managing angular masks of galaxy surveys}

As galaxy surveys become larger and more complex, keeping track of
the completeness, magnitude limit, and other survey parameters as
a function of direction on the sky becomes an increasingly challenging
computational task. For example, typical angular masks of SDSS contain
about $N=300{,}000$ distinct spherical polygons. Managing masks with
such large numbers of polygons becomes intractably slow, particularly
for tasks that run in time $\mathcal{O}\left(N^{2}\right)$ with a
naive algorithm, such as finding which polygons overlap each other.
In chapter~\ref{chap:mangle}, we present a {}``divide-and-conquer''
solution to this challenge: we first split the angular mask into predefined
regions called {}``pixels,'' such that each polygon is in only one
pixel, and then perform further computations, such as checking for
overlap, on the polygons within each pixel separately. This reduces
$\mathcal{O}\left(N^{2}\right)$ tasks to $\mathcal{O}\left(N\right)$,
and also reduces the important task of determining in which polygon(s)
a point on the sky lies from $\mathcal{O}\left(N\right)$ to $\mathcal{O}\left(1\right)$,
resulting in significant computational speedup. 

Additionally, we present a method to efficiently convert any angular
mask to and from the popular \emph{HEALPix} format. This method can
be generically applied to convert to and from any desired spherical
pixelization. We have implemented these techniques in a new version
of the \noun{mangle} software package, which is freely available at
\texttt{\url{http://space.mit.edu/home/tegmark/mangle/}}, along with
complete documentation and example applications. These new methods
should prove quite useful to the astronomical community, and since
\noun{mangle} is a generic tool for managing angular masks on a sphere,
it has the potential to benefit terrestrial mapmaking applications
as well.

\subsection{Relative galaxy bias}

Differences in clustering properties between galaxy subpopulations
complicate the cosmological interpretation of the galaxy power spectrum,
but can also provide insights about the physics underlying galaxy
formation. Chapter~\ref{chap:bias} details a study of the nature
of this relative clustering in which we perform a counts-in-cells
analysis of galaxies in SDSS to measure the relative bias between
pairs of galaxy subsamples of different luminosities and colors. We
use a generalized $\chi^{2}$ test to determine if the relative bias
between each pair of subsamples is consistent with the simplest deterministic
linear bias model, and we also use a maximum likelihood technique
to further understand the nature of the relative bias between each
pair. 

We find that the simple, deterministic model is a good fit for the
luminosity-dependent bias on scales above $\sim2\, h^{-1}\mathrm{Mpc}$,
which is good news for using magnitude-limited surveys for cosmology.
However, the color-dependent bias shows evidence for stochasticity
and/or non-linearity which increases in strength toward smaller scales,
in agreement with previous studies of stochastic bias. Also, confirming
hints seen in earlier work, the luminosity-dependent bias for red
galaxies is significantly different from that of blue galaxies: both
luminous and dim red galaxies have higher bias than moderately bright
red galaxies, whereas the biasing of blue galaxies is not strongly
luminosity-dependent. These results can be used to constrain galaxy
formation models and also to quantify how the color and luminosity
selection of a galaxy survey can impact measurements of the cosmological
matter power spectrum.

\chapter{Methods for rapidly processing angular masks of next-generation galaxy surveys}
\label{chap:mangle}

\begin{quote}
This chapter is adapted from the paper ``Methods for rapidly processing angular masks of next-generation galaxy surveys'' by Molly E.~C. Swanson, Max Tegmark, Andrew J.~S. Hamilton, and J. Colin Hill, which has been accepted for publication in the \emph{Monthly Notices of the Royal Astronomical Society} \citep{2007arXiv0711.4352S}.
\end{quote}

\section{Introduction}
Over the past few decades, galaxy surveys have provided a wealth of
information about the large-scale structure of our Universe, and the
next generation of surveys currently being planned promises to provide
even more insight. In order to realize the full potential of upcoming
surveys, it is essential to avoid unnecessary errors and approximations
in the way they are analyzed. The tremendous volumes of data produced
by these new surveys will shrink statistical uncertainty to unprecedented
levels, and in order to take advantage of this we must ensure that
the systematic uncertainties can keep pace. The purpose of this chapter
is to maximize the scientific utility of next-generation surveys by
providing methods for processing angular masks as rapidly and accurately
as possible.

Angular masks of a galaxy survey are functions of direction on the
sky that model the survey completeness, magnitude limit, seeing, dust
extinction, or other parameters that vary across the sky. The earliest
galaxy redshift surveys -- the first Center for Astrophysics redshift
survey (CfA1; \citealt{1983ApJS...52...89H}) and the first Southern
Sky Redshift Survey (SSRS1; \citealt{1991ApJS...75..935D}) -- had
simple angular masks defined by boundaries in declination and Galactic
latitude. The next generation of surveys -- IRAS \citep{1992ApJS...83...29S}
and PCSz \citep{2000MNRAS.317...55S} -- had somewhat more complex
masks, with some regions of high contamination excluded from the survey.

The present generation of surveys -- the Two Degree Field Galaxy Redshift
Survey (2dFGRS; \citealt{2001MNRAS.328.1039C,2003astro.ph..6581C})
and the Sloan Digital Sky Survey (SDSS; \citealt{2000AJ....120.1579Y})
-- consist of photometric surveys that identify galaxies and measure
their angular positions combined with spectroscopic surveys that measure
a redshift for each galaxy to determine its distance from us. Angular
masks are useful for describing parameters for both photometric and
spectroscopic surveys -- for example, seeing and magnitude limit are
key parameters to model in photometric surveys, and the survey completeness
-- i.e., the fraction of photometrically selected target galaxies
for which a spectrum has been measured -- is vital for analyzing spectroscopic
surveys.

The angular masks of SDSS and 2dFGRS consist of circular fields defined
by the spectroscopic plates of the redshift survey superimposed on
an angular mask of the parent photometric survey. 2dFGRS uses the
Automatic Plate Measurement (APM) survey (Maddox et al.~1990a,b, 1996)
\nocite{1990MNRAS.243..692M,1990MNRAS.246..433M,1996MNRAS.283.1227M}
as its parent photometric survey and covers approximately 1500 square
degrees. The APM angular mask consists of 269 $5^{\circ}\times5^{\circ}$
photographic plates which are drilled with numerous holes to avoid
bright stars, satellite trails, plate defects, and so forth. Combining
the photographic plates, holes, and spectroscopic fields gives a total
of 3525 polygons that define the spectroscopic angular mask of 2dFGRS.

The SDSS covers a larger area on the sky -- 5740 square degrees in
Data Release 5 (DR\%; \citealt{2007ApJS..172..634A}) -- and has a
yet more complicated angular mask than 2dFGRS. The SDSS photometric
survey is done by drift-scanning: each scan across the sky covers
six long, narrow scanlines, and the gaps between these lines are filled
in with a second scan slightly offset from the first, producing a
{}``stripe'' about $2.5^{\circ}$ wide assembled from 12 scanlines.
In addition to the fairly intricate pattern produced by this scanning
strategy, there are nearly 250,000 holes masked out of the photometric
survey for various reasons, plus the circular $3^{\circ}$ spectroscopic
fields. Combining all of these elements produces an angular mask for
the spectroscopic survey that contains 340,351 polygons.

To accurately manage the 2dFGRS and SDSS angular masks, \citet{2004MNRAS.349..115H}
developed a suite of general-purpose software called \noun{mangle},
which performs several important procedures on angular masks using
computational methods detailed in \citet{1993ApJ...406L..47H,1993ApJ...417...19H}.
This software has proved to be a valuable resource to the astronomical
community: it has been used in several analyses of galaxy survey data
\citep{2002MNRAS.335..887T,2004ApJ...606..702T,2005MNRAS.361.1287M,2005PASJ...57..709H,2005ApJ...633...11P,2005ApJ...635..990C,2007ApJ...658..898P,2007PASJ...59...93N,2007AJ....133.2222S,2007ApJ...664..608W,2007arXiv0707.3445T}.
Additionally, it was used extensively in the preparation of the New
York University Value-Added Galaxy Catalog (NYU-VAGC; Blanton et al.~2005b), \nocite{2005AJ....129.2562B}
which has been used as the basis for almost all publications on large-scale
structure by the SDSS collaboration.

However, many of functions in the original version of \noun{mangle}
run in $\mathcal{O}\left(N^{2}\right)$ time, which becomes quite
computationally challenging as the size and complexity of surveys
continues to increase -- computations involving the SDSS mask can
take several months of CPU time. In this chapter we present new algorithms
that can process complicated angular masks such as SDSS dramatically
faster with no loss of accuracy. Our method is based on splitting
an angular mask into pixels, reducing the processing time to $\mathcal{O}\left(N\right)$
by adding an $\mathcal{O}\left(N\log N\right)$ preprocessing step. 
Similar methods based on hierarchical spatial subdivisions have been
found to be useful in the field of computational geometry
(see, e.g., \citealt{bigbluebook}),
but have not previously been applied to angular masks in an astronomy context.

This will be especially useful for next-generation surveys, such as
the Dark Energy Survey (DES; \citealt{2005astro.ph.10346T}) and surveys
done with the Wide-Field Multi-Object Spectrograph (WFMOS; \citealt{2006PhRvD..74f3525Y,wfmos}),
the Panoramic Survey Telescope and Rapid Response System (Pan-STARRS;
\citealt{2004SPIE.5489...11K}), and the Large Synoptic Survey Telescope
(LSST; \citealt{2002SPIE.4836...10T,2004AAS...20510802S,2006AIPC..870...44T,LSST}).
DES, Pan-STARRS, and LSST will perform photometric surveys and use
techniques for estimating redshifts based on the photometric information;
WFMOS will perform spectroscopic surveys using one of the upcoming
photometric surveys for target selection. The methods we present here
are useful for both photometric and spectroscopic surveys -- keeping
track of factors such as seeing and dust extinction could prove particularly
important for photometric redshift determinations \citep{2007MNRAS.375...68C,2008ApJ...674..768O,2007arXiv0711.1059B}.

\begin{figure*}
\includegraphics[width=1\textwidth]{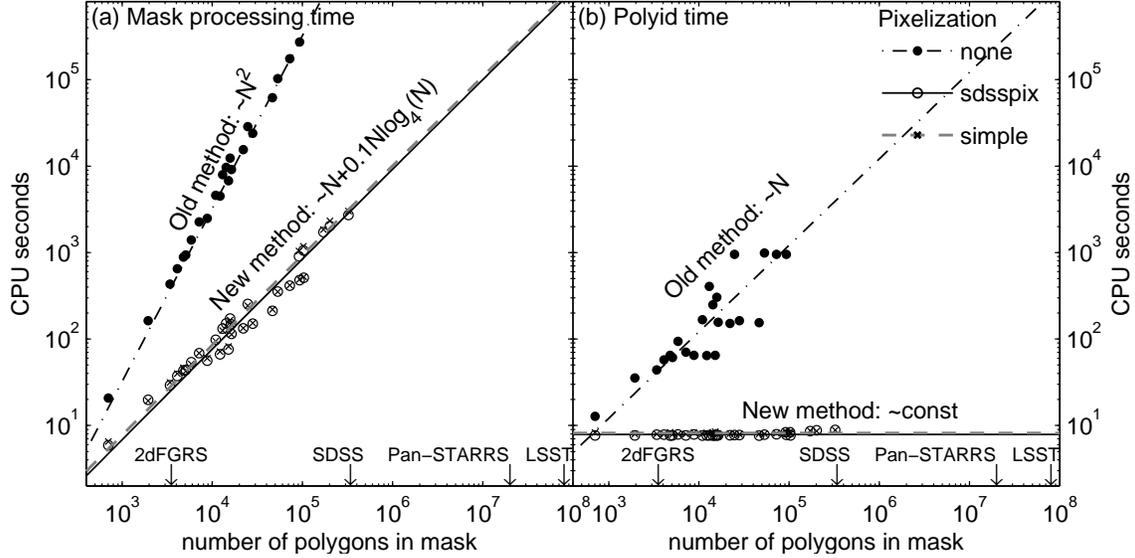}

\caption[Speed trials for a series of portions
of the SDSS DR5 mask with and without pixelization.]{\label{fig:speed_fractions}Speed trials for a series of portions
of the SDSS DR5 mask with and without pixelization. (a) Time required
for pixelization, snapping, balkanization and unification of the mask.
(b) Time required to identify in which polygon each of the ${\sim}400{,}000$
SDSS DR5 galaxies lies. Each set of trials is fitted with a power
law to show how the processing time scales with the number of polygons
$N$. Also shown on the x-axis are the number of polygons in the 2dFGRS
mask, the SDSS DR5 mask, and conservative estimates for the Pan-STARRS
and LSST large-scale structure masks based on scaling up 2dFGRS.}
\end{figure*}

The proposed large-scale structure survey to be produced by Pan-STARRS
will cover ${\sim}30{,}000$ square degrees in 5 wavelength bands,
with each field being observed ${\sim}50$ times such that the images
can be co-added. A naive scaling up of the 2dFGRS area and number
of polygons gives an estimate of ${\sim}2\times10^7$ polygons for
the final Pan-STARRS mask. Similarly, the LSST large-scale structure
survey will cover ${\sim}20{,}000$ square degrees in 6 bands, with
${\sim}200$ co-added images, suggesting ${\sim}8\times10^7$ polygons
for the LSST mask. The need for an improvement in mask processing
speed is clearly illustrated in Fig.~\ref{fig:speed_fractions}:
with the old algorithms, the projected processing time for the LSST
mask would be over 6000 years. With our new method, this time is reduced
by a factor of ${\sim}24{,}000$ to just ten days.

In addition to developing faster algorithms for processing angular
masks, we have also integrated the \noun{mangle} utilities with \emph{HEALPix},
a widely used tool for discretizing the celestial sphere \citep{2005ApJ...622..759G}.
The methods used by \noun{mangle} are complementary to \emph{HEALPix}:
\noun{mangle} is best used for functions that are piecewise-constant
in distinct regions of the sky, such as the completeness of a galaxy
survey. In contrast, \emph{HEALPix} is optimal for describing functions
that are continuously varying across the sky, such as the cosmic microwave
background (CMB) or the amount of extinction due to galactic dust.
The ability to convert rapidly between these two formats allows for
easy comparison of these two types of data without the unnecessary
approximation inherent in discretizing an angular mask. Furthermore,
converting a mask into \emph{HEALPix} format allows users to take
advantage of pre-existing \emph{HEALPix} tools for rapidly computing
spherical harmonics. 

The spectacular surveys on the horizon are preparing to generate massive,
powerful datasets that will be made publicly available -- this in
turn necessitates powerful and intuitive general-purpose tools that
assist the community to do science with this avalanche of data. We
provide a such tools with this new generation of the \noun{mangle}
software and describe these new tools here. However, this chapter is
not a software manual (a manual is provided on the \noun{mangle} website)
but rather a description of the underlying algorithms. These tools
have been utilized in recent analyses of SDSS data (\citealt{2006PhRvD..74l3507T},
discussed in \S\ref{sec:cosmological-parameters};
\citealt{2008MNRAS.385.1635S}, discussed in Chapter~\ref{chap:bias});
we are now making them public so others can use them as well.

The outline of this chapter is as follows: in \S\ref{sec:Mangle-terminology}
we give an overview of the terminology we use to describe angular
masks and the basic tasks we wish to perform, and in \S\ref{sec:Speedup:-pixelization}
we detail our algorithms for accelerating these tasks and quantify
their speed. We describe our methods for integrating \noun{mangle}
with \emph{HEALPix} in \S\ref{sec:Unification-with-HEALPix} and
summarize in \S\ref{sec:Summary}.


\section{\noun{Mangle} terminology}

\label{sec:Mangle-terminology}

The process of defining an angular mask of a galaxy survey in a generic
way requires a set of standardized terminology. We use the terminology
from \citet{2004MNRAS.349..115H} and present a summary of it here.
Our formal definition of an angular mask is a union of an arbitrary
number of weighted angular regions bounded by arbitrary numbers of
edges. The restrictions on the mask are

\begin{figure*}
\includegraphics[width=.488\textwidth]{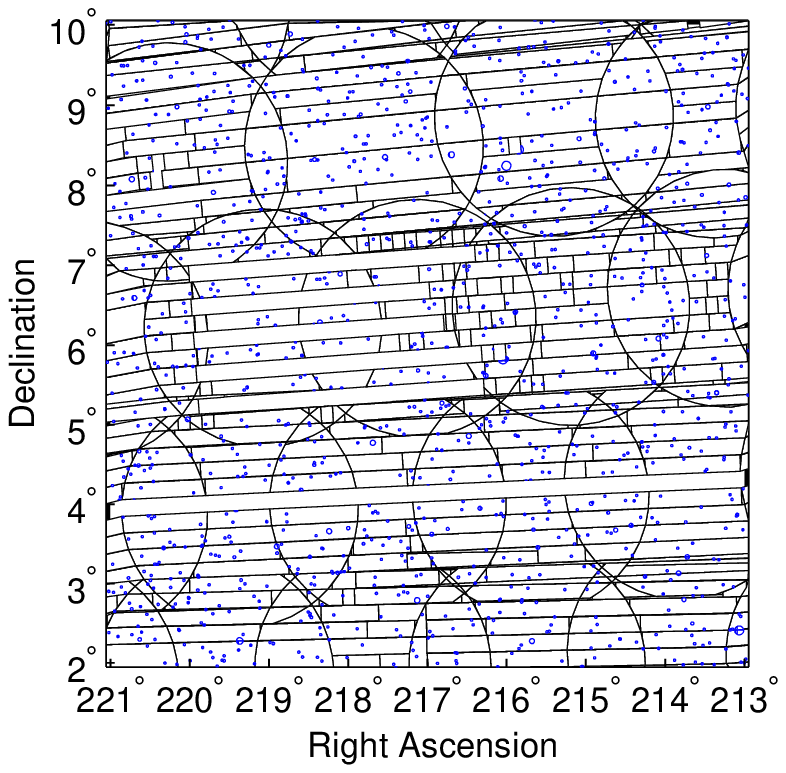}\includegraphics[width=.512\textwidth]{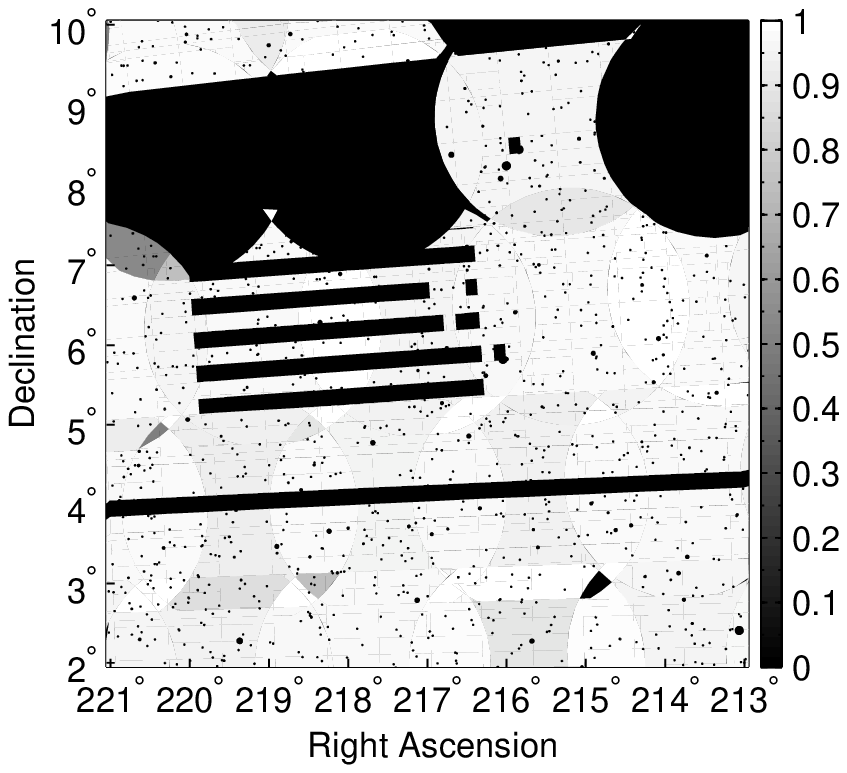}

\caption[A portion of
the SDSS DR5 angular mask.]{\label{fig:examplemask}\label{fig:examplemask_final}A portion of
the SDSS DR5 angular mask (Blanton et al.~2005b; \citealt{2007ApJS..172..634A}).
Left: Polygons defining the mask: spectroscopic plates and lines delineating
different scans and spectroscopic plates are shown in black, and holes
in the mask are shown in blue/gray. Right: Processed version of the
mask, shaded according to survey completeness.}
\end{figure*}

\begin{enumerate}
\item that each edge must be part of some circle on the sphere (but not
necessarily a great circle), and
\item that the weight within each subregion of the mask must be constant.
\end{enumerate}
This definition does not cover every theoretical possibility of how
a piecewise-constant function on a sphere could be defined, but in
practice it is sufficiently broad to accommodate the design of essentially
any galaxy survey. Furthermore, as we discuss in detail in \S\ref{sec:Unification-with-HEALPix},
a curvilinear angular region (such as a \emph{HEALPix} pixel) can
be well-approximated by segments of circles at high resolution. As
an example of a typical angular mask, we show a portion of the SDSS
angular mask from Data Release 5 (DR5; \citealt{2007ApJS..172..634A})
in Fig.~\ref{fig:examplemask}. 

The fundamental building block of an angular mask is the spherical
polygon, which is defined as a region bounded by edges that are part
of a circle on the sphere. An angular mask is thus the union of arbitrarily
weighted non-overlapping polygons. For convenience, we provide an
updated version of a table from \citet{2004MNRAS.349..115H} in Table~\ref{tab:mangleterms}
containing the definitions of key terms used in this chapter. 
\begin{table*}
\begin{minipage}[c][110mm]{100mm}%

\caption{\label{tab:mangleterms}Definitions of Terms, in Alphabetical Order}

\begin{tabular}{@{}lp{127mm}}
Term &
Definition \tabularnewline
\hline
boundary &
A set of edges bounding a polygon. \tabularnewline
cap &
A spherical disk, a region above a circle on the unit sphere. \tabularnewline
circle &
A line of constant latitude with respect to some arbitrary polar axis
on the unit sphere. \tabularnewline
edge &
An edge is part of a circle. A polygon is enclosed by its edges. \tabularnewline
great circle &
A line of zero latitude with respect to some arbitrary polar axis
on the unit sphere. A great circle is a circle, but a circle is not
necessarily a great circle. \tabularnewline
mask &
The union of an arbitrary number of weighted polygons. \tabularnewline
pixel&
A special polygon that specifies some predefined region on the sky
as part of a scheme for discretizing the unit sphere. Once a mask
has been pixelized, each polygon in the mask is guaranteed to overlap
with exactly one pixel.\tabularnewline
polygon &
The intersection of an arbitrary number of caps. \tabularnewline
rectangle &
A special kind of polygon, a rectangular polygon bounded by lines
of constant longitude and latitude. \tabularnewline
vertex &
A point of intersection of two circles. A vertex of a polygon is a
point where two of its edges meet. \tabularnewline
weight &
The weight assigned to a polygon. The spherical harmonics of a mask
are the sum of the spherical harmonics of its polygons, each weighted
according to its weight. A weight of 1 is the usual weight. A weight
of 0 signifies an empty polygon, a hole. In general the weight may
be some arbitrary positive or negative real number. \tabularnewline
\hline
\end{tabular}%
\end{minipage}%

\end{table*}

The basic procedure used by the \noun{mangle} software to process
an angular mask consists of the following steps, which are described
in greater detail in \citet{2004MNRAS.349..115H}:

\begin{enumerate}
\item Snap
\item Balkanize
\item Weight
\item Unify
\end{enumerate}
The snapping step identifies edges of polygons that are nearly coincident
and snaps them together so the edges line up exactly. This is necessary
because nearly-coincident edges can cause significant numerical issues
in later computations. In practice, this situation occurs when two
polygons in a survey are intended to abut perfectly, but are prevented
from doing so by roundoff errors or numerical imprecision in the mask
definition. There are several tunable tolerances that control how
close two edges can be before they get snapped together -- these can
be adjusted based on how precisely the polygons defining the mask
are specified.

\begin{figure*}
\includegraphics[width=1\textwidth]{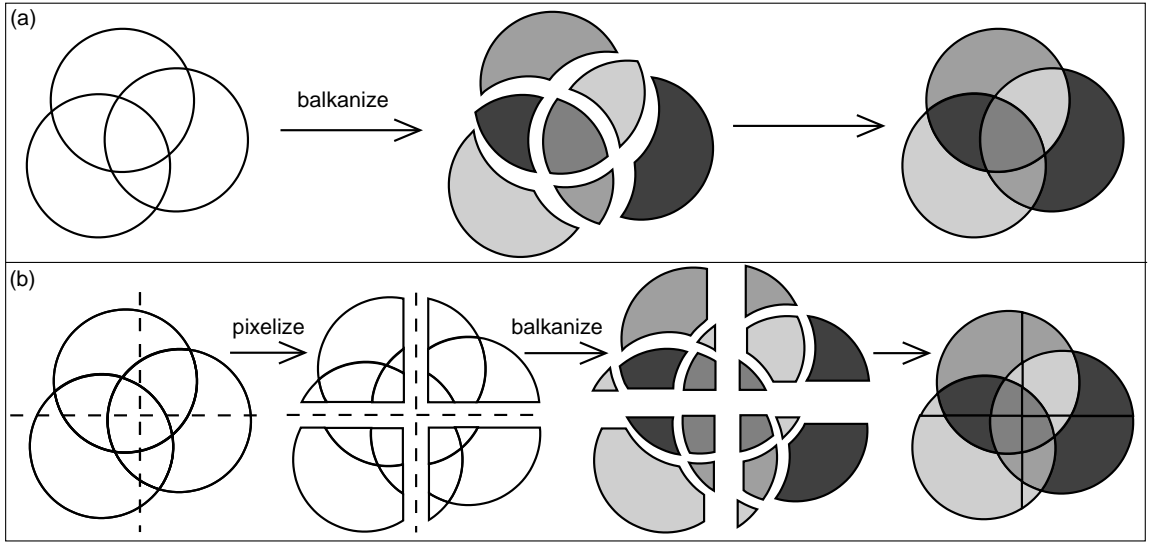}

\caption{\label{fig:pix_concept}A cartoon illustrating the process of balkanization
(a) with no pixelization, (b) with pixelization.}
\end{figure*}

Balkanization is the process of resolving a mask into a set of non-overlapping
polygons. It checks for overlap between each pair of polygons in the
mask, and if the polygons overlap, it fragments them into non-overlapping
pieces. After this is completed, it identifies polygons having disconnected
geometry and subdivides them into connected parts. The basic concept
of balkanization is illustrated in the top panel of Fig.~\ref{fig:pix_concept}.
The purpose of this procedure is to define all of the distinct regions
on the sky in which the piecewise-constant function we are intending
to model might take on a different value. For example, the 2dFGRS
and SDSS spectroscopic surveys generate masks containing many overlapping
circles defining each spectroscopic field observed. The SDSS spectrographs
can observe 640 objects in each field, so if there are more than 640
desired target galaxies in the field, they might not all be observed.
For example, one field may have spectra for 80 per cent of the targets,
and a neighboring field may have 90 per cent, but in the region where
they overlap all of the targets may have been observed. This is how
the survey completeness is determined and illustrates why balkanization
is necessary.

After the mask is balkanized, weights are assigned to each polygon,
representing the value of the survey completeness (or any other desired
parameter) in that region. The way this is done depends on how this
information is provided for a given survey. For example, the 2dFGRS
mask software by Peder Norberg and Shaun Cole\footnote{\url{http://magnum.anu.edu.au/~TDFgg/Public/Release/Masks}}
provides a function that takes an angular position on the sky and
returns the completeness, the magnitude limit, the photographic plate
number, and the value of the parameter $\mu$ (described in \citealt{2001MNRAS.328.1039C})
at that location. This information can be imported into \noun{mangle}
by producing a list of the midpoints of each of the polygons in the
mask, applying the 2dFGRS software to calculate the value of the desired
function at each midpoint, and assigning these weights to the appropriate
polygon by using the {}``weight'' routine in \noun{mangle}. For
the SDSS mask, the files provided by the NYU-VAGC\footnote{\url{http://sdss.physics.nyu.edu/vagc}}
(Blanton et al.~2005b) already include weights for the survey
completeness, so this step is not needed.

The final step of the processing is unification, which discards polygons
with zero weight and combines neighboring polygons that have the
same weight. While not strictly necessary, this procedure clears out
unneeded clutter and makes subsequent calculations more efficient.

After an angular mask has been processed in this fashion, it can be
used for function evaluation: i.e., given a point on the sky, determine
in which one of the non-overlapping polygons it lies, and then get
the weight of that polygon to obtain the value of the function at
the input point. It can also be used for creating a random sample
of points with the same selection function as the survey, calculating
Data-Random $\left\langle DR\right\rangle $ and Random-Random $\left\langle RR\right\rangle $
angular integrals, and computing the spherical harmonics of the mask.
The \noun{mangle} software provides utilities for all of these tasks.

\section{Speedup: pixelization}

\label{sec:Speedup:-pixelization}

The tasks of snapping, balkanization, and unification all require
comparing pairs of polygons in the mask. The brute-force method to
accomplish this is simply to compare each polygon with every other
polygon, which is what the original version of \noun{mangle} did.
This naive algorithm is $\mathcal{O}\left(N^{2}\right)$, which is
easily sufficient for masks such as the 2dFGRS mask, with an $N$
of a few thousand polygons. However, the SDSS mask has about 100 times
as many polygons, and points to the need for a cleverer approach.
The method we present here is a divide-and-conquer approach we dub
{}``pixelization,'' which processes the mask so that each polygon
needs to be compared with only a few nearby polygons.

\subsection{Pixelization concept}

\label{sub:Pixelization-concept}

The underlying concept of pixelization is as follows: before performing
any snapping, balkanization, or unification, divide the mask into
predefined regions called {}``pixels'' and split each polygon along
the pixel boundaries such that each polygon is only in one pixel.
Then for following tasks, polygons need only be compared with other
polygons in the same pixel.

The process of balkanization with and without pixelization is illustrated
in Fig.~\ref{fig:pix_concept} for three overlapping polygons. The
top panel shows the unpixelized version: balkanization checks for
overlap between each pair of polygons, and then fragments it into
seven non-overlapping polygons. 

The bottom panel shows the same process but using pixelization as
a first step. First, each polygon is divided along the pixel boundaries,
shown by the dotted lines. At this point, each of the four pixels
shown has three polygons in it: the intersections between that pixel
and each of the original three polygons. Then balkanization is performed
within each pixel: the three polygons in the upper left pixel are
split into five non-overlapping polygons, and so forth, yielding a
final set of 18 non-overlapping polygons. In this illustrative example,
pixelization increases the complexity of the process, but in general
it replaces the $\mathcal{O}\left(N^{2}\right)$ algorithm for balkanization
with one that is roughly $\mathcal{O}\left(M\left(N/M\right)^{2}\right)$
for $N$ polygons and $M$ pixels, which is roughly $\mathcal{O}\left(N\right)$
if $M\sim N$. For large, complicated masks such as SDSS, this speeds
up the processing time by a factor of ${\sim}1200$. 

Once a mask has been pixelized, the important task of determining
in which polygon(s) a given point lies is sped up greatly as well:
one merely has to calculate in which pixel the point lies, and then
test if the point is in each polygon within that pixel. For a typical
pixelization scheme, the appropriate pixel number for a given point
can be found with a simple formula, i.e. an $\mathcal{O}\left(1\right)$
calculation, so pixelization reduces the $\mathcal{O}\left(N\right)$
algorithm of testing every polygon in the mask to $\mathcal{O}\left(N/M\right)$.
This means that with pixelization, this task does not depend on the
total number of polygons in the mask at all if $M\sim N$. 

It is important to note that the pixelization procedure makes no approximations
to the original mask -- the pixels are used simply as a tool to determine
which polygons are close to each other, not as a means of discretizing
the mask itself. A mask that has been balkanized after pixelization
contains all of the same information as it would without pixelization,
except that it took a tiny fraction of the time to produce. Furthermore,
unification can be applied across the whole mask rather than within
each pixel, which effectively unpixelizes the mask if desired. Thus
there are essentially no drawbacks to using our pixelization procedure.

\subsection{Pixelization schemes}

\label{sub:Pixelization-schemes}

\subsubsection{Simple scheme}

The most straightforward means of pixelizing the sky is to use lines
of equal azimuth and elevation as the pixel boundaries. The azimuth
and elevation typically correspond to celestial coordinates -- right
ascension (RA) and declination (Dec) -- in a survey mask. In this
scheme, the whole sky is split into quadrants along the equator and
prime meridian to form the lowest resolution of pixelization. In \noun{mangle}
this is defined as resolution 1. 

To pixelize to higher resolutions, each pixel is split into four child
pixels, with the boundaries at the midpoints of azimuth and of cos(elevation)
within the pixel. This creates pixels with equal area. Thus resolution
2 consists of each of the four resolution 1 quadrants split into four
pixels each, and so forth. This procedure produces a hierarchical
pixel structure known in computer science terminology as a quadtree \citep{markbook}: 
there are $4^{r}$ pixels in this scheme at resolution
$r$, and each of these pixels has four child pixels at resolution
$r+1$ and $r$ parent pixels, one at each lower resolution. Resolution
0 is defined as being the whole sky. 

\begin{figure*}
\includegraphics[width=1\textwidth]{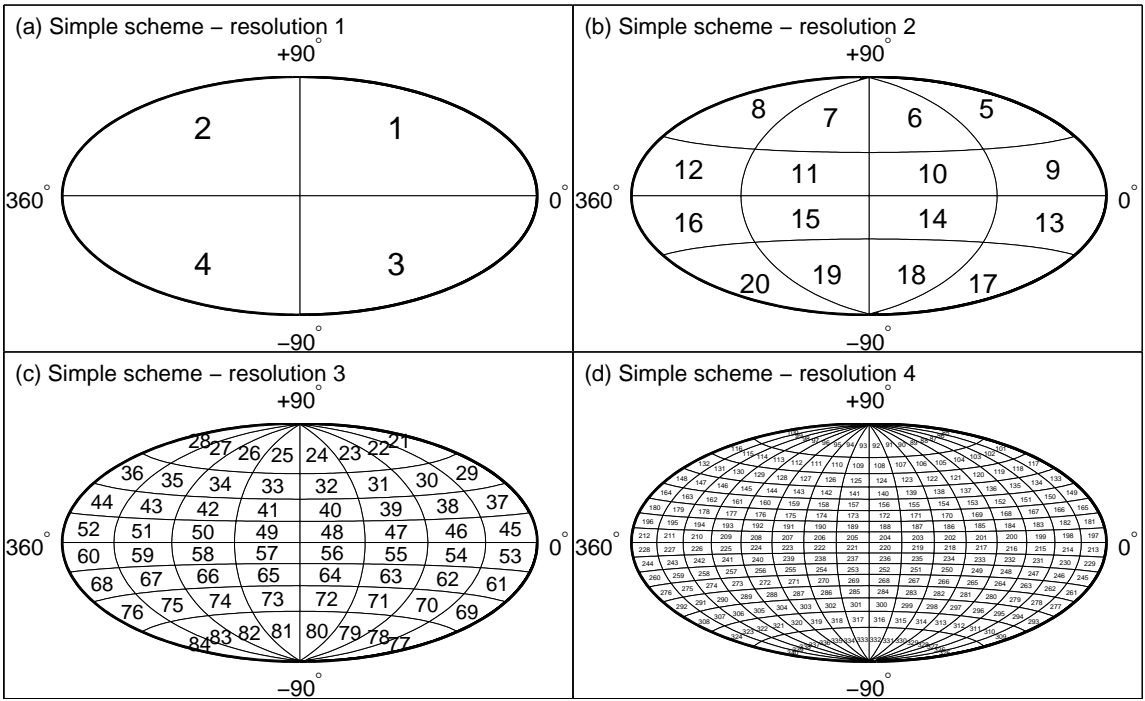}

\caption[The full sky pixelized with the simple pixelization scheme.]{\label{fig:simple_pix}The full sky (shown in a Hammer-Aitoff projection
in celestial coordinates) pixelized with the simple pixelization scheme
at the four lowest resolutions.}
\end{figure*}

In \noun{mangle}, the pixels of the simple scheme are numbered as
follows: the whole sky is pixel 0, the four quadrants of resolution
1 are pixels 1, 2, 3, and 4, resolution 2 is pixels 5-20, and so forth.
Thus just one number specifies both the resolution and the pixel location.
At each resolution, the pixels are numbered in a ring pattern, starting
from an elevation of $90^{\circ}$ and along increasing azimuth for
each ring of equal elevation. The simple scheme pixels at the four
lowest resolutions are shown in Fig.~\ref{fig:simple_pix}.
\begin{figure*}
\includegraphics[width=1\textwidth]{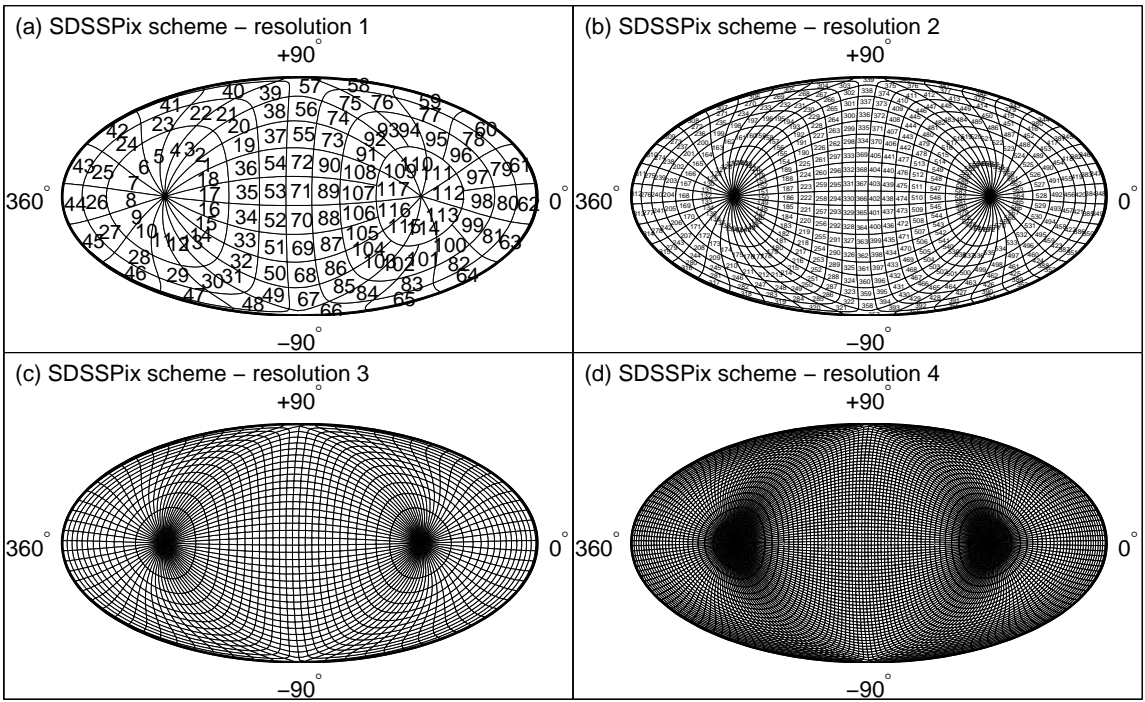}

\caption[The full sky pixelized with the SDSSPix pixelization
scheme.]{\label{fig:sdsspix}The full sky (shown in a Hammer-Aitoff projection
in celestial coordinates) pixelized with the SDSSPix pixelization
scheme at the four lowest resolutions.}
\end{figure*}

\subsubsection{SDSSPix scheme}

Alternatively, the pixelization can be done such that it is more closely
aligned with the mask of a given survey. In particular, a pixelization
scheme called SDSSPix\footnotemark\footnotetext{\url{http://lahmu.phyast.pitt.edu/~scranton/SDSSPix/}}
has been developed for use with the SDSS geometry. Like the simple
scheme, SDSSPix is a hierarchical, equal-area pixelization scheme,
and it is based on the SDSS survey coordinates $\lambda$ and $\eta$.
As described in \citet{2002AJ....123..485S}, SDSS survey coordinates
form a spherical coordinate system rotated relative to the celestial
coordinate system. The poles are located at $\rmn{RA}=95^{\circ},\rmn{Dec}=0^{\circ}$
and $\rmn{RA}=275^{\circ},\rmn{Dec}=0^{\circ}$ (J2000), which are
strategically located outside the SDSS covered area and in the galactic
plane. $\eta$ is the azimuthal angle, with lines of constant $\eta$
being great circles perpendicular to the survey equator, and $\lambda$
is the elevation angle, with lines of constant $\lambda$ being small
circles parallel to survey equator. $\lambda=0^{\circ},\eta=0^{\circ}$
is located at $\rmn{RA}=185^{\circ},\rmn{Dec}={32.5}^{\circ}$ with
$\eta$ increasing northward. This configuration has been chosen such
that the stripes produced by the SDSS scanning pattern lie along lines
of constant $\eta$. 

The SDSSPix base resolution is defined by 36 divisions in the $\eta$
direction (equally spaced in $\eta$) and 13 in the $\lambda$ direction
(equally spaced in $\cos\lambda$), for a total of 468 equal-area
pixels. These divisions are chosen such that at a special resolution
level (called the {}``superpixel'' resolution), there is exactly
one pixel across each SDSS stripe. Finally, as in the simple scheme,
higher resolutions are achieved by hierarchically subdividing each
pixel at a given resolution into four smaller pixels. 

SDSSPix has been included in \noun{mangle} by incorporating several
routines from the SDSSPix software package available online.\footnotemark[\value{footnote}]
The numbering scheme in the \noun{mangle} implementation of SDSSPix
differs somewhat from the internal SDSSPix numbering: in \noun{mangle},
the entire sky is pixel 0, as in the simple scheme, but it has 117
child pixels (instead of 4), numbered from 1 to 117. These pixels,
which comprise resolution 1, are not part of the official SDSSPix
scheme, but are created by combining sets of 4 pixels from what is
defined as resolution 2 in \textsc{mangle}. The resolution 2 pixels
are the official SDSSPix base resolution pixels, and are numbered
from 118 to 585. Higher resolutions are constructed through the standard
SDSSPix hierarchical division of each pixel into 4 child pixels. As
in the simple scheme, the pixel number identifies both the resolution
and the pixel position. The superpixel resolution described above
is defined as resolution 5 in \noun{mangle} and contains a total of
7488 pixels.

\noun{Mangle} typically uses RA and Dec as its internal azimuth and
elevation coordinates, so the SDSSPix pixels are constructed as rectangles
in $\eta-\lambda$ coordinates and then rotated into celestial coordinates.
The SDSSPix pixels at the four lowest resolutions are shown in Fig.~\ref{fig:sdsspix}.

\subsubsection{Other schemes}

\label{sub:Other-schemes}

The implementation of pixelization in \noun{mangle} is designed to
be flexible: it is simple for users to add their own scheme as well.
The pixelization uses only four basic routines:

\begin{enumerate}
\item get\_pixel: Given a pixel number, return a polygon representing that
pixel.
\item which\_pixel: Given a point on the sky and a resolution, return the
pixel number containing the point.
\item get\_child\_pixels: Given a pixel number, return the numbers of its
child pixels.
\item get\_parent\_pixels: Given a pixel number, return the numbers of its
parent pixels.
\end{enumerate}
Adding a new pixelization scheme simply requires creating appropriate
versions of these four routines. Note that it also requires that the
pixels can be represented as polygons -- this is \emph{not} strictly
the case for the \emph{HEALPix} pixels, as discussed further in \S\ref{sec:Unification-with-HEALPix}.

\subsection{Pixelization algorithm}

\label{sub:Pixelization-algorithm}

The purpose of pixelization is to speed up the processing of angular
masks, which means that the pixelization itself must be done with
a clever, speedy algorithm or nothing will be gained. The naive algorithm
is to search through all of the polygons in the mask for those that
overlap that pixel. This is $\mathcal{O}\left(NM\right)$ for $N$
polygons and $M$ pixels, and is not sufficient for our purposes.

Our fast pixelization algorithm is a recursive method that takes advantage
of the hierarchical nature of the pixelization schemes. The method
works as follows:

\begin{enumerate}
\item Start with all the mask polygons that are in pixel $i$.
\item Create polygons for each child pixel of pixel $i$ at the next resolution
level.
\item Split the mask polygons in pixel $i$ along the child pixel boundaries
such that each polygon lies within one child pixel.
\item Repeat steps 1-3 for the mask polygons in each child pixel until desired
stopping point is reached.
\end{enumerate}
Starting this with pixel $i=0$, i.e., the whole sky, will pixelize
the entire mask in $\mathcal{O}\left(N\log M\right)$ time. Thus the
pixelization does not add too much overhead time to the overall mask
processing.

\begin{figure*}
\includegraphics[width=.5\textwidth]{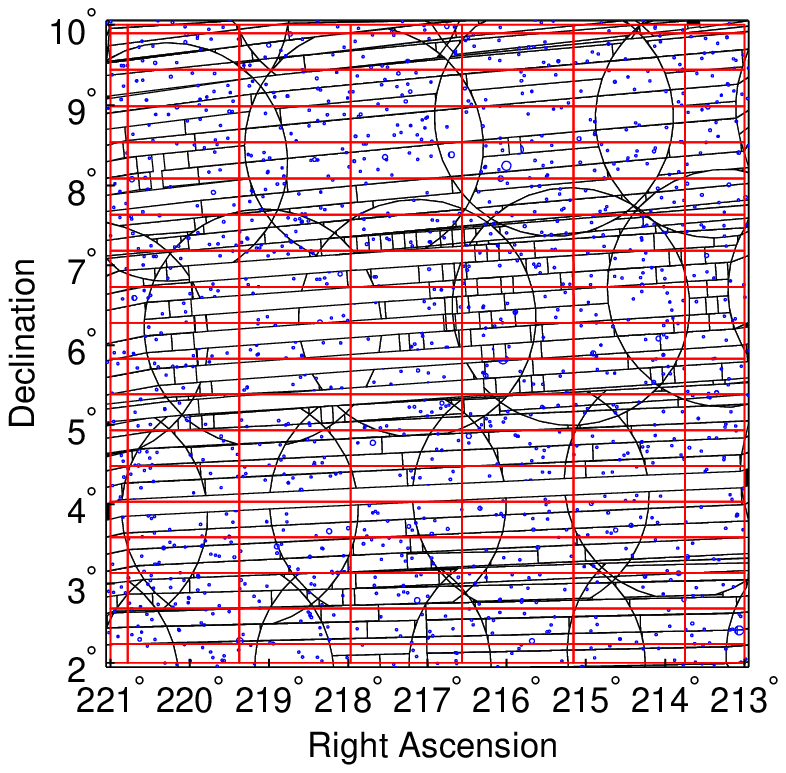}\includegraphics[width=.5\textwidth]{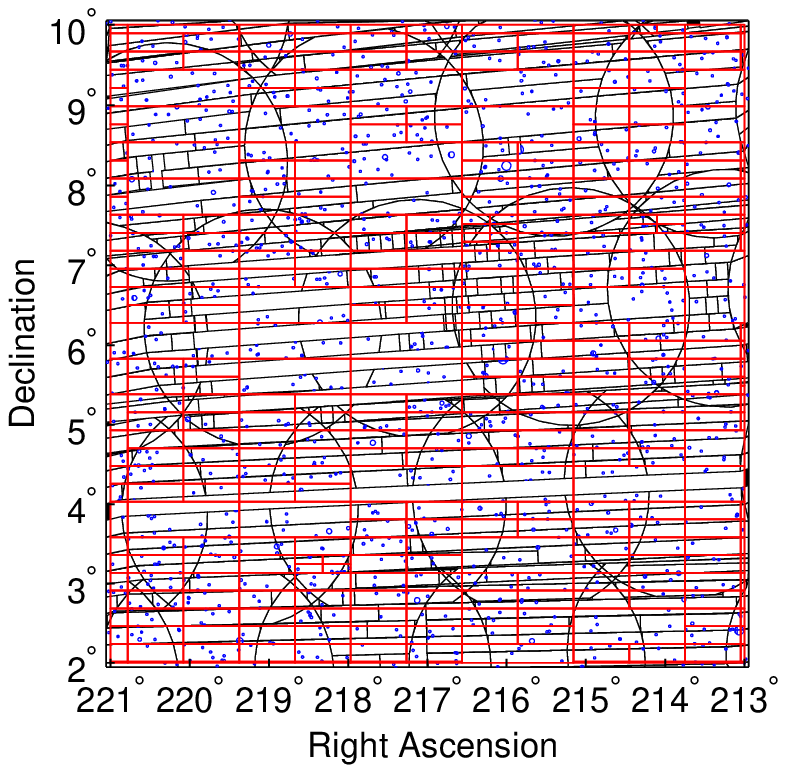}

\caption[Left: portion of SDSS mask from
Fig.~\ref{fig:examplemask} pixelized to a fixed resolution with the simple pixelization scheme.  Right: the same mask pixelized with the
simple pixelization scheme using the adaptive resolution method.]{\label{fig:examplemask_pixelized}Left: portion of SDSS mask from
Fig.~\ref{fig:examplemask} pixelized to resolution 8 ($4^{8}$ total
pixels on the sky) with the simple pixelization scheme. Pixel boundaries
are shown in red/light gray. Right: the same mask pixelized with the
simple pixelization scheme using the adaptive resolution method with
a maximum of 30 polygons per pixel.}
\end{figure*}

There are two different methods for choosing the desired stopping
point of the pixelization. The simplest method is to stop at a fixed
resolution, such that the entire mask is pixelized with pixels of
the same size. The example mask from Fig.~\ref{fig:examplemask}
is shown in the left panel of Fig.~\ref{fig:examplemask_pixelized}
pixelized to a fixed resolution in the simple scheme.

Alternatively, the stopping condition can be chosen to be a maximum
number of polygons allowed in each pixel: if there are more than $N_{\rmn{max}}$
polygons in a given pixel, continue the recursion and divide those
polygons into pixels at the next resolution level. This results in
an adaptively pixelized mask, where higher resolutions are automatically
used in regions of the mask that are more complicated. This method
is especially useful for masks with varying degrees of complexity
in different areas. The example mask from Fig.~\ref{fig:examplemask}
is shown again in the right panel of Fig.~\ref{fig:examplemask_pixelized}
pixelized adaptively with $N_{\rmn{max}}=30$.

The implementation of pixelization in \noun{mangle} allows users to
choose either of these methods and to select values for the fixed
resolution level or for $N_{\rmn{max}}$.

\subsection{Speed trials}

In order to choose optimal default values for the maximum resolution
and $N_{\rmn{max}}$ as well as to demonstrate the dramatic improvements
in speed that pixelization provides, we have conducted a series of
speed trials of the new \noun{mangle} software.

There are two procedures we are interested in optimizing: firstly,
basic processing of a mask detailed in \S\ref{sec:Mangle-terminology}
involving pixelization, snapping, balkanization, and unification,
and secondly, the use of the final mask to identify in which polygon
a given point lies. We carried out trials of these two procedures
using both the simple and SDSSPix pixelization schemes described in
\S\ref{sub:Pixelization-schemes}. For each of these schemes, we
tested both the fixed and adaptive resolution methods for stopping
the pixelization algorithm described in \S\ref{sub:Pixelization-algorithm}.
For the fixed resolution method, we measured the time for several
different values of the maximum resolution, and for the adaptive resolution
method, we measured the time as a function of $N_{\rmn{max}}$. 

\begin{figure*}
\includegraphics[width=1\textwidth]{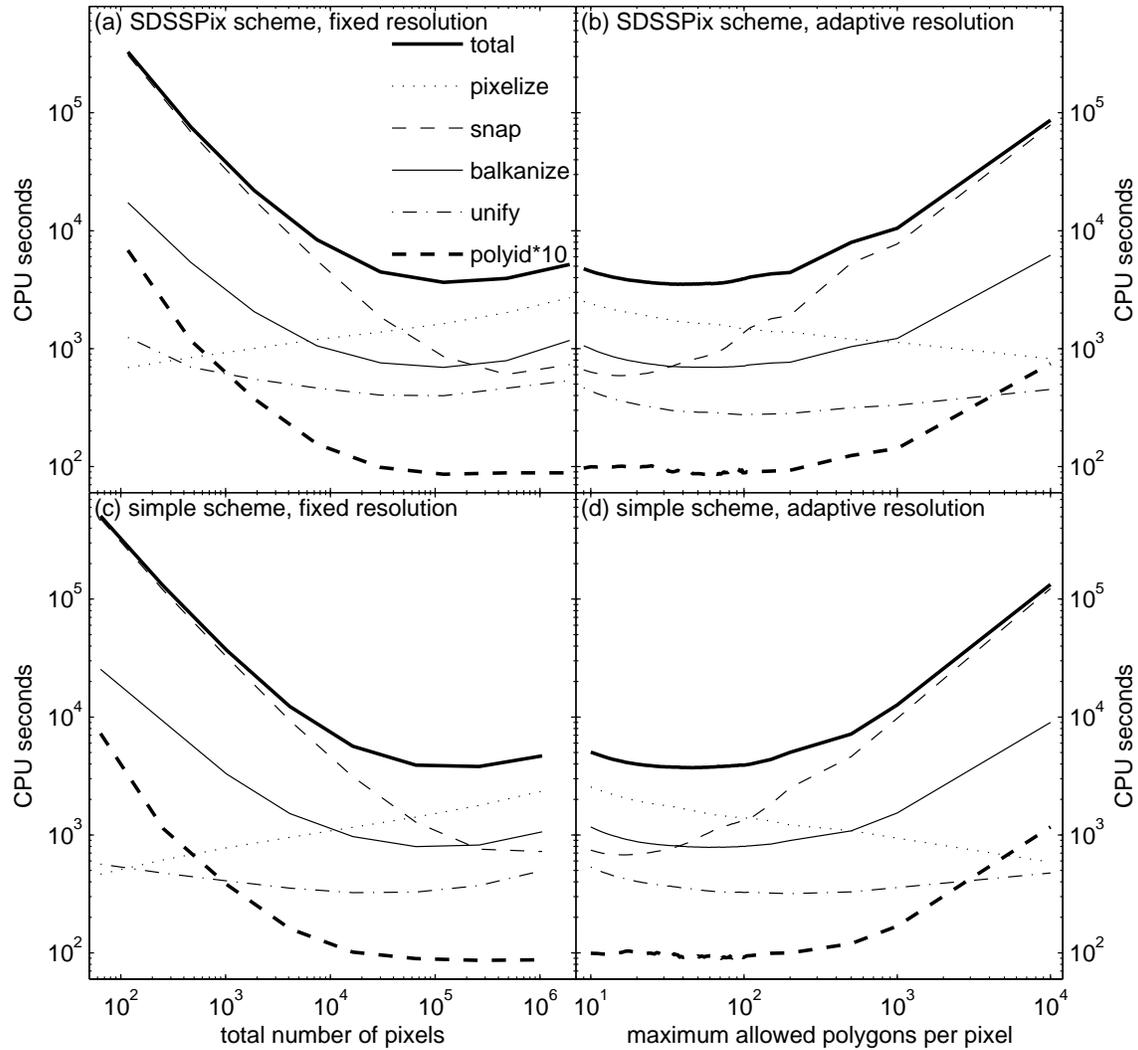}

\caption[Time required to process the full SDSS DR5
mask with different choices of pixelization schemes and methods.]{\label{fig:speedtrials}Time required to process the full SDSS DR5
mask with different choices of pixelization schemes and methods. The
{}``total'' curves are the sum of the pixelization, snapping, balkanization,
and unification curves. Also shown is the time required to identify
in which polygon each of the ${\sim}400{,}000$ SDSS DR5 galaxies
lies (polyid), scaled up by a factor of 10. The plots on the left
show time vs. the number of pixels at a fixed resolution. The plots
on the right show time vs. $N_{\rmn{max}}$, the maximum number of
polygons in allowed in each pixel when pixelizing with the adaptive
resolution method.}
\end{figure*}

The results are shown in Fig.~\ref{fig:speedtrials}. (Note that
the overhead time for reading and writing files and doing general
setup has been subtracted -- the times shown here are just for the
primary operations.) From the fixed resolution trials, we see that
the optimal resolution choice for the SDSS mask is the one that has
approximately $10^{5}$ total pixels on the sky for both the simple
and SDSSPix schemes. This corresponds to resolution 9 for the simple
scheme and resolution 6 for SDSSPix. When using the adaptive method,
the choice for the maximum number of polygons allowed in each pixel
that gives the fastest processing is $N_{\rmn{max}}=40$ for the simple
scheme and $N_{\rmn{max}}=46$ for SDSSPix. Overall, the fastest choice
(by a slight margin) for the SDSS mask is using the SDSSPix scheme
with adaptive resolution. It is also interesting to note that different
\noun{mangle} processes have different optimum values -- for example,
snapping is fastest when there are fewer polygons in each pixel compared
to balkanization.

The impact of our pixelization algorithm is most clearly demonstrated
by Fig.~\ref{fig:speed_fractions}. This shows the processing time
and polygon identification time for a series of selected portions
of the SDSS mask as a function of the total number of polygons, both
with and without pixelization -- again, overhead time has been subtracted
here. Pixelization clearly gives an improvement in speed that becomes
increasingly significant for larger numbers of polygons. 

To quantify this, we fit theoretical models to each series of trials.
Without pixelization, the processing time for snapping, balkanization,
and unification is well-fitted by $AN^{2}$ where $A$ is a free parameter
with a best-fitted value of $A=3.2\times10^{-5}$ CPU seconds. With
pixelization (and using the adaptive resolution method with the optimal
choice for $N_{\rmn{max}}$) this is reduced substantially. We fit
the times using theoretical models based on the formulas in \S\ref{sub:Pixelization-concept}
assuming $M\sim N$. We fit the pixelization time with $BN\log_{4}N$
and the snapping, balkanization, and unification time with $CN$ where
$B$ and $C$ are free parameters. The simple scheme gives $\left(B,\, C\right)=\left(0.0054,\,4.7\times10^{-4}\right)$
and the SDSSPix scheme gives $\left(B,\, C\right)=\left(0.0045,\,4.8\times10^{-4}\right)$
(units are all in CPU seconds) -- thus for processing the SDSS mask,
the SDSSPix scheme is slightly faster than the simple scheme. Our
fits give $C/B=0.1$, so the overall processing time scales like $\mathcal{O}\left(N+0.1N\log_{4}N\right)$. 

These fitted curves are shown in Fig.~\ref{fig:speed_fractions}
and allow us to extrapolate estimates for processing masks with larger
numbers of polygons. For the full SDSS DR5 mask containing about 300,000
polygons, pixelization reduces the processing time by a factor of
${\sim}1200$. The improvement for future surveys will be even more
dramatic: the mask for the co-added LSST survey might contain ${\sim}10^8$
polygons -- pixelization reduces the processing time by a factor of
${\sim}24{,}000$.

The time for identifying in which polygon each SDSS galaxy lies is
significantly sped up as well: without pixelization it is fitted by
$DN$ where $D=0.012$ CPU seconds, although the scatter is rather
large due to dependence on which polygons are used. With pixelization,
it is well-fitted by a constant (8.2 CPU seconds for the simple scheme,
7.9 CPU seconds for the SDSSPix scheme) -- thus the time for polygon
identification does not depend on how many polygons are in the mask
if the number of pixels used is chosen to be roughly proportional to the 
number of polygons.
For the SDSS mask, the time to identify the polygons for all of the
SDSS galaxies is reduced by a factor of nearly 500.

\section{Unification with \emph{HEALPix} and other pixelized tools}

\label{sec:Unification-with-HEALPix}

\emph{HEALPix} \citep{2005ApJ...622..759G,1999astro.ph..5275G} is
a hierarchical, equal-area, isolatitudinal pixelization scheme for
the sphere motivated by computational challenges in analyzing CMB
data \citep{1999elss.conf...37G,1999CoScE...1...21B}. Its base resolution
consists of 12 equal-area pixels; to generate higher resolutions,
each pixel at a given resolution is hierarchically subdivided into
four smaller pixels, and it represents an interesting class of spherical
projections \citep{2007MNRAS.381..865C}. \emph{}Resolution in \emph{HEALPix}
is defined in terms of $N_{\rmn{side}}$, the number of divisions
along the side of a base-resolution pixel required to reach the desired
resolution. Because of the hierarchical definition of the higher resolution
pixels, $N_{\rmn{side}}$ is always a power of 2, and the total number
of pixels at a given resolution is $12N_{\rmn{side}}^{2}$.

\emph{HEALPix} is a very useful scheme in that it allows for fast
and accurate astrophysical computations by means of appropriately
discretizing functions on the sphere to high resolution \citep{1998astro.ph..3317W,2001A&A...374..358D}.
In particular, \emph{HEALPix} includes routines for fast computations
of spherical harmonics \citep{2002ApJ...567....2H,2001PhRvD..64h3003W,2001ApJ...561L..11S,2001PhRvD..64h3001D}.
It is widely used in the analysis of cosmic microwave background data
from WMAP \citep{2007ApJS..170..377S} and has recently been used
to approximate galaxy survey masks as well \citep{2007ApJ...657..645P}. 

Combining \emph{HEALPix} with \noun{mangle} is useful because it facilitates
comparisons between galaxy survey data and the CMB as is done in experiments
measuring the integrated Sachs-Wolfe effect \citep{2005PhRvD..72d3525P,2006PhRvD..74f3520G,2007MNRAS.377.1085R},
the Sunyaev-Zel'dovich effect \citep{2004MNRAS.347L..67M,2006ApJ...651..643R},
etc., and can also be used for generating masks that block out regions
of high dust extinction from galaxy surveys (\citealt{1998ApJ...500..525S},
dust map available in \emph{HEALPix} format online%
\footnote{\protect\url{http://lambda.gsfc.nasa.gov/product/foreground/ebv_map.cfm} %
}).

In general, it can be applied to any task requiring comparison between
a piecewise-constant function on the sphere to a continuous function
sampled on a discretized spherical grid. Converting an angular mask
into \emph{HEALPix} format also allows for rapid computations of approximate
spherical harmonics of the mask using the existing \emph{HEALPix}
tools. 

The implementation of the \emph{HEALPix} scheme in \textsc{mangle}
consists of two components:

\begin{enumerate}
\item A new polygon format, {}``healpix\_weight'', which allows the user
to input a list of weights corresponding to each \emph{HEALPix} pixel
at a given $N_{\rmn{side}}$ parameter; 
\item A new utility, {}``rasterize'', which essentially allows the user
to pixelize a mask against the \emph{HEALPix} pixels, by means of
a different technique than the pixelization method described in \S\ref{sec:Speedup:-pixelization}. 
\end{enumerate}
Together, these new features allow effective two-way conversion between
the \emph{HEALPix} specifications and those of \textsc{mangle}.

\subsection{Importing \emph{HEALPix} maps into \noun{mangle}}

\label{sub:Importing-HEALPix-files}

The structure of the healpix\_weight format is quite simple: an input
file consists of a list of numbers corresponding to the weight of
each \emph{HEALPix} pixel at a given $N_{\rmn{side}}$ parameter,
using the nested numbering scheme described in \citet{2005ApJ...622..759G}.
In addition, the definition of the $N_{\rmn{side}}$ parameter is
extended to include 0, which corresponds to a single pixel covering
the entire sphere. \textsc{mangle} constructs polygons approximately
equivalent to the \emph{HEALPix} pixels through the following procedure:

\begin{enumerate}
\item The exact azimuth and elevation of each vertex of a given pixel are
calculated using the \emph{HEALPix} utility {}``pix2vec\_nest''; 
\item The exact azimuth and elevation of each vertex of the four child pixels
of the current pixel are calculated using the same utility, four of
which are the midpoints of the edges of the current pixel; 
\item The four vertices of the given pixel are combined with the four midpoints
to construct the current pixel. Each edge is defined by the circle
that passes through the two vertices and the midpoint, using the {}``edges''
format described in \citet{2004MNRAS.349..115H}.
\item To eliminate spurious antipodal pieces of the polygon defining the
pixel, a fifth cap is constructed whose axis coordinates are the exact
center of the current pixel and whose radius is $10^{-6}$ radians
greater than the distance from the center of the pixel to any of the
four vertices---this cap thus encloses the entire pixel. 
\end{enumerate}
A similar technique could be applied to incorporate other pixelization
schemes not exactly described by spherical polygons as well. It is
important to note that the pixels constructed in this manner are not
exactly equivalent to the actual \emph{HEALPix} pixels -- while the
hierarchical and isolatitudinal properties of the \emph{HEALPix} pixels
are preserved, the equal area property is not. For example, at $N_{\rmn{side}}=1$,
the areas of the approximate pixels differ on average from the actual
area by about $0.08\%$. 

However, as the $N_{\rmn{side}}$ parameter increases, this difference
decreases rapidly: at $N_{\rmn{side}}=512$, the average difference
from the actual area is $0.000002\%$. The boundaries of the both
the actual \emph{HEALPix} pixels and our circles approximating them
become straight lines in the flat-sky approximation, i.e., when the
pixel size is much less than 1 radian. Thus for the resolutions at
which \emph{HEALPix} is typically used, the difference is totally
negligible. 

In applications involving integrations over the sphere, such as calculating
spherical harmonics, the slight area differences between the pixels
can be corrected for in a straightforward manner since the area of
each pixel is known (and can be easily extracted using \noun{mangle}'s
{}``area'' format): simply multiply the value of the function in
each pixel by the area of that pixel divided by the average pixel
area, and then the \emph{HEALPix} spherical harmonics routines will
be exact. Furthermore, the paranoid user can get higher precision
by rasterizing to a higher resolution and then using the {}``ud\_grade''
\emph{HEALPix} utility to obtain results for lower resolutions.

\subsection{Exporting polygon files as \emph{HEALPix} maps}

The idea behind the rasterization method used to convert polygon files
into \emph{HEALPix} maps is very similar to that of pixelization:
to split up the polygons that comprise a mask using a given set of
pixels, such that afterward each polygon lies in only one pixel. However,
rasterization is somewhat different, in that afterward the converse
statement also holds: each pixel contains only one polygon, namely,
itself. In particular, rasterization uses an arbitrary user-defined
spherical pixelization as the pre-determined scheme against which
to split up the polygons in a given input mask. In general, the user-defined
{}``rasterizer'' pixels may be from any pixelization of the sphere,
but the method was originally developed for use with the approximate
\emph{HEALPix} pixels described in the previous section. From a practical
standpoint, it would be simplest to implement any spherical pixelization
that can be exactly represented as spherical polygons by following
the steps in \S\ref{sub:Other-schemes}; however, for pixelization
schemes that do not possess this property (such as \emph{HEALPix}),
rasterization provides an alternate method of implementation.

The final product of rasterization is a polygon file in which each polygon
corresponds to one of the rasterizer pixels (either the approximate
\emph{HEALPix} pixels or a user-defined set of pixels) and its weight
is the area-averaged weight of the input angular mask within that
pixel. This output can easily be converted into a FITS file read by
the \emph{HEALPix} software using a simple script provided with \noun{mangle}.
\begin{figure*}
\includegraphics[width=.5\textwidth]{manglefig2b}\includegraphics[width=.5\textwidth]{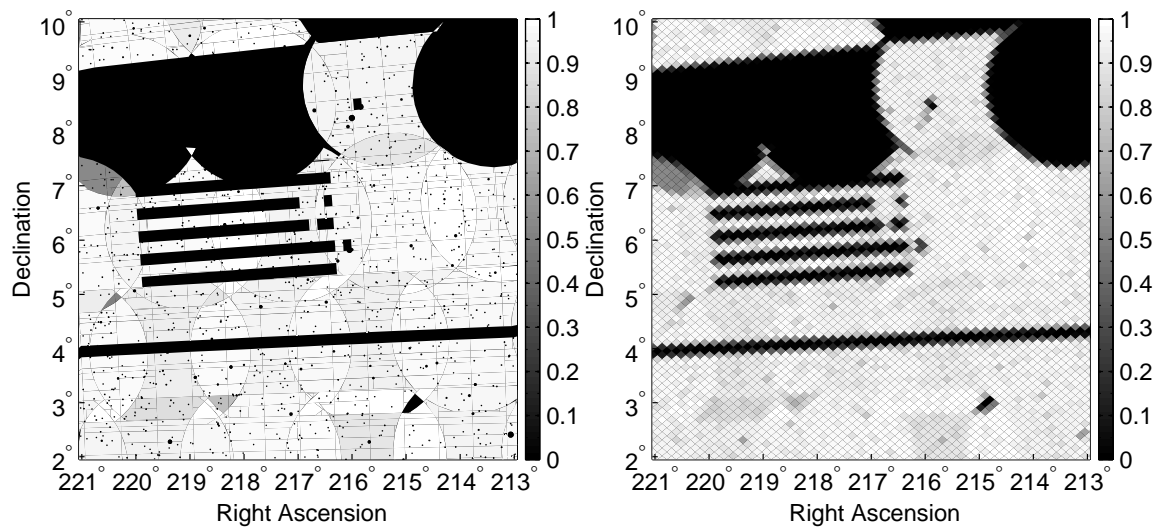}

\caption[Left: Portion of SDSS mask as shown in Fig.~\ref{fig:examplemask}.
Right: Portion of SDSS mask from Fig.~\ref{fig:examplemask} as approximated
by \emph{HEALPix} pixels.]{\label{fig:rasterized}Left: Portion of SDSS mask as shown in Fig.~\ref{fig:examplemask}.
Right: Portion of SDSS mask from Fig.~\ref{fig:examplemask} as approximated
by \emph{HEALPix} pixels, rasterized to $N_{side}=512$.}
\end{figure*}

\begin{figure*}
\begin{center}
\includegraphics[width=11cm]{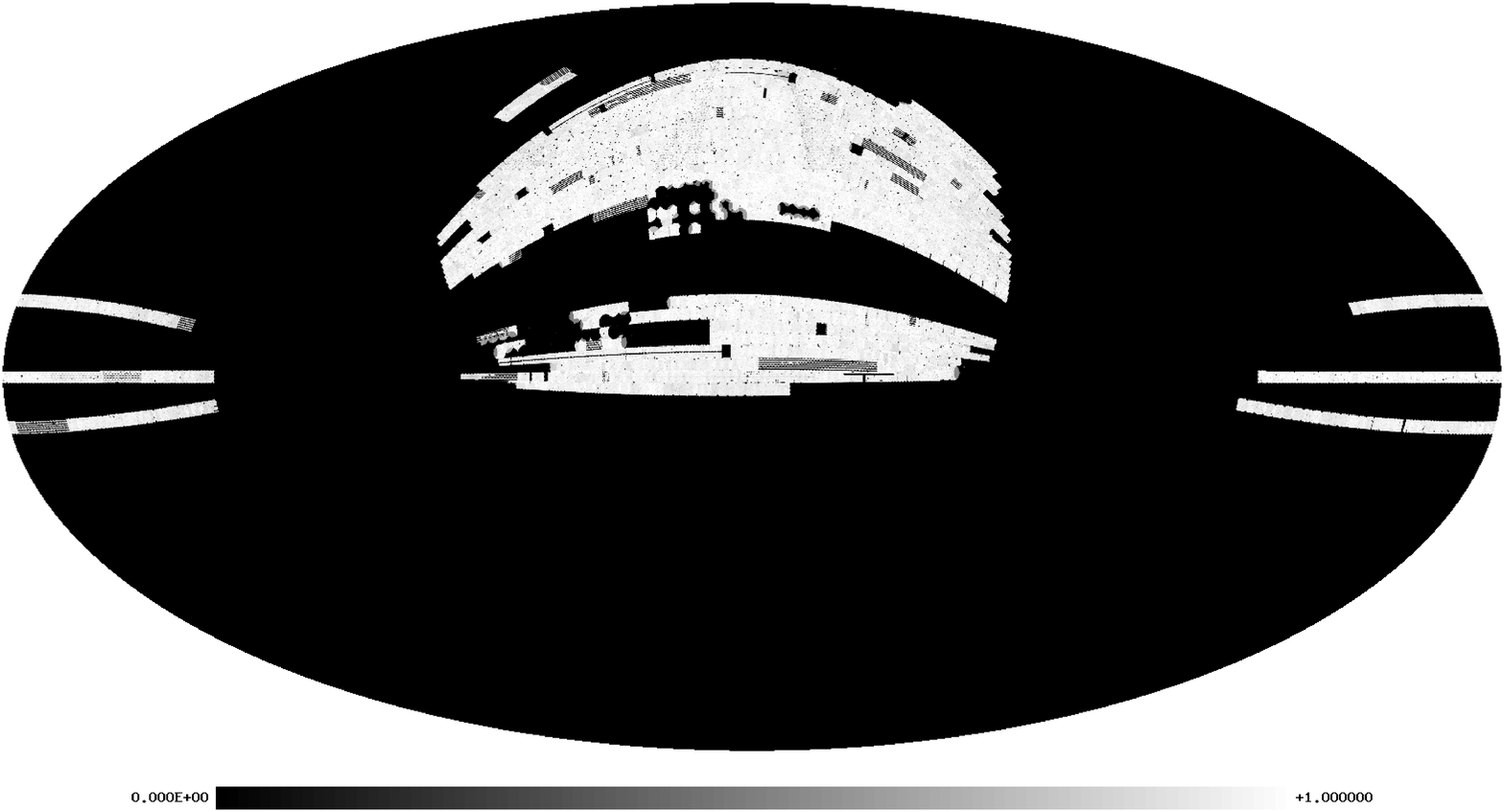}

\includegraphics[width=11cm]{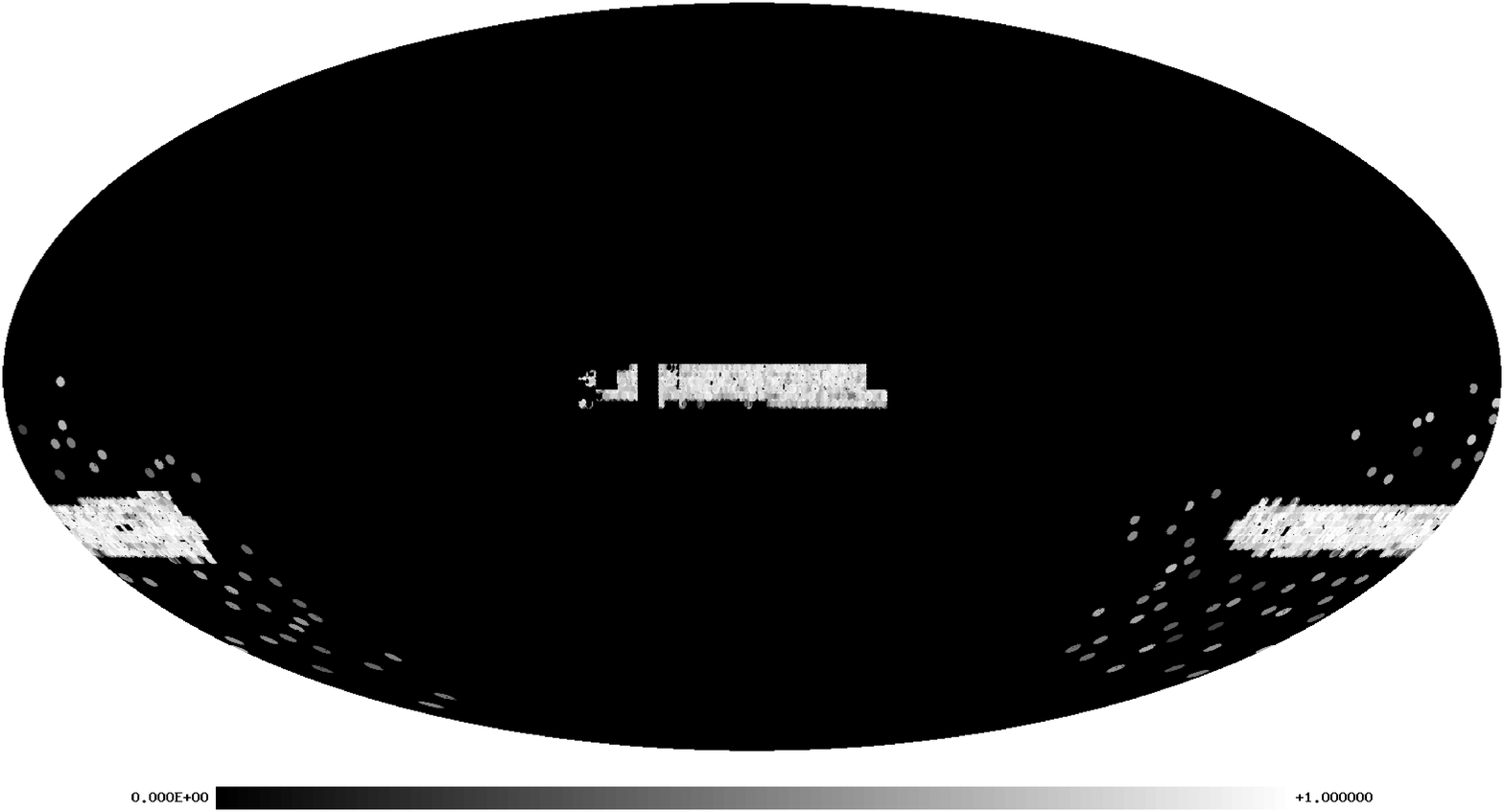}

\includegraphics[width=11cm]{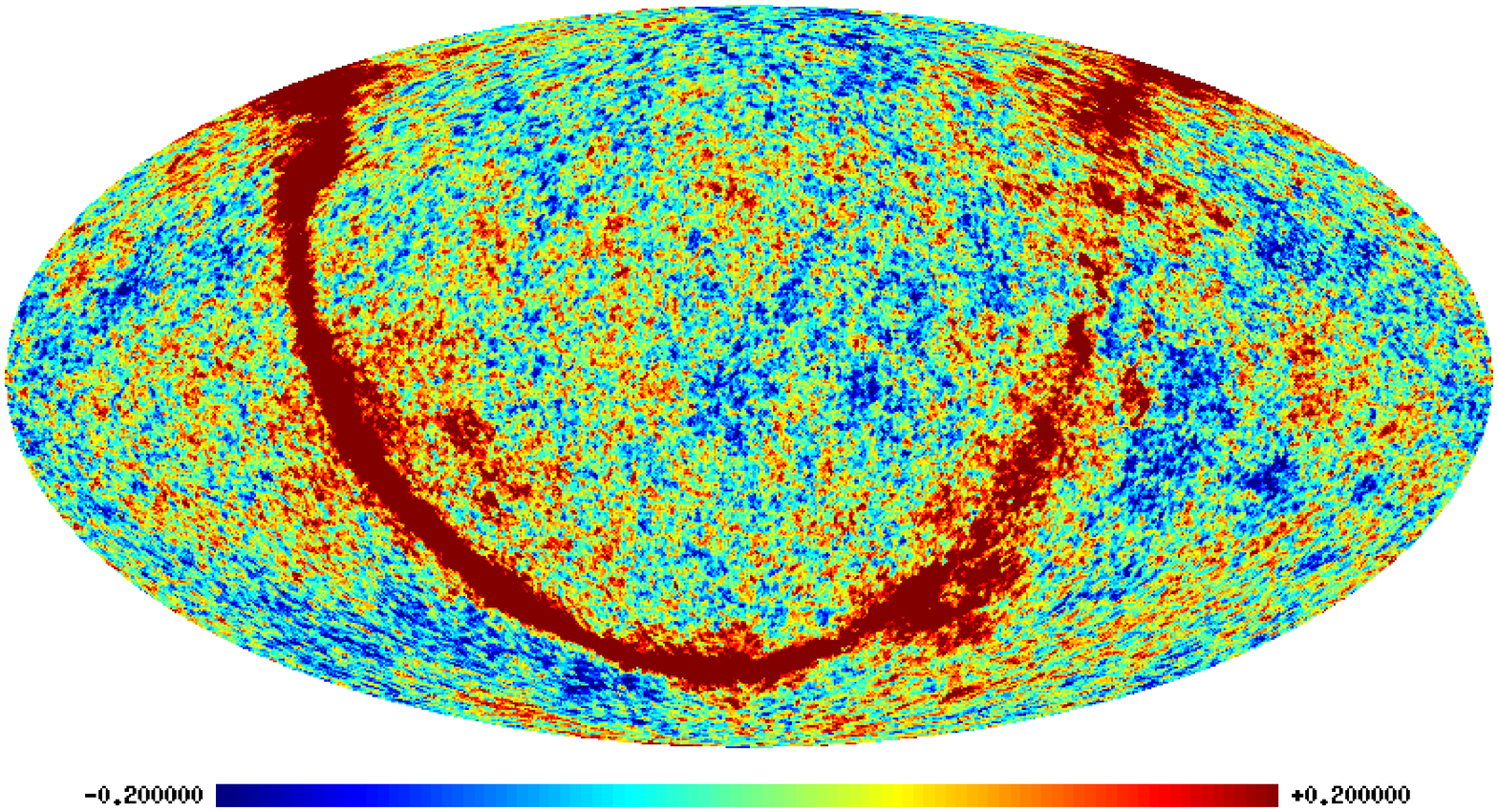}
\end{center}

\caption[Top: The SDSS DR5 completeness mask,
rasterized and plotted using HEALPix routines. Middle: The final 2dFGRS
completeness mask, rasterized
and plotted using \emph{HEALPix} routines. Bottom: CMB temperature
difference map measured by WMAP channel 4.]{\label{fig:masksandcmb}The new version of \noun{mangle} can output
angular masks in \emph{HEALPix} format, which allows for easy comparisons
to CMB and other sky map data and allows users to take advantage of
existing \emph{HEALPix} tools. Top: The SDSS DR5 completeness mask,
rasterized and plotted using HEALPix routines. Middle: The final 2dFGRS
completeness mask as determined in \citet{Hamilton_in_prep}, rasterized
and plotted using \emph{HEALPix} routines. Bottom: CMB temperature
difference map measured by WMAP channel 4, with units in mK. \citet{2007ApJS..170..377S}
All three maps are shown at $N_{side}=512$ in celestial coordinates.}
\end{figure*}

Rasterization consists of the following steps:

\begin{enumerate}
\item Calculate the area of each rasterizer pixel;
\item Compute the area of the intersection of each input mask polygon with
each rasterizer (e.g., \emph{HEALPix}) pixel;
\item Calculate the area-averaged weight within each rasterizer pixel. 
\end{enumerate}
Note that, as with snapping, balkanization, and unification, this
procedure is greatly accelerated by pixelizing both the rasterizer
polygons and the polygons defining the mask to the same resolution
using one of the pixelization schemes described in \S\ref{sub:Pixelization-schemes},
e.g. simple or SDSSPix. Then step (ii) in the above procedure then
involves comparing only polygons in the same simple/SDSSPix pixel.
Our example mask from Fig.~\ref{fig:examplemask} (duplicated in
the left panel of Fig.~ \ref{fig:rasterized}) is shown in the right
panel of Fig.~\ref{fig:rasterized} rasterized with \emph{HEALPix}
pixels at $N_{\rmn{side}}=512$. The ability to convert between \noun{mangle}
and \emph{HEALPix} formats allows for straightforward comparisons
between different types of functions defined on the sphere, e.g. angular
masks of galaxy surveys and the CMB, as illustrated in Fig.~\ref{fig:masksandcmb}.
%
%

\section{Summary}

\label{sec:Summary}

As technologies for surveying the sky continue to improve, the process
of managing the angular masks of galaxy surveys grows ever more complicated.
The primary purpose of this chapter has been to present a set of dramatically
faster algorithms for completing these tasks. 

These algorithms are based on dividing the sky into regions called
{}``pixels'' and performing key operations only within each pixel
rather than across the entire sky. The pixelization is based on a hierarchical
subdivision of the sky -- this produces a quadtree data structure that keeps
track of which polygons are nearby each other. The preprocessing step of
pixelization is $\mathcal{O}\left(N\log N\right)$ for a mask with $N$ polygons
and ${\sim}N$ pixels,
and it reduces
the mask processing time from $\mathcal{O}\left(N^{2}\right)$
to $\mathcal{O}\left(N\right)$.
Furthermore, it reduces
the time required to locate a point within a polygon from $\mathcal{O}\left(N\right)$
to $\mathcal{O}\left(1\right)$.
 
This method is exact, i.e., it
does not make a discrete approximation to the mask, 
and it takes only a tiny fraction of the computation time.
It accelerates the processing of the SDSS mask
by a factor of about 1200 and 
reduces the time to locate all of the
galaxies in the SDSS mask by a factor of nearly 500.
It will provide even more dramatic gains
for future surveys: processing time for the LSST large-scale structure
mask could be reduced by a factor of about 24,000.

We have also described a method for converting between masks described
by spherical polygons and sky maps in the \emph{HEALPix} format commonly
used by CMB and large-scale structure experiments. This provides a
convenient way to work with both piecewise-constant functions on the
sky such as the completeness of a galaxy survey and continuously varying
sky maps such as the CMB temperature. Converting angular masks into \emph{HEALPix}
format also allows users to take advantage of existing \emph{HEALPix} tools
for rapidly computing spherical harmonics.

All of the new algorithms and features detailed here have been integrated
into the \noun{mangle} software suite, which is available for free
download at \texttt{\url{http://space.mit.edu/home/tegmark/mangle/}}.
This updated software package should prove increasingly useful in the
coming years, especially as next-generation surveys such as DES, WFMOS,
Pan-STARRS, and LSST get underway.


\chapter{SDSS galaxy clustering: luminosity \& color dependence and stochasticity}
\label{chap:bias}

\begin{quote}
This chapter is adapted from the paper ``SDSS galaxy clustering: luminosity and colour dependence and stochasticity'' by Molly E.~C. Swanson, Max Tegmark, Michael Blanton, and Idit Zehavi, which was previously published in the \emph{Monthly Notices of the Royal Astronomical Society} 385, pp. 1635-1655 \citep{2008MNRAS.385.1635S}.
\end{quote}

\section{Introduction}

\label{Introduction} 
In order to use galaxy surveys to study the large-scale distribution
of matter, the relation between the galaxies and the underlying matter
-- known as the \emph{galaxy bias} -- must be understood. Developing
a detailed understanding of this bias is important for two reasons:
bias is a key systematic uncertainty in the inference of cosmological
parameters from galaxy surveys, and it also has implications for galaxy
formation theory.

Since it is difficult to measure the dark matter distribution directly,
we can gain insight by studying \emph{relative bias}, i.e., the relation
between the spatial distributions of different galaxy subpopulations.
There is a rich body of literature on this subject tracing back many
decades (see, e.g., \citealt{1931ApJ....74...43H,1976ApJ...208...13D,1988ApJ...331L..59H,1988ApJ...333L..45W,1994ApJ...431..569P,1995ApJ...442..457L}),
and been studied extensively in recent years as well, both theoretically
\citep{2001MNRAS.325.1359S,2003MNRAS.340..771V,2005MNRAS.363..337C,2005astro.ph.11773S,2007ApJ...659..877T}
and observationally. Such studies have established that biasing depends
on the type of galaxy under consideration -- for example, early-type,
red galaxies are more clustered than late-type, blue galaxies \citep{1997ApJ...489...37G,2002MNRAS.332..827N,2003MNRAS.344..847M,2005MNRAS.356..456C,2006MNRAS.368...21L,2007MNRAS.379.1562C},
and luminous galaxies are more clustered than dim galaxies \citep{1998AJ....115..869W,2001MNRAS.328...64N,2004ApJ...606..702T,2005ApJ...630....1Z,2005PhRvD..71d3511S,2006MNRAS.369...68S}.
Since different types of galaxies do not exactly trace each other,
it is thus impossible for them all to be exact tracers of the underlying
matter distribution.

More quantitatively, the luminosity dependence of bias has been measured
in the 2 Degree Field Galaxy Redshift Survey (2dFGRS; \citealt{2001MNRAS.328.1039C})
\citep{2001MNRAS.328...64N,2002MNRAS.332..827N} and in the Sloan
Digital Sky Survey (SDSS; \citealt{2000AJ....120.1579Y,2002AJ....123..485S})
\citep{2004ApJ...606..702T,2005ApJ...630....1Z,2006MNRAS.368...21L}
as well as other surveys, and it is generally found that luminous
galaxies are more strongly biased, with the difference becoming more
pronounced above $L_{*}$, the characteristic luminosity of a galaxy
in the Schechter luminosity function \citep{1976ApJ...203..297S}.

These most recent studies measured the bias from ratios of correlation
functions or power spectra. The variances of clustering estimators
like correlation functions and power spectra are well-known to be
the sum of two physically separate contributions: Poisson shot noise
(due to the sampling of the underlying continuous density field with
a finite number of galaxies) and sample variance (due to the fact
that only a finite spatial volume is probed). On the large scales
most relevant to cosmological parameter studies, sample variance dominated
the aforementioned 2dFGRS and SDSS measurements, and therefore dominated
the error bars on the inferred bias.

This sample variance is easy to understand: if the power spectrum
of distant luminous galaxies is measured to be different than that
of nearby dim galaxies, then part of this measured bias could be due
to the nearby region happening to be more/less clumpy than the distant
one. In this chapter, we will eliminate this annoying sample variance
by comparing how different galaxies cluster in the \textit{same} region
of space, extending the counts-in-cells work of \citet{1999ApJ...518L..69T},
\citet{2000ApJ...544...63B}, and \citet{2005MNRAS.356..247W} and
the correlation function work of \citet{2001MNRAS.328...64N}, \citet{2002MNRAS.332..827N},
\citet{2005ApJ...630....1Z}, and \citet{2006MNRAS.368...21L}. Here
we use the counts-in-cells technique: we divide the survey volume
into roughly cubical cells and compare the number of galaxies of each
type within each cell. This yields a local, point-by-point measure
of the relative bias rather than a global one as in the correlation
function method. In other words, by comparing two galaxy density fields
directly in real space, including the phase information that correlation
function and power spectrum estimators discard, we are able to provide
substantially sharper bias constraints.


This local approach also enables us to quantify so-called stochastic
bias \citep{1998ApJ...504..601P,1998ApJ...500L..79T,1999ApJ...520...24D,1999ApJ...525..543M}.
It is well-known that the relation between galaxies and dark matter
or between two different types of galaxies is not necessarily deterministic
-- galaxy formation processes that depend on variables other than
the local matter density give rise to stochastic bias as described
in \citet{1998ApJ...504..601P}, \citet{1998ApJ...500L..79T}, \citet{1999ApJ...520...24D},
and \citet{1999ApJ...525..543M}. Evidence for stochasticity in the
relative bias between early-type and late-type galaxies has been presented
in \citet{2005MNRAS.356..247W}, \citet{2005MNRAS.356..456C}, \citet{1999ApJ...518L..69T},
and \citet{2000ApJ...544...63B}. Additionally, \citet{2007A&A...461..861S}
finds evidence for stochastic bias between galaxies and dark matter
via weak lensing. The time evolution of such stochastic bias has been
modeled in \citet{1998ApJ...500L..79T} and was recently updated
in \citet{2005A&A...430..827S}. Stochasticity is even predicted in
the relative bias between virialized clumps of dark matter (halos)
and the linearly-evolved dark matter distribution \citep{2002MNRAS.333..730C,2004MNRAS.355..129S}.
Here we aim to test whether stochasticity is necessary for modeling
the luminosity-dependent or the color-dependent relative bias.

In this chapter, we study the relative bias as a function of scale using
a simple stochastic biasing model by comparing pairs of SDSS galaxy
subsamples in cells of varying size. Such a study is timely for two
reasons. First of all, the galaxy power spectrum has recently been
measured to high precision on large scales with the goal of constraining
cosmology \citep{2004ApJ...606..702T,2006PhRvD..74l3507T,2007MNRAS.374.1527B,2007MNRAS.378..852P}.
As techniques continue to improve and survey volumes continue to grow,
it is necessary to reduce systematic uncertainties in order to keep
pace with shrinking statistical uncertainties. A detailed understanding
of complications due to the dependence of galaxy bias on scale, luminosity,
and color will be essential for making precise cosmological inferences
with the next generation of galaxy redshift surveys \citep{2004MNRAS.347..645P,2005ApJ...625..613A,2007ApJ...657..645P,2007ApJ...659....1Z,2007MNRAS.379.1195M,2007PhRvD..75h3510K}.

Secondly, a great deal of theoretical progress on models of galaxy
formation has been made in recent years, and 2dFGRS and SDSS contain
a large enough sample of galaxies that we can now begin to place robust
and detailed observational constraints on these models. The framework
known as the halo model \citep{2000MNRAS.318..203S} (see \citealt{2002PhR...372....1C}
for a comprehensive review) provides the tools needed to make comparisons
between theory and observations. The halo model assumes that all galaxies
form in dark matter halos, so the galaxy distribution can be modeled
by first determining the halo distribution -- either analytically
\citep{1998MNRAS.297..692C,2006PhRvD..74j3512M,2007PhRvD..75f3512S}
or using $N$-body simulations \citep{2003MNRAS.341.1311S,2004ApJ...609...35K,2006astro.ph..7463K}
-- and then populating the halos with galaxies. This second step
can be done using semi-analytical galaxy formation models \citep{2001MNRAS.320..289S,2003ApJ...593....1B,2006MNRAS.365...11C,2006RPPh...69.3101B}
or with a statistical approach using a model for the halo occupation
distribution (HOD) \citep{2000MNRAS.318.1144P,2002ApJ...575..587B,2005PhRvD..71f3001S}
or conditional luminosity function (CLF) \citep{2003MNRAS.339.1057Y,2003MNRAS.340..771V}
which prescribes how galaxies populate halos.

Although there are some concerns that the halo model does not capture
all of the relevant physics \citep{2006ApJ...638L..55Y,2006astro.ph..1090C,2007MNRAS.377L...5G},
it has been applied successfully in a number of different contexts
\citep{2003MNRAS.339..410S,2005MNRAS.361..415C,2006ApJ...647..737T,2006MNRAS.369...68S}.
The correlation between a galaxy's environment (i.e., the local density
of surrounding galaxies) and its color and luminosity (\citealt{2004ApJ...601L..29H};
Blanton et al.~2005a)\nocite{2005ApJ...629..143B} has been interpreted
in the context of the halo model \citep{2005ApJ...629..625B,2006ApJ...645..977B,2006MNRAS.372.1749A},
and \citet{2003MNRAS.340..771V} and \citet{2005MNRAS.363..337C}
make predictions for the bias as a function of galaxy type and luminosity
using the CLF formalism. Additionally, \citet{2005ApJ...630....1Z},
\citet{2003MNRAS.346..186M}, and \citet{2005MNRAS.364.1327A} use
correlation function methods to study the luminosity and color dependence
of galaxy clustering, and interpret the results using the Halo Occupation
Distribution (HOD) framework. The analysis presented here is complementary
to this body of work in that the counts-in-cells method is sensitive
to larger scales, uses a different set of assumptions, and compares
the two density fields directly in each cell rather than comparing
ratios of their second moments. The halo model provides a natural
framework in which to interpret the luminosity and color dependence
of galaxy biasing statistics we measure here.

The rest of this chapter is organized as follows: \S\ref{SDSS-Galaxy-Data}
describes our galaxy data, and \S\ref{sub:Overlapping-Volume-Limited-Samples}~and~\S\ref{sub:Counts-in-Cells-Methodology}
describe the construction of our galaxy samples and the partition
of the survey volume into cells. In \S\ref{sub:Relative-Bias-Framework}
we outline our relative bias framework, and in \S\ref{sub:The-Null-buster-Test}~and~\S\ref{sub:Maximum-Likelihood-Method}
we describe our two main analysis methods. We present our results
in \S\ref{Results} and conclude with a qualitative interpretation
of our results in the halo model context in \S\ref{Conclusions}.

\section{SDSS Galaxy Data}

\label{SDSS-Galaxy-Data} The SDSS \citep{2000AJ....120.1579Y,2002AJ....123..485S}
uses a mosaic CCD camera \citep{1998AJ....116.3040G} on a dedicated
telescope \citep{2006AJ....131.2332G} to image the sky in five photometric
bandpasses denoted $u$, $g$, $r$, $i$ and $z$ \citep{1996AJ....111.1748F}.
After astrometric calibration \citep{2003AJ....125.1559P}, photometric
data reduction \citep{2001ASPC..238..269L}, and photometric calibration
\citep{2001AJ....122.2129H,2002AJ....123.2121S,2004AN....325..583I,2006AN....327..821T},
galaxies are selected for spectroscopic observations. To a good approximation,
the main galaxy sample consists of all galaxies with $r$-band apparent
Petrosian magnitude $r<17.77$ after correction for reddening as per
\citet{1998ApJ...500..525S}; there are about 90 such galaxies per
square degree, with a median redshift of 0.1 and a tail out to $z\sim0.25$.
Galaxy spectra are also measured for the Luminous Red Galaxy sample
\citep{2001AJ....122.2267E}, which is not used in this chapter. These
targets are assigned to spectroscopic plates of diameter $2.98^{\circ}$
by an adaptive tiling algorithm (Blanton et al.~2003b)\nocite{2003AJ....125.2276B}
and observed with a pair of CCD spectrographs \citep{2004SPIE.5492.1411U},
after which the spectroscopic data reduction and redshift determination
are performed by automated pipelines. The rms galaxy redshift errors
are of order $30\,\rmn{km\, s^{-1}}$ for main galaxies, hence negligible
for the purposes of the present chapter.

Our analysis is based on 380,614 main galaxies (the `safe0' cut) from
the 444,189 galaxies in the 5th SDSS data release (`DR5') \citep{2007ApJS..172..634A},
processed via the SDSS data repository at New York University (Blanton
et al.~2005b)\nocite{2005AJ....129.2562B}. The details of how these
samples were processed and modeled are given in Appendix A of \citet{2004ApJ...606..702T}
and in \citet{2005ApJ...633..560E}. The bottom line is that each
sample is completely specified by three entities:

\begin{enumerate}
\item The galaxy positions (RA, Dec, and comoving redshift space distance
$r$ for each galaxy) ;
\item The radial selection function $\bar{n}\left(r\right)$, which gives
the expected (not observed) number density of galaxies as a function
of distance ;
\item The angular selection function $\bar{n}\left(\hat{\bmath{r}}\right)$,
which gives the completeness as a function of direction in the sky,
specified in a set of spherical polygons \citep{2004MNRAS.349..115H}. 
\end{enumerate}
The three-dimensional selection functions of our samples are separable,
i.e., simply the product $\bar{n}\left(\bmath{r}\right)=\bar{n}\left(\hat{\bmath{r}}\right)\bar{n}\left(r\right)$
of an angular and a radial part; here $r\equiv|\bmath{r}|$ and $\hat{\bmath{r}}\equiv\bmath{r}/r$
are the radial comoving distance and the unit vector corresponding
to the position $\bmath{r}$. The volume-limited samples used in this
chapter are constructed so that their radial selection function $\bar{n}\left(r\right)$
is constant over a range of $r$ and zero elsewhere. The effective
sky area covered is $\Omega\equiv\int\bar{n}\left(\hat{\bmath{r}}\right)d\Omega\approx5183$
square degrees, and the typical completeness $\bar{n}\left(\hat{\bmath{r}}\right)$
exceeds 90 per cent. The conversion from redshift $z$ to comoving
distance was made for a flat $\Lambda\rmn{CDM}$ cosmological model
with $\Omega_{m}=0.25$. 
Additionally, we make a first-order correction for redshift space
distortions by applying the finger-of-god compression algorithm described
in \citet{2004ApJ...606..702T} with a threshold density of $\delta_{c}=200$.

\section{Analysis Methods}

\label{Analysis-Methods}

\subsection{\label{sub:Overlapping-Volume-Limited-Samples}Overlapping Volume-Limited
Samples}

The basic technique used in this chapter is pairwise comparison of the
three-dimensional density fields of galaxy samples with different
colors and luminosities. As in \citet{2005ApJ...630....1Z}, we focus
on these two properties (as opposed to morphological type, spectral
type, or surface brightness) for two reasons: they are straightforward
to measure from SDSS data, and recent work (Blanton et al.~2005a)
has found that luminosity and color is the pair of properties that
is most predictive of the local overdensity. Since color and spectral
type are strongly correlated, our study of the color dependence of
bias probes similar physics as studies using spectral type \citep{1999ApJ...518L..69T,2000ApJ...544...63B,2002MNRAS.332..827N,2005MNRAS.356..247W,2005MNRAS.356..456C}.

Our base sample of SDSS galaxies (`safe0') has an $r$-band apparent
magnitude range of $14.5<r<17.5$. Following the method used in \citet{2004ApJ...606..702T},
we created a series of volume-limited samples containing galaxies
in different luminosity ranges. These samples are defined by selecting
a range of absolute magnitude $M_{\rmn{luminous}}<M_{^{0.1}r}<M_{\rmn{dim}}$
and defining a redshift range such that the near limit has $M_{^{0.1}r}=M_{\rmn{luminous}}$,
$r=14.5$ and the far limit has $M_{^{0.1}r}=M_{\rmn{dim}}$, $r=17.5$.
Thus by discarding all galaxies outside the redshift range, we are
left with a sample with a uniform radial selection function $\bar{n}\left(r\right)$
that contains all of the galaxies in the given absolute magnitude
range in the volume defined by the redshift limits. Here $M_{^{0.1}r}$
is defined as the absolute magnitude in the $r$-band shifted to a
redshift of $z=0.1$ (Blanton et al.~2003a)\nocite{2003AJ....125.2348B}.
\begin{figure}
\begin{center}%
\includegraphics[width=0.4746\textwidth]{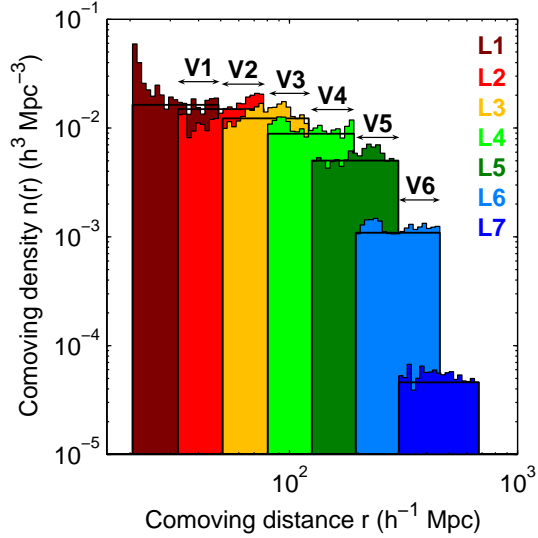}
\end{center}%
\caption[Histogram of the comoving number density of the volume-limited samples.]{\label{fig:vlim_hist}Histogram of the comoving number density (after
finger-of-god compression) of the volume-limited samples L1-L7. The
cuts used to define these samples are shown in Table~\ref{tab:cuts}.
Note that the radial selection function $\bar{n}\left(r\right)$ is
uniform over the allowed range for each sample. Arrows indicate volumes
V1-V6 where neighboring volume-limited samples overlap.}
\end{figure}

Our volume-limited samples are labeled L1 through L7, with L1 being
the dimmest and L7 being the most luminous. Figure~\ref{fig:vlim_hist}
shows a histogram of the comoving galaxy density $n\left(r\right)$
for L1-L7. The cuts used to make these samples are shown in Table~\ref{tab:cuts}.%
\begin{table*}
\begin{center}

\caption{\label{tab:cuts}Summary of cuts used to create luminosity-binned
volume-limited samples.}

\begin{tabular}{cccc}
\hline 
Luminosity-binned&
&
&
\tabularnewline
volume-limited&
&
&
Comoving number\tabularnewline
samples&
Absolute magnitude&
Redshift&
density $\bar{n}$ $\left(h^{3}{\rmn{Mpc}}^{-3}\right)$\tabularnewline
\hline 
L1&
$-17<M_{^{0.1}r}<-16$&
$0.007<z<0.016$&
$\left(1.63\pm0.05\right)\times10^{-2}$\tabularnewline
L2&
$-18<M_{^{0.1}r}<-17$&
$0.011<z<0.026$&
$\left(1.50\pm0.03\right)\times10^{-2}$\tabularnewline
L3&
$-19<M_{^{0.1}r}<-18$&
$0.017<z<0.041$&
$\left(1.23\pm0.01\right)\times10^{-2}$\tabularnewline
L4&
$-20<M_{^{0.1}r}<-19$&
$0.027<z<0.064$&
$\left(8.86\pm0.05\right)\times10^{-3}$\tabularnewline
L5&
$-21<M_{^{0.1}r}<-20$&
$0.042<z<0.100$&
$\left(5.02\pm0.02\right)\times10^{-3}$\tabularnewline
L6&
$-22<M_{^{0.1}r}<-21$&
$0.065<z<0.152$&
$\left(1.089\pm0.005\right)\times10^{-3}$\tabularnewline
L7&
$-23<M_{0.1_{r}}<-22$&
$0.101<z<0.226$&
$\left(4.60\pm0.06\right)\times10^{-5}$\tabularnewline
\hline
\end{tabular}

\end{center}
\end{table*}

Each sample overlaps spatially only with the samples in neighboring
luminosity bins -- since the apparent magnitude range spans three
magnitudes and the absolute magnitude ranges for each bin span one
magnitude, the far redshift limit of a given luminosity bin is approximately
equal to the near redshift limit of the bin two notches more luminous.
(It is not precisely equal due to evolution and K-corrections.)

The regions where neighboring volume-limited samples overlap provide
a clean way to select data for studying the luminosity-dependent bias.
By using only the galaxies in the overlapping region from each of
the two neighboring luminosity bins, we obtain two sets of objects
(one from the dimmer bin and one from more luminous bin) whose selection
is volume-limited and redshift-independent. Furthermore, since they
occupy the same volume, they are correlated with the same underlying
matter distribution, which eliminates uncertainty due to sample variance
and removes potential systematic effects due to sampling different
volume sizes \citep{2005A&A...443...11J}. We label the overlapping
volume regions V1 through V6, where V1 is defined as the overlap between
L1 and L2, and so forth. The redshift ranges for V1-V6 are shown Table~\ref{tab:vols}.%
\begin{table}
\begin{center}
\caption{\label{tab:vols}Overlapping volumes in which neighboring luminosity
bins are compared.}

\begin{tabular}{ccc}
\hline 
Pairwise comparison &
Overlapping&
\tabularnewline
(overlapping) volumes&
bins&
Redshift\tabularnewline
\hline 
V1&
L1 \& L2&
$0.011<z<0.016$\tabularnewline
V2&
L2 \& L3&
$0.017<z<0.026$\tabularnewline
V3&
L3 \& L4&
$0.027<z<0.041$\tabularnewline
V4&
L4 \& L5&
$0.042<z<0.064$\tabularnewline
V5&
L5 \& L6&
$0.065<z<0.100$\tabularnewline
V6&
L6 \& L7&
$0.101<z<0.152$\tabularnewline
\hline
\end{tabular}
\end{center}
\end{table}

\begin{figure}
\begin{center}%
\includegraphics[width=0.4746\textwidth]{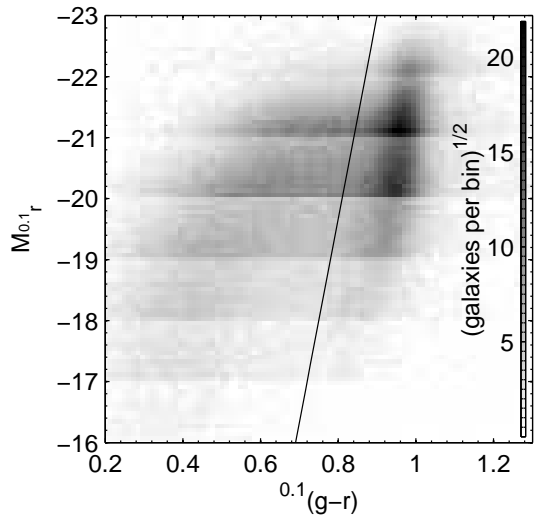}
\end{center}%
\caption[Color-magnitude diagram showing the number
density distribution of the galaxies in the volume-limited samples.]{\label{fig:colormag}Color-magnitude diagram showing the number
density distribution of the galaxies in the volume-limited samples.
The shading scale has a square-root stretch, with darker areas indicating
regions of higher density. The line shows the color cut of $^{0.1}\left(g-r\right)=0.9-0.03\left(M_{^{0.1}r}+23\right)$.
We refer to galaxies falling to the left of this line as blue and
ones falling to the right of the line as red. }
\end{figure}

To study the color dependence of the bias, we further divide each
sample into red galaxies and blue galaxies. Figure~\ref{fig:colormag}
shows the galaxy distribution of our volume-limited samples on a color-magnitude
diagram. The sharp boundaries between the different horizontal slices
are due to the differences in density and total volume sampled in
each luminosity bin. This diagram illustrates the well-known color
bimodality, with the redder galaxies falling predominantly in a region
commonly known as the E-S0 ridgeline \citep{2004ApJ...615L.101B,2006MNRAS.370..721M,2006MNRAS.373..469B}.
To separate the E-S0 ridgeline from the rest of the population, we
use the same magnitude-dependent color cut as in \citet{2005ApJ...630....1Z}:
we define galaxies with $^{0.1}\left(g-r\right)<0.9-0.03\left(M_{^{0.1}r}+23\right)$
to be blue and galaxies on the other side of this line to be red.%

\begin{figure*}
\includegraphics[width=1\textwidth]{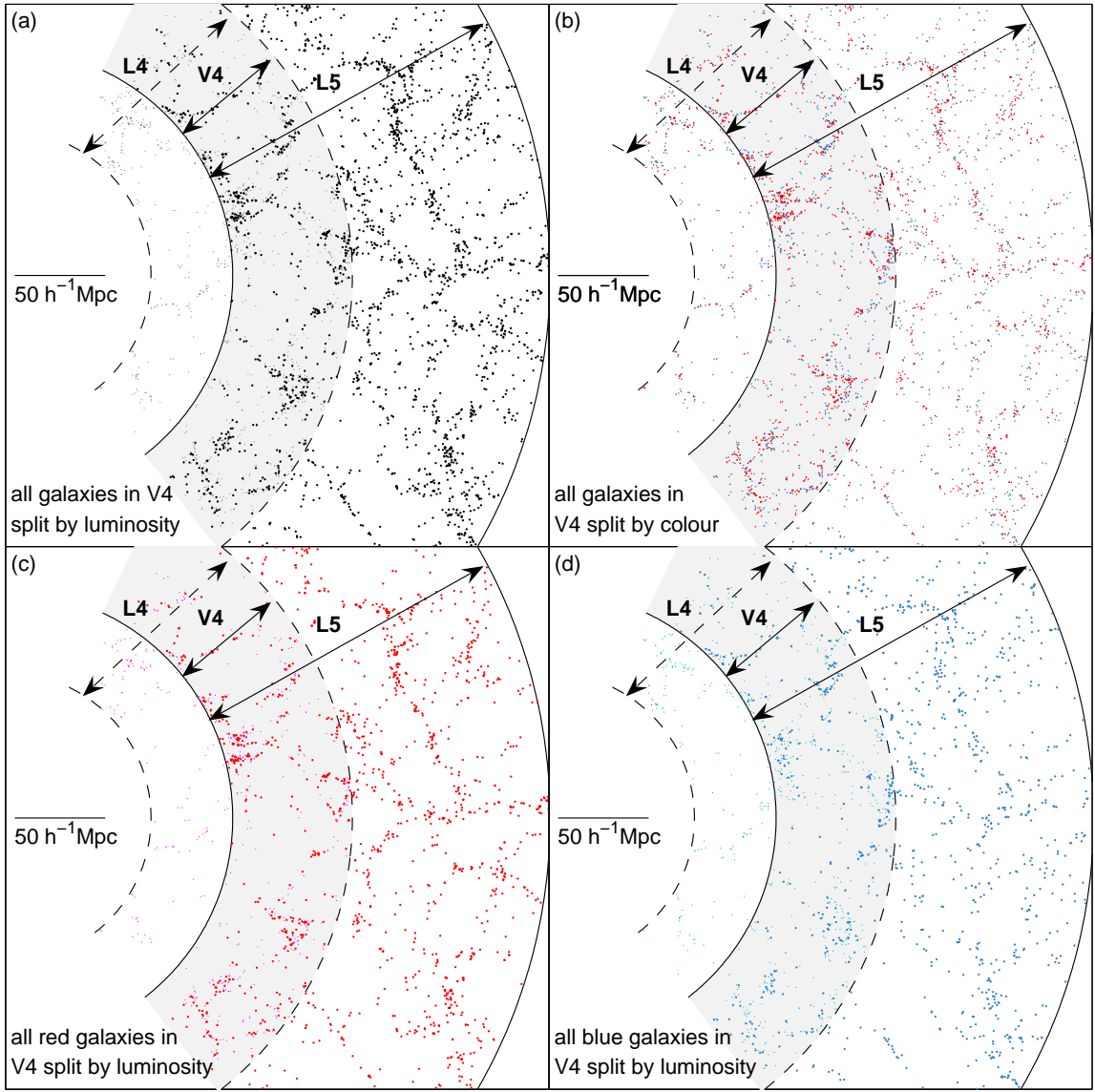}

\caption[Galaxy distributions
plotted in comoving spatial coordinates for a radial slice, showing the four different types of pairwise comparisons made.]{\label{fig:4pairwise}Galaxy distributions (after finger-of-god compression)
plotted in comoving spatial coordinates for a radial slice of the
volume-limited samples L4 (smaller dots, radial boundaries denoted
by dashed lines) and L5 (larger dots, radial boundaries denoted by
solid lines), which overlap in volume V4. Four different types of
pairwise comparisons are illustrated: (a) luminous galaxies (L5) vs.
dim galaxies (L4), (b) red galaxies vs. blue galaxies (both in V4),
(c) luminous red galaxies (L5) vs. dim red galaxies (L4), and (d)
luminous blue galaxies (L5) vs. dim blue galaxies (L4). The shaded
regions denote the volume in which the two sets of galaxies are compared.
A simple visual inspection shows that the different samples of galaxies
being compared generally appear to cluster in the same physical locations
-- one key question we aim to answer here is if these observed correlations
can be described with a simple linear bias model.}
\end{figure*}

In each volume V1-V6, we make four separate pairwise comparisons:
luminous galaxies vs. dim galaxies, red galaxies vs. blue galaxies,
luminous red galaxies vs. dim red galaxies, and luminous blue galaxies
vs. dim blue galaxies. The luminous vs. dim comparisons measures the
relative bias between galaxies in neighboring luminosity bins, and
from this we can extract the luminosity dependence of the bias for
all galaxies combined and for red and blue galaxies separately. The
red vs. blue comparison measures the color-dependent bias. This set
of four different types of pairwise comparisons is illustrated in
Fig.~\ref{fig:4pairwise} for V4, and the number of galaxies in each
sample being compared is shown in Table~\ref{tab:counts}.%
\begin{table}
\begin{center}
\caption{\label{tab:counts}Number of galaxies in each sample being compared.}

\begin{tabular}{c|ff||ff}
\hline 
&
\multicolumn{2}{c||}{All split by luminosity}&
\multicolumn{2}{c}{All split by color}\tabularnewline
&
\multicolumn{1}{c}{Luminous}&
\multicolumn{1}{c||}{Dim}&
\multicolumn{1}{c}{Red}&
\multicolumn{1}{c}{Blue}\tabularnewline
\hline 
V1&
427&
651&
125&
953\tabularnewline
V2&
2102&
2806&
1117&
3791\tabularnewline
V3&
6124&
8273&
5147&
9250\tabularnewline
V4&
12122&
23534&
17144&
18512\tabularnewline
V5&
11202&
53410&
37472&
27140\tabularnewline
V6&
1784&
38920&
27138&
13566\tabularnewline
\hline
\hline 
&
\multicolumn{2}{c||}{Red split by luminosity}&
\multicolumn{2}{c}{Blue split by luminosity}\tabularnewline
&
\multicolumn{1}{c}{Red luminous}&
\multicolumn{1}{c||}{Red dim}&
\multicolumn{1}{c}{Blue luminous}&
\multicolumn{1}{c}{Blue dim}\tabularnewline
\hline 
V1&
72&
53&
355&
598\tabularnewline
V2&
620&
497&
1482&
2309\tabularnewline
V3&
2797&
2350&
3327&
5923\tabularnewline
V4&
6848&
10296&
5274&
13238\tabularnewline
V5&
7514&
29958&
3688&
23452\tabularnewline
V6&
1451&
25687&
333&
13233\tabularnewline
\hline
\end{tabular}
\end{center}
\end{table}

\subsection{\label{sub:Counts-in-Cells-Methodology}Counts-in-Cells Methodology}

To compare the different pairs of galaxy samples, we perform a counts-in-cells
analysis: we divide each comparison volume into roughly cubical cells
and use the number of galaxies of each type in each cell as the primary
input to our statistical analysis. This method is complementary to
studies based on the correlation function since it involves point-by-point
comparison of the two density fields and thus provides a more direct
test of the local deterministic linear bias hypothesis. We probe scale
dependence by varying the size of the cells.

\begin{figure*}
\includegraphics[width=1\textwidth]{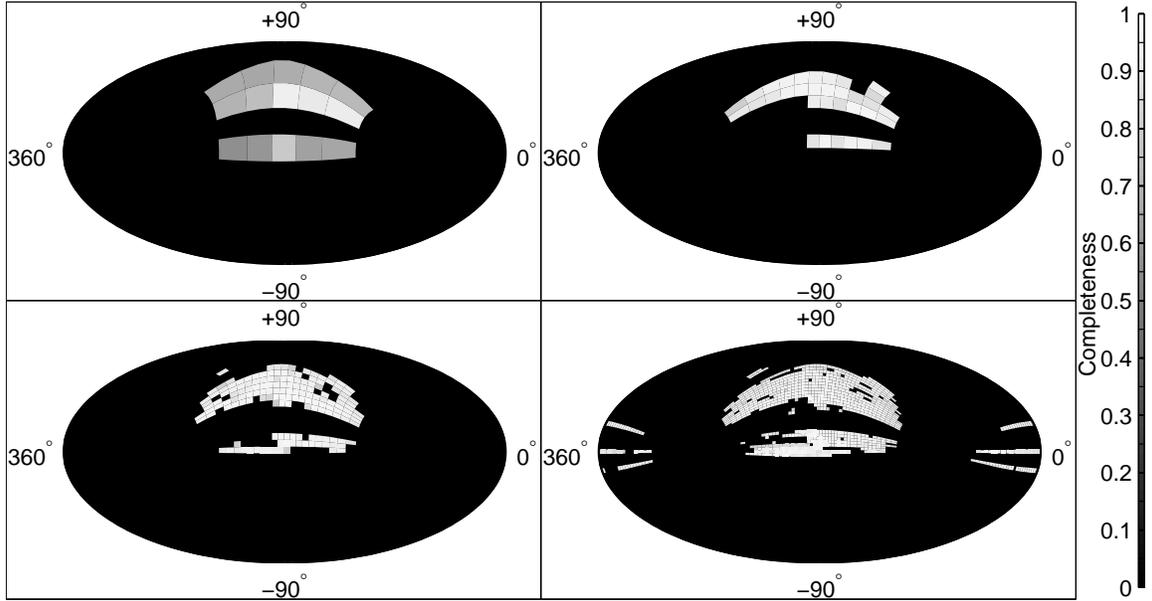}

\caption[The SDSS DR5 angular mask pixelized at the
four different resolutions used to partition the survey into cells.
]{\label{fig:angularcells}The SDSS DR5 angular mask pixelized at the
four different resolutions used to partition the survey into cells,
shown in Hammer-Aitoff projection in equatorial coordinates. Shading
indicates completeness level: 0 per cent is black, 100 per cent is
white.}
\end{figure*}

To create our cells, we first divide the sky into two-dimensional
`pixels' at four different angular resolutions using the SDSSPix pixelization
scheme%
\footnote{See http://lahmu.phyast.pitt.edu/~scranton/SDSSPix/%
} as implemented by the updated version of the angular mask processing
software \noun{mangle} discussed in Chapter~\ref{chap:mangle}
\citep{2004MNRAS.349..115H,2007arXiv0711.4352S}.
The angular selection function $\bar{n}\left(\hat{\bmath{r}}\right)$
is averaged over each pixel to obtain the completeness. To reduce
the effects of pixels on the edge of the survey area or in regions
affected by internal holes in the survey, we apply a cut on pixel
completeness: we only use pixels with a completeness higher than 80
per cent (50 per cent for the lowest angular resolution). Figure~\ref{fig:angularcells}
shows the pixelized SDSS angular mask at our four different resolutions,
including only the pixels that pass our completeness cut. The different
angular resolutions have 15, 33, 157, and 901 of these angular pixels
respectively. At the lowest resolution, each pixel covers 353 square
degrees, and the angular area of the pixels decreases by a factor
of $1/4$ at each resolution level, yielding pixels covering 88, 22,
and 5 square degrees at the three higher resolutions.%

To produce three-dimensional cells from our pixels, we divide each
comparison volume into radial shells of equal volume. We choose the
number of radial subdivisions at each angular resolution in each comparison
volume such that our cells are approximately cubical, i.e., the radial
extent of a cell is approximately equal to its transverse (angular)
extent. This procedure makes cells that are not quite perfect cubes
-- there is some slight variation in the cell shapes, with cells on
the near edge of the volume slightly elongated radially and cells
on the far edge slightly flattened. We state all of our results as
a function of cell size $\elll$, defined as the cube root of the
cell volume. At the lowest resolution, there is just 1 radial shell
for each volume; at the next resolution, we have 3 radial shells for
volumes V4 and V5 and 2 radial shells for the other volumes. There
are 5 radial shells at the second highest resolution, and 10 at the
highest.

\begin{figure*}
\includegraphics[width=1\textwidth]{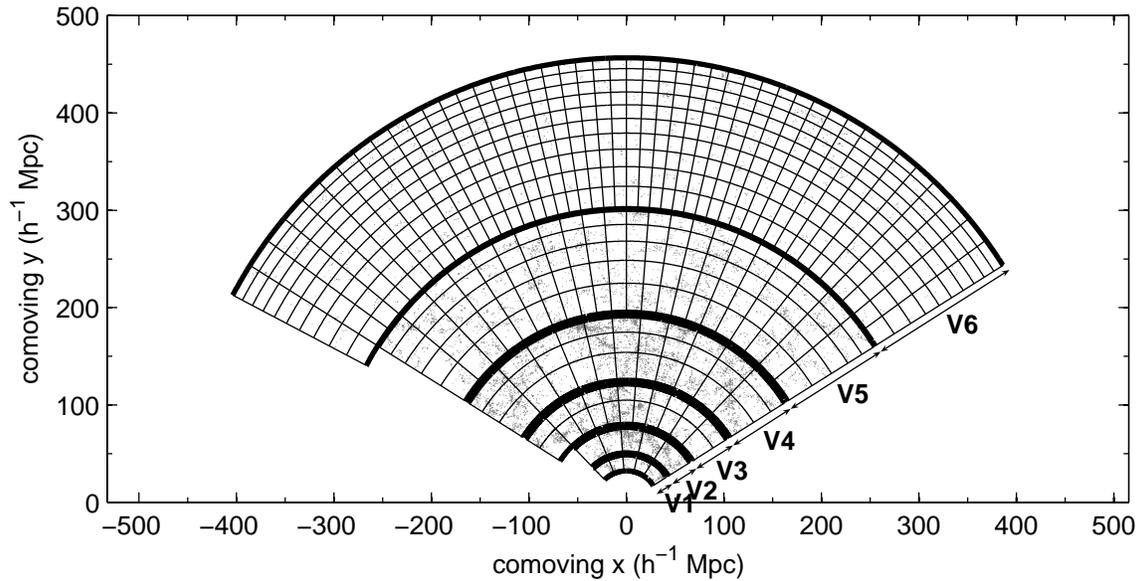}

\caption[A radial slice of the SDSS survey volume
divided into cells.]{\label{fig:radialcells}A radial slice of the SDSS survey volume
divided into cells of size $\sim20{\, h}^{-1}\rmn{Mpc}$ with the
galaxies in each cell (after finger-of-god compression) shown in grey.}
\end{figure*}

Since each comparison volume is at a different distance from us, the
angular geometry gives us cells of different physical size in each
of the volumes. At the lowest resolution, where there is only one
shell in each volume, the cell size is $14\, h^{-1}\rmn{Mpc}$ in
V1 and $134\, h^{-1}\rmn{Mpc}$ in V6. At the highest resolution,
the cell size is $1.7\, h^{-1}\rmn{Mpc}$ in V1 and $16\, h^{-1}\rmn{Mpc}$
in V6. Figure~\ref{fig:radialcells} shows the cells in each volume
V1-V6 that are closest to a size of $\sim20{\, h}^{-1}\rmn{Mpc}$,
the range in which the length scales probed by the different volumes
overlap. (These are the cells used to produce the results shown in
Fig.~\ref{fig:lum-bias}.) %

\subsection{\label{sub:Relative-Bias-Framework}Relative Bias Framework}

Our task is to quantify the relationship between two fractional overdensity
fields $\delta_{1}\left(\bmath{x}\right)\equiv\rho_{1}\left(\bmath{x}\right)/\bar{\rho}_{1}-1$
and $\delta_{2}\left(\bmath{x}\right)\equiv\rho_{2}\left(\bmath{x}\right)/\bar{\rho}_{2}-1$
representing two different types of objects. This framework is commonly
used with types (1,2) representing (dark matter, galaxies), or as
in \citet{2000ApJ...544...63B}, \citet{2005MNRAS.356..247W}, and
\citet{2005MNRAS.356..456C}, (early-type galaxies, late-type galaxies).
Here we use it to represent (more luminous galaxies, dimmer galaxies)
or (red galaxies, blue galaxies) to compare the samples described
in \S\ref{sub:Overlapping-Volume-Limited-Samples}. Galaxies are
of course discrete objects, and as customary, we use the continuous
field $\rho_{\alpha}\left(\bmath{x}\right)$ (where $\alpha$=1 or
2) to formally refer to the intensity of the Poisson point process
involved in distributing the type $\alpha$ galaxies, as described
in more detail below.

The simplest (and frequently assumed) relationship between $\delta_{1}$
and $\delta_{2}$ is linear deterministic bias: \begin{equation}
\delta_{2}\left(\bmath{x}\right)=b_{\rmn{lin}}\delta_{1}\left(\bmath{x}\right),\label{eq:linearbias}\end{equation}
 where $b_{\rmn{lin}}$ is a constant parameter. This model cannot
hold in all cases -- note that it can give negative densities if $b_{\rmn{lin}}>1$
-- but is typically a reasonable approximation on cosmologically large
length scales where the density fluctuations $\delta_{1}\ll1$, as
is the case for the measurements of the large scale power spectrum
recently used to constrain cosmological parameters \citep{2004ApJ...606..702T,2005MNRAS.362..505C,2006PhRvD..74l3507T}.
This simple model is likely to break down on length scales small enough
to be affected by galaxy formation physics, and such a breakdown could
potentially impact the cosmological interpretation of medium- and
small-scale power spectrum results from galaxy surveys.

More complicated models allow for non-linearity and stochasticity,
as described in detail in \citet{1999ApJ...520...24D}: \begin{equation}
\delta_{2}\left(\bmath{x}\right)=b\left[\delta_{1}\left(\bmath{x}\right)\right]\delta_{1}\left(\bmath{x}\right)+\epsilon\left(\bmath{x}\right),\label{eq:nonlinearbias}\end{equation}
 where the bias $b$ is now a (typically slightly non-linear) function
of $\delta_{1}$. The stochasticity is represented by a random field
$\epsilon$ -- allowing for stochasticity removes the restriction
implied by deterministic models that the peaks of $\delta_{1}$ and
$\delta_{2}$ must coincide spatially. Stochasticity is basically
the scatter in the relationship between the two density fields due
to physical variables besides the local matter density. Non-local
galaxy formation processes can also give rise to stochasticity, as
discussed in \citet{1999ApJ...525..543M}.

The theoretical model we use to describe the galaxy distribution is
based on the idea that our universe is one realization drawn from
an ensemble of all possible universes that have a given set of cosmological
parameters. Any measurement we make in our universe can be described
as a random variable over this ensemble of universes. The value we
observe is a realization of this random variable. Since we only have
one universe to observe, we cannot make repeated observations to draw
different values from the distribution. However, we can make theoretical
predictions for the probability distribution for a random variable,
and then perform statistical tests to see if it is plausible that
we would observe the particular value we do if our model is correct.
This is the basic idea behind the statistical tests detailed in \S\ref{sub:The-Null-buster-Test}
and \S\ref{sub:Maximum-Likelihood-Method}.

Our statistical model is composed of two separate random processes.
The first component is the overdensity field $\delta\left(\bmath{x}\right)$,
which is a realization of a zero-mean random field with a power spectrum
predicted by a set of cosmological parameters, as described in \S\ref{sub:randomfields}.
Here, as in \S\ref{sub:randomfields}, we use regular angle brackets
$\left\langle \right\rangle $ to denote expectation values over the
probability distribution for this random field, e.g., a zero-mean
random field $\delta\left(\bmath{x}\right)$ has $\left\langle \delta\left(\bmath{x}\right)\right\rangle =0$.
The second component is a random point process which describes the
locations of galaxies. This is typically taken to be a Poisson point
process, which means that the probability of finding $m$ galaxies
in a bounded region $A$ is given by \begin{equation}
P\left(m\:\mathrm{galaxies\: in\:}A\right)=\frac{\left[\Lambda\left(A\right)\right]^{m}}{m!}e^{-\Lambda\left(A\right)},\label{eq:poissonprocess}\end{equation}
where \begin{equation}
\Lambda\left(A\right)\equiv\int_{A}\lambda\left(\bmath{x}\right)d^{3}x.\label{eq:poissonintensity}\end{equation}
The function $\lambda\left(\bmath{x}\right)$ is the intensity function
of the point process. For describing the galaxy distribution, we use
$\lambda\left(\bmath{x}\right)=\bar{n}\left(\bmath{x}\right)\left(1+\delta\left(\bmath{x}\right)\right)$,
where $\bar{n}\left(\bmath{x}\right)$ is the selection function of
the survey as defined in \S\ref{SDSS-Galaxy-Data} -- in other words,
the expected galaxy density if there were no clustering. Expectation
values over this Poisson point process will be denoted by bold angle
brackets $\pmb{\bm{\langle\rangle}}$.

Our basic observational measurement is $\underline{N}_{\alpha}^{(i)}$,
the number of galaxies of type $\alpha$ in cell $i$ observed in
our SDSS data sample. We will denote quantities calculated using our
SDSS counts-in-cells data -- i.e., realizations of random variables
in our universe -- with underlines. Non-underlined quantities indicate
the random variables themselves, for which we compute theoretical
expectation values based on the statistical model described above.
For example, taking the expectation value over the Poisson point process
for type $\alpha$ galaxies in cell $i$ gives \begin{equation}
\pmb{\bm{\langle}}N_{\alpha}^{(i)}\pmb{\bm{\rangle}}=\bar{N}_{\alpha}^{(i)}\left(1+\delta_{\alpha}^{(i)}\right)\label{eq:poiss_expectation}\end{equation}
where $\delta_{\alpha}^{(i)}$ is the volume average of $\delta_{\alpha}\left(\bmath{x}\right)$
over cell $i$ and $\bar{N}_{\alpha}^{(i)}$ is the integral of the
selection function $\bar{n}\left(\bmath{x}\right)$ over cell $i$.
Note that aside from slight variations in the angular selection function,
$\bar{N}_{\alpha}^{(i)}$ is the same for all cells $i$ -- it is
simply the expected number of type $\alpha$ galaxies in cell $i$
in the absence of any clustering. Taking the expectation over the
random field realization as well gives

\begin{equation}
\left\langle \pmb{\bm{\langle}}N_{\alpha}^{(i)}\pmb{\bm{\rangle}}\right\rangle =\bar{N}_{\alpha}^{(i)}.\label{eq:full_expectation}\end{equation}

\begin{sloppypar}We estimate the overdensity of galaxies of type
$\alpha$ in cell $i$ by \begin{equation}
\underline{g}_{\alpha}^{(i)}\equiv\frac{\underline{N}_{\alpha}^{(i)}-\bar{N}_{\alpha}^{(i)}}{\bar{N}_{\alpha}^{(i)}}.\label{eq:countsincells}\end{equation}
 The $n$-dimensional vectors \begin{equation}
{\bmath{\underline{g}}}_{\alpha}\equiv\left(\begin{array}{c}
\underline{g}_{\alpha}^{(1)}\\
\vdots\\
\underline{g}_{\alpha}^{(n)}\end{array}\right)\label{eq:g_vec1}\end{equation}
 contain the counts-in-cells data to which we apply the statistical
analyses in \S\ref{sub:The-Null-buster-Test}~and~\S\ref{sub:Maximum-Likelihood-Method}.
\end{sloppypar}

${\bmath{\underline{g}}}_{\alpha}$ is the realization of the random
variable $\bmath{g}_{\alpha}$ for our particular universe that we
measure from the SDSS data. $ $The covariance matrix of $\bmath{g}_{\alpha}$,
which will be the same for any universe drawn from the ensemble defined
by our cosmological model, is given by\begin{equation}
\left\langle \pmb{\bm{\langle}}g_{\alpha}^{(i)}g_{\beta}^{(j)}\pmb{\bm{\rangle}}\right\rangle =\left\langle \delta_{\alpha}^{(i)}\delta_{\beta}^{(j)}\right\rangle +\delta_{\alpha\beta}{\mathbfss{N}}_{\alpha}^{ij},\label{eq:gdef}\end{equation}
where $\delta_{\alpha\beta}$ is a Kronecker delta and ${\mathbfss{N}}_{\alpha}$
is known as the shot noise covariance matrix. The presence of $\delta_{\alpha\beta}$
implies that we are assuming that type 1 galaxies and type 2 galaxies
are distributed via \emph{independent} point processes, so their shot
noise is uncorrelated -- the correlations between type 1 and type
2 galaxies are encoded in the relationship between between $\delta_{1}\left(\bmath{x}\right)$
and $\delta_{2}\left(\bmath{x}\right)$ as in equations \eqref{eq:linearbias}
and \eqref{eq:nonlinearbias}. Although one might expect the fact
that the counts of type 1 and type 2 galaxies in a cell is constrained
to be equal to the total number of galaxies in the cell could induce
correlations in the shot noise, we do not explicitly use the combined
total count in our analyses -- uncorrelated shot noise is thus a reasonable
assumption.

For Poisson point processes, we have \begin{equation}
{\mathbfss{N}}_{\alpha}^{ij}\equiv\delta_{ij}/\bar{N}_{\alpha}^{(i)}.\label{eq:shotnoise}\end{equation}
For comparing pairs of different types of galaxies, we construct the
$2n$-dimensional data vector \begin{equation}
\bmath{\underline{g}}\equiv\left(\begin{array}{c}
{\bmath{\underline{g}}}_{1}\\
{\bmath{\underline{g}}}_{2}\end{array}\right),\label{eq:g_vec}\end{equation}
 which, under our theoretical statistical model described above, has
covariance matrix \begin{equation}
\mathbfss{C}\equiv\left\langle \pmb{\bm{\langle}}\bmath{g}{\bmath{g}}^{T}\pmb{\bm{\rangle}}\right\rangle =\mathbfss{S}+\mathbfss{N},\label{eq:cov_matrix}\end{equation}
 with\begin{equation}
\mathbfss{N}\equiv\left(\begin{array}{cc}
{\mathbfss{N}}_{1} & 0\\
0 & {\mathbfss{N}}_{2}\end{array}\right),\qquad\mathbfss{S}\equiv\left(\begin{array}{cc}
{\mathbfss{S}}_{11} & {\mathbfss{S}}_{12}\\
{\mathbfss{S}}_{12} & {\mathbfss{S}}_{22}\end{array}\right),\label{eq:N_and_S}\end{equation}
 and the elements of the matrix $\mathbfss{S}$ given by

\begin{equation}
{\mathbfss{S}}_{\alpha\beta}^{ij}=\left\langle \delta_{\alpha}^{(i)}\delta_{\beta}^{(j)}\right\rangle .\label{eq:s_matrices}\end{equation}
 $\mathbfss{N}$ is a diagonal matrix, which indicates that the shot
noise is uncorrelated between different cells and also between type
1 and type 2 galaxies within the same cell. 

Regarding the matrix $\mathbfss{S}$, other counts-in-cells analyses
often assume that the cosmological correlations between different
cells can be ignored, i.e., $\left\langle \delta_{\alpha}^{(i)}\delta_{\beta}^{(j)}\right\rangle =0$
unless $i=j$. Here we account for cosmological correlations by computing
the elements of $\mathbfss{S}$ using the best-fitting $\Lambda$CDM
matter power spectrum as we will now describe in detail. The power
spectrum $P_{\alpha\beta}\left(\bmath{k}\right)$ is defined as $\left\langle \hat{\delta}_{\alpha}\left(\bmath{k}\right)\hat{\delta}_{\beta}\left({\bmath{k}}^{\prime}\right)^{\dagger}\right\rangle =\left(2\pi\right)^{3}\delta^{D}\left(\bmath{k}-{\bmath{k}}^{\prime}\right)P_{\alpha\beta}\left(\bmath{k}\right)$,
where $\hat{\delta}_{\alpha}\left(\bmath{k}\right)\equiv\int e^{-i\bmath{k}\cdot\bmath{x}}\delta_{\alpha}\left(\bmath{x}\right)d^{3}\bmath{x}$
is the Fourier transform of the overdensity field. $P_{11}\left(\bmath{k}\right)$
and $P_{22}\left(\bmath{k}\right)$ are the power spectra of type
1 and 2 galaxies respectively, and $P_{12}\left(\bmath{k}\right)$
is the cross spectrum between type 1 and 2 galaxies. We assume isotropy
and homogeneity, so that $P_{\alpha\beta}\left(\bmath{k}\right)$
is a function only of $k\equiv\left|\bmath{k}\right|$, and rewrite
the galaxy power spectra in terms of the matter power spectrum $P\left(k\right)$:

\begin{eqnarray}
P_{11}\left(k\right) & = & b_{1}\left(k\right)^{2}P\left(k\right)\nonumber \\
P_{12}\left(k\right) & = & {b_{1}\left(k\right)b}_{2}\left(k\right)r_{12}\left(k\right)P\left(k\right)\nonumber \\
P_{22}\left(k\right) & = & b_{2}\left(k\right)^{2}P\left(k\right),\label{eq:Pks}\end{eqnarray}
 which defines the functions $b_{1}\left(k\right)$, $b_{2}\left(k\right)$,
and $r_{12}\left(k\right)$.

To calculate $\left\langle \delta_{\alpha}^{(i)}\delta_{\beta}^{(j)}\right\rangle $
exactly, we need to convolve $\delta_{\alpha}(\bmath{x})$ with a
filter representing cell $i$ and $\delta_{\beta}(\bmath{x})$ with
a filter representing cell $j$. This is complicated since our cells,
while all roughly cubical, have slightly different shapes. We therefore
use an approximation of a spherical top hat smoothing filter with
radius $R$: $w\left(r,R\right)\equiv3/(4\pi R^{3})\Theta(R-r)$ with
the Fourier transform given by \begin{equation}
\hat{w}\left(k,R\right)=\frac{3}{\left(kR\right)^{3}}\left[\sin\left(kR\right)-kR\cos\left(kR\right)\right].\end{equation}
 $R$ is chosen so that the effective scale corresponds to cubes with
side length $\elll$: $R=\sqrt{5/12}\elll$, where $\elll$ is the
cell size defined in \S\ref{sub:Counts-in-Cells-Methodology}. (See
p. 500 in \citealt{1999coph.book.....P} for derivation of the $\sqrt{5/12}$
factor.) This gives

\begin{equation}
\left\langle \delta_{\alpha}^{(i)}\delta_{\beta}^{(j)}\right\rangle =\frac{1}{2\pi^{2}}\int_{0}^{\infty}\frac{\sin\left(kr_{ij}\right)}{kr_{ij}}P_{\alpha\beta}\left(k\right)\left|\hat{w}\left(k,R\right)\right|^{2}k^{2}{\rm {d}k,}\label{eq:s_elements}\end{equation}
 where $r_{ij}$ is the distance between the centers of cells $i$
and $j$. The kernel of this integrand -- meaning everything besides
$P_{\alpha\beta}\left(k\right)$ here -- typically peaks at $k\sim1/R$
and is only non-negligible in a range of $\Delta\log_{10}k\sim1$.
Assuming that the functions $b_{1}\left(k\right)$, $b_{2}\left(k\right)$,
and $r_{12}\left(k\right)$ vary slowly with $k$ over this range,
they can be approximated by their values at $k_{\rmn{peak}}\equiv1/R=\sqrt{12/5}/\elll$
and pulled outside the integral, allowing us to write

\begin{equation}
\mathbfss{S}=\sigma_{1}^{2}\left(\elll\right)\left[\begin{array}{cc}
{\mathbfss{S}}_{M} & b_{\rmn{rel}}\left(\elll\right)r_{\rmn{rel}}\left(\elll\right){\mathbfss{S}}_{M}\\
b_{\rmn{rel}}\left(\elll\right)r_{\rmn{rel}}\left(\elll\right){\mathbfss{S}}_{M} & b_{\rmn{rel}}\left(\elll\right)^{2}{\mathbfss{S}}_{M}\end{array}\right]\label{eq:smatrix}\end{equation}
 where $\sigma_{1}^{2}\left(\elll\right)\equiv b_{1}\left(k_{\rmn{peak}}\right)^{2}$,
$b_{\rmn{rel}}\left(\elll\right)\equiv b_{2}\left(k_{\rmn{peak}}\right)/b_{1}\left(k_{\rmn{peak}}\right)$,
$r_{\rmn{rel}}\left(\elll\right)\equiv r_{12}\left(k_{\rmn{peak}}\right)$,
and ${\mathbfss{S}}_{M}$ is the correlation matrix for the underlying
matter density:

\begin{equation}
{\mathbfss{S}}_{M}^{ij}=\frac{1}{2\pi^{2}}\int_{0}^{\infty}\frac{\sin\left(kr_{ij}\right)}{kr_{ij}}P\left(k\right)\left|\hat{w}\left(k,R\right)\right|^{2}k^{2}{\rm {d}k.}\label{eq:sd_elements}\end{equation}
 For the matter power spectrum $P\left(k\right)$, we use the fitting
formula from \citet{1999A&A...347..799N} with the best-fitting `vanilla'
parameters from \citet{2004PhRvD..69j3501T} and apply the non-linear
transformation of \citet{2003MNRAS.341.1311S}. In principle, using
$P\left(k\right)$ as determined previously from SDSS data could introduce
some model dependence in our results. However, our tests in \S\ref{sub:Uncorrelated-signal-matrix}
using an uncorrelated (i.e., diagonal) signal matrix indicate that
the particular choice of the matter power spectrum has no significant
effect on the results, so this is not a problem in practice.

Our primary parameters are the relative bias factor $b_{\rmn{rel}}\left(\elll\right)$,
the relative cross-correlation coefficient $r_{\rmn{rel}}\left(\elll\right)$,
and the overall normalization $\sigma_{1}^{2}\left(\elll\right)$.
The only assumptions we have made in defining these parameters are
homogeneity, isotropy, and that $b_{1}\left(k\right)$, $b_{2}\left(k\right)$,
and $r_{12}\left(k\right)$ vary slowly in $k$. 
These parameters are closely related to those in the biasing models
specified in equations~\eqref{eq:linearbias}~and~\eqref{eq:nonlinearbias}:
If linear deterministic biasing holds, then $b_{\rmn{rel}}=b_{\rmn{lin}}$
and $r_{\rmn{rel}}=1$, 
and the addition of either non-linearity or stochasticity will give
$r_{\rmn{rel}}<1$. 
As discussed in \citet{1999ApJ...525..543M}, stochasticity is expected
to vanish in Fourier space (i.e., $r_{12}\left(k\right)=1$) on large
scales where the density fluctuations are small, but scale dependence
of $b_{1}\left(k\right)$ and $b_{2}\left(k\right)$ can still give
rise to stochasticity in real space. We will measure the parameters
$b_{\rmn{rel}}\left(\elll\right)$ and $r_{\rmn{rel}}\left(\elll\right)$
as a function of scale, thus testing whether the bias is scale dependent
and determining the range of scales on which linear biasing holds.

\subsection{\label{sub:The-Null-buster-Test}The Null-buster Test}

Can the relative bias between dim and luminous galaxies or between
red and blue galaxies be explained by simple linear deterministic
biasing? To address this question, we use the so-called null-buster
test described in \citet{1999ApJ...518L..69T}. For a pair of different
types of galaxies, we calculate an $n$-dimensional difference map
vector \begin{equation}
{\bmath{\underline{g}}}_{\Delta}\equiv{\bmath{\underline{g}}}_{2}-f{\bmath{\underline{g}}}_{1}\label{eq:diffmap}\end{equation}
for a range of values of $f$. (Note that ${\bmath{\underline{g}}}_{\Delta}$
contains a model parameter $f$ even though it is constructed from
observed quantities.) If equation~\eqref{eq:linearbias} holds and
$f=b_{\rmn{lin}}$, then the density fluctuations cancel and ${\bmath{\underline{g}}}_{\Delta}$
will contain only shot noise, with a covariance matrix $\left\langle \pmb{\bm{\langle}}{\bmath{g}}_{\Delta}{\bmath{g}}_{\Delta}^{T}\pmb{\bm{\rangle}}\right\rangle ={\mathbfss{N}}_{\Delta}\equiv{\mathbfss{N}}_{2}+f^{2}{\mathbfss{N}}_{1}$
-- this is our null hypothesis. Alternatively, if equation~\eqref{eq:linearbias}
does not hold, the covariance matrix is instead given by $\left\langle \pmb{\bm{\langle}}{\bmath{g}}_{\Delta}{\bmath{g}}_{\Delta}^{T}\pmb{\bm{\rangle}}\right\rangle ={\mathbfss{N}}_{\Delta}+{\mathbfss{S}}_{\Delta}$,
where ${\mathbfss{S}}_{\Delta}$ is some residual signal. We want
to construct a test to confirm or refute the null hypothesis of deterministic
linear bias. We aim to distinguish between the following two cases:

\begin{itemize}
\item Null hypothesis $H_{0}$: $\left\langle \pmb{\bm{\langle}}{\bmath{g}}_{\Delta}{\bmath{g}}_{\Delta}^{T}\pmb{\bm{\rangle}}\right\rangle ={\mathbfss{N}}_{\Delta}$
for some value of $f$.
\item Alternate hypothesis $H_{1}$: $\left\langle \pmb{\bm{\langle}}{\bmath{g}}_{\Delta}{\bmath{g}}_{\Delta}^{T}\pmb{\bm{\rangle}}\right\rangle ={\mathbfss{N}}_{\Delta}+{\mathbfss{S}}_{\Delta}$
even when $f$ is chosen to minimize the contribution of ${\mathbfss{S}}_{\Delta}$.
\end{itemize}
One valid test of $H_{0}$ would be a $\chi^{2}$ test -- that is,
compute a $\chi^{2}$ statistic assuming $H_{0}$ is true (so $\underline{\chi}^{2}\equiv{\bmath{\underline{g}}}_{\Delta}^{T}{\mathbfss{N}}_{\Delta}^{-1}{\bmath{\underline{g}}}_{\Delta}$),
minimize $\chi^{2}$ with respect to the parameter $f$, and then
check to see whether $\underline{\chi_{\rmn{min}}^{2}}$, the value
of $ $$\underline{\chi}^{2}$ at the minimum point, indicates a good
model fit. For a system with $N$ degrees of freedom where $N\gg1$,
the distribution of $\chi^{2}$ can be approximated as a Gaussian
distribution with mean $N$ and variance $2N$ -- here $N=n-1$ (the
number of cells $n$ minus 1 fitted parameter $f$). Thus $\left(\underline{\chi_{\rmn{min}}^{2}}-N\right)/\sqrt{2N}$
is a measure of the number of `sigmas' the observed $\underline{\chi_{\rmn{min}}^{2}}$
value lies away from the mean expected value if $H_{0}$ is true.
The usual convention is to consider the model a good fit if $\left|\left(\underline{\chi_{\rmn{min}}^{2}}-N\right)/\sqrt{2N}\right|<2$
, i.e., the data is consistent with the null hypothesis at the $2\sigma$
level.

However, if we have some knowledge of the form that a deviation from
the null hypothesis is likely to take (that is, a model for the residual
signal ${\mathbfss{S}}_{\Delta}$), we can construct a statistic that
is more sensitive than $\chi^{2}$ for ruling out the null hypothesis
$H_{0}$ in the event that the alternate hypothesis $H_{1}$ is true.
Here we follow \citet{1998ApJ...500L..79T} and \citet{1999ApJ...519..513T}
and generalize the $\chi^{2}$ test statistic by constructing a quadratic
test statistic $\underline{q}\equiv{\bmath{\underline{g}}}_{\Delta}^{T}\mathbfss{E}{\bmath{\underline{g}}}_{\Delta}$
for some matrix $\mathbfss{E}$. We want to choose $\mathbfss{E}$
such that our test will rule out $H_{0}$ with maximum significance
if $H_{1}$ is true. Our generalized version of $\left(\underline{\chi_{\rmn{min}}^{2}}-N\right)/\sqrt{2N}$,
the significance level (i.e. the number of `sigmas') at which we can
rule out $H_{0}$, is given by\begin{equation}
\underline{\nu}\equiv\frac{\underline{q}-\left\langle \pmb{\bm{\langle}}q|H_{0}\pmb{\bm{\rangle}}\right\rangle }{\Delta q}\label{eq:genchisquared}\end{equation}
where $\left(\Delta q\right)^{2}\equiv\left\langle \pmb{\bm{\langle}}q^{2}|H_{0}\pmb{\bm{\rangle}}\right\rangle -\left\langle \pmb{\bm{\langle}}q|H_{0}\pmb{\bm{\rangle}}\right\rangle ^{2}$
is the variance of $q$ given that $H_{0}$ is true. The desired matrix
$\mathbfss{E}$ maximizes\begin{equation}
\left\langle \pmb{\bm{\langle}}\nu|H_{1}\pmb{\bm{\rangle}}\right\rangle =\frac{\left\langle \pmb{\bm{\langle}}q|H_{1}\pmb{\bm{\rangle}}\right\rangle -\left\langle \pmb{\bm{\langle}}q|H_{0}\pmb{\bm{\rangle}}\right\rangle }{\Delta q},\label{eq:nullbuster_tominimize}\end{equation}
the expected value of $\nu$ if the alternate hypothesis $H_{1}$
holds. The matrix $\mathbfss{E}$ satisfying this requirement is derived
in \citet{1999ApJ...519..513T} to be $\mathbfss{E}\propto{\mathbfss{N}}_{\Delta}^{-1}{\mathbfss{S}}_{\Delta}{\mathbfss{N}}_{\Delta}^{-1}$,
which we insert into equation~\eqref{eq:genchisquared} to give our
null-buster statistic:

\begin{equation}
\underline{\nu}=\frac{{\bmath{\underline{g}}}_{\Delta}^{T}{\mathbfss{N}}_{\Delta}^{-1}{\mathbfss{S}}_{\Delta}{\mathbfss{N}}_{\Delta}^{-1}{\bmath{\underline{g}}}_{\Delta}-\rmn{Tr}\left({\mathbfss{N}}_{\Delta}^{-1}{\mathbfss{S}}_{\Delta}\right)}{\left[\rmn{2\, Tr}\left({\mathbfss{N}}_{\Delta}^{-1}{\mathbfss{S}}_{\Delta}{\mathbfss{N}}_{\Delta}^{-1}{\mathbfss{S}}_{\Delta}\right)\right]^{1/2}}.\label{eq:nullbuster}\end{equation}
 This can be interpreted as the significance level (i.e. the number
of `sigmas') at which we can rule out the null hypothesis $H_{0}$
of deterministic linear bias. As detailed in \citet{1999ApJ...519..513T},
this test assumes that the Poisson shot noise contribution can be
approximated as Gaussian but makes no other assumptions about the
probability distribution of ${\bmath{\underline{g}}}_{\Delta}$. It
is a valid test for any choice of ${\mathbfss{S}}_{\Delta}$ and reduces
to a standard $\chi^{2}$ test if ${\mathbfss{S}}_{\Delta}={\mathbfss{N}}_{\Delta}$,
but it rules out $H_{0}$ with maximum significance in the case where
$H_{1}$ is true, i.e., where ${\mathbfss{S}}_{\Delta}$ is the true
residual signal.

Using equations~\eqref{eq:N_and_S},~\eqref{eq:s_matrices},~and~\eqref{eq:smatrix},
the covariance matrix of ${\bmath{g}}_{\Delta}$ can be written as
\begin{equation}
\left\langle \pmb{\bm{\langle}}{\bmath{g}}_{\Delta}{\bmath{g}}_{\Delta}^{T}\pmb{\bm{\rangle}}\right\rangle =\sigma_{1}^{2}\left(f^{2}-2b_{{\rm \rmn{rel}}}r_{\rmn{rel}}f+b_{\rmn{rel}}^{2}\right){\mathbfss{S}}_{M}+{\mathbfss{N}}_{\Delta},\end{equation}
 where ${\mathbfss{S}}_{M}$ is given by equation~\eqref{eq:sd_elements}.
We use ${\mathbfss{S}}_{\Delta}={\mathbfss{S}}_{M}$ in equation~\eqref{eq:nullbuster}
(note that $\nu$ is independent of the normalization of $\mathbfss{S}$,
which scales out) since deviations from linear deterministic bias
are likely to be correlated with large-scale structure. 

To apply the null-buster test, we compute $\underline{\nu}$ as a
function of $f$ and then minimize it. If the minimum value $\underline{\nu_{\rmn{min}}}>2$,
we rule out linear deterministic bias at $>2\sigma$. If the null
hypothesis cannot be ruled out and we choose to accept it as an accurate
description of the data, we can take $r_{\rmn{rel}}$ to be equal
to unity and then use the value of $f$ that gives $\underline{\nu_{\rmn{min}}}$
as a measure of $b_{\rmn{rel}}$.

We calculate the uncertainty on $b_{\rmn{rel}}$ using two different
methods. The first method makes use of the fact that $\nu$ is generalized
$\chi^{2}$ statistic: the uncertainty on $b_{\rmn{rel}}$, is given
by the range in $f$ that gives $\sqrt{2N}\left(\underline{\nu}-\underline{\nu_{\rmn{min}}}\right)\le1$,
where $N$ is the number of degrees of freedom (equal to the number
of cells minus 1 fitted parameter). This is a generalization of the
standard $\Delta\chi^{2}=1$ uncertainty since $\nu$ is a generalization
of $\left(\underline{\chi_{\rmn{min}}^{2}}-N\right)/\sqrt{2N}$. The
second method uses jackknife resampling, which is described in \S\ref{sub:Jackknifes}
along with a comparison of the two methods. We present all of our
results derived from the null-buster test using the uncertainties
from the jackknife method.

\subsection{\label{sub:Maximum-Likelihood-Method}Maximum Likelihood Method}

In addition to the null-buster test, we use a maximum likelihood analysis
to determine the parameters $b_{\rmn{rel}}$ and $r_{\rmn{rel}}$.
Our method is a generalization of the maximum likelihood method used
in previous work, accounting for correlations between different cells
but making a somewhat different set of assumptions.

In \citet{2000ApJ...544...63B}, \citet{2005MNRAS.356..247W}, \citet{2005MNRAS.356..456C},
the probability of observing $N_{1}$ galaxies of type 1 and $N_{2}$
galaxies of type 2 in a given cell is expressed as

\begin{eqnarray}
\lefteqn{P\left(N_{1},N_{2}\right)=\int_{-1}^{\infty}\int_{-1}^{\infty}\rmn{Poiss}\left[N_{1},\bar{N}_{1}\left(1+\delta_{1}\right)\right]}\nonumber \\
 &  & \times\rmn{Poiss}\left[N_{2},\bar{N}_{2}\left(1+\delta_{2}\right)\right]f\left({\delta_{1},\delta}_{2},\alpha\right)\rmn{d}\delta_{1}\rmn{d}\delta_{2},\label{eq:wild}\end{eqnarray}
 where $f\left({\delta_{1},\delta}_{2},\alpha\right)$ is the joint
probability distribution of $\delta_{1}$ and $\delta_{2}$ in one
cell, $\alpha$ represents a set of parameters which depend on the
biasing model, and $\rmn{Poiss}\left(N,\lambda\right)\equiv\lambda^{N}e^{-\lambda}/N!$
is the Poisson probability to observe $N$ objects given a mean value
$\lambda$. The likelihood function for $n$ cells is given a set
of observations $\underline{N}_{1}^{(i)}$ and $\underline{N}_{2}^{(i)}$
is then given by \begin{equation}
\mathcal{L}\left(\alpha\right)\equiv\prod_{i=1}^{n}P\left(\underline{N}_{1}^{(i)},\underline{N}_{2}^{(i)}\right),\label{eq:uncorr_L}\end{equation}
 which is minimized with respect to the parameters $\alpha$. This
treatment makes two assumptions: it neglects correlations between
different cells and it assumes that the galaxy discreteness is Poissonian.
These assumptions greatly simplify the computation of $\mathcal{L}$,
but are understood to be approximations to the true process. Cosmological
correlations are known to exist on large scales, although their impact
on counts-in-cells analyses has been argued to be small \citep{1995ApJ...438...49B,2005MNRAS.356..456C}.
Semi-analytical modeling \citep{2001MNRAS.322..901S,2002ApJ...575..587B},
$N$-body simulation \citep{2002MNRAS.333..730C,2004ApJ...609...35K},
and smoothed particle hydrodynamic simulation \citep{2003ApJ...593....1B}
investigations suggest that the probability distribution for galaxies/halos
is sub-Poissonian in some regimes, and in fact non-Poissonian behavior
is implied by observations as well \citep{2003MNRAS.339.1057Y,2005MNRAS.356..247W}.

\begin{sloppypar} Dropping these two assumptions, we can write a
more general expression for the likelihood function for $n$ cells:
\begin{eqnarray}
\lefteqn{{\mathcal{L}\left(\alpha,\beta\right)\equiv P\left(\underline{N}_{1}^{\left(1\right)},\ldots,\underline{N}_{1}^{\left(n\right)},\underline{N}_{2}^{\left(1\right)},\ldots,\underline{N}_{2}^{\left(n\right)}\right)=}}\nonumber \\
 &  & \int_{-1}^{\infty}\ldots\int_{-1}^{\infty}\left[\prod_{i=1}^{n}P_{g}\left(\underline{N}_{1}^{\left(i\right)},\bar{N}_{1}^{\left(i\right)}\left(1+\delta_{1}^{\left(i\right)}\right),\beta\right)\right]\nonumber \\
 &  & \times\left[\prod_{j=1}^{n}P_{g}\left(\underline{N}_{2}^{\left(j\right)},\bar{N}_{2}^{\left(j\right)}\left(1+\delta_{2}^{\left(j\right)}\right),\beta\right)\right]\nonumber \\
 &  & \times f\left(\delta_{1}^{\left(1\right)},\ldots,\delta_{1}^{\left(n\right)},\delta_{2}^{\left(1\right)},\ldots,\delta_{2}^{\left(n\right)},\alpha\right)\nonumber \\
 &  & \times\rmn{d}\delta_{1}^{\left(1\right)}\ldots\rmn{d}\delta_{1}^{\left(n\right)}\rmn{d}\delta_{2}^{\left(1\right)}\ldots\rmn{d}\delta_{2}^{\left(n\right)},\label{eq:corr_L}\end{eqnarray}
 where $\rmn{Poiss}\left(N,\lambda\right)$ has been replaced with
a generic probability $P_{g}\left(N,\lambda,\beta\right)$ for the
galaxy distribution parameterized by some parameters $\beta$ and
$f\left(\delta_{1}^{\left(1\right)},\ldots,\delta_{1}^{\left(n\right)},\delta_{2}^{\left(1\right)},\ldots,\delta_{2}^{\left(n\right)},\alpha\right)$
is a joint probability distribution relating $\delta_{1}$ and $\delta_{2}$
in all cells. In practice, this would be prohibitively difficult to
calculate as it involves $2n$ integrations \citep{1997astro.ph.12074D},
and would require a reasonable parameterized form for $P_{g}\left(N,\lambda,\beta\right)$
as well as $f\left(\delta_{1}^{\left(1\right)},\ldots,\delta_{1}^{\left(n\right)},\delta_{2}^{\left(1\right)},\ldots,\delta_{2}^{\left(n\right)},\alpha\right)$.
\end{sloppypar}

In this chapter, we take a simpler approach and approximate the probability
distribution for our data vector $\bmath{g}$ to be Gaussian with
the covariance matrix $\mathbfss{C}$ as defined by equations~\eqref{eq:cov_matrix},~\eqref{eq:N_and_S}~and~\eqref{eq:smatrix},
and use this to define our likelihood function in terms of the parameters
$\sigma_{1}^{2}$, $b_{\rmn{rel}}$, and $r_{\rmn{rel}}$:

\begin{eqnarray}
\lefteqn{{\mathcal{L}\left(\sigma_{1}^{2},b_{\rmn{rel}},r_{\rmn{rel}}\right)\equiv P\left(\underline{g}_{1}^{\left(1\right)},\ldots,\underline{g}_{1}^{\left(n\right)},\underline{g}_{2}^{\left(1\right)},\ldots,\underline{g}_{2}^{\left(n\right)}\right)}}\nonumber \\
 &  & =\frac{1}{\left(2\pi\right)^{n}\left|\mathbfss{C}\right|^{1/2}}\exp\left(-\frac{1}{2}{\bmath{\underline{g}}}^{\dagger}{\mathbfss{C}}^{-1}\bmath{\underline{g}}\right).\label{eq:likelihood}\end{eqnarray}
 Note that this includes the shot noise since $\mathbfss{C}=\mathbfss{S}+\mathbfss{N}$,
and is not precisely equivalent to assuming that $P_{g}$ and $f$
in equation~\eqref{eq:corr_L} are Gaussian.

For $r_{\rmn{rel}}$ values of $\left|r_{\rmn{rel}}\right|>1$, the
matrix $\mathbfss{C}$ is singular, and thus the likelihood function
cannot be computed. Hence this analysis method automatically incorporates
the constraint that $\left|r_{\rmn{rel}}\right|\le1$, which is physically
expected for a cross-correlation coefficient.

To determine the best fit values of our parameters for each pairwise
comparison, we maximize $2\ln\mathcal{L}\left(\sigma_{1}^{2},b_{\rmn{rel}},r_{\rmn{rel}}\right)$
with respect to $\sigma_{1}^{2}$, $b_{\rmn{rel}}$, and $r_{\rmn{rel}}$.
Since our method of comparing pairs of galaxy samples primarily probes
the relative biasing between the two types of galaxies, it is not
particularly sensitive to $\sigma_{1}^{2}$, which represents the
bias of type 1 galaxies relative to the dark matter power spectrum
used in equation~\eqref{eq:sd_elements}. Thus we marginalize over
$\sigma_{1}^{2}$ and calculate the uncertainty on $b_{\rmn{rel}}$
and $r_{\rmn{rel}}$ using the $\Delta\left(2\ln\mathcal{L}\right)=1$
contour in the $b_{\rmn{rel}}$-$r_{\rmn{rel}}$ plane. This procedure
is discussed in more detail in \S\ref{sub:Likelihood-contour-plots}.

\section{Results}

\label{Results}

\subsection{\label{sub:Null-buster-Results}Null-buster Results}

To test the deterministic linear bias model, we apply the null-buster
test described in \S\ref{sub:The-Null-buster-Test} to the
pairs of galaxy samples described in \S\ref{sub:Overlapping-Volume-Limited-Samples}.
For studying the luminosity-dependent bias, we use the galaxies in
the more luminous bin as the type 1 galaxies and the dimmer bin as
the type 2 galaxies for each pair of neighboring luminosity bins,
and repeat this in each volume V1-V6. We do this for all galaxies
and also for red and blue galaxies separately. For the color dependence,
we use red galaxies as type 1 and blue galaxies as type 2, and again
repeat this in each volume. To determine the scale dependence, we
repeat all of these tests for four different values of the cell size
$\elll$ as described in \S\ref{sub:Counts-in-Cells-Methodology}.

\subsubsection{\label{sub:linear-deterministic}Is the bias linear and deterministic?}

\begin{figure*}
\includegraphics[width=1\textwidth]{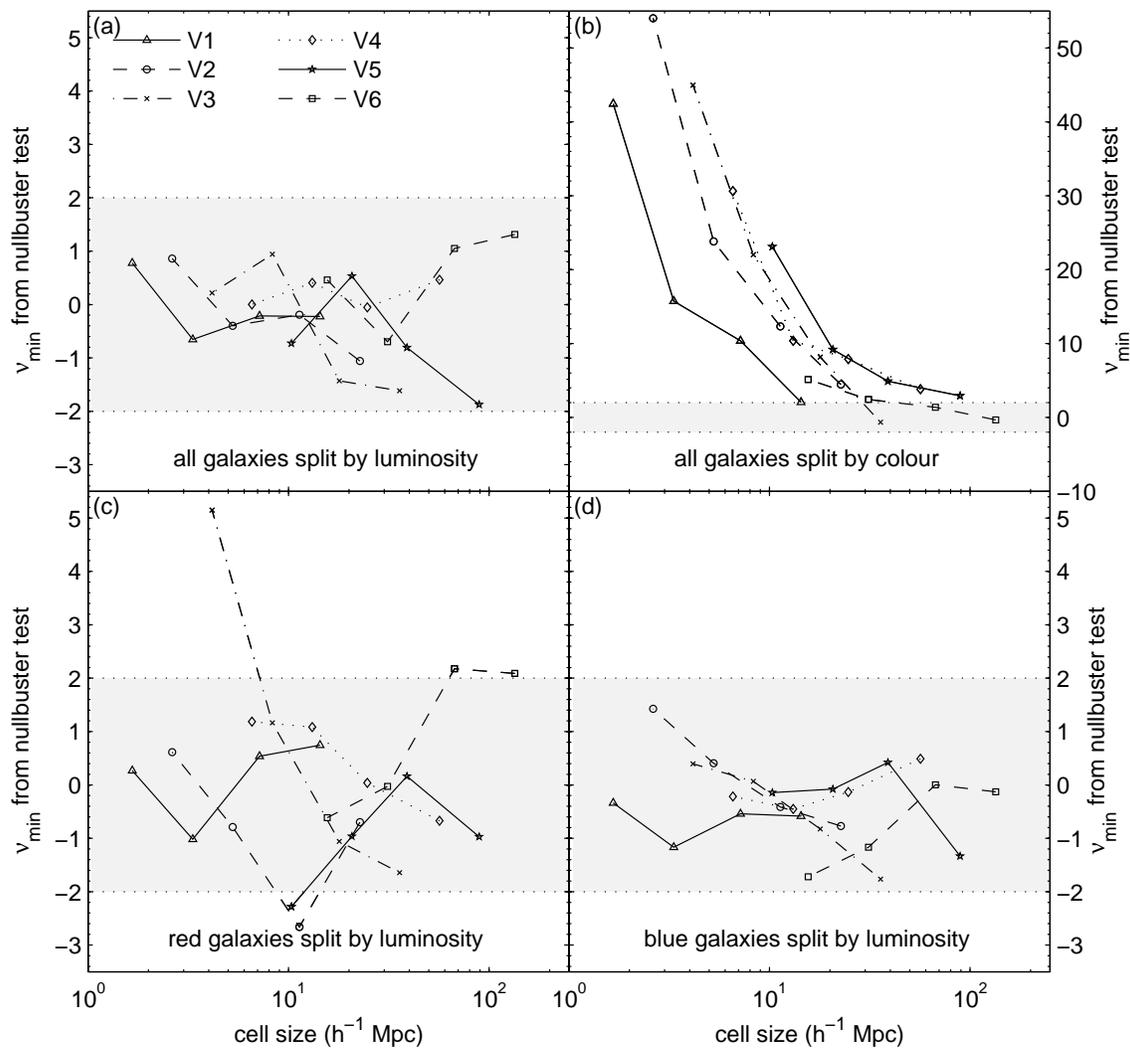}

\caption[Null-buster results for pairwise comparisons.]{\label{fig:nullbusterplot}Null-buster results for pairwise comparisons.
$\underline{\nu_{\rmn{min}}}$ measures the number of sigmas at which deterministic
linear biasing can be ruled out as a model of relative bias between
the two samples being compared. Shaded areas indicate $\left|\underline{\nu_{\rmn{min}}}\right|<2$,
where data is consistent with the null hypothesis at the $2\sigma$
level. Four different types of pairwise comparison are illustrated:
(a) luminous vs. dim, (b) red vs. blue, (c) luminous red vs. dim red,
and (d) luminous blue vs. dim blue. The different symbols denote the
different comparison volumes V1-V6. The luminosity-dependent bias
(a, c, d) is consistent with deterministic linear biasing but color-dependent
bias (b) is not.}
\end{figure*}

The results are plotted in Fig.~\ref{fig:nullbusterplot}, which
shows the minimum value of the null-buster test statistic $\underline{\nu_{\rmn{min}}}$
vs. cell size $\elll$. According to this test, deterministic linear
biasing is in fact an excellent fit for the luminosity-dependent bias:
nearly all $\underline{\nu_{\rmn{min}}}$ fall within $\left|\underline{\nu_{\rmn{min}}}\right|<2$,
indicating consistency with the null hypothesis at the $2\sigma$
level. (There are a few exceptions in the case of the red galaxies,
the largest being $\underline{\nu_{\rmn{min}}}\sim5$ for the smallest cell size
in V3.) For color-dependent bias, however, deterministic linear biasing
is ruled out quite strongly, especially at smaller scales.

The cases where the null hypothesis survives are quite noteworthy,
since this implies that essentially all of the large clustering signal
that is present in the data (and is visually apparent in Fig.~\ref{fig:4pairwise})
is common to the two galaxy samples and can be subtracted out. For
example, for the V5 luminosity split at the highest angular resolution
($\elll=10\, h^{-1}\rmn{Mpc}$), clustering signal is detected at
$953\sigma$ in the faint sample ($\underline{\nu}(f)\approx953$ for $f=0$)
and at $255\sigma$ in the bright sample ($\underline{\nu}(f)\approx255$ for
$f=\infty$), yet the weighted difference of the two maps is consistent
with mere shot noise ($\underline{\nu}(0.88)\approx-0.63$). This also shows that
no luminosity-related systematic errors afflict the sample selection
even at that low level.%

\subsubsection{\label{sub:scale-dependent}Is the bias independent of scale?}

\begin{figure*}
\includegraphics[width=1\textwidth]{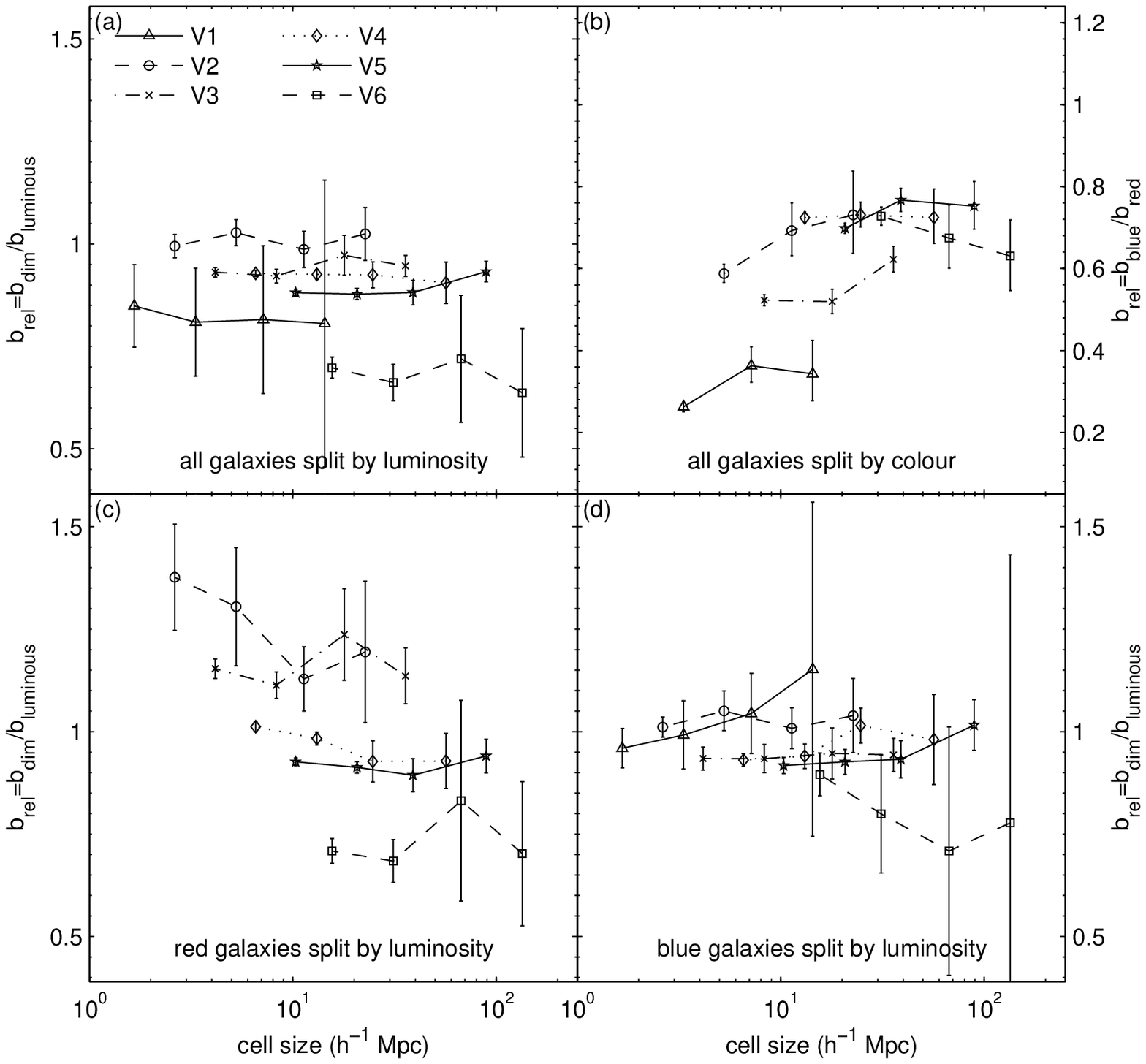}

\caption[Relative bias $b_{\rmn{rel}}$ between pairwise
samples.]{\label{fig:biasplot}Relative bias $b_{\rmn{rel}}$ between pairwise
samples. (a) luminous vs. dim, (b) red vs. blue, (c) luminous red
vs. dim red, and (d) luminous blue vs. dim blue, revealing no significant
scale dependence of luminosity- or color-dependent bias. The $b_{\rmn{rel}}$
values shown for luminosity dependent splittings (a), (c), and (d)
were computed with the null-buster analysis, those shown for the color-dependent
splitting (b) were computed with the likelihood analysis. The different
symbols denote the different comparison volumes V1-V6.}
\end{figure*}

For the luminosity-dependent bias, we use the value of $f$ that gives
$\underline{\nu_{\rmn{min}}}$ as a measure of $b_{\rmn{rel}}$, the relative
bias between two neighboring luminosity bins. Since deterministic
linear bias is ruled out in the case of the color-dependent bias,
we instead use the value of $b_{\rmn{rel}}$ from the likelihood analysis
here. We find that although the value of $b_{\rmn{rel}}$ depends
on luminosity, it does not appear to depend strongly on scale, as
can be seen in Fig.~\ref{fig:biasplot}: in all plots the curves
appear roughly horizontal. To test this `chi-by-eye' inference of
scale independence quantitatively, we applied a simple $\chi^{2}$
fit on the four data points (or three in the color-dependent case)
in each volume using a one-parameter model: a horizontal line with
a constant value of $b_{\mathrm{rel}}$. For this fit we use covariance
matrices derived from jackknife resampling, as discussed in \S\ref{sub:Jackknifes}.
We define this model to be a good fit if the goodness-of-fit value
(the probability that a $\chi^{2}$ as poor as as the value calculated
should occur by chance, as defined in \citealt{1992nrca.book.....P})
exceeds 0.01.

We find that this model of no scale dependence is a good fit for all
data sets plotted in Fig.~\ref{fig:biasplot}. We therefore find
no evidence that the luminosity- or color-dependent bias is scale-dependent
on the scales we probe here$\left(2-160\, h^{-1}\rmn{Mpc}\right)$.
This implies that recent cosmological parameter analyses which use
only measurements on scales $\ga60{\, h}^{-1}\rmn{Mpc}$ (e.g., \citealt{2006MNRAS.366..189S,2006PhRvD..74l3507T,2007ApJS..170..377S})
are probably justified in assuming scale independence of luminosity-dependent
bias. %

In comparison to previous work \citep{2005ApJ...630....1Z,2006MNRAS.368...21L},
it is perhaps surprising to see as little scale dependence as we do
-- \citet{2006MNRAS.368...21L} find the luminosity-dependent bias
to vary with scale (see their fig.~4), in contrast to what we find
here. The measurement of luminosity-dependent bias in \citet{2005ApJ...630....1Z}
agrees more closely with our observation of scale independence, but
their their fig.~10 indicates that we might expect to see scale dependence
of the luminosity-dependent bias in the most luminous samples. However,
we measure the bias in our most luminous samples (in V6) at $16-134\, h^{-1}\rmn{Mpc}$,
well above the range probed in \citet{2005ApJ...630....1Z}, so there
is no direct conflict here. Additionally, fig.~13 of \citet{2005ApJ...630....1Z}
and fig.~10 of \citet{2006MNRAS.368...21L} show that correlation
functions of red and blue galaxies have significantly different slopes,
implying that the color-dependent bias should be strongly scale-dependent
on $0.1-10\, h^{-1}\rmn{Mpc}$ scales. However, the points $>1\, h^{-1}\rmn{Mpc}$
in these plots (the range comparable to the scales we probe here)
do not appear strongly scale dependent, so our results are not inconsistent
with these correlation function measurements. This interpretation
is further supported by recent work \citep{2007ApJ...664..608W} that
finds correlation functions for different luminosities and colors
to be roughly parallel above $\sim1\, h^{-1}\rmn{Mpc}$.

\subsubsection{\label{sub:luminosity-dependent}How bias depends on luminosity}

Our next step is to calculate the relative bias parameter $b/b_{*}$
(the bias relative to $L_{*}$ galaxies) as a function of luminosity
by combining the measured values of $b_{\rmn{rel}}$ between the different
pairs of luminosity bins. This function has been measured previously
using SDSS power spectra \citep{2004ApJ...606..702T} at length scales
of $\sim60\, h^{-1}\rmn{Mpc}$ as well as SDSS \citep{2005ApJ...630....1Z,2006MNRAS.368...21L,2007ApJ...664..608W}
and 2dFGRS \citep{2001MNRAS.328...64N} correlation functions at length
scales of $\sim1\, h^{-1}\rmn{Mpc}$ -- here we measure it at length
scales of $\sim20\, h^{-1}\rmn{Mpc}$. 

The bias of each luminosity bin relative to the central bin L4 is
given by

\begin{eqnarray}
 &  & \frac{b_{1}}{b_{4}}=b_{12}b_{23}b_{34},\qquad\frac{b_{2}}{b_{4}}=b_{23}b_{34},\qquad\frac{b_{3}}{b_{4}}=b_{34},\nonumber \\
 &  & \frac{b_{7}}{b_{4}}=\frac{1}{b_{45}b_{56}b_{67}},\qquad\frac{b_{6}}{b_{4}}=\frac{1}{b_{45}b_{56}},\qquad\frac{b_{5}}{b_{4}}=\frac{1}{b_{45}},\label{eq:lum_bias}\end{eqnarray}
 where $b_{\alpha\beta}$ denotes the measured value of $b_{\rmn{rel}}$
between luminosity bins L$\alpha$ and L$\beta$ using all galaxies
and $b_{\alpha}$ denotes the bias of galaxies in luminosity bin L$\alpha$
relative to the dark matter. For each pairwise comparison, we choose
the value of $b_{\rmn{rel}}$ calculated at the resolution where the
cell size is closest to $20\, h^{-1}\rmn{Mpc}$, as illustrated in
Fig.~\ref{fig:radialcells}. (Since we see no evidence for scale
dependence of $b_{\rmn{rel}}$ for the luminosity-dependent bias,
this choice does not strongly influence the results.)

To compute the error bars on $b_{\alpha}/b_{4}$, we rewrite equation~\eqref{eq:lum_bias}
as a linear matrix equation using the logs of the bias values:

\begin{equation}
\left(\begin{array}{cccccc}
1 & -1 & 0 & 0 & 0 & 0\\
0 & 1 & -1 & 0 & 0 & 0\\
0 & 0 & 1 & 0 & 0 & 0\\
0 & 0 & 0 & -1 & 0 & 0\\
0 & 0 & 0 & 1 & -1 & 0\\
0 & 0 & 0 & 0 & 1 & -1\end{array}\right)\left(\begin{array}{c}
\log{b_{1}/b_{4}}\\
\log{b_{2}/b_{4}}\\
\log{b_{3}/b_{4}}\\
\log{b_{5}/b_{4}}\\
\log{b_{6}/b_{4}}\\
\log{b_{7}/b_{4}}\end{array}\right)=\left(\begin{array}{c}
\log{b_{12}}\\
\log{b_{23}}\\
\log{b_{34}}\\
\log{b_{45}}\\
\log{b_{56}}\\
\log{b_{67}}\end{array}\right),\label{eq:lum_matrix}\end{equation}
 or $\mathbfss{A}{\bmath{b}}_{\rmn{log}}={\bmath{b}}_{\rmn{log,rel}}$,
where ${\bmath{b}}_{\rmn{log,rel}}$ is a vector of the log of our
relative bias measurements $b_{\alpha\beta}$, ${\bmath{b}}_{\rmn{log}}$
is a vector of the log of the bias values $b_{\alpha}/b_{4}$, and
$\mathbfss{A}$ is the matrix relating them. We determine the covariance
matrix $\mathbf{\Sigma}_{\rmn{rel}}$ of ${\bmath{b}}_{\rmn{rel}}$
(a vector of the relative bias measurements $b_{\alpha\beta}$) from
the jackknife resampling described in \S\ref{sub:Jackknifes},
and then compute the covariance matrix $\mathbf{\Sigma}_{\rmn{log,rel}}$
of ${\bmath{b}}_{\rmn{log,rel}}$ by

\begin{equation}
\mathbf{\Sigma}_{\rmn{log,rel}}=\left({\mathbfss{B}}_{\rmn{rel}}^{T}\right)^{-1}\mathbf{\Sigma}_{\rmn{rel}}{\mathbfss{B}}_{\rmn{rel}}^{-1}\label{eq:cov_logrel}\end{equation}
 where ${\mathbfss{B}}_{\rmn{rel}}\equiv\rmn{diag}\left({\bmath{b}}_{\rmn{rel}}\right)$.
We invert equation~\eqref{eq:lum_matrix} to give ${\bmath{b}}_{\rmn{log}}$:

\begin{equation}
{\bmath{b}}_{\rmn{log}}={\mathbfss{A}}^{-1}{\bmath{b}}_{\rmn{log,rel}},\label{eq:invert}\end{equation}
 with the covariance matrix for ${\bmath{b}}_{\rmn{log}}$ given by

\begin{equation}
\mathbf{\Sigma}_{\rmn{log}}=\left({\mathbfss{A}}^{T}\mathbf{\Sigma}_{\rmn{log,rel}}^{-1}\mathbfss{A}\right)^{-1}.\label{eq:cov_log}\end{equation}

We then fit our data with the model used in \citet{2001MNRAS.328...64N}:
$b\left(M\right)/b_{*}=a_{1}+a_{2}\left(L/L_{*}\right)$, parameterized
by $\bmath{a}\equiv\left(a_{1},\, a_{2}\right)$. Here $M$ is the
central absolute magnitude of the bin, $L$ is the corresponding luminosity,
and $M_{*}=-20.83$. We use a weighted least-squares fit that is linear
in the parameters $\left(a_{1},\, a_{2}\right)$ -- that is, we solve
the matrix equation\begin{equation}
\left(\begin{array}{c}
b_{1}/b_{4}\\
b_{2}/b_{4}\\
b_{3}/b_{4}\\
b_{4}/b_{4}\\
b_{5}/b_{4}\\
b_{6}/b_{4}\end{array}\right)=\left(\begin{array}{cc}
1 & L_{1}/L_{4}\\
1 & L_{2}/L_{4}\\
1 & L_{3}/L_{4}\\
1 & L_{4}/L_{4}\\
1 & L_{5}/L_{4}\\
1 & L_{6}/L_{4}\end{array}\right)\left(\begin{array}{c}
a_{1}\\
a_{2}\end{array}\right),\label{eq:model_matrix}\end{equation}
 or $\bmath{b}=\mathbfss{X}\bmath{a}$, where $\bmath{b}$ is a vector
of the bias values $b_{\alpha}/b_{4}$ and $\mathbfss{X}$ is the
matrix representing our model. We solve for $\bmath{a}$ using

\begin{equation}
\bmath{a}=\left({\mathbfss{X}}^{T}\mathbf{\Sigma}^{-1}\mathbfss{X}\right)^{-1}{\mathbfss{X}}^{T}\mathbf{\Sigma}^{-1}\bmath{b}.\label{eq:best_fit_LS}\end{equation}
 Here $\mathbf{\Sigma}$ is the covariance matrix of $\bmath{b}$,
given by\begin{equation}
\mathbf{\Sigma}=\mathbfss{B}\mathbf{\Sigma}_{\rmn{log}}{\mathbfss{B}}^{T},\label{eq:cov_notlog}\end{equation}
 where $\mathbf{\Sigma}_{\rmn{log}}$ is given by equation~\eqref{eq:cov_log}
and $\mathbfss{B}\equiv\rmn{diag}\left(\bmath{b}\right)$. This procedure
gives us the best-fitting values for the parameters $a_{1}$ and $a_{2}$,
accounting for the correlations between the data points that are induced
we compute the bias values $\bmath{b}$ from our relative bias measurements.
We then normalize the model such that $b\left(M_{*}\right)/b_{*}=1$.%
\begin{figure}
\begin{center}%
\includegraphics[width=0.4746\textwidth]{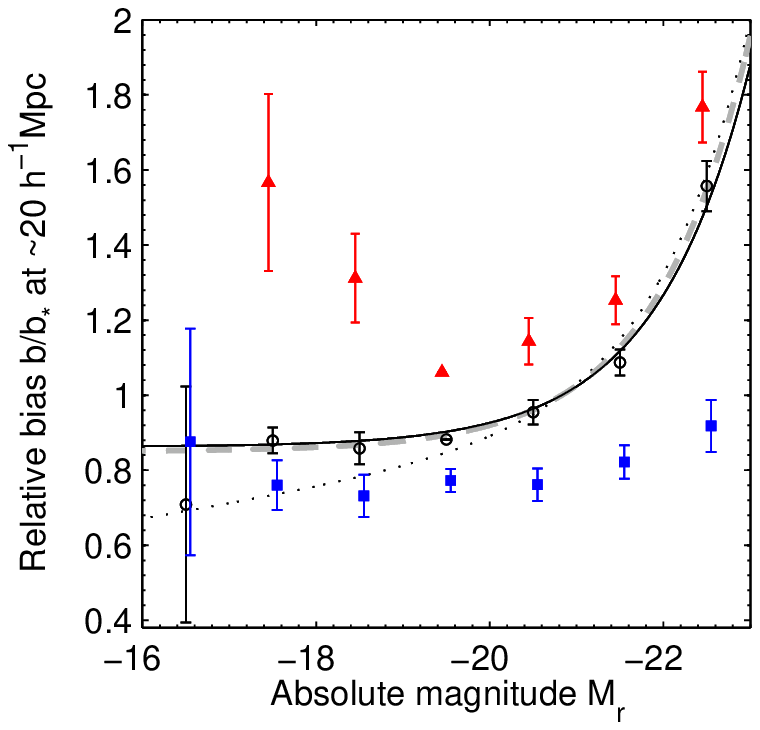}
\end{center}%
\caption[Luminosity dependence of bias for all, red, and blue galaxies.]{\label{fig:lum-bias}Luminosity dependence of bias for all (open
circles), red (solid triangles), and blue (solid squares) galaxies
at a cell size of $\sim20\, h^{-1}\rmn{Mpc}$ from null-buster results.
The solid line is a model fit to the all-galaxy data points, the dotted
line shows the model from \citet{2004ApJ...606..702T}, and the grey
dashed line shows the model from \citet{2001MNRAS.328...64N}. The
\citet{2001MNRAS.328...64N} model has been computed using the SDSS
$r$-band value of $M_{*}=-20.83$.}
\end{figure}

Figure~\ref{fig:lum-bias} shows a plot of $b/b_{*}$ vs. $M$: results
for all galaxies are plotted with black open circles, our best-fitting
model is shown by the solid line, the best-fitting model from \citet{2001MNRAS.328...64N}
is shown by the grey dashed line,and the best-fitting model from \citet{2004ApJ...606..702T}
is shown by the dotted line. The error bars represent the diagonal
elements of $\mathbf{\Sigma}$ from equation~\eqref{eq:cov_notlog}.
Our model, with $\left(a_{1},a_{2}\right)=\left(0.862,\,0.138\right)$,
agrees extremely well with the model from \citet{2001MNRAS.328...64N},
with $\left(a_{1},a_{2}\right)=\left(0.85,\,0.15\right)$. This agreement
is quite remarkable since we use data from a different survey and
analyze it with a completely different technique. 

\begin{figure*}
\includegraphics[width=1\textwidth]{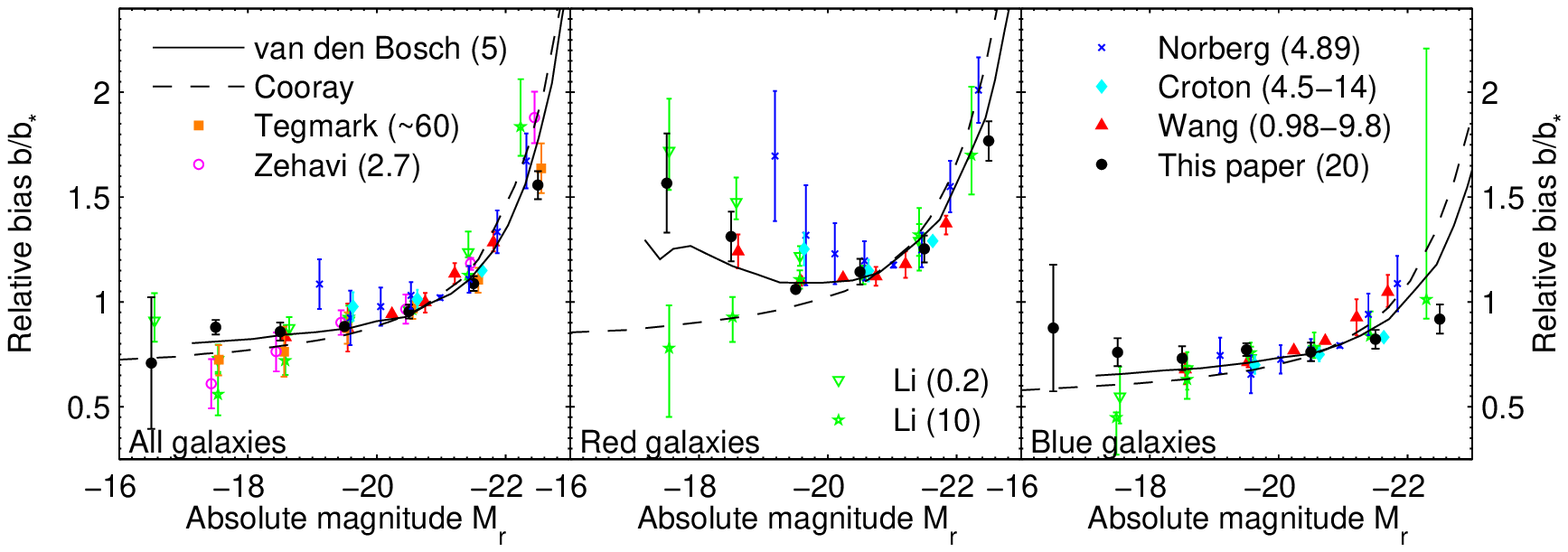}

\caption[Comparison to previous results for the
luminosity dependence of bias for all, red, and blue galaxies.]{\label{fig:comparison_bvmag}Comparison to previous results for the
luminosity dependence of bias for all, red, and blue galaxies. \citet{2002MNRAS.332..827N,2005ApJ...630....1Z,2006MNRAS.368...21L},
and \citet{2007ApJ...664..608W} use correlation function measurements,
\citet{2004ApJ...606..702T} use the power spectrum, and \citet{2007MNRAS.379.1562C}
use counts in cells. To better illustrate the similarities and differences
in the trends as a function of luminosity, we have normalized all
measurements to match our results using the bin closest to $M_{*}=-20.83$.
The error bars shown are all relative: they do not include uncertainties
due to the normalization. Numbers in parentheses denote the scale
in $h^{-1}\rmn{Mpc}$ at which the measurements were done. Also shown
are theoretical models from \citet{2003MNRAS.340..771V} (we show
their model B as a representative example) and \citet{2005MNRAS.363..337C}
-- these are also normalized to match our results at $M_{*}$.}
\end{figure*}

A comparison of our results with previous measurements is shown in
Fig.~\ref{fig:comparison_bvmag} in the left panel. %
 In order to compare our SDSS results with results from 2dFGRS \citep{2002MNRAS.332..827N,2007MNRAS.379.1562C},
we have added a constant factor of $-1.13$ to their quoted values
for $M_{b_{J}}-5\log_{10}h$ in order to line up the value of $M_{*}$
used in \citet{2002MNRAS.332..827N} ($M_{b_{J}}-5\log_{10}h=-19.7$)
with the value used here ($M_{^{01.}r}=-20.83$). Note that this is
necessarily a rough correction since the magnitude in the different
bands varies depending on the spectrum of each galaxy, but this method
provides a reasonable means of comparing the different results. This
plot shows excellent agreement over a wide range of scales, lending
further support to our conclusion that the luminosity-dependent bias
is independent of scale.

We also use equation~\eqref{eq:invert} to calculate $b/b_{*}$ vs.
$M$ for red and blue galaxies separately. To plot the points for
red, blue, and all galaxies on the same $b/b_{*}$ vs. $M$ plot,
we need to determine their relative normalizations. Applying equation~\eqref{eq:invert}
to the red and blue galaxies gives $b_{\alpha,\rmn{red}}/b_{4,\rmn{red}}$
and $b_{\alpha,\rmn{blue}}/b_{4,\rmn{blue}}$, so to normalize the
red-galaxy and blue-galaxy data points to the all-galaxy data points
in Fig.~\ref{fig:lum-bias}, we need to calculate \begin{equation}
\frac{b_{\alpha,\rmn{red}}}{b_{*,\rmn{all}}}=\frac{b_{4,\rmn{all}}}{b_{*,\rmn{all}}}\frac{b_{4,\rmn{red}}}{b_{4,\rmn{all}}}\frac{b_{\alpha,\rmn{red}}}{b_{4,\rmn{red}}}\label{eq:red_norm}\end{equation}
and

\begin{equation}
\frac{b_{\alpha,\rmn{blue}}}{b_{*,\rmn{all}}}=\frac{b_{4,\rmn{all}}}{b_{*,\rmn{all}}}\frac{b_{4,\rmn{red}}}{b_{4,\rmn{all}}}\frac{b_{4,\rmn{blue}}}{b_{4,\rmn{red}}}\frac{b_{\alpha,\rmn{blue}}}{b_{4,\rmn{blue}}}.\label{eq:blue_norm}\end{equation}

The factor $b_{4,\rmn{all}}/b_{*,\rmn{all}}$ is simply the normalization
factor chosen for the above model to give $b\left(M_{*}\right)/b_{*}=1$.
To determine $b_{4,\rmn{red}}/b_{4,\rmn{all}}$, we use best-fitting
values of $\sigma_{1}^{2}$ from the likelihood analysis described
in \S\ref{sub:Maximum-Likelihood-Method} at the resolution
with cell sizes closest to $20\, h^{-1}\rmn{Mpc}$: $\sigma_{1}$
from the comparison of dimmer and more luminous galaxies in V3 gives
$b_{4,\rmn{all}}$, and similarly $\sigma_{1}$ from the comparison
of blue and red galaxies in L4 gives $b_{4,\rmn{red}}$, so\begin{equation}
\frac{b_{4,\rmn{red}}}{b_{4,\rmn{all}}}=\left(\frac{\sigma_{1,\rmn{red\, vs.\, blue\, L4}}^{2}}{\sigma_{1,\rmn{lum\, vs.\, dim\, V3}}^{2}}\right)^{1/2}.\label{eq:red_v_all}\end{equation}
The blue points are then normalized relative to the red points using
$b_{4,\rmn{blue}}/b_{4,\rmn{red}}$ equal to the measured value of
$b_{\rmn{rel}}$ from the likelihood comparison of blue and red galaxies
in L4. Thus the shapes of the red and blue curves are determined using
the luminosity-dependent bias from the null-buster analysis, but their
normalization uses information from the likelihood analysis as well.


Splitting the luminosity dependence of the bias by color reveals
some interesting features. The bias of the blue galaxies shows only
a weak dependence on luminosity, and both luminous ($M\sim-22$) and
dim ($M\sim-17$) red galaxies have slightly higher bias than moderately
bright ($M\sim-20\sim M_{*}$) red galaxies. The previously observed
luminosity dependence of bias, with a weak dependence dimmer than
$L_{*}$ and a strong increase above $L_{*}$, is thus quite sensitive
to the color selection: the lower luminosity bins contain mostly
blue galaxies and thus show weak luminosity dependence, whereas the
more luminous bins are dominated by red galaxies which drive the observed
trend of more luminous galaxies being more strongly biased. It is
instructive to compare these results with the mean local overdensity
in color-magnitude space, as in fig.~2 of Blanton et al.~(2005a).
Although our bias measurements are necessarily much coarser, it can
be seen that the bias is strongest where the overdensity is largest,
as has been seen previously \citep{2006MNRAS.372.1749A}.

Comparisons of our results to other measurements of luminosity-dependent
bias for red and blue galaxies are shown in Fig.~\ref{fig:comparison_bvmag}
in the middle and right panels. Indications of the differing trends
for red and blue galaxies have been observed in previous work: an
early hint of the upturn in the bias for dim red galaxies was seen
in \citet{2002MNRAS.332..827N}, and recent results \citep{2007ApJ...664..608W}
also indicate higher bias for dim red galaxies at scales $>1\, h^{-1}\rmn{Mpc}$.
However, there is some inconsistency between these results compared
to \citet{2005ApJ...630....1Z} and \citet{2006MNRAS.368...21L} regarding
the dim red galaxies: they find that dim red galaxies exhibit the
strongest clustering on scales $<1\, h^{-1}\rmn{Mpc}$ and luminous
red galaxies exhibit the strongest clustering on larger scales, as
can be seen from the green points in Fig.~\ref{fig:comparison_bvmag}.
This is shown in fig.~14 of \citet{2005ApJ...630....1Z} and fig.~11
of \citet{2006MNRAS.368...21L}. However, we find the dim red galaxies
to have higher bias than $L_{*}$ red galaxies at all the scales we
probe ($2-40\, h^{-1}\rmn{Mpc}$ in this case). This upturn of the
bias for dim red galaxies is present in the halo model-based theoretical
curves from \citet{2003MNRAS.340..771V}, although not in the theoretical
curves from \citet{2005MNRAS.363..337C}. Note also that \citet{2003MNRAS.340..771V}
use the data from \citet{2002MNRAS.332..827N} to constrain their
models so the agreement between the theory and data should be interpreted
with some caution.

Recent measurements of higher order clustering statistics \citep{2007MNRAS.379.1562C}
find the same trends in the clustering strengths of red and blue galaxies,
although they indicate that their linear bias measurement (which should
be comparable to ours) shows the opposite trends -- little luminosity
dependence for red galaxies and a slight monotonic increase for blue
galaxies. However, their luminosity range is much narrower than ours
so the trends are less clear, and placing their data points on Fig.~\ref{fig:comparison_bvmag}
shows that they are in good agreement with our results. 

Previous studies \citep{2002MNRAS.332..827N,2006MNRAS.368...21L,2007ApJ...664..608W}
have also reported a somewhat stronger luminosity dependence of blue
galaxy clustering than we have measured here. As can be seen in Fig.~\ref{fig:comparison_bvmag},
\citet{2002MNRAS.332..827N} and \citet{2007ApJ...664..608W} measure
slightly higher bias for luminous blue galaxies, and \citep{2006MNRAS.368...21L}
measure slightly lower bias for dim blue galaxies. Although the quantitative
disagreement is fairly small, the qualitative trends of the previous
studies imply that the bias of blue galaxies increases with luminosity,
as opposed to our measurement which indicates a lack of luminosity
dependence.

\subsection{\label{sub:Likelihood-Results}Likelihood Results}

To study the luminosity dependence, color dependence and stochasticity
of bias in more detail, we also apply the maximum likelihood method
described in \S\ref{sub:Maximum-Likelihood-Method} to all
of the same pairs of samples used in the null-buster test. Due to
constraints on computing power and memory, we perform these calculations
for only three values of the cell size $\elll$ rather than four,
dropping the highest resolution (smallest cell size) shown in Fig.~\ref{fig:angularcells}.
The likelihood analysis makes a few additional assumptions, but provides
a valuable cross-check and also a measurement of the parameter $r_{\rmn{rel}}$
which encodes the stochasticity and non-linearity of the relative
bias.

For each pair of samples, the likelihood function given in equation~\eqref{eq:likelihood}
is maximized with respect to the parameters $\sigma_{1}^{2}$, $b_{\rmn{rel}}$,
and $r_{\rmn{rel}}$ and marginalized over $\sigma_{1}^{2}$ to determine
the best-fitting values of $b_{\rmn{rel}}$ and $r_{\rmn{rel}}$,
with uncertainties defined by the $\Delta\left(2\ln\mathcal{L}\right)=1$
contour in the $b_{\rmn{rel}}$-$r_{\rmn{rel}}$ plane. As we discuss
in \S\ref{sub:Comparison-with-null-buster}, the values of
$b_{\rmn{rel}}$ found here are consistent with those determined using
the null-buster test. 
%
\begin{figure*}
\includegraphics[width=1\textwidth]{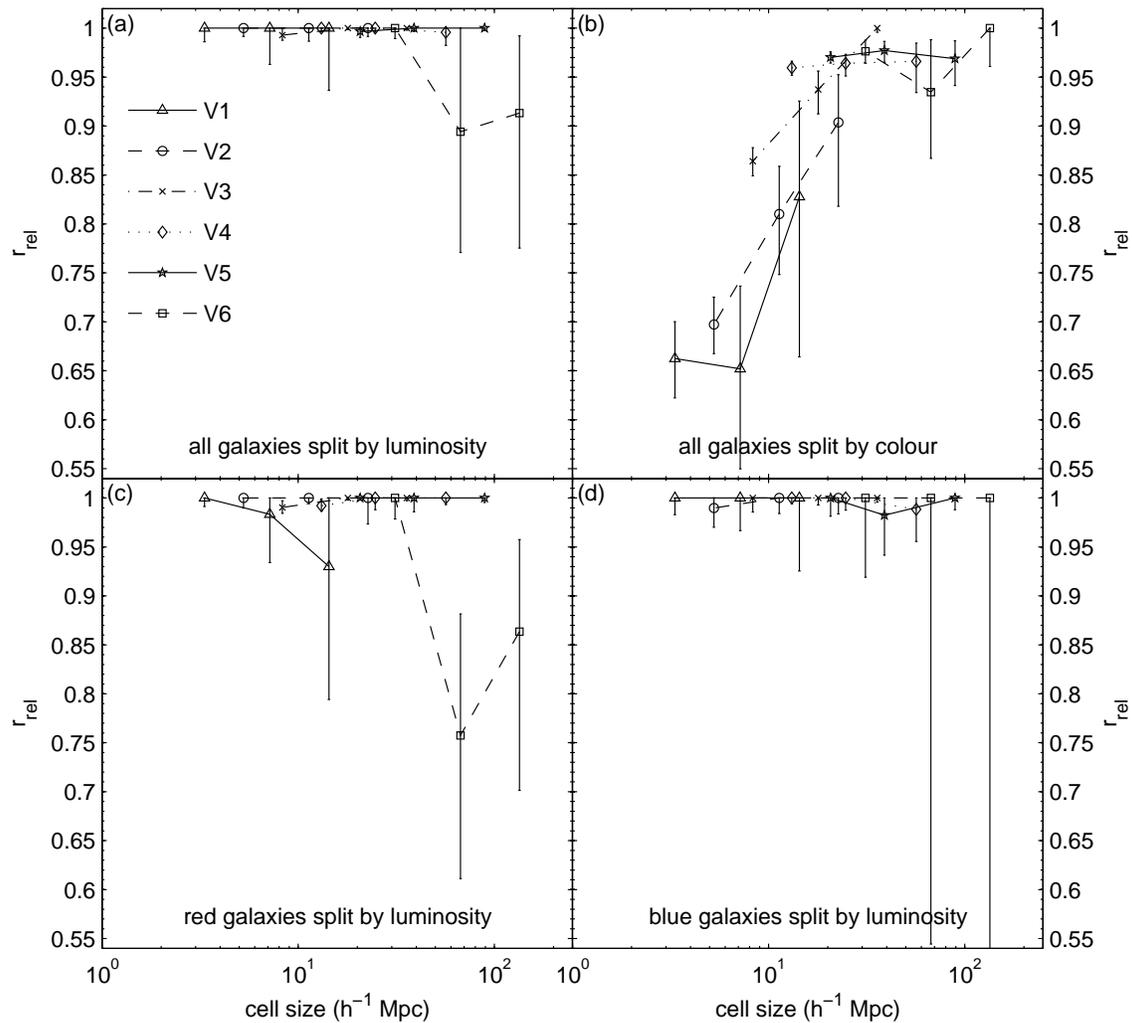}

\caption[The best-fitting values of the relative cross-correlation
coefficient $r_{\rmn{rel}}$ between pairwise samples.]{\label{fig:rplot}The best-fitting values of the relative cross-correlation
coefficient $r_{\rmn{rel}}$ between pairwise samples. Four different
types of pairwise comparison are illustrated: (a) luminous vs. dim,
(b) red vs. blue, (c) luminous red vs. dim red, and (d) luminous blue
vs. dim blue. The different symbols denote the different comparison
volumes V1-V6. }
\end{figure*}

Figure~\ref{fig:rplot} shows the best-fitting values of $r_{\rmn{rel}}$
as a function of cell size $\elll$. For the comparisons between neighboring
luminosity bins, the results are consistent with $r_{\rmn{rel}}=1$.
On the other hand, the comparisons between red and blue galaxies give
$r_{\rmn{rel}}<1$, with smaller cell sizes $\elll$ giving smaller
values of $r_{\rmn{rel}}$. This confirms the null-buster result that
the luminosity-dependent bias can be accurately modeled using simple
deterministic linear bias but color-dependent bias demands a more
complicated model. Also, $r_{\rmn{rel}}$ for the color-dependent
bias is seen to depend on scale but not strongly on luminosity. In
contrast, $b_{\rmn{rel}}$ (both in the null-buster and likelihood
analyses) depends on luminosity but not on scale.

To summarize, we find that the simple, deterministic model is a good
fit for the luminosity-dependent bias, but the color-dependent bias
shows evidence for stochasticity and/or non-linearity which increases
in strength towards smaller scales. These results are consistent with
previous detections of stochasticity/non-linearity in spectral-type-dependent
bias \citep{1999ApJ...518L..69T,2000ApJ...544...63B,2005MNRAS.356..456C},
and also agree with \citep{2005MNRAS.356..247W} which measures significant
stochasticity between galaxies of different color or spectral type,
but not between galaxies of different luminosities. 

\begin{figure}
\begin{center}%
\includegraphics[width=.4746\textwidth]{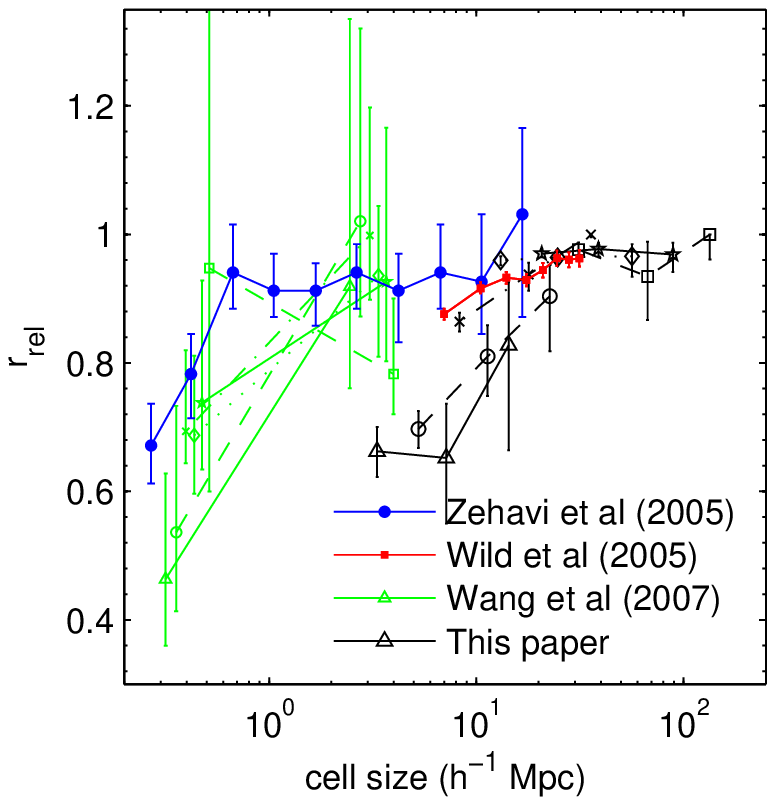}
\end{center}%
\caption[Comparison of relative cross-correlation
coefficient $r_{\rmn{rel}}$ between red and blue galaxies as measured
with different techniques.]{\label{fig:comparison_r}Comparison of relative cross-correlation
coefficient $r_{\rmn{rel}}$ between red and blue galaxies as measured
with different techniques. The points from \citet{2005ApJ...630....1Z}
are extracted from cross-correlation measurements between red and
blue galaxies in SDSS with $M_{^{0.1}r}<-21$, and the \citet{2005MNRAS.356..247W}
points are from a counts-in-cells analysis using all 2dFGRS galaxies.
Our results and the \citet{2007ApJ...664..608W} results (also from
SDSS) are separated by luminosity -- symbols are the same as in Fig.~\ref{fig:rplot},
and for the \citet{2007ApJ...664..608W} results, open triangles denote
their dimmest bin ($-19<M_{^{01.}r}<-18$) and open squares denote
their most luminous bin ($-23<M_{^{01.}r}<-21.5$). The length scales
used in \citet{2007ApJ...664..608W} are averages over small scales
($0.16-0.98\, h^{-1}\rmn{Mpc}$) and large scales ($0.98-9.8\, h^{-1}\rmn{Mpc}$)
-- points here are shown in the middle of these ranges and offset
for clarity.}
\end{figure}

We compare our results for $r_{\rmn{rel}}$ for red and blue galaxies
to previous results in Fig.~\ref{fig:comparison_r}. This shows good
agreement between our results and those of \citet{2005MNRAS.356..247W}
($r_{\rmn{lin}}$ from their fig.~11), implying that these results
are quite robust since our analysis uses a different data set, employs
different methods, and makes different assumptions. %

For the results from cross-correlation measurements, however, the
agreement is not as clear. \citet{2005ApJ...630....1Z} find that
the cross-correlation between red and blue galaxies (their fig.~24),
indicates that $r_{\rmn{rel}}$ is consistent with 1 on scales $>1\, h^{-1}\rmn{Mpc}$.
However, it is not clear that this result disagrees with ours, as
their result is for luminous galaxies ($M_{^{0.1}r}<-21$) and and
we do not see a strong indication of $r_{\rmn{rel}}<1$ for our V6
sample ($23<M_{^{0.1}r}<-21$). 

More recent cross-correlation measurements \citep{2007ApJ...664..608W}
do find evidence for stochasticity/non-linearity between red and blue
galaxies at scales $<1\, h^{-1}\rmn{Mpc}$ and also show an indication
that dimmer galaxies have slightly lower values of $r_{\rmn{rel}}$.
Note also that the method of calculating $r_{\rmn{rel}}$ by taking
ratios of cross- and auto-correlation functions as used for \citet{2005ApJ...630....1Z}
and \citet{2007ApJ...664..608W} does not automatically incorporate
the constraint that $\left|r_{\rmn{rel}}\right|\le1$ as our analysis
does, so their error bars are allowed to extend above $r_{\rmn{rel}}=1$
in Fig.~\ref{fig:comparison_r}.

Overall, the counts-in-cells measurements (this chapter, \citealt{2005MNRAS.356..247W})
show stronger evidence for stochasticity/non-linearity at larger scales
than the cross-correlation measurements \citep{2005ApJ...630....1Z,2007ApJ...664..608W},
indicating either that there might be some slight systematic variation
between the two methods or that the counts-in-cells method is more
sensitive to these effects.

\section{Conclusions}

\label{Conclusions} To shed further light on how galaxies trace matter,
we have quantified how different types of galaxies trace each other.
We have analyzed the relative bias between pairs of volume-limited
galaxy samples of different luminosities and colors using counts-in-cells
at varying length scales. This method is most sensitive to length
scales between those probed by correlation function and power spectrum
methods, and makes point-by-point comparisons of the density fields
rather than using ratios of moments, thereby eliminating sample variance
and obtaining a local rather than global measure of the bias. We applied
a null-buster test on each pair of subsamples to determine if the
relative bias was consistent with 
deterministic linear biasing, and we also performed a maximum-likelihood
analysis to find the best-fitting parameters for a simple stochastic
biasing model.

\subsection{Biasing results}

Our primary results are:

\begin{enumerate}
\item The luminosity-dependent bias for red galaxies is significantly different
from that of blue galaxies: the bias of blue galaxies shows only a
weak dependence on luminosity, whereas both luminous and dim red galaxies
have higher bias than moderately bright ($L_{*}$) red galaxies. 
\item Both of our analysis methods indicate that the simple, deterministic
model is a good fit for the luminosity-dependent bias, but that the
color-dependent bias is more complicated, showing strong evidence
for stochasticity and/or non-linearity on scales $\la10h^{-1}$ Mpc. 
\item The luminosity-dependent bias is consistent with being scale-independent
over the range of scales probed here $\left(2-160\, h^{-1}\rmn{Mpc}\right)$.
The color-dependent bias depends on luminosity but not on scale,
while the cross-correlation coefficient $r_{\rmn{rel}}$ depends on
scale but not strongly on luminosity, giving smaller $r_{\rmn{rel}}$
values at smaller scales. 
\end{enumerate}
These results are encouraging from the perspective of using galaxy
clustering to measure cosmological parameters: simple scale-independent
linear biasing appears to be a good approximation on the $\ga60{\, h}^{-1}\rmn{Mpc}$
scales used in many recent cosmological studies (e.g., \citet{2006MNRAS.366..189S,2006PhRvD..74l3507T,2007ApJS..170..377S}).
However, further quantification of small residual effects will be
needed to do full justice to the precision of next-generation data
sets on the horizon. Moreover, our results regarding color sensitivity
suggest that more detailed bias studies are worthwhile for luminous
red galaxies, which have emerged a powerful cosmological probe because
of their visibility at large distances and near-optimal number density
\citep{2001AJ....122.2267E,2005ApJ...633..560E,2006PhRvD..74l3507T},
since color cuts are involved in their selection.

\subsection{Implications for galaxy formation}

What can these results tell us about galaxy formation in the context
of the halo model? First of all, as discussed in \citet{2005ApJ...630....1Z},
the large bias of the faint red galaxies can be explained by the fact
that such galaxies tend to be satellites in high mass halos, which
are more strongly clustered than low mass halos. Previous studies
have found that central galaxies in low-mass halos are preferentially
blue, central galaxies in high mass halos tend to be red, and that
the luminosity of the central galaxy is strongly correlated with the
halo mass \citep{2005MNRAS.358..217Y,2005ApJ...633..791Z}. Our observed
lack of luminosity dependence of the bias for blue galaxies would
then be a reflection of the correlation between luminosity and halo
mass being weaker for blue galaxies than for red ones. 
Additional work is needed to study this quantitatively and compare
it with theoretical predictions from galaxy formation models.

The detection of stochasticity between red and blue galaxies may imply
that red and blue galaxies tend to live in different halos -- a study
of galaxy groups in SDSS \citep{2006MNRAS.366....2W} recently presented
evidence supporting this, but this is at odds with the cross-correlation
measurement in \citet{2005ApJ...630....1Z}, which implies that blue
and red galaxies are well-mixed within halos. The fact that the stochasticity
is strongest at small scales suggests that this effect is due to the
1-halo term, i.e., arising from pairs of galaxies in the same halo,
although some amount of stochasticity persists even for large scales.
However, the halo model implications for stochasticity have not been
well-studied to date. 





In summary, our results on galaxy biasing and future work along these
lines should be able to deepen our understanding of both cosmology
(by quantifying systematic uncertainties) and galaxy formation.


\begin{subappendices}
\section{Consistency Checks}

\subsection{Alternate null-buster analyses}

In order to test the robustness of our results against various systematic
effects, we have repeated the null-buster analysis with four different
modifications: splitting the galaxy samples randomly, offsetting the
pixel positions, using galaxy positions without applying the finger-of-god
compression algorithm, and ignoring the cosmological correlations
between neighboring cells.

\subsubsection{Randomly split samples}

The null-buster test assumes that the Poissonian shot noise for each
type of galaxy in each pixel is uncorrelated -- i.e., that the matrix
$\mathbfss{N}$ in equation~\eqref{eq:N_and_S} is diagonal -- and
that this shot noise can be approximated as Gaussian. To test the
impact of these assumptions on our results, we repeated the null-buster
analysis using randomly split galaxy samples rather than splitting
by luminosity or color. For each volume V1-V6, we created two samples
by generating a uniformly distributed random number for each galaxy
and assigning it to sample 1 for numbers $>0.5$ and sample 2 otherwise.

\begin{figure*}
\includegraphics[width=1\textwidth]{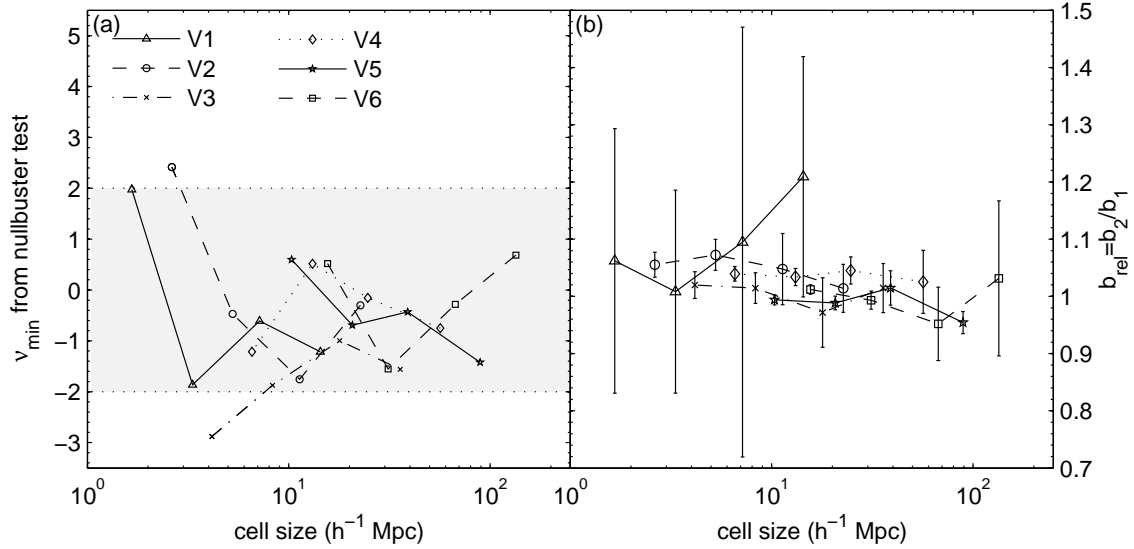}

\caption{\label{fig:randomplot}Null-buster results for randomly split samples.}
\end{figure*}

If the null-buster test is accurate, we expect the pairwise comparison
for the randomly split samples to be consistent with deterministic
linear bias with $b_{\rmn{rel}}=1$. The results are shown in Fig.~\ref{fig:randomplot}
-- we find that, indeed, deterministic linear bias is not ruled out,
with nearly all of the $\nu_{\rmn{min}}$ points falling within $\pm2$.
Furthermore, the measured values of $b_{\rmn{rel}}$ are seen to be
consistent with 1. Thus, we detect no systematic effects due to the
null-buster assumptions.

\subsubsection{Offset pixel positions}

\label{sub:Offset-pixel-positions}

To test if our results are stable against the pixelization chosen,
particularly at large scales where we have a small number of cells,
we shifted the locations on the sky of the angular pixels defining
the cells by half a pixel width in declination. Applying the null-buster
analysis to the offset pixels reveals no significant differences from
the original analysis: the luminosity-dependent comparisons for all,
red, and blue galaxies are still consistent with deterministic linear
bias, and the color-dependent comparison still shows strong evidence
for stochasticity and/or nonlinearity, especially at smaller scales.

For the all-, red-, and blue-galaxy, luminosity-dependent comparisons,
we also compared the measured values of $b_{\rmn{rel}}$ from the
offset analysis with the original analysis: we took the difference
between the two measured values in each volume at each resolution
and divided this by the larger of the error bars on the two analyses
to determine the number of sigmas by which the two analyses differ.
In order to be conservative, we did not add the error bars from the
two analyses in quadrature, since this would overestimate the error
on the difference if they are correlated and this would make our test
less robust. Note that a fully proper treatment would necessitate
accounting for the correlations between the errors from each analysis,
which we have not done -- the discussions in the this and the following
sections are meant only to serve as a crude reality check.

For the error bars on the original analysis, we used the jackknife
uncertainties described in \S\ref{sub:Jackknifes}, and for
the offset analysis error bars we use the generalized $\chi^{2}$
uncertainties described in \S\ref{sub:The-Null-buster-Test}
computed from the offset results. (This is because we did not perform
jackknife resampling for the offset case or the other modified analyses.) 

The results show good agreement: out of a total of 72 measured $b_{\rmn{rel}}$
values, only 4 differ by more than $2\sigma$ (all galaxies in V3
and V5 at the second-smallest cell size, at $2.6\sigma$ and $3.2\sigma$
respectively, and the red galaxies in V3 and V5 at the second-smallest
cell size, at $2.2\sigma$and $3.3\sigma$ respectively). As a rough
test for systematic trends, we also counted the number of measurements
for which the measured value of $b_{\rmn{rel}}$ is larger in each
analysis. We found that in 38 cases the value from the offset analysis
was larger, and in 34 cases the value from the original analysis was
larger, indicating no systematic trends in the deviations.

\subsubsection{No finger-of-god compression}

Our analysis used the finger-of-god compression algorithm from \citet{2004ApJ...606..702T}
with a threshold density of $\delta_{c}=200$. This gives a first-order
correction for redshift space distortions, but it complicates comparisons
to other analyses which work purely in redshift space (e.g. \citealt{2005MNRAS.356..247W})
or use projected correlation functions (e.g. \citealt{2005ApJ...630....1Z}),
particularly at small scales $\left(\lesssim10\, h^{-1}\rmn{Mpc}\right)$
where the effects of virialized galaxy clusters could be significant.
To test the sensitivity of our results to this correction, we repeated
the null-buster analysis with no finger-of-god compression. 

The results show excellent agreement with the original analysis --
the smallest-scale measurements for the color-dependent comparison
only rule out deterministic linear bias at 30 sigma rather than 40,
but the conclusions remain the same. Additionally, we compared the
measured $b_{\rmn{rel}}$ values to the original analysis as in 
\S\ref{sub:Offset-pixel-positions} and find all 72 measurements to
be within $2\sigma$. In 25 cases, the analysis without finger-of-god
compression gave a larger $b_{\rmn{rel}}$ value, and in 47 cases
the original analysis gave a larger value. This indicates that there
might be a very slight tendency to underestimate $b_{\rmn{rel}}$
if fingers-of-god are not accounted for, but the effect is quite small
and well within our error bars. Thus, the finger-of-god compression
has no substantial impact on our results.

\subsubsection{Uncorrelated signal matrix}
\label{sub:Uncorrelated-signal-matrix}
The null-buster test requires a choice of residual signal matrix ${\mathbfss{S}}_{\Delta}$
-- our analysis uses a signal matrix derived from the matter power
spectrum, thus accounting for cosmological correlations between neighboring
cells. However, these correlations are commonly assumed to be negligible
in other counts-in-cells analyses \citep{2000ApJ...544...63B,2005MNRAS.356..247W,2005MNRAS.356..456C}.
To test the sensitivity to the choice of ${\mathbfss{S}}_{\Delta}$,
we repeated the analysis using ${\mathbfss{S}}_{\Delta}$ equal to
the identity matrix.

Again, we find the results to agree well with the original analysis
and lead to the same conclusions. When comparing the measured $b_{\rmn{rel}}$
values to the original analysis as in \S\ref{sub:Offset-pixel-positions},
we find only 4 out of 72 points differing by more than $2\sigma$
(all galaxies in V5 at the smallest cell size, at $-2.6\sigma$, red
galaxies in V4 at the second-smallest and smallest cell size, at $2.2\sigma$
and $2.8\sigma$ respectively, and blue galaxies in V4 at the smallest
cell size, at $-3.1\sigma$). In 39 cases the value of $b_{\rmn{rel}}$
is larger with the uncorrelated signal matrix, and in 33 cases $b_{\rmn{rel}}$
is larger in the original analysis, indicating no strong systematic
effects. Thus we expect our results to be directly comparable to other
counts-in-cells analyses done without accounting for cosmological
correlations.

\section{Uncertainty Calculations}

\subsection{Jackknife uncertainties for null-buster analysis}

\label{sub:Jackknifes}

\begin{figure*}
\includegraphics[width=1\textwidth]{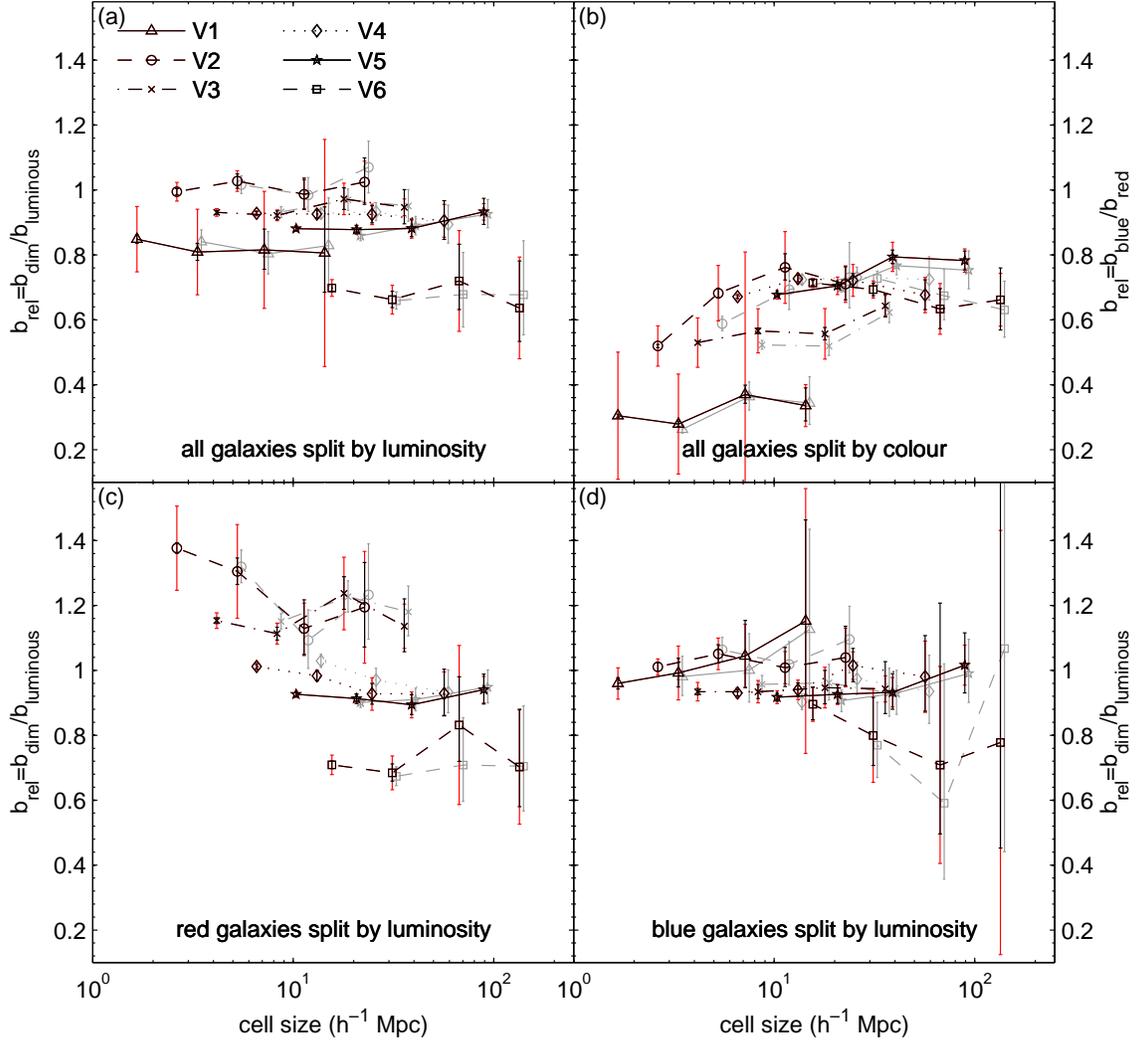}

\caption[Comparison of jackknife and generalized $\chi^{2}$
uncertainties on $b_{\rmn{rel}}$ from the null-buster analysis.]{\label{fig:likelihoodvnullbuster}Comparison of two methods for calculating
uncertainties on $b_{\rmn{rel}}$ from the null-buster analysis: jackknife
resampling (red) and the generalized $\chi^{2}$ method (black). Also
shown are the results for $b_{\rmn{rel}}$ from the likelihood analysis
(grey).}
\end{figure*}
We use jackknife resampling to calculate the uncertainties for the
null-buster analysis. The concept is as follows: divide area covered
on the sky into $N$ spatially contiguous regions, and then repeat
the analysis $N$ times, omitting each of the $N$ regions in turn.
The covariance matrix for the measured parameters is then estimated
by \begin{equation}
\mathbf{\Sigma}_{\rmn{rel}}^{ij}=\frac{N-1}{N}\sum_{k=1}^{N}\left(b_{\rmn{rel},k}^{i}-\overline{b_{\rmn{rel}}^{i}}\right)\left(b_{\rmn{rel},k}^{j}-\overline{b_{\rmn{rel}}^{j}}\right)\label{eq:jackknife}\end{equation}
where superscripts $i$ and $j$ denote measurements of $b_{\rmn{rel}}$
in different volumes and at different scales, $b_{\rmn{rel},k}^{i}$
denotes the value of $b_{\rmn{rel}}^{i}$ with the $k$th jackknife
region omitted, and $\overline{b_{\rmn{rel}}^{i}}$ is the average
over all $N$ values of $b_{\rmn{rel},k}^{i}$.

For our analysis, we use the 15 pixels at our lowest resolution (upper
left panel in Fig.~\ref{fig:angularcells}) as the jackknife regions.
However, since we use a looser completeness cut at the lowest resolution,
two of these pixels cover an area that is not used at higher resolutions.
Thus we chose not to include these two pixels in our jackknifes since
they do not omit much (or any) area at the higher resolutions. Thus
our jackknife resampling has $N=13$. This technique allows us to
estimate the uncertainties on all of our $b_{\rmn{rel}}$ measurements
as well as the covariance matrix quantifying the correlations between
them. We use these covariances in the model-fitting done in \S\ref{sub:scale-dependent}~and~\S\ref{sub:luminosity-dependent}.

Figure~\ref{fig:likelihoodvnullbuster} shows the uncertainties on
$b_{\rmn{rel}}$ as calculated from jackknife resampling compared
to those calculated with the generalized $\chi^{2}$ method described
in \S\ref{sub:The-Null-buster-Test}. Overall, the two methods
agree well, but the jackknife method gives larger uncertainties at
the smallest scales and in volume V1. The reason for the large jackknife
uncertainties in volume V1 is because it is significantly smaller
than the other volumes, and it is small enough that omitting a cell
containing just one large cluster can have a substantial effect on
the measured value of $b_{\rmn{rel}}$. Thus the large uncertainties
in V1 reflect the effects of sampling a small volume. Since there
are so few dim red galaxies, these effects are particularly egregious
for the measurements of luminosity-dependent bias of red galaxies
in V1. Thus, based on the jackknife results, we elected to not use
V1 in our analysis of the red galaxies.

\subsection{Likelihood uncertainties}

\label{sub:Likelihood-contour-plots}

\subsubsection{Likelihood contours}

\begin{figure*}
\includegraphics[width=1\textwidth]{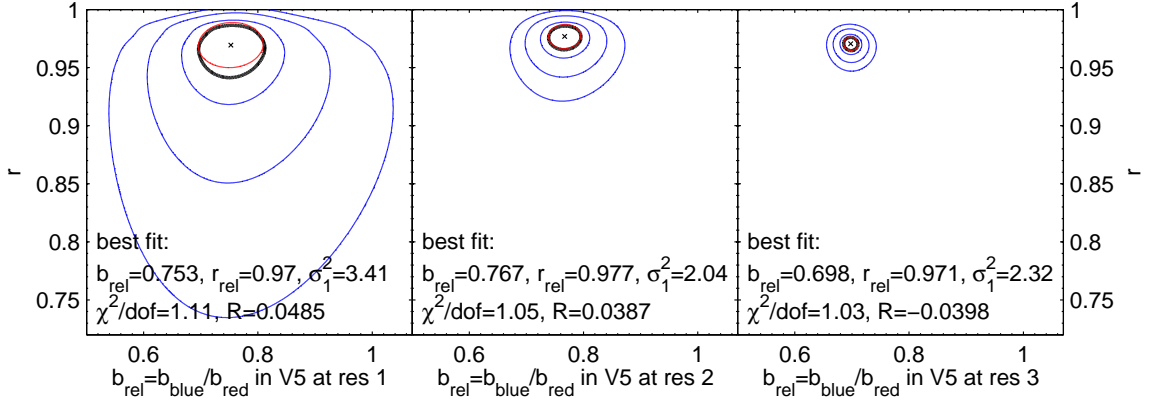}

\caption[Typical contour plots of $\Delta\left(2\ln\mathcal{L}\right)$
for volume V5 for three different resolutions.]{\label{fig:likelihoodexamples}Typical contour plots of $\Delta\left(2\ln\mathcal{L}\right)$
for volume V5 for three different resolutions corresponding (from
left- to right-hand-side) to cell sizes of 89, 39, and 21 $h^{-1}\rmn{Mpc}$.
Blue contours denote the $1$, $2$ and $3\sigma$ two-dimensional
confidence regions, and black contours denote the $1\sigma$ one-dimensional
confidence region used for computing error bars on $b_{\rmn{rel}}$
and $r_{\rmn{rel}}$. The red contours denote the error ellipse calculated
from the second-order approximation to $2\ln\mathcal{L}$ at the best-fit
point, marked with a $\times$.}
\end{figure*}
As described in \S\ref{sub:Maximum-Likelihood-Method}, we
calculate the uncertainty on $b_{\rmn{rel}}$ and $r_{\rmn{rel}}$
for the likelihood method using the $\Delta\left(2\ln\mathcal{L}\right)=1$
contour in the $b_{\rmn{rel}}$-$r_{\rmn{rel}}$ plane after marginalizing
over $\sigma_{1}^{2}$. This means that for each comparison volume
and at each resolution, we calculate $\mathcal{L}$ from equation~\eqref{eq:likelihood}
over a grid of $b_{\rmn{rel}}$ and $r_{\rmn{rel}}$ values and maximize
$2\ln\mathcal{L}$ with respect to $\sigma_{1}^{2}$ at each grid
point. This gives us a 2-dimensional likelihood function, which we
then maximize to find the best-fit values for $b_{\rmn{rel}}$ and
$r_{\rmn{rel}}$. Uncertainties are calculated using the function
\begin{equation}
\Delta\left(2\ln\mathcal{L}\left(b_{\rmn{rel},}r_{\rmn{rel}}\right)\right)\equiv2\ln\mathcal{L}\left(b_{\rmn{rel},}^{\rmn{max}}r_{\rmn{rel}}^{\rmn{max}}\right)-2\ln\mathcal{L}\left(b_{\rmn{rel},}r_{\rmn{rel}}\right).\label{eq:contour}\end{equation}
Typical contour plots of this function for volume V5 at each of the
three cell sizes used are shown in Fig.~\ref{fig:likelihoodexamples}.

We define 1- and 2-dimensional confidence regions using the standard
procedures detailed in \citet{1992nrca.book.....P}, using $\Delta\left(2\ln\mathcal{L}\right)$
as an equivalent to $\Delta\chi^{2}$: the $1\sigma$ (68.3\%) 1-dimensional
confidence region is given by $\Delta\left(2\ln\mathcal{L}\right)=1$,
so we define our error bars on $b_{\rmn{rel}}$ and $r_{\rmn{rel}}$
by projecting the $\Delta\left(2\ln\mathcal{L}\right)=1$ contour
(shown in black in Fig.~\ref{fig:likelihoodexamples}) onto the $b_{\rmn{rel}}$
and $r_{\rmn{rel}}$ axes. For illustrative purposes we also show
the $1\sigma$, $2\sigma$, and $3\sigma$ (68.3\%, 95.4\%, and
99.73\%) 2-dimensional confidence regions in these plots, given by
$\Delta\left(2\ln\mathcal{L}\right)=2.30$, $6.17$, and $11.8$ respectively.

To check the goodness of fit, we also compute an effective value of
$\chi^{2}$: \begin{equation}
\chi_{\rmn{eff}}^{2}\equiv-2\ln\mathcal{L}-\ln\left|\mathbfss{C}\right|-2n\ln\left(2\pi\right),\label{eq:chi_eff}\end{equation}
where $\mathbfss{C}$ is given by equation~\eqref{eq:cov_matrix}
and $n$ is the number of cells. If our model is a good fit, the value
of $\chi_{\rmn{eff}}^{2}$ at the best fit parameter values should
be close to the number of degrees of freedom, given by $\rmn{dof}=2n-2$
($2n$ data points for type 1 and 2 galaxies in each cell minus 2
parameters $b_{\rmn{rel}}$ and $r_{\rmn{rel}}$). We calculated $\chi_{\rmn{eff}}^{2}/\rmn{dof}$
for each volume and resolution and found they all lie quite close
to 1, ranging from a minimum value of $0.678$ to a maximum value
of $1.11$. Thus this test indicates our model is a good fit.

The uncertainties on $b_{\rmn{rel}}$ and $r_{\rmn{rel}}$ could perhaps
be calculated more accurately using jackknife resampling as we did
for the null-buster case; however, repeating the analysis for each
jackknife sample is computationally prohibitive since performing all
the calculations for just one likelihood analysis took several months
of CPU time.

\subsubsection{$b_{\rmn{rel}}$-$r_{\rmn{rel}}$ covariance matrices}

Alternatively, we can calculate the uncertainties using the parameter
covariance matrix at the best fit parameter values, as is commonly
done in $\chi^{2}$ analyses. The Hessian matrix of second derivatives
is given by \begin{equation}
\mathbfss{H}\equiv\left(\begin{array}{cc}
\frac{d^{2}\left(2\ln\mathcal{L}\right)}{db_{\rmn{rel}}^{2}} & \frac{d^{2}\left(2\ln\mathcal{L}\right)}{db_{\rmn{rel}}dr_{\rmn{rel}}}\\
\frac{d^{2}\left(2\ln\mathcal{L}\right)}{db_{\rmn{rel}}dr_{\rmn{rel}}} & \frac{d^{2}\left(2\ln\mathcal{L}\right)}{dr_{\rmn{rel}}^{2}}\end{array}\right)\label{eq:hessian}\end{equation}
and the parameter covariance matrix is given by \begin{equation}
{\mathbfss{C}}_{\rmn{param}}\equiv2{\mathbfss{H}}^{-1}=\left(\begin{array}{cc}
\sigma_{b_{\rmn{rel}}}^{2} & \sigma_{{b_{\rmn{rel}}r}_{\rmn{rel}}}^{2}\\
\sigma_{{b_{\rmn{rel}}r}_{\rmn{rel}}}^{2} & \sigma_{r_{\rmn{rel}}}^{2}\end{array}\right).\label{eq:param_cov}\end{equation}
Thus the uncertainties are given by $\sigma_{b_{\rmn{rel}}}^{2}$
and $\sigma_{r_{\rmn{rel}}}^{2}$ with this method. This is equivalent
to approximating the likelihood function $2\ln\mathcal{L}$ with its
second-order Taylor series about the best-fit point, and it defines
an error ellipse that approximates the $\Delta\left(2\ln\mathcal{L}\right)=1$
contour. These error ellipses are shown in Fig.~\ref{fig:likelihoodexamples}
in red, and are seen to be in close agreement with the true $\Delta\left(2\ln\mathcal{L}\right)=1$
contours.

This method also allows us to measure the correlation between $b_{\rmn{rel}}$
and $r_{\rmn{rel}}$ by calculating the correlation coefficient, given
by \begin{equation}
R\equiv\frac{\sigma_{{b_{\rmn{rel}}r}_{\rmn{rel}}}^{2}}{\left(\sigma_{b_{\rmn{rel}}}^{2}\sigma_{r_{\rmn{rel}}}^{2}\right)^{1/2}}.\label{eq:corr_coeff}\end{equation}
$R$ will fall between -1 (perfectly anti-correlated) and 1 (perfectly
correlated). Effectively this measures the tilt of the error ellipse
in the $b_{\rmn{rel}}$-$r_{\rmn{rel}}$ plane. Overall we find the
values of $R$ to be quite small -- typically $\left|R\right|\sim0.05$
-- indicating no large correlations between $b_{\rmn{rel}}$ and $r_{\rmn{rel}}$.
Out of the 72 points we calculate, only 6 have $\left|R\right|>0.3$.
The cases with the largest $R$ values are for blue galaxies in volume
V6, where the uncertainties are quite large due to the small number
of bright blue galaxies and the error ellipses are not good approximations
to the likelihood contours anyway -- thus these few cases with large
$R$ are not overly concerning. 


\subsubsection{Comparison with null-buster results}

\label{sub:Comparison-with-null-buster}

Finally, we compare the results for $b_{\rmn{rel}}$ from the likelihood
method to the results from the null-buster analysis in Fig.~\ref{fig:likelihoodvnullbuster},
with the likelihood points shown in grey. As can be seen in this plot,
the likelihood and null-buster values for $b_{\rmn{rel}}$ agree within
the uncertainties, even for the color-dependent bias where the null-buster
values are not necessarily accurate since deterministic linear bias
is ruled out. Thus our two analysis methods are in excellent agreement
with each other.

\end{subappendices}
\cleardoublepage
\addtocontents{toc}{\vspace{2\baselineskip}}
\phantomsection
\part*{}
\chapter{Conclusions}
\label{chap:conclusions}

We set out in this thesis to explore the connections between particle physics and astrophysics by using the tools of particle physics to do astronomy, a.k.a. ``Astrophysics Underground,'' and by using the tools of astronomy to do particle physics, a.k.a. ``Particle Physics in the Sky.''  What have we accomplished in these endeavors?

In ``Astrophysics Underground,'' we performed a search for extremely high-energy upward-going muons in the Super-K neutrino detector and set an upper limit on the flux of neutrinos coming from astrophysical sources such as AGNs. Have we really learned anything about astrophysics here?  Yes, albeit rather indirectly -- our upper limit on the astrophysical neutrino flux tells us that AGNs are not producing an unexpectedly large amount of neutrinos. 
However, much more time and energy was devoted to thinking about the particle physics technical issues than about the primary astrophysical science goal.
We developed a way to use previously ignored information from the Super-K detector, studied how muons propagate through rock and water, and did a detailed analysis of various particle physics phenomena contributing to the background flux. 
Ultimately, most of what we learned was really about particle physics.  

This same effect was even more true in ``Particle Physics in the Sky.''  Here we developed a powerful software tool to manage angular masks of the next generation of galaxy surveys and analyzed the details of relative bias between galaxies of different luminosities and colors. These projects contribute to cosmology by providing techniques to better use galaxy surveys as a cosmological tool. In particular, gaining a better understanding of bias is key for reducing the systematic uncertainties inherent in galaxy survey cosmology. Have we learned any particle physics?  Again, yes, though again rather indirectly, in that we have contributed to the overall cosmological programs to understand dark matter and dark energy. What we really learned about is the physics of galaxies -- our results on how different types of galaxies cluster differently most directly shed light on the complex physical processes governing galaxy formation and evolution. It may appear overly optimistic to claim that we have really done any particle physics here, but it is important to remember that our understanding of cosmology from galaxy surveys is only as good as our understanding of bias, and if we hope to gain further insight about particle physics from the next generation of cosmological measurements, it is essential that we understand the physics of galaxies at a deeper level.

One particularly interesting direction for future research is to use relative bias measurements as a guide for refining halo model methods for putting galaxies in dark matter halos, as discussed in \S\ref{sub:Halo-model}. This can help us gain a better understanding of the physics that links galaxy properties to halo mass, and provide a better way of modeling of complicated bias. Furthermore, we can combined this with advances in nonlinear modeling of structure formation in order to further refine galaxy surveys as a tool for doing cosmology. This makes the biggest difference at length scales slightly smaller than where the power spectrum is typically measured, a regime which is particularly important for setting cosmological bounds on the neutrino mass. Such investigations will also help reduce systematic uncertainties on measuring the baryon oscillation length scale as a standard ruler for studying dark energy. Thus, even if we have not addressed them directly, the particle physics science goals are clearly in sight.

It is perhaps a natural outcome of studying the science questions of one field with the techniques of another that one gets more immersed in the area of the technique. This is borne out in the traditional physics department divisions as well -- the work on neutrino astrophysics presented here was done within a particle physics division, and in general is done by scientists who call themselves particle physicists rather than astronomers.  Likewise, the work on cosmology was done within an astrophysics division and is more often grouped with astronomy, not high energy physics.  In many ways this makes sense -- after all, if a scientist at such an interface spends most of their time thinking about the physics of the techniques they are using, it is perfectly logical to surround oneself with experts in that area.  However, more direct interaction between scientists on either side of the fence can potentially be extremely beneficial, as is happening at new interdisciplinary research centers across the world.

In conclusion, we have hopefully provided valuable tools for upcoming experiments in both neutrino astronomy and cosmology. These lines of research represent areas where the traditionally separate fields of particle physics and astrophysics overlap and have produced much fruitful interaction between physicists exploring the largest and smallest scales in our universe. The work in this thesis makes contributions to the exciting research in progress at the interface of these two rich fields.


\addcontentsline{toc}{chapter}{Bibliography}
\phantomsection
\bibliography{thesis,bias,bias1,mangle,mangle1,apj-jour,superk,proposal,halo}

\begin{thebibliography}{}

\bibitem[\protect\citeauthoryear{{Abazajian} et~al.}{{Abazajian}
  et~al.}{2005}]{2005ApJ...625..613A}
{Abazajian}, K. et~al. (2005).
\newblock {Cosmology and the Halo Occupation Distribution from Small-Scale
  Galaxy Clustering in the Sloan Digital Sky Survey}.
\newblock \apj~625, 613--620.

\bibitem[\protect\citeauthoryear{{Abbas} and {Sheth}}{{Abbas} and
  {Sheth}}{2005}]{2005MNRAS.364.1327A}
{Abbas}, U. and R.~K. {Sheth} (2005).
\newblock {The environmental dependence of clustering in hierarchical models}.
\newblock \mnras~364, 1327--1336.

\bibitem[\protect\citeauthoryear{{Abbas} and {Sheth}}{{Abbas} and
  {Sheth}}{2006}]{2006MNRAS.372.1749A}
{Abbas}, U. and R.~K. {Sheth} (2006).
\newblock {The environmental dependence of galaxy clustering in the Sloan
  Digital Sky Survey}.
\newblock \mnras~372, 1749--1754.

\bibitem[\protect\citeauthoryear{{Abe} et~al.}{{Abe}
  et~al.}{2006}]{2006ApJ...652..198A}
{Abe}, K. et~al. (2006).
\newblock {High-Energy Neutrino Astronomy Using Upward-going Muons in
  Super-Kamiokande I}.
\newblock \apj~652, 198--205.

\bibitem[\protect\citeauthoryear{{Abell}}{{Abell}}{1959}]{1959ASPL....8..121A}
{Abell}, G.~O. (1959).
\newblock {The National Geographic Society-Palomar Observatory Sky Survey}.
\newblock Leaflet of the Astronomical Society of the Pacific~8, 121--+.

\bibitem[\protect\citeauthoryear{{Abell}}{{Abell}}{1961}]{1961AJ.....66..607A}
{Abell}, G.~O. (1961).
\newblock {Evidence regarding second-order clustering of galaxies and
  interactions between clusters of galaxies}.
\newblock \aj~66, 607--+.

\bibitem[\protect\citeauthoryear{{Abraham} et~al.}{{Abraham}
  et~al.}{2007}]{auger_agns}
{Abraham}, J. et~al. (2007).
\newblock {Correlation of the Highest-Energy Cosmic Rays with Nearby
  Extragalactic Objects}.
\newblock Science~318, 938--943.

\bibitem[\protect\citeauthoryear{{Achterberg} et~al.}{{Achterberg}
  et~al.}{2007}]{2007PhRvD..76d2008A}
{Achterberg}, A. et~al. (2007).
\newblock {Multiyear search for a diffuse flux of muon neutrinos with
  AMANDA-II}.
\newblock \prd~76(4), 042008--+.

\bibitem[\protect\citeauthoryear{{Adelman-McCarthy} et~al.}{{Adelman-McCarthy}
  et~al.}{2006}]{2006ApJS..162...38A}
{Adelman-McCarthy}, J.~K. et~al. (2006).
\newblock {The Fourth Data Release of the Sloan Digital Sky Survey}.
\newblock \apjs~162, 38--48.

\bibitem[\protect\citeauthoryear{{Adelman-McCarthy} et~al.}{{Adelman-McCarthy}
  et~al.}{2007}]{2007ApJS..172..634A}
{Adelman-McCarthy}, J.~K. et~al. (2007).
\newblock {The Fifth Data Release of the Sloan Digital Sky Survey}.
\newblock \apjs~172, 634--644.

\bibitem[\protect\citeauthoryear{{Aggouras} et~al.}{{Aggouras}
  et~al.}{2005}]{2005ICRC....5...91A}
{Aggouras}, G. et~al. (2005).
\newblock {Operation, Performance and Measurements with the NESTOR Test
  Detector}.
\newblock Volume~5 of Proceedings of the 29th International Cosmic Ray
  Conference,  91--+.

\bibitem[\protect\citeauthoryear{{Ahrens} et~al.}{{Ahrens}
  et~al.}{2004}]{2004NIMPA.524..169A}
{Ahrens}, J. et~al. (2004).
\newblock {Muon track reconstruction and data selection techniques in AMANDA}.
\newblock Nuclear Instruments and Methods in Physics Research A~524, 169--194.

\bibitem[\protect\citeauthoryear{{Albrecht}, {Bernstein}, {Cahn}, {Freedman},
  {Hewitt}, {Hu}, {Huth}, {Kamionkowski}, {Kolb}, {Knox}, {Mather}, {Staggs},
  and {Suntzeff}}{{Albrecht} et~al.}{2006}]{2006astro.ph..9591A}
{Albrecht}, A., G.~{Bernstein}, R.~{Cahn}, W.~L. {Freedman}, J.~{Hewitt},
  W.~{Hu}, J.~{Huth}, M.~{Kamionkowski}, E.~W. {Kolb}, L.~{Knox}, J.~C.
  {Mather}, S.~{Staggs}, and N.~B. {Suntzeff} (2006).
\newblock {Report of the Dark Energy Task Force}.
\newblock astro-ph/0609591.

\bibitem[\protect\citeauthoryear{{Ambrosio} et~al.}{{Ambrosio}
  et~al.}{2003}]{2003APh....19....1M}
{Ambrosio}, M. et~al. (2003).
\newblock {Search for diffuse neutrino flux from astrophysical sources with
  MACRO}.
\newblock Astropart. Phys.~19, 1--13.

\bibitem[\protect\citeauthoryear{{Ashie} et~al.}{{Ashie}
  et~al.}{2005}]{2005PhRvD..71k2005A}
{Ashie}, Y. et~al. (2005).
\newblock {Measurement of atmospheric neutrino oscillation parameters by
  Super-Kamiokande I}.
\newblock \prd~71(11), 112005--+.

\bibitem[\protect\citeauthoryear{{Baldry}, {Balogh}, {Bower}, {Glazebrook},
  {Nichol}, {Bamford}, and {Budavari}}{{Baldry}
  et~al.}{2006}]{2006MNRAS.373..469B}
{Baldry}, I.~K., M.~L. {Balogh}, R.~G. {Bower}, K.~{Glazebrook}, R.~C.
  {Nichol}, S.~P. {Bamford}, and T.~{Budavari} (2006).
\newblock {Galaxy bimodality versus stellar mass and environment}.
\newblock \mnras~373, 469--483.

\bibitem[\protect\citeauthoryear{{Balogh}, {Baldry}, {Nichol}, {Miller},
  {Bower}, and {Glazebrook}}{{Balogh} et~al.}{2004}]{2004ApJ...615L.101B}
{Balogh}, M.~L., I.~K. {Baldry}, R.~{Nichol}, C.~{Miller}, R.~{Bower}, and
  K.~{Glazebrook} (2004).
\newblock {The Bimodal Galaxy Color Distribution: Dependence on Luminosity and
  Environment}.
\newblock \apjl~615, L101--L104.

\bibitem[\protect\citeauthoryear{{Banerji}, {Abdalla}, {Lahav}, and
  {Lin}}{{Banerji} et~al.}{2007}]{2007arXiv0711.1059B}
{Banerji}, M., F.~B. {Abdalla}, O.~{Lahav}, and H.~{Lin} (2007).
\newblock {Photometric Redshifts for the Dark Energy Survey and VISTA and
  Implications for Large Scale Structure}.
\newblock arXiv:0711.1059.

\bibitem[\protect\citeauthoryear{{Barwick} et~al.}{{Barwick}
  et~al.}{2006}]{2006PhRvL..96q1101B}
{Barwick}, S.~W. et~al. (2006).
\newblock {Constraints on Cosmic Neutrino Fluxes from the Antarctic Impulsive
  Transient Antenna Experiment}.
\newblock Physical Review Letters~96(17), 171101--+.

\bibitem[\protect\citeauthoryear{{Baugh}}{{Baugh}}{2006}]{2006RPPh...69.3101B}
{Baugh}, C.~M. (2006).
\newblock {A primer on hierarchical galaxy formation: the semi-analytical
  approach}.
\newblock Reports of Progress in Physics~69, 3101--3156.

\bibitem[\protect\citeauthoryear{{Benabed} and {van Waerbeke}}{{Benabed} and
  {van Waerbeke}}{2004}]{2004PhRvD..70l3515B}
{Benabed}, K. and L.~{van Waerbeke} (2004).
\newblock {Constraining dark energy evolution with gravitational lensing by
  large scale structures}.
\newblock \prd~70(12), 123515--+.

\bibitem[\protect\citeauthoryear{{Berghaus} et~al.}{{Berghaus}
  et~al.}{2007}]{2007arXiv0712.4406B}
{Berghaus}, P. et~al. (2007).
\newblock {Status and Results from AMANDA/IceCube}.
\newblock arXiv:0712.4406.

\bibitem[\protect\citeauthoryear{{Berlind}, {Blanton}, {Hogg}, {Weinberg},
  {Dav{\'e}}, {Eisenstein}, and {Katz}}{{Berlind}
  et~al.}{2005}]{2005ApJ...629..625B}
{Berlind}, A.~A., M.~R. {Blanton}, D.~W. {Hogg}, D.~H. {Weinberg},
  R.~{Dav{\'e}}, D.~J. {Eisenstein}, and N.~{Katz} (2005).
\newblock {Interpreting the Relationship between Galaxy Luminosity, Color, and
  Environment}.
\newblock \apj~629, 625--632.

\bibitem[\protect\citeauthoryear{{Berlind} and {Weinberg}}{{Berlind} and
  {Weinberg}}{2002}]{2002ApJ...575..587B}
{Berlind}, A.~A. and D.~H. {Weinberg} (2002).
\newblock {The Halo Occupation Distribution: Toward an Empirical Determination
  of the Relation between Galaxies and Mass}.
\newblock \apj~575, 587--616.

\bibitem[\protect\citeauthoryear{{Berlind}, {Weinberg}, {Benson}, {Baugh},
  {Cole}, {Dav{\'e}}, {Frenk}, {Jenkins}, {Katz}, and {Lacey}}{{Berlind}
  et~al.}{2003}]{2003ApJ...593....1B}
{Berlind}, A.~A., D.~H. {Weinberg}, A.~J. {Benson}, C.~M. {Baugh}, S.~{Cole},
  R.~{Dav{\'e}}, C.~S. {Frenk}, A.~{Jenkins}, N.~{Katz}, and C.~G. {Lacey}
  (2003).
\newblock {The Halo Occupation Distribution and the Physics of Galaxy
  Formation}.
\newblock \apj~593, 1--25.

\bibitem[\protect\citeauthoryear{{Bionta}, {Blewitt}, {Bratton}, {Caspere}, and
  {Ciocio}}{{Bionta} et~al.}{1987}]{1987PhRvL..58.1494B}
{Bionta}, R.~M., G.~{Blewitt}, C.~B. {Bratton}, D.~{Caspere}, and A.~{Ciocio}
  (1987).
\newblock {Observation of a neutrino burst in coincidence with supernova 1987A
  in the Large Magellanic Cloud}.
\newblock Physical Review Letters~58, 1494--1496.

\bibitem[\protect\citeauthoryear{{Biviano}}{{Biviano}}{2000}]{2000cucg.confE..%
.1B}
{Biviano}, A. (2000).
\newblock {From Messier to Abell: 200 Years of Science with Galaxy Clusters}.
\newblock astro-ph/0010409.

\bibitem[\protect\citeauthoryear{{Blake}, {Collister}, {Bridle}, and
  {Lahav}}{{Blake} et~al.}{2007}]{2007MNRAS.374.1527B}
{Blake}, C., A.~{Collister}, S.~{Bridle}, and O.~{Lahav} (2007).
\newblock {Cosmological baryonic and matter densities from 600000 SDSS luminous
  red galaxies with photometric redshifts}.
\newblock \mnras~374, 1527--1548.

\bibitem[\protect\citeauthoryear{{Blanton}}{{Blanton}}{2000}]{2000ApJ...544...%
63B}
{Blanton}, M. (2000).
\newblock {How Stochastic Is the Relative Bias between Galaxy Types?}
\newblock \apj~544, 63--80.

\bibitem[\protect\citeauthoryear{{Blanton} et~al.}{{Blanton}
  et~al.}{005b}]{2005AJ....129.2562B}
{Blanton}, M.~R. et~al. ({2005b}).
\newblock {New York University Value-Added Galaxy Catalog: A Galaxy Catalog
  Based on New Public Surveys}.
\newblock \aj~129, 2562--2578.

\bibitem[\protect\citeauthoryear{{Blanton}, {Brinkmann}, {Csabai}, {Doi},
  {Eisenstein}, {Fukugita}, {Gunn}, {Hogg}, and {Schlegel}}{{Blanton}
  et~al.}{003a}]{2003AJ....125.2348B}
{Blanton}, M.~R., J.~{Brinkmann}, I.~{Csabai}, M.~{Doi}, D.~{Eisenstein},
  M.~{Fukugita}, J.~E. {Gunn}, D.~W. {Hogg}, and D.~J. {Schlegel} ({2003a}).
\newblock {Estimating Fixed-Frame Galaxy Magnitudes in the Sloan Digital Sky
  Survey}.
\newblock \aj~125, 2348--2360.

\bibitem[\protect\citeauthoryear{{Blanton}, {Eisenstein}, {Hogg}, {Schlegel},
  and {Brinkmann}}{{Blanton} et~al.}{005a}]{2005ApJ...629..143B}
{Blanton}, M.~R., D.~{Eisenstein}, D.~W. {Hogg}, D.~J. {Schlegel}, and
  J.~{Brinkmann} ({2005a}).
\newblock {Relationship between Environment and the Broadband Optical
  Properties of Galaxies in the Sloan Digital Sky Survey}.
\newblock \apj~629, 143--157.

\bibitem[\protect\citeauthoryear{{Blanton}, {Eisenstein}, {Hogg}, and
  {Zehavi}}{{Blanton} et~al.}{2006}]{2006ApJ...645..977B}
{Blanton}, M.~R., D.~{Eisenstein}, D.~W. {Hogg}, and I.~{Zehavi} (2006).
\newblock {The Scale Dependence of Relative Galaxy Bias: Encouragement for the
  ``Halo Model'' Description}.
\newblock \apj~645, 977--985.

\bibitem[\protect\citeauthoryear{{Blanton}, {Lin}, {Lupton}, {Maley}, {Young},
  {Zehavi}, and {Loveday}}{{Blanton} et~al.}{003b}]{2003AJ....125.2276B}
{Blanton}, M.~R., H.~{Lin}, R.~H. {Lupton}, F.~M. {Maley}, N.~{Young},
  I.~{Zehavi}, and J.~{Loveday} ({2003b}).
\newblock {An Efficient Targeting Strategy for Multiobject Spectrograph
  Surveys: the Sloan Digital Sky Survey ``Tiling'' Algorithm}.
\newblock \aj~125, 2276--2286.

\bibitem[\protect\citeauthoryear{{Bond}, {Cole}, {Efstathiou}, and
  {Kaiser}}{{Bond} et~al.}{1991}]{1991ApJ...379..440B}
{Bond}, J.~R., S.~{Cole}, G.~{Efstathiou}, and N.~{Kaiser} (1991).
\newblock {Excursion set mass functions for hierarchical Gaussian
  fluctuations}.
\newblock \apj~379, 440--460.

\bibitem[\protect\citeauthoryear{{Bond}, {Crittenden}, {Jaffe}, and
  {Knox}}{{Bond} et~al.}{1999}]{1999CoScE...1...21B}
{Bond}, J.~R., R.~G. {Crittenden}, A.~H. {Jaffe}, and L.~{Knox} (1999).
\newblock {Computing challenges of the cosmic microwave background.}
\newblock Comput.~Sci.~Eng.~1, 21--35.

\bibitem[\protect\citeauthoryear{{Bottai} and {Perrone}}{{Bottai} and
  {Perrone}}{2001}]{2001NIMPA.459..319B}
{Bottai}, S. and L.~{Perrone} (2001).
\newblock {Simulation of UHE muons propagation for GEANT3}.
\newblock Nucl. Instrum. Methods Phys. Res., Sect. A~459, 319--325.

\bibitem[\protect\citeauthoryear{{Bowman}, {Morales}, and {Hewitt}}{{Bowman}
  et~al.}{2007}]{2007ApJ...661....1B}
{Bowman}, J.~D., M.~F. {Morales}, and J.~N. {Hewitt} (2007).
\newblock {Constraints on Fundamental Cosmological Parameters with Upcoming
  Redshifted 21 cm Observations}.
\newblock \apj~661, 1--9.

\bibitem[\protect\citeauthoryear{{Broadhurst}, {Taylor}, and
  {Peacock}}{{Broadhurst} et~al.}{1995}]{1995ApJ...438...49B}
{Broadhurst}, T.~J., A.~N. {Taylor}, and J.~A. {Peacock} (1995).
\newblock {Mapping cluster mass distributions via gravitational lensing of
  background galaxies}.
\newblock \apj~438, 49--61.

\bibitem[\protect\citeauthoryear{{Bugaev}, {Naumov}, {Sinegovskii}, and
  {Zaslavskaia}}{{Bugaev} et~al.}{1989}]{1989NCimC..12...41B}
{Bugaev}, E.~V., V.~A. {Naumov}, S.~I. {Sinegovskii}, and E.~S. {Zaslavskaia}
  (1989).
\newblock {Prompt leptons in cosmic rays}.
\newblock Nuovo Cimento C~12, 41--73.

\bibitem[\protect\citeauthoryear{{Bullock}, {Kolatt}, {Sigad}, {Somerville},
  {Kravtsov}, {Klypin}, {Primack}, and {Dekel}}{{Bullock}
  et~al.}{2001}]{2001MNRAS.321..559B}
{Bullock}, J.~S., T.~S. {Kolatt}, Y.~{Sigad}, R.~S. {Somerville}, A.~V.
  {Kravtsov}, A.~A. {Klypin}, J.~R. {Primack}, and A.~{Dekel} (2001).
\newblock {Profiles of dark haloes: evolution, scatter and environment}.
\newblock \mnras~321, 559--575.

\bibitem[\protect\citeauthoryear{{Calabretta} and {Roukema}}{{Calabretta} and
  {Roukema}}{2007}]{2007MNRAS.381..865C}
{Calabretta}, M.~R. and B.~F. {Roukema} (2007).
\newblock {Mapping on the HEALPix grid}.
\newblock \mnras~381, 865--872.

\bibitem[\protect\citeauthoryear{{Carpenter}}{{Carpenter}}{1938}]{1938ApJ....8%
8..344C}
{Carpenter}, E.~F. (1938).
\newblock {Some Characteristics of Associated Galaxies. I. a. Density
  Restriction in the Metagalaxy.}
\newblock \apj~88, 344--+.

\bibitem[\protect\citeauthoryear{{Casas-Miranda}, {Mo}, {Sheth}, and
  {Boerner}}{{Casas-Miranda} et~al.}{2002}]{2002MNRAS.333..730C}
{Casas-Miranda}, R., H.~J. {Mo}, R.~K. {Sheth}, and G.~{Boerner} (2002).
\newblock {On the distribution of haloes, galaxies and mass}.
\newblock \mnras~333, 730--738.

\bibitem[\protect\citeauthoryear{{Catelan}, {Lucchin}, {Matarrese}, and
  {Porciani}}{{Catelan} et~al.}{1998}]{1998MNRAS.297..692C}
{Catelan}, P., F.~{Lucchin}, S.~{Matarrese}, and C.~{Porciani} (1998).
\newblock {The bias field of dark matter haloes}.
\newblock \mnras~297, 692--712.

\bibitem[\protect\citeauthoryear{{Christensen}, {Meyer}, {Knox}, and
  {Luey}}{{Christensen} et~al.}{2001}]{2001CQGra..18.2677C}
{Christensen}, N., R.~{Meyer}, L.~{Knox}, and B.~{Luey} (2001).
\newblock {Bayesian methods for cosmological parameter estimation from cosmic
  microwave background measurements }.
\newblock Classical and Quantum Gravity~18, 2677--2688.

\bibitem[\protect\citeauthoryear{{Coil} et~al.}{{Coil}
  et~al.}{2008}]{2008ApJ...672..153C}
{Coil}, A.~L. et~al. (2008).
\newblock {The DEEP2 Galaxy Redshift Survey: Color and Luminosity Dependence of
  Galaxy Clustering at $z \sim 1$}.
\newblock \apj~672, 153--176.

\bibitem[\protect\citeauthoryear{{Cole} et~al.}{{Cole}
  et~al.}{2005}]{2005MNRAS.362..505C}
{Cole}, S. et~al. (2005).
\newblock {The 2dF Galaxy Redshift Survey: power-spectrum analysis of the final
  data set and cosmological implications}.
\newblock \mnras~362, 505--534.

\bibitem[\protect\citeauthoryear{{Colless} et~al.}{{Colless}
  et~al.}{2001}]{2001MNRAS.328.1039C}
{Colless}, M. et~al. (2001).
\newblock {The 2dF Galaxy Redshift Survey: spectra and redshifts}.
\newblock \mnras~328, 1039--1063.

\bibitem[\protect\citeauthoryear{{Colless} et~al.}{{Colless}
  et~al.}{2003}]{2003astro.ph..6581C}
{Colless}, M. et~al. (2003).
\newblock {The 2dF Galaxy Redshift Survey: Final Data Release}.
\newblock astro-ph/0306581.

\bibitem[\protect\citeauthoryear{{Collister} et~al.}{{Collister}
  et~al.}{2007}]{2007MNRAS.375...68C}
{Collister}, A. et~al. (2007).
\newblock {MegaZ-LRG: a photometric redshift catalogue of one million SDSS
  luminous red galaxies}.
\newblock \mnras~375, 68--76.

\bibitem[\protect\citeauthoryear{{Collister} and {Lahav}}{{Collister} and
  {Lahav}}{2005}]{2005MNRAS.361..415C}
{Collister}, A.~A. and O.~{Lahav} (2005).
\newblock {Distribution of red and blue galaxies in groups: an empirical test
  of the halo model}.
\newblock \mnras~361, 415--427.

\bibitem[\protect\citeauthoryear{{Conrad}, {Botner}, {Hallgren}, and {P{\'e}rez
  de Los Heros}}{{Conrad} et~al.}{2003}]{2003PhRvD..67a2002C}
{Conrad}, J., O.~{Botner}, A.~{Hallgren}, and C.~{P{\'e}rez de Los Heros}
  (2003).
\newblock {Including systematic uncertainties in confidence interval
  construction for Poisson statistics}.
\newblock \prd~67(1), 012002--+.

\bibitem[\protect\citeauthoryear{{Conroy}, {Coil}, {White}, {Newman}, {Yan},
  {Cooper}, {Gerke}, {Davis}, and {Koo}}{{Conroy}
  et~al.}{2005}]{2005ApJ...635..990C}
{Conroy}, C., A.~L. {Coil}, M.~{White}, J.~A. {Newman}, R.~{Yan}, M.~C.
  {Cooper}, B.~F. {Gerke}, M.~{Davis}, and D.~C. {Koo} (2005).
\newblock {The DEEP2 Galaxy Redshift Survey: The Evolution of Void Statistics
  from $z \sim 1$ to $z \sim 0$}.
\newblock \apj~635, 990--1005.

\bibitem[\protect\citeauthoryear{{Conway} et~al.}{{Conway}
  et~al.}{2005}]{2005MNRAS.356..456C}
{Conway}, E. et~al. (2005).
\newblock {The 2dF Galaxy Redshift Survey: the nature of the relative bias
  between galaxies of different spectral type}.
\newblock \mnras~356, 456--474.

\bibitem[\protect\citeauthoryear{{Conway}}{{Conway}}{2002}]{Conway:2002}
{Conway}, J. (2002).
\newblock {Efficiency Uncertainties: A Bayesian Prescription}.
\newblock Technical Report {CDF/PUB/5894}, CDF, Batavia, IL.
\newblock
  \url{http://www-cdf.fnal.gov/physics/statistics/statistics_recommendations.h%
tml}.

\bibitem[\protect\citeauthoryear{{Cooray}}{{Cooray}}{2005}]{2005MNRAS.363..337%
C}
{Cooray}, A. (2005).
\newblock {A divided Universe: red and blue galaxies and their preferred
  environments}.
\newblock \mnras~363, 337--352.

\bibitem[\protect\citeauthoryear{{Cooray}}{{Cooray}}{2006}]{2006astro.ph..1090%
C}
{Cooray}, A. (2006).
\newblock {Are galaxy properties only determined by the dark matter halo mass?}
\newblock astro-ph/0601090.

\bibitem[\protect\citeauthoryear{{Cooray} and {Sheth}}{{Cooray} and
  {Sheth}}{2002}]{2002PhR...372....1C}
{Cooray}, A. and R.~{Sheth} (2002).
\newblock {Halo models of large scale structure}.
\newblock \physrep~372, 1--129.

\bibitem[\protect\citeauthoryear{{Cousins} and {Highland}}{{Cousins} and
  {Highland}}{1992}]{1992NIMPA.320..331C}
{Cousins}, R.~D. and V.~L. {Highland} (1992).
\newblock {Incorporating systematic uncertainties into an upper limit}.
\newblock Nucl. Instrum. Methods Phys. Res., Sect. A~320, 331--335.

\bibitem[\protect\citeauthoryear{{Crocce} and {Scoccimarro}}{{Crocce} and
  {Scoccimarro}}{2008}]{2008PhRvD..77b3533C}
{Crocce}, M. and R.~{Scoccimarro} (2008).
\newblock {Nonlinear evolution of baryon acoustic oscillations}.
\newblock \prd~77(2), 023533--+.

\bibitem[\protect\citeauthoryear{{Crook}, {Huchra}, {Martimbeau}, {Masters},
  {Jarrett}, and {Macri}}{{Crook} et~al.}{2007}]{2007ApJ...655..790C}
{Crook}, A.~C., J.~P. {Huchra}, N.~{Martimbeau}, K.~L. {Masters}, T.~{Jarrett},
  and L.~M. {Macri} (2007).
\newblock {Groups of Galaxies in the Two Micron All Sky Redshift Survey}.
\newblock \apj~655, 790--813.

\bibitem[\protect\citeauthoryear{{Croton}, {Norberg}, {Gazta{\~n}aga}, and
  {Baugh}}{{Croton} et~al.}{2007}]{2007MNRAS.379.1562C}
{Croton}, D.~J., P.~{Norberg}, E.~{Gazta{\~n}aga}, and C.~M. {Baugh} (2007).
\newblock {Statistical analysis of galaxy surveys - III. The non-linear
  clustering of red and blue galaxies in the 2dFGRS}.
\newblock \mnras~379, 1562--1570.

\bibitem[\protect\citeauthoryear{{Croton}, {Springel}, {White}, {De Lucia},
  {Frenk}, {Gao}, {Jenkins}, {Kauffmann}, {Navarro}, and {Yoshida}}{{Croton}
  et~al.}{2006}]{2006MNRAS.365...11C}
{Croton}, D.~J., V.~{Springel}, S.~D.~M. {White}, G.~{De Lucia}, C.~S. {Frenk},
  L.~{Gao}, A.~{Jenkins}, G.~{Kauffmann}, J.~F. {Navarro}, and N.~{Yoshida}
  (2006).
\newblock {The many lives of active galactic nuclei: cooling flows, black holes
  and the luminosities and colours of galaxies}.
\newblock \mnras~365, 11--28.

\bibitem[\protect\citeauthoryear{{da Costa}}{{da
  Costa}}{1999}]{1999elss.conf...87D}
{da Costa}, L. (1999).
\newblock {Galaxy redshift surveys: 20 years later (invited review)}.
\newblock In A.~J. {Banday}, R.~K. {Sheth}, and L.~N. {da Costa} (Eds.),
  Evolution of Large Scale Structure: From Recombination to Garching,  87--+.

\bibitem[\protect\citeauthoryear{{da Costa}, {Pellegrini}, {Davis}, {Meiksin},
  {Sargent}, and {Tonry}}{{da Costa} et~al.}{1991}]{1991ApJS...75..935D}
{da Costa}, L.~N., P.~S. {Pellegrini}, M.~{Davis}, A.~{Meiksin}, W.~L.~W.
  {Sargent}, and J.~L. {Tonry} (1991).
\newblock {Southern Sky Redshift Survey - The catalog}.
\newblock \apjs~75, 935--964.

\bibitem[\protect\citeauthoryear{{Dark Energy Survey Collaboration}}{{Dark
  Energy Survey Collaboration}}{2005}]{2005astro.ph.10346T}
{Dark Energy Survey Collaboration} (2005).
\newblock {The Dark Energy Survey}.
\newblock astro-ph/0510346.

\bibitem[\protect\citeauthoryear{{Davis} et~al.}{{Davis}
  et~al.}{2003}]{2003SPIE.4834..161D}
{Davis}, M. et~al. (2003).
\newblock {Science Objectives and Early Results of the DEEP2 Redshift Survey}.
\newblock In P.~{Guhathakurta} (Ed.), Discoveries and Research Prospects from
  6- to 10-Meter-Class Telescopes II, Volume 4834 of Proceedings of the SPIE,
  161--172.

\bibitem[\protect\citeauthoryear{{Davis} and {Geller}}{{Davis} and
  {Geller}}{1976}]{1976ApJ...208...13D}
{Davis}, M. and M.~J. {Geller} (1976).
\newblock {Galaxy Correlations as a Function of Morphological Type}.
\newblock \apj~208, 13--19.

\bibitem[\protect\citeauthoryear{{Davis}}{{Davis}}{1996}]{1996NuPhS..48..284D}
{Davis}, R. (1996).
\newblock {A review of measurements of the solar neutrino flux and their
  variation.}
\newblock Nuclear Physics B Proceedings Supplements~48, 284--298.

\bibitem[\protect\citeauthoryear{{Davis}}{{Davis}}{2003}]{2003RvMP...75..985D}
{Davis}, R. (2003).
\newblock {Nobel Lecture: A half-century with solar neutrinos}.
\newblock Rev. Mod. Phys.~75, 985--994.

\bibitem[\protect\citeauthoryear{{de Berg}, {van Kreveld}, {Overmars}, and
  {Schwarzkopf}}{{de Berg} et~al.}{2000}]{markbook}
{de Berg}, M., M.~{van Kreveld}, M.~{Overmars}, and O.~{Schwarzkopf} (2000).
\newblock {Computational Geometry: Algorithms and Applications}, Chapter~14,
  291--298.
\newblock New York: Springer.

\bibitem[\protect\citeauthoryear{{de Oliveira-Costa}}{{de
  Oliveira-Costa}}{2005}]{2005ASPC..343..485D}
{de Oliveira-Costa}, A. (2005).
\newblock {The Cosmic Microwave Background and its Polarization}.
\newblock In A.~{Adamson}, C.~{Aspin}, C.~{Davis}, and T.~{Fujiyoshi} (Eds.),
  Astronomical Polarimetry: Current Status and Future Directions, Volume 343 of
  Astronomical Society of the Pacific Conference Series,  485--+.

\bibitem[\protect\citeauthoryear{{de Vaucouleurs}}{{de
  Vaucouleurs}}{1958}]{1958AJ.....63..253D}
{de Vaucouleurs}, G. (1958).
\newblock {Further evidence for a local super-cluster of galaxies: rotation and
  expansion}.
\newblock \aj~63, 253--+.

\bibitem[\protect\citeauthoryear{{Dekel} and {Lahav}}{{Dekel} and
  {Lahav}}{1999}]{1999ApJ...520...24D}
{Dekel}, A. and O.~{Lahav} (1999).
\newblock {Stochastic Nonlinear Galaxy Biasing}.
\newblock \apj~520, 24--34.

\bibitem[\protect\citeauthoryear{{Desai} et~al.}{{Desai}
  et~al.}{2003}]{2003ICRC....3.1673D}
{Desai}, S. et~al. (2003).
\newblock {Study of Upward Showering Muons in Super-Kamiokande}.
\newblock Volume~3 of Proceedings of the 28th International Cosmic Ray
  Conference,  1673.

\bibitem[\protect\citeauthoryear{{Desai} et~al.}{{Desai}
  et~al.}{2004}]{2004PhRvD..70h3523D}
{Desai}, S. et~al. (2004).
\newblock {Search for dark matter WIMPs using upward through-going muons in
  Super-Kamiokande}.
\newblock \prd~70(8), 083523--+.

\bibitem[\protect\citeauthoryear{{Desai} et~al.}{{Desai}
  et~al.}{2008}]{2008APh....29...42D}
{Desai}, S. et~al. (2008).
\newblock {Study of TeV neutrinos with upward showering muons in
  Super-Kamiokande}.
\newblock Astroparticle Physics~29, 42--54.

\bibitem[\protect\citeauthoryear{{Dodelson}}{{Dodelson}}{2003}]{2003moco.book.%
....D}
{Dodelson}, S. (2003).
\newblock {Modern cosmology}.
\newblock Amsterdam: Academic Press.

\bibitem[\protect\citeauthoryear{{Dodelson}, {Hui}, and {Jaffe}}{{Dodelson}
  et~al.}{1997}]{1997astro.ph.12074D}
{Dodelson}, S., L.~{Hui}, and A.~{Jaffe} (1997).
\newblock {Likelihood Analysis of Galaxy Surveys}.
\newblock astro-ph/9712074.

\bibitem[\protect\citeauthoryear{{Dolag}, {Borgani}, {Schindler}, {Diaferio},
  and {Bykov}}{{Dolag} et~al.}{2008}]{2008SSRv..tmp...26D}
{Dolag}, K., S.~{Borgani}, S.~{Schindler}, A.~{Diaferio}, and A.~M. {Bykov}
  (2008).
\newblock {Simulation Techniques for Cosmological Simulations}.
\newblock Space Science Reviews~134, 229--268.

\bibitem[\protect\citeauthoryear{{Dor{\'e}}, {Knox}, and {Peel}}{{Dor{\'e}}
  et~al.}{2001}]{2001PhRvD..64h3001D}
{Dor{\'e}}, O., L.~{Knox}, and A.~{Peel} (2001).
\newblock {CMB power spectrum estimation via hierarchical decomposition}.
\newblock \prd~64(8), 083001--+.

\bibitem[\protect\citeauthoryear{{Dor{\'e}}, {Teyssier}, {Bouchet}, {Vibert},
  and {Prunet}}{{Dor{\'e}} et~al.}{2001}]{2001A&A...374..358D}
{Dor{\'e}}, O., R.~{Teyssier}, F.~R. {Bouchet}, D.~{Vibert}, and S.~{Prunet}
  (2001).
\newblock {MAPCUMBA: A fast iterative multi-grid map-making algorithm for CMB
  experiments}.
\newblock \aap~374, 358--370.

\bibitem[\protect\citeauthoryear{{Doughty}, {Shane}, and {Wood}}{{Doughty}
  et~al.}{1974}]{1974SouSt..25..107D}
{Doughty}, N.~A., C.~D. {Shane}, and F.~B. {Wood} (1974).
\newblock {The all-sky sixteenth magnitude photographic sky surveys of Mount
  John University Observatory and Lick Observatory.}
\newblock Southern Stars~25, 107--127.

\bibitem[\protect\citeauthoryear{{Dziewonski}}{{Dziewonski}}{1989}]{earthmodel}
{Dziewonski}, A. (1989).
\newblock {The Encyclopedia of Solid Earth Geophysics}, Chapter {``Earth
  Structure, Global''},  331.
\newblock New York: Van Nostrand Reinhold.

\bibitem[\protect\citeauthoryear{{Eisenstein} et~al.}{{Eisenstein}
  et~al.}{2001}]{2001AJ....122.2267E}
{Eisenstein}, D.~J. et~al. (2001).
\newblock {Spectroscopic Target Selection for the Sloan Digital Sky Survey: The
  Luminous Red Galaxy Sample}.
\newblock \aj~122, 2267--2280.

\bibitem[\protect\citeauthoryear{{Eisenstein} et~al.}{{Eisenstein}
  et~al.}{2005}]{2005ApJ...633..560E}
{Eisenstein}, D.~J. et~al. (2005).
\newblock {Detection of the Baryon Acoustic Peak in the Large-Scale Correlation
  Function of SDSS Luminous Red Galaxies}.
\newblock \apj~633, 560--574.

\bibitem[\protect\citeauthoryear{{Eisenstein} and {Hu}}{{Eisenstein} and
  {Hu}}{1999}]{1999ApJ...511....5E}
{Eisenstein}, D.~J. and W.~{Hu} (1999).
\newblock {Power Spectra for Cold Dark Matter and Its Variants}.
\newblock \apj~511, 5--15.

\bibitem[\protect\citeauthoryear{{Eisenstein}, {Seo}, and {White}}{{Eisenstein}
  et~al.}{2007}]{2007ApJ...664..660E}
{Eisenstein}, D.~J., H.-J. {Seo}, and M.~{White} (2007).
\newblock {On the Robustness of the Acoustic Scale in the Low-Redshift
  Clustering of Matter}.
\newblock \apj~664, 660--674.

\bibitem[\protect\citeauthoryear{{Elgar{\o}y} and {Lahav}}{{Elgar{\o}y} and
  {Lahav}}{2006}]{2006PhST..127..105E}
{Elgar{\o}y}, {\O}. and O.~{Lahav} (2006).
\newblock {Sub-eV upper limits on neutrino masses from cosmology}.
\newblock Physica Scripta Volume T~127, 105--106.

\bibitem[\protect\citeauthoryear{{Falcke} et~al.}{{Falcke}
  et~al.}{2007}]{2007HiA....14..386F}
{Falcke}, H.~D. et~al. (2007).
\newblock {A very brief description of LOFAR the Low Frequency Array}.
\newblock Highlights of Astronomy~14, 386--387.

\bibitem[\protect\citeauthoryear{{Feldman} and {Cousins}}{{Feldman} and
  {Cousins}}{1998}]{1998PhRvD..57.3873F}
{Feldman}, G.~J. and R.~D. {Cousins} (1998).
\newblock {Unified approach to the classical statistical analysis of small
  signals}.
\newblock \prd~57, 3873--3889.

\bibitem[\protect\citeauthoryear{{Feng}}{{Feng}}{2008}]{2008arXiv0801.1334F}
{Feng}, J.~L. (2008).
\newblock {Collider Physics and Cosmology}.
\newblock arXiv:0801.1334.

\bibitem[\protect\citeauthoryear{{Fermi}}{{Fermi}}{1949}]{fermi}
{Fermi}, E. (1949).
\newblock {On the Origin of Cosmic Acceleration}.
\newblock Phys. Rev.~75, 1169--1174.

\bibitem[\protect\citeauthoryear{{Frieman}, {Turner}, and {Huterer}}{{Frieman}
  et~al.}{2008}]{2008arXiv0803.0982F}
{Frieman}, J., M.~{Turner}, and D.~{Huterer} (2008).
\newblock {Dark Energy and the Accelerating Universe}.
\newblock arXiv:0803.0982.

\bibitem[\protect\citeauthoryear{{Fukuda} et~al.}{{Fukuda}
  et~al.}{2003}]{2003NIMPA.501..418T}
{Fukuda}, S. et~al. (2003).
\newblock {The Super-Kamiokande detector}.
\newblock Nucl. Instrum. Methods Phys. Res., Sect. A~501, 418--462.

\bibitem[\protect\citeauthoryear{{Fukuda} et~al.}{{Fukuda}
  et~al.}{1998}]{1998PhRvL..81.1562F}
{Fukuda}, Y. et~al. (1998).
\newblock {Evidence for Oscillation of Atmospheric Neutrinos}.
\newblock \prl~81, 1562--1567.

\bibitem[\protect\citeauthoryear{{Fukuda} et~al.}{{Fukuda}
  et~al.}{1999}]{1999PhRvL..82.2644F}
{Fukuda}, Y. et~al. (1999).
\newblock {Measurement of the Flux and Zenith-Angle Distribution of Upward
  Throughgoing Muons by Super-Kamiokande}.
\newblock \prl~82, 2644--2648.

\bibitem[\protect\citeauthoryear{{Fukugita}, {Ichikawa}, {Gunn}, {Doi},
  {Shimasaku}, and {Schneider}}{{Fukugita} et~al.}{1996}]{1996AJ....111.1748F}
{Fukugita}, M., T.~{Ichikawa}, J.~E. {Gunn}, M.~{Doi}, K.~{Shimasaku}, and
  D.~P. {Schneider} (1996).
\newblock {The Sloan Digital Sky Survey Photometric System}.
\newblock \aj~111, 1748--+.

\bibitem[\protect\citeauthoryear{{Gaisser}}{{Gaisser}}{1990}]{1990cup..book...%
..G}
{Gaisser}, T.~K. (1990).
\newblock {Cosmic rays and particle physics}.
\newblock New York: Cambridge University Press.

\bibitem[\protect\citeauthoryear{{Gaisser}, {Halzen}, and {Stanev}}{{Gaisser}
  et~al.}{1995}]{1995PhR...258..173G}
{Gaisser}, T.~K., F.~{Halzen}, and T.~{Stanev} (1995).
\newblock {Particle astrophysics with high energy neutrinos}.
\newblock \physrep~258, 173--236.

\bibitem[\protect\citeauthoryear{{Gandhi}, {Quigg}, {Reno}, and
  {Sarcevic}}{{Gandhi} et~al.}{1996}]{1996APh.....5...81G}
{Gandhi}, R., C.~{Quigg}, M.~H. {Reno}, and I.~{Sarcevic} (1996).
\newblock {Ultrahigh-energy neutrino interactions}.
\newblock Astropart. Phys.~5, 81--110.

\bibitem[\protect\citeauthoryear{{Gao} and {White}}{{Gao} and
  {White}}{2007}]{2007MNRAS.377L...5G}
{Gao}, L. and S.~D.~M. {White} (2007).
\newblock {Assembly bias in the clustering of dark matter haloes}.
\newblock \mnras~377, L5--L9.

\bibitem[\protect\citeauthoryear{{Gelmini}, {Gondolo}, and
  {Varieschi}}{{Gelmini} et~al.}{2000a}]{2000PhRvD..61e6011G}
{Gelmini}, G., P.~{Gondolo}, and G.~{Varieschi} (2000a).
\newblock {Prompt atmospheric neutrinos and muons: Dependence on the gluon
  distribution function}.
\newblock \prd~61(5), 056011--+.

\bibitem[\protect\citeauthoryear{{Gelmini}, {Gondolo}, and
  {Varieschi}}{{Gelmini} et~al.}{2000b}]{2000PhRvD..61c6005G}
{Gelmini}, G., P.~{Gondolo}, and G.~{Varieschi} (2000b).
\newblock {Prompt atmospheric neutrinos and muons: NLO versus LO QCD
  predictions}.
\newblock \prd~61(3), 036005--+.

\bibitem[\protect\citeauthoryear{{Gelmini}, {Gondolo}, and
  {Varieschi}}{{Gelmini} et~al.}{2003}]{2003PhRvD..67a7301G}
{Gelmini}, G., P.~{Gondolo}, and G.~{Varieschi} (2003).
\newblock {Measuring the prompt atmospheric neutrino flux with down-going muons
  in neutrino telescopes}.
\newblock \prd~67(1), 017301--+.

\bibitem[\protect\citeauthoryear{{Giannantonio}, {Crittenden}, {Nichol},
  {Scranton}, {Richards}, {Myers}, {Brunner}, {Gray}, {Connolly}, and
  {Schneider}}{{Giannantonio} et~al.}{2006}]{2006PhRvD..74f3520G}
{Giannantonio}, T., R.~G. {Crittenden}, R.~C. {Nichol}, R.~{Scranton}, G.~T.
  {Richards}, A.~D. {Myers}, R.~J. {Brunner}, A.~G. {Gray}, A.~J. {Connolly},
  and D.~P. {Schneider} (2006).
\newblock {High redshift detection of the integrated Sachs-Wolfe effect}.
\newblock \prd~74(6), 063520--+.

\bibitem[\protect\citeauthoryear{{Gilks}, {Richardson}, and
  {Spiegelhalter}}{{Gilks} et~al.}{1996}]{Gilks96}
{Gilks}, W.~R., S.~{Richardson}, and D.~J. {Spiegelhalter} (1996).
\newblock {Markov Chain Monte Carlo in Practice}.
\newblock London: Chapman \& Hall.

\bibitem[\protect\citeauthoryear{{Glazebrook} et~al.}{{Glazebrook}
  et~al.}{2006}]{wfmos}
{Glazebrook}, K. et~al. (2006).
\newblock Wfmos detf white paper.
\newblock \url{http://home.fnal.gov/~rocky/DETF/Dey.pdf}.

\bibitem[\protect\citeauthoryear{{Gluck}, {Reya}, and {Vogt}}{{Gluck}
  et~al.}{1995}]{1995ZPhyC..67..433G}
{Gluck}, M., E.~{Reya}, and A.~{Vogt} (1995).
\newblock {Dynamical Parton Distributions of the Proton and Small X Physics}.
\newblock Z. Phys., C~67, 433--+.

\bibitem[\protect\citeauthoryear{{Gondolo}}{{Gondolo}}{2004}]{2004astro.ph..30%
64G}
{Gondolo}, P. (2004).
\newblock {Introduction to Non-Baryonic Dark Matter}.
\newblock astro-ph/0403064.

\bibitem[\protect\citeauthoryear{{Goodman} and {O'Rourke}}{{Goodman} and
  {O'Rourke}}{2004}]{bigbluebook}
{Goodman}, J.~E. and J.~{O'Rourke} (Eds.) (2004).
\newblock {Handbook of Discrete and Computational Geometry},  778 and 871--872.
\newblock New York: Chapman \& Hall/CRC.

\bibitem[\protect\citeauthoryear{{G{\'o}rski} et~al.}{{G{\'o}rski}
  et~al.}{1999}]{1999elss.conf...37G}
{G{\'o}rski}, K.~M. et~al. (1999).
\newblock {Analysis issues for large CMB data sets}.
\newblock In A.~J. {Banday}, R.~K. {Sheth}, and L.~N. {da Costa} (Eds.),
  Evolution of Large Scale Structure: From Recombination to Garching,  37--+.

\bibitem[\protect\citeauthoryear{{G{\'o}rski}, {Hivon}, {Banday}, {Wandelt},
  {Hansen}, {Reinecke}, and {Bartelmann}}{{G{\'o}rski}
  et~al.}{2005}]{2005ApJ...622..759G}
{G{\'o}rski}, K.~M., E.~{Hivon}, A.~J. {Banday}, B.~D. {Wandelt}, F.~K.
  {Hansen}, M.~{Reinecke}, and M.~{Bartelmann} (2005).
\newblock {HEALPix: A Framework for High-Resolution Discretization and Fast
  Analysis of Data Distributed on the Sphere}.
\newblock \apj~622, 759--771.

\bibitem[\protect\citeauthoryear{{Gorski}, {Wandelt}, {Hansen}, {Hivon}, and
  {Banday}}{{Gorski} et~al.}{1999}]{1999astro.ph..5275G}
{Gorski}, K.~M., B.~D. {Wandelt}, F.~K. {Hansen}, E.~{Hivon}, and A.~J.
  {Banday} (1999).
\newblock {The HEALPix Primer}.
\newblock astro-ph/9905275.

\bibitem[\protect\citeauthoryear{{Greisen}}{{Greisen}}{1966}]{GZK1}
{Greisen}, K. (1966).
\newblock {End to the Cosmic-Ray Spectrum?}
\newblock \prl~16, 748--750.

\bibitem[\protect\citeauthoryear{{Gunn} et~al.}{{Gunn}
  et~al.}{1998}]{1998AJ....116.3040G}
{Gunn}, J.~E. et~al. (1998).
\newblock {The Sloan Digital Sky Survey Photometric Camera}.
\newblock \aj~116, 3040--3081.

\bibitem[\protect\citeauthoryear{{Gunn} et~al.}{{Gunn}
  et~al.}{2006}]{2006AJ....131.2332G}
{Gunn}, J.~E. et~al. (2006).
\newblock {The 2.5 m Telescope of the Sloan Digital Sky Survey}.
\newblock \aj~131, 2332--2359.

\bibitem[\protect\citeauthoryear{{Guzzo}, {Strauss}, {Fisher}, {Giovanelli},
  and {Haynes}}{{Guzzo} et~al.}{1997}]{1997ApJ...489...37G}
{Guzzo}, L., M.~A. {Strauss}, K.~B. {Fisher}, R.~{Giovanelli}, and M.~P.
  {Haynes} (1997).
\newblock {Redshift-Space Distortions and the Real-Space Clustering of
  Different Galaxy Types}.
\newblock \apj~489, 37--+.

\bibitem[\protect\citeauthoryear{{Haiman} et~al.}{{Haiman}
  et~al.}{2005}]{2005astro.ph..7013H}
{Haiman}, Z. et~al. (2005).
\newblock {An X-ray Galaxy Cluster Survey for Investigations of Dark Energy}.
\newblock astro-ph/0507013.

\bibitem[\protect\citeauthoryear{{Halzen} and {Hooper}}{{Halzen} and
  {Hooper}}{2002}]{2002RPPh...65.1025H}
{Halzen}, F. and D.~{Hooper} (2002).
\newblock {High-energy neutrino astronomy: the cosmic ray connection.}
\newblock Reports of Progress in Physics~65, 1025--1078.

\bibitem[\protect\citeauthoryear{{Halzen} and {O'Murchadha}}{{Halzen} and
  {O'Murchadha}}{2008}]{2008arXiv0802.0887H}
{Halzen}, F. and A.~{O'Murchadha} (2008).
\newblock {Neutrinos from Auger Sources}.
\newblock arXiv:0802.0887.

\bibitem[\protect\citeauthoryear{{Hamilton}}{{Hamilton}}{1988}]{1988ApJ...331L%
..59H}
{Hamilton}, A.~J.~S. (1988).
\newblock {Evidence for biasing in the CfA survey}.
\newblock \apjl~331, L59--L62.

\bibitem[\protect\citeauthoryear{{Hamilton}}{{Hamilton}}{1993a}]{1993ApJ...406%
L..47H}
{Hamilton}, A.~J.~S. (1993a).
\newblock {Omega from the anisotropy of the redshift correlation function in
  the IRAS 2 Jansky survey}.
\newblock \apjl~406, L47--L50.

\bibitem[\protect\citeauthoryear{{Hamilton}}{{Hamilton}}{1993b}]{1993ApJ...417%
...19H}
{Hamilton}, A.~J.~S. (1993b).
\newblock {Toward Better Ways to Measure the Galaxy Correlation Function}.
\newblock \apj~417, 19--+.

\bibitem[\protect\citeauthoryear{{Hamilton}}{{Hamilton}}{1998}]{1998ASSL..231.%
.185H}
{Hamilton}, A.~J.~S. (1998).
\newblock {Linear Redshift Distortions: a Review}.
\newblock In D.~{Hamilton} (Ed.), The Evolving Universe, Volume 231 of
  Astrophysics and Space Science Library,  185--+.

\bibitem[\protect\citeauthoryear{{Hamilton} and {Tegmark}}{{Hamilton} and
  {Tegmark}}{2004}]{2004MNRAS.349..115H}
{Hamilton}, A.~J.~S. and M.~{Tegmark} (2004).
\newblock {A scheme to deal accurately and efficiently with complex angular
  masks in galaxy surveys}.
\newblock \mnras~349, 115--128.

\bibitem[\protect\citeauthoryear{{Hamilton}, {Tegmark}, and
  {Swanson}}{{Hamilton} et~al.}{2008}]{Hamilton_in_prep}
{Hamilton}, A.~J.~S., M.~{Tegmark}, and M.~E.~C. {Swanson} (2008).
\newblock in preparation.

\bibitem[\protect\citeauthoryear{{Hannestad}, {Mirizzi}, {Raffelt}, and
  {Wong}}{{Hannestad} et~al.}{2008}]{2008arXiv0803.1585H}
{Hannestad}, S., A.~{Mirizzi}, G.~G. {Raffelt}, and Y.~Y.~Y. {Wong} (2008).
\newblock {Cosmological constraints on neutrino plus axion hot dark matter:
  Update after WMAP-5}.
\newblock arXiv:0803.1585.

\bibitem[\protect\citeauthoryear{{Hastings}}{{Hastings}}{1970}]{Hastings}
{Hastings}, W.~K. (1970).
\newblock {Monte Carlo sampling methods using Markov chains and their
  applications}.
\newblock Biometrika~57, 97.

\bibitem[\protect\citeauthoryear{{Hikage}, {Matsubara}, {Suto}, {Park},
  {Szalay}, and {Brinkmann}}{{Hikage} et~al.}{2005}]{2005PASJ...57..709H}
{Hikage}, C., T.~{Matsubara}, Y.~{Suto}, C.~{Park}, A.~S. {Szalay}, and
  J.~{Brinkmann} (2005).
\newblock {Fourier Phase Analysis of SDSS Galaxies}.
\newblock \pasj~57, 709--718.

\bibitem[\protect\citeauthoryear{{Hill}}{{Hill}}{2003}]{2003PhRvD..67k8101H}
{Hill}, G.~C. (2003).
\newblock {Comment on ``Including systematic uncertainties in confidence
  interval construction for Poisson statistics''}.
\newblock \prd~67(11), 118101--+.

\bibitem[\protect\citeauthoryear{{Hinshaw} et~al.}{{Hinshaw}
  et~al.}{2008}]{2008arXiv0803.0732H}
{Hinshaw}, G. et~al. (2008).
\newblock {Five-Year Wilkinson Microwave Anisotropy Probe (WMAP) Observations:
  Data Processing, Sky Maps, and Basic Results}.
\newblock arXiv:0803.0732.

\bibitem[\protect\citeauthoryear{{Hirata}, {Kajita}, {Koshiba}, {Nakahata}, and
  {Oyama}}{{Hirata} et~al.}{1987}]{1987PhRvL..58.1490H}
{Hirata}, K., T.~{Kajita}, M.~{Koshiba}, M.~{Nakahata}, and Y.~{Oyama} (1987).
\newblock {Observation of a neutrino burst from the supernova SN1987A}.
\newblock Physical Review Letters~58, 1490--1493.

\bibitem[\protect\citeauthoryear{{Hivon}, {G{\'o}rski}, {Netterfield}, {Crill},
  {Prunet}, and {Hansen}}{{Hivon} et~al.}{2002}]{2002ApJ...567....2H}
{Hivon}, E., K.~M. {G{\'o}rski}, C.~B. {Netterfield}, B.~P. {Crill},
  S.~{Prunet}, and F.~{Hansen} (2002).
\newblock {MASTER of the Cosmic Microwave Background Anisotropy Power Spectrum:
  A Fast Method for Statistical Analysis of Large and Complex Cosmic Microwave
  Background Data Sets}.
\newblock \apj~567, 2--17.

\bibitem[\protect\citeauthoryear{{Hogg} et~al.}{{Hogg}
  et~al.}{2004}]{2004ApJ...601L..29H}
{Hogg}, D.~W. et~al. (2004).
\newblock {The Dependence on Environment of the Color-Magnitude Relation of
  Galaxies}.
\newblock \apjl~601, L29--L32.

\bibitem[\protect\citeauthoryear{{Hogg}, {Finkbeiner}, {Schlegel}, and
  {Gunn}}{{Hogg} et~al.}{2001}]{2001AJ....122.2129H}
{Hogg}, D.~W., D.~P. {Finkbeiner}, D.~J. {Schlegel}, and J.~E. {Gunn} (2001).
\newblock {A Photometricity and Extinction Monitor at the Apache Point
  Observatory}.
\newblock \aj~122, 2129--2138.

\bibitem[\protect\citeauthoryear{{Honda}, {Kajita}, {Kasahara}, and
  {Midorikawa}}{{Honda} et~al.}{2004}]{2004PhRvD..70d3008H}
{Honda}, M., T.~{Kajita}, K.~{Kasahara}, and S.~{Midorikawa} (2004).
\newblock {New calculation of the atmospheric neutrino flux in a
  three-dimensional scheme}.
\newblock \prd~70(4), 043008--+.

\bibitem[\protect\citeauthoryear{{Hu}}{{Hu}}{2005}]{2005ASPC..339..215H}
{Hu}, W. (2005).
\newblock {Dark Energy Probes in Light of the CMB}.
\newblock In S.~C. {Wolff} and T.~R. {Lauer} (Eds.), Observing Dark Energy,
  Volume 339 of Astronomical Society of the Pacific Conference Series,  215--+.

\bibitem[\protect\citeauthoryear{{Hu}, {DeDeo}, and {Vale}}{{Hu}
  et~al.}{2007}]{2007NJPh....9..441H}
{Hu}, W., S.~{DeDeo}, and C.~{Vale} (2007).
\newblock {Cluster mass estimators from CMB temperature and polarization
  lensing}.
\newblock New Journal of Physics~9, 441--+.

\bibitem[\protect\citeauthoryear{{Hu}, {Fukugita}, {Zaldarriaga}, and
  {Tegmark}}{{Hu} et~al.}{2001}]{2001ApJ...549..669H}
{Hu}, W., M.~{Fukugita}, M.~{Zaldarriaga}, and M.~{Tegmark} (2001).
\newblock {Cosmic Microwave Background Observables and Their Cosmological
  Implications}.
\newblock \apj~549, 669--680.

\bibitem[\protect\citeauthoryear{{Hubble}}{{Hubble}}{1934}]{1934CMWCI.485....1%
H}
{Hubble}, E. (1934).
\newblock {No. 485. The distribution of extra-galactic nebulae.}
\newblock Contributions from the Mount Wilson Observatory / Carnegie
  Institution of Washington~485, 1--69.

\bibitem[\protect\citeauthoryear{{Hubble} and {Humason}}{{Hubble} and
  {Humason}}{1931}]{1931ApJ....74...43H}
{Hubble}, E. and M.~L. {Humason} (1931).
\newblock {The Velocity-Distance Relation among Extra-Galactic Nebulae}.
\newblock \apj~74, 43--+.

\bibitem[\protect\citeauthoryear{{Huchra} et~al.}{{Huchra}
  et~al.}{2005}]{2005ASPC..329..135H}
{Huchra}, J. et~al. (2005).
\newblock {The 2MASS Redshift Survey and Low Galactic Latitude Large-Scale
  Structure}.
\newblock In Nearby Large-Scale Structures and the Zone of Avoidance,
  Astronomical Society of the Pacific Conference Series, Volume 329,  135--+.

\bibitem[\protect\citeauthoryear{{Huchra}, {Davis}, {Latham}, and
  {Tonry}}{{Huchra} et~al.}{1983}]{1983ApJS...52...89H}
{Huchra}, J., M.~{Davis}, D.~{Latham}, and J.~{Tonry} (1983).
\newblock {A survey of galaxy redshifts. IV - The data}.
\newblock \apjs~52, 89--119.

\bibitem[\protect\citeauthoryear{{Humason}, {Mayall}, and {Sandage}}{{Humason}
  et~al.}{1956}]{1956AJ.....61...97H}
{Humason}, M.~L., N.~U. {Mayall}, and A.~R. {Sandage} (1956).
\newblock {Redshifts and magnitudes of extragalactic nebulae.}
\newblock \aj~61, 97--162.

\bibitem[\protect\citeauthoryear{{IceCube Collaboration}}{{IceCube
  Collaboration}}{2007}]{2007arXiv0711.0353T}
{IceCube Collaboration} (2007).
\newblock {The IceCube Collaboration: contributions to the 30th International
  Cosmic Ray Conference (ICRC 2007)}.
\newblock arXiv:0711.0353.

\bibitem[\protect\citeauthoryear{{Ikeda} et~al.}{{Ikeda}
  et~al.}{2007}]{2007ApJ...669..519I}
{Ikeda}, M. et~al. (2007).
\newblock {Search for Supernova Neutrino Bursts at Super-Kamiokande}.
\newblock \apj~669, 519--524.

\bibitem[\protect\citeauthoryear{{Ivezi{\'c}} et~al.}{{Ivezi{\'c}}
  et~al.}{2004}]{2004AN....325..583I}
{Ivezi{\'c}}, {\v Z}. et~al. (2004).
\newblock {SDSS data management and photometric quality assessment}.
\newblock Astronomische Nachrichten~325, 583--589.

\bibitem[\protect\citeauthoryear{{Joyce}, {Sylos Labini}, {Gabrielli},
  {Montuori}, and {Pietronero}}{{Joyce} et~al.}{2005}]{2005A&A...443...11J}
{Joyce}, M., F.~{Sylos Labini}, A.~{Gabrielli}, M.~{Montuori}, and
  L.~{Pietronero} (2005).
\newblock {Basic properties of galaxy clustering in the light of recent results
  from the Sloan Digital Sky Survey}.
\newblock \aap~443, 11--16.

\bibitem[\protect\citeauthoryear{{Kaiser}}{{Kaiser}}{2004}]{2004SPIE.5489...11%
K}
{Kaiser}, N. (2004).
\newblock {Pan-STARRS: a wide-field optical survey telescope array}.
\newblock In J.~M. {Oschmann}, Jr. (Ed.), Ground-based Telescopes, Volume 5489
  of Proceedings of the SPIE,  11--22.

\bibitem[\protect\citeauthoryear{{Kappes} et~al.}{{Kappes}
  et~al.}{2007}]{2007arXiv0711.0563K}
{Kappes}, A. et~al. (2007).
\newblock {KM3NeT: A Next Generation Neutrino Telescope in the Mediterranean
  Sea}.
\newblock arXiv:0711.0563.

\bibitem[\protect\citeauthoryear{{Katz}}{{Katz}}{2004}]{2004EPJC...33S.971K}
{Katz}, U.~F. (2004).
\newblock {Status of the ANTARES project}.
\newblock Eur. Phys. J. C~33, 971--974.

\bibitem[\protect\citeauthoryear{{Katz}}{{Katz}}{2006}]{2006NIMPA.567..457K}
{Katz}, U.~F. (2006).
\newblock {KM3NeT: Towards a $\mathrm{km}^{3}$ Mediterranean neutrino
  telescope}.
\newblock Nuclear Instruments and Methods in Physics Research A~567, 457--461.

\bibitem[\protect\citeauthoryear{{Kernan} and {Krauss}}{{Kernan} and
  {Krauss}}{1995}]{1995NuPhB.437..243K}
{Kernan}, P.~J. and L.~M. {Krauss} (1995).
\newblock {Updated limits on the electron neutrino mass and large angle
  oscillations from SN 1987A.}
\newblock Nuclear Physics B~437, 243--256.

\bibitem[\protect\citeauthoryear{{Kolb} and {Turner}}{{Kolb} and
  {Turner}}{1990}]{1990eaun.book.....K}
{Kolb}, E.~W. and M.~S. {Turner} (1990).
\newblock {The Early Universe}.
\newblock New York: Addison-Wesley.

\bibitem[\protect\citeauthoryear{{Kolb}}{{Kolb}}{2007}]{2007RPPh...70.1583K}
{Kolb}, R. (2007).
\newblock {A thousand invisible cords binding astronomy and high-energy
  physics}.
\newblock Reports of Progress in Physics~70, 1583--1595.

\bibitem[\protect\citeauthoryear{{Komatsu} et~al.}{{Komatsu}
  et~al.}{2008}]{2008arXiv0803.0547K}
{Komatsu}, E. et~al. (2008).
\newblock {Five-Year Wilkinson Microwave Anisotropy Probe (WMAP) Observations:
  Cosmological Interpretation}.
\newblock arXiv:0803.0547.

\bibitem[\protect\citeauthoryear{{Koshiba}}{{Koshiba}}{2003}]{2003RvMP...75.10%
11K}
{Koshiba}, M. (2003).
\newblock {Nobel Lecture: Birth of neutrino astrophysics}.
\newblock Rev. Mod. Phys.~75, 1011--1020.

\bibitem[\protect\citeauthoryear{{Kravtsov}}{{Kravtsov}}{2006}]{2006astro.ph..%
7463K}
{Kravtsov}, A.~V. (2006).
\newblock {Modeling Galaxy Clustering with Cosmological Simulations}.
\newblock astro-ph/0607463.

\bibitem[\protect\citeauthoryear{{Kravtsov}, {Berlind}, {Wechsler}, {Klypin},
  {Gottl{\"o}ber}, {Allgood}, and {Primack}}{{Kravtsov}
  et~al.}{2004}]{2004ApJ...609...35K}
{Kravtsov}, A.~V., A.~A. {Berlind}, R.~H. {Wechsler}, A.~A. {Klypin},
  S.~{Gottl{\"o}ber}, B.~{Allgood}, and J.~R. {Primack} (2004).
\newblock {The Dark Side of the Halo Occupation Distribution}.
\newblock \apj~609, 35--49.

\bibitem[\protect\citeauthoryear{{Kristiansen}, {Elgar{\o}y}, and
  {Dahle}}{{Kristiansen} et~al.}{2007}]{2007PhRvD..75h3510K}
{Kristiansen}, J.~R., {\O}.~{Elgar{\o}y}, and H.~{Dahle} (2007).
\newblock {Using the cluster mass function from weak lensing to constrain
  neutrino masses}.
\newblock \prd~75(8), 083510--+.

\bibitem[\protect\citeauthoryear{{Lesgourgues} and {Pastor}}{{Lesgourgues} and
  {Pastor}}{2006}]{2006PhR...429..307L}
{Lesgourgues}, J. and S.~{Pastor} (2006).
\newblock {Massive neutrinos and cosmology}.
\newblock \physrep~429, 307--379.

\bibitem[\protect\citeauthoryear{{Lewis} and {Bridle}}{{Lewis} and
  {Bridle}}{2002}]{2002PhRvD..66j3511L}
{Lewis}, A. and S.~{Bridle} (2002).
\newblock {Cosmological parameters from CMB and other data: A Monte Carlo
  approach}.
\newblock \prd~66(10), 103511--+.

\bibitem[\protect\citeauthoryear{{Li}, {Kauffmann}, {Jing}, {White},
  {B{\"o}rner}, and {Cheng}}{{Li} et~al.}{2006}]{2006MNRAS.368...21L}
{Li}, C., G.~{Kauffmann}, Y.~P. {Jing}, S.~D.~M. {White}, G.~{B{\"o}rner}, and
  F.~Z. {Cheng} (2006).
\newblock {The dependence of clustering on galaxy properties}.
\newblock \mnras~368, 21--36.

\bibitem[\protect\citeauthoryear{{Lipari} and {Stanev}}{{Lipari} and
  {Stanev}}{1991}]{1991PhRvD..44.3543L}
{Lipari}, P. and T.~{Stanev} (1991).
\newblock {Propagation of multi-TeV muons}.
\newblock \prd~44, 3543--3554.

\bibitem[\protect\citeauthoryear{{Lobashev}}{{Lobashev}}{2003}]{2003NuPhA.719.%
.153L}
{Lobashev}, V.~M. (2003).
\newblock {The search for the neutrino mass by direct method in the tritium
  beta-decay and perspectives of study it in the project KATRIN}.
\newblock Nuclear Physics A~719, 153--160.

\bibitem[\protect\citeauthoryear{{Loveday}, {Maddox}, {Efstathiou}, and
  {Peterson}}{{Loveday} et~al.}{1995}]{1995ApJ...442..457L}
{Loveday}, J., S.~J. {Maddox}, G.~{Efstathiou}, and B.~A. {Peterson} (1995).
\newblock {The Stromlo-APM redshift survey. 2: Variation of galaxy clustering
  with morphology and luminosity}.
\newblock \apj~442, 457--468.

\bibitem[\protect\citeauthoryear{{LSST Collaboration}}{{LSST
  Collaboration}}{2006}]{LSST}
{LSST Collaboration} (2006).
\newblock {LSST DETF White Paper}.
\newblock \url{http://www.lsst.org/Science/docs/LSST_DETF_Whitepaper.pdf}.

\bibitem[\protect\citeauthoryear{{Lupton}, {Gunn}, {Ivezi{\'c}}, {Knapp}, and
  {Kent}}{{Lupton} et~al.}{2001}]{2001ASPC..238..269L}
{Lupton}, R., J.~E. {Gunn}, Z.~{Ivezi{\'c}}, G.~R. {Knapp}, and S.~{Kent}
  (2001).
\newblock {Astronomical Data Analysis Software and Systems X}.
\newblock In F.~R. {Harnden}, Jr., F.~A. {Primini}, and H.~E. {Payne} (Eds.),
  ASP Conf.~Ser., Volume 238, San Francisco,  269--+. Astron.~Soc.~Pac.

\bibitem[\protect\citeauthoryear{{Maddox}, {Efstathiou}, and
  {Sutherland}}{{Maddox} et~al.}{990a}]{1990MNRAS.246..433M}
{Maddox}, S.~J., G.~{Efstathiou}, and W.~J. {Sutherland} ({1990a}).
\newblock {The APM Galaxy Survey - Part Two - Photometric Corrections}.
\newblock \mnras~246, 433--+.

\bibitem[\protect\citeauthoryear{{Maddox}, {Efstathiou}, and
  {Sutherland}}{{Maddox} et~al.}{1996}]{1996MNRAS.283.1227M}
{Maddox}, S.~J., G.~{Efstathiou}, and W.~J. {Sutherland} (1996).
\newblock {The APM Galaxy Survey - III. an analysis of systematic errors in the
  angular correlation function and cosmological implications}.
\newblock \mnras~283, 1227--1263.

\bibitem[\protect\citeauthoryear{{Maddox}, {Efstathiou}, {Sutherland}, and
  {Loveday}}{{Maddox} et~al.}{990b}]{1990MNRAS.243..692M}
{Maddox}, S.~J., G.~{Efstathiou}, W.~J. {Sutherland}, and J.~{Loveday}
  ({1990b}).
\newblock {The APM galaxy survey. I - APM measurements and star-galaxy
  separation}.
\newblock \mnras~243, 692--712.

\bibitem[\protect\citeauthoryear{{Madgwick} et~al.}{{Madgwick}
  et~al.}{2003}]{2003MNRAS.344..847M}
{Madgwick}, D.~S. et~al. (2003).
\newblock {The 2dF Galaxy Redshift Survey: galaxy clustering per spectral
  type}.
\newblock \mnras~344, 847--856.

\bibitem[\protect\citeauthoryear{{Magliocchetti} and
  {Porciani}}{{Magliocchetti} and {Porciani}}{2003}]{2003MNRAS.346..186M}
{Magliocchetti}, M. and C.~{Porciani} (2003).
\newblock {The halo distribution of 2dF galaxies}.
\newblock \mnras~346, 186--198.

\bibitem[\protect\citeauthoryear{{Mandelbaum}, {Hirata}, {Seljak}, {Guzik},
  {Padmanabhan}, {Blake}, {Blanton}, {Lupton}, and {Brinkmann}}{{Mandelbaum}
  et~al.}{2005}]{2005MNRAS.361.1287M}
{Mandelbaum}, R., C.~M. {Hirata}, U.~{Seljak}, J.~{Guzik}, N.~{Padmanabhan},
  C.~{Blake}, M.~R. {Blanton}, R.~{Lupton}, and J.~{Brinkmann} (2005).
\newblock {Systematic errors in weak lensing: application to SDSS galaxy-galaxy
  weak lensing}.
\newblock \mnras~361, 1287--1322.

\bibitem[\protect\citeauthoryear{{Mannheim}, {Protheroe}, and
  {Rachen}}{{Mannheim} et~al.}{2001}]{Mannheim:1998wp}
{Mannheim}, K., R.~J. {Protheroe}, and J.~P. {Rachen} (2001).
\newblock {Cosmic ray bound for models of extragalactic neutrino production}.
\newblock \prd~63(2), 023003--+.

\bibitem[\protect\citeauthoryear{{Mart{\'{\i}}nez} and
  {Saar}}{{Mart{\'{\i}}nez} and {Saar}}{2002}]{2002sgd..book.....M}
{Mart{\'{\i}}nez}, V.~J. and E.~{Saar} (2002).
\newblock {Statistics of the Galaxy Distribution}.
\newblock New York: Chapman \& Hall/CRC.

\bibitem[\protect\citeauthoryear{{Mateus}, {Sodr{\'e}}, {Cid Fernandes},
  {Stasi{\'n}ska}, {Schoenell}, and {Gomes}}{{Mateus}
  et~al.}{2006}]{2006MNRAS.370..721M}
{Mateus}, A., L.~{Sodr{\'e}}, R.~{Cid Fernandes}, G.~{Stasi{\'n}ska},
  W.~{Schoenell}, and J.~M. {Gomes} (2006).
\newblock {Semi-empirical analysis of Sloan Digital Sky Survey galaxies - II.
  The bimodality of the galaxy population revisited}.
\newblock \mnras~370, 721--737.

\bibitem[\protect\citeauthoryear{{Mather} et~al.}{{Mather}
  et~al.}{1994}]{1994ApJ...420..439M}
{Mather}, J.~C. et~al. (1994).
\newblock {Measurement of the cosmic microwave background spectrum by the COBE
  FIRAS instrument}.
\newblock \apj~420, 439--444.

\bibitem[\protect\citeauthoryear{{Matsubara}}{{Matsubara}}{1999}]{1999ApJ...52%
5..543M}
{Matsubara}, T. (1999).
\newblock {Stochasticity of Bias and Nonlocality of Galaxy Formation: Linear
  Scales}.
\newblock \apj~525, 543--553.

\bibitem[\protect\citeauthoryear{{McDonald}}{{McDonald}}{2006}]{2006PhRvD..74j%
3512M}
{McDonald}, P. (2006).
\newblock {Clustering of dark matter tracers: Renormalizing the bias
  parameters}.
\newblock \prd~74(10), 103512--+.

\bibitem[\protect\citeauthoryear{{Metropolis}, {Rosenbluth}, {Rosenbluth},
  {Teller}, and {Teller}}{{Metropolis} et~al.}{1953}]{1953JChPh..21.1087M}
{Metropolis}, N., A.~W. {Rosenbluth}, M.~N. {Rosenbluth}, A.~H. {Teller}, and
  E.~{Teller} (1953).
\newblock {Equation of State Calculations by Fast Computing Machines}.
\newblock \jcp~21, 1087--1092.

\bibitem[\protect\citeauthoryear{{Mo} and {White}}{{Mo} and
  {White}}{1996}]{1996MNRAS.282..347M}
{Mo}, H.~J. and S.~D.~M. {White} (1996).
\newblock {An analytic model for the spatial clustering of dark matter haloes}.
\newblock \mnras~282, 347--361.

\bibitem[\protect\citeauthoryear{{M{\"o}ller}, {Kitzbichler}, and
  {Natarajan}}{{M{\"o}ller} et~al.}{2007}]{2007MNRAS.379.1195M}
{M{\"o}ller}, O., M.~{Kitzbichler}, and P.~{Natarajan} (2007).
\newblock {Strong lensing statistics in large, $z \lesssim 0.2$, surveys: bias
  in the lens galaxy population}.
\newblock \mnras~379, 1195--1208.

\bibitem[\protect\citeauthoryear{{Myers}, {Shanks}, {Outram}, {Frith}, and
  {Wolfendale}}{{Myers} et~al.}{2004}]{2004MNRAS.347L..67M}
{Myers}, A.~D., T.~{Shanks}, P.~J. {Outram}, W.~J. {Frith}, and A.~W.
  {Wolfendale} (2004).
\newblock {Evidence for an extended Sunyaev-Zel'dovich effect in WMAP data}.
\newblock \mnras~347, L67--L72.

\bibitem[\protect\citeauthoryear{{Navarro}, {Frenk}, and {White}}{{Navarro}
  et~al.}{1996}]{1996ApJ...462..563N}
{Navarro}, J.~F., C.~S. {Frenk}, and S.~D.~M. {White} (1996).
\newblock {The Structure of Cold Dark Matter Halos}.
\newblock \apj~462, 563--+.

\bibitem[\protect\citeauthoryear{{Neyman} and {Scott}}{{Neyman} and
  {Scott}}{1952}]{1952ApJ...116..144N}
{Neyman}, J. and E.~L. {Scott} (1952).
\newblock {A Theory of the Spatial Distribution of Galaxies.}
\newblock \apj~116, 144--+.

\bibitem[\protect\citeauthoryear{{Nishimichi}, {Kayo}, {Hikage}, {Yahata},
  {Taruya}, {Jing}, {Sheth}, and {Suto}}{{Nishimichi}
  et~al.}{2007}]{2007PASJ...59...93N}
{Nishimichi}, T., I.~{Kayo}, C.~{Hikage}, K.~{Yahata}, A.~{Taruya}, Y.~P.
  {Jing}, R.~K. {Sheth}, and Y.~{Suto} (2007).
\newblock {Bispectrum and Nonlinear Biasing of Galaxies: Perturbation Analysis,
  Numerical Simulation, and SDSS Galaxy Clustering}.
\newblock \pasj~59, 93--106.

\bibitem[\protect\citeauthoryear{{Norberg} et~al.}{{Norberg}
  et~al.}{2001}]{2001MNRAS.328...64N}
{Norberg}, P. et~al. (2001).
\newblock {The 2dF Galaxy Redshift Survey: luminosity dependence of galaxy
  clustering}.
\newblock \mnras~328, 64--70.

\bibitem[\protect\citeauthoryear{{Norberg} et~al.}{{Norberg}
  et~al.}{2002}]{2002MNRAS.332..827N}
{Norberg}, P. et~al. (2002).
\newblock {The 2dF Galaxy Redshift Survey: the dependence of galaxy clustering
  on luminosity and spectral type}.
\newblock \mnras~332, 827--838.

\bibitem[\protect\citeauthoryear{{Novosyadlyj}, {Durrer}, and
  {Lukash}}{{Novosyadlyj} et~al.}{1999}]{1999A&A...347..799N}
{Novosyadlyj}, B., R.~{Durrer}, and V.~N. {Lukash} (1999).
\newblock {An analytic approximation of MDM power spectra in four dimensional
  parameter space}.
\newblock \aap~347, 799--808.

\bibitem[\protect\citeauthoryear{{Oyaizu}, {Lima}, {Cunha}, {Lin}, {Frieman},
  and {Sheldon}}{{Oyaizu} et~al.}{2008}]{2008ApJ...674..768O}
{Oyaizu}, H., M.~{Lima}, C.~E. {Cunha}, H.~{Lin}, J.~{Frieman}, and E.~S.
  {Sheldon} (2008).
\newblock {A Galaxy Photometric Redshift Catalog for the Sloan Digital Sky
  Survey Data Release 6}.
\newblock \apj~674, 768--783.

\bibitem[\protect\citeauthoryear{{Padmanabhan} et~al.}{{Padmanabhan}
  et~al.}{2007}]{2007MNRAS.378..852P}
{Padmanabhan}, N. et~al. (2007).
\newblock {The clustering of luminous red galaxies in the Sloan Digital Sky
  Survey imaging data}.
\newblock \mnras~378, 852--872.

\bibitem[\protect\citeauthoryear{{Padmanabhan}, {Hirata}, {Seljak}, {Schlegel},
  {Brinkmann}, and {Schneider}}{{Padmanabhan}
  et~al.}{2005}]{2005PhRvD..72d3525P}
{Padmanabhan}, N., C.~M. {Hirata}, U.~{Seljak}, D.~J. {Schlegel},
  J.~{Brinkmann}, and D.~P. {Schneider} (2005).
\newblock {Correlating the CMB with luminous red galaxies: The integrated
  Sachs-Wolfe effect}.
\newblock \prd~72(4), 043525--+.

\bibitem[\protect\citeauthoryear{{Padmanabhan}}{{Padmanabhan}}{1993}]{1993sfu.%
.book.....P}
{Padmanabhan}, T. (1993).
\newblock {Structure Formation in the Universe}.
\newblock New York: Cambridge University Press.

\bibitem[\protect\citeauthoryear{{Park}, {Choi}, {Vogeley}, {Gott}, and
  {Blanton}}{{Park} et~al.}{2007}]{2007ApJ...658..898P}
{Park}, C., Y.-Y. {Choi}, M.~S. {Vogeley}, J.~R.~I. {Gott}, and M.~R. {Blanton}
  (2007).
\newblock {Environmental Dependence of Properties of Galaxies in the Sloan
  Digital Sky Survey}.
\newblock \apj~658, 898--916.

\bibitem[\protect\citeauthoryear{{Park}, {Choi}, {Vogeley}, {Gott}, {Kim},
  {Hikage}, {Matsubara}, {Park}, {Suto}, and {Weinberg}}{{Park}
  et~al.}{2005}]{2005ApJ...633...11P}
{Park}, C., Y.-Y. {Choi}, M.~S. {Vogeley}, J.~R.~I. {Gott}, J.~{Kim},
  C.~{Hikage}, T.~{Matsubara}, M.-G. {Park}, Y.~{Suto}, and D.~H. {Weinberg}
  (2005).
\newblock {Topology Analysis of the Sloan Digital Sky Survey. I. Scale and
  Luminosity Dependence}.
\newblock \apj~633, 11--22.

\bibitem[\protect\citeauthoryear{{Park}, {Vogeley}, {Geller}, and
  {Huchra}}{{Park} et~al.}{1994}]{1994ApJ...431..569P}
{Park}, C., M.~S. {Vogeley}, M.~J. {Geller}, and J.~P. {Huchra} (1994).
\newblock {Power spectrum, correlation function, and tests for luminosity bias
  in the CfA redshift survey}.
\newblock \apj~431, 569--585.

\bibitem[\protect\citeauthoryear{{Pasquali}, {Reno}, and {Sarcevic}}{{Pasquali}
  et~al.}{1999}]{1999PhRvD..59c4020P}
{Pasquali}, L., M.~H. {Reno}, and I.~{Sarcevic} (1999).
\newblock {Lepton fluxes from atmospheric charm}.
\newblock \prd~59(3), 034020--+.

\bibitem[\protect\citeauthoryear{{Peacock}}{{Peacock}}{1999}]{1999coph.book...%
..P}
{Peacock}, J.~A. (1999).
\newblock {Cosmological Physics}.
\newblock New York: Cambridge University Press.

\bibitem[\protect\citeauthoryear{{Peacock} and {Dodds}}{{Peacock} and
  {Dodds}}{1994}]{1994MNRAS.267.1020P}
{Peacock}, J.~A. and S.~J. {Dodds} (1994).
\newblock {Reconstructing the Linear Power Spectrum of Cosmological Mass
  Fluctuations}.
\newblock \mnras~267, 1020--+.

\bibitem[\protect\citeauthoryear{{Peacock} and {Heavens}}{{Peacock} and
  {Heavens}}{1990}]{1990MNRAS.243..133P}
{Peacock}, J.~A. and A.~F. {Heavens} (1990).
\newblock {Alternatives to the Press-Schechter cosmological mass function}.
\newblock \mnras~243, 133--143.

\bibitem[\protect\citeauthoryear{{Peacock} and {Smith}}{{Peacock} and
  {Smith}}{2000}]{2000MNRAS.318.1144P}
{Peacock}, J.~A. and R.~E. {Smith} (2000).
\newblock {Halo occupation numbers and galaxy bias}.
\newblock \mnras~318, 1144--1156.

\bibitem[\protect\citeauthoryear{{Pen}}{{Pen}}{1998}]{1998ApJ...504..601P}
{Pen}, U.-L. (1998).
\newblock {Reconstructing Nonlinear Stochastic Bias from Velocity Space
  Distortions}.
\newblock \apj~504, 601--+.

\bibitem[\protect\citeauthoryear{{Percival} et~al.}{{Percival}
  et~al.}{2007}]{2007ApJ...657..645P}
{Percival}, W.~J. et~al. (2007).
\newblock {The Shape of the Sloan Digital Sky Survey Data Release 5 Galaxy
  Power Spectrum}.
\newblock \apj~657, 645--663.

\bibitem[\protect\citeauthoryear{{Percival}, {Cole}, {Eisenstein}, {Nichol},
  {Peacock}, {Pope}, and {Szalay}}{{Percival}
  et~al.}{2007}]{2007MNRAS.381.1053P}
{Percival}, W.~J., S.~{Cole}, D.~J. {Eisenstein}, R.~C. {Nichol}, J.~A.
  {Peacock}, A.~C. {Pope}, and A.~S. {Szalay} (2007).
\newblock {Measuring the Baryon Acoustic Oscillation scale using the Sloan
  Digital Sky Survey and 2dF Galaxy Redshift Survey}.
\newblock \mnras~381, 1053--1066.

\bibitem[\protect\citeauthoryear{{Percival}, {Verde}, and {Peacock}}{{Percival}
  et~al.}{2004}]{2004MNRAS.347..645P}
{Percival}, W.~J., L.~{Verde}, and J.~A. {Peacock} (2004).
\newblock {Fourier analysis of luminosity-dependent galaxy clustering}.
\newblock \mnras~347, 645--653.

\bibitem[\protect\citeauthoryear{{Perlmutter} et~al.}{{Perlmutter}
  et~al.}{1999}]{1999ApJ...517..565P}
{Perlmutter}, S. et~al. (1999).
\newblock {Measurements of Omega and Lambda from 42 High-Redshift Supernovae}.
\newblock \apj~517, 565--586.

\bibitem[\protect\citeauthoryear{{Peskin}}{{Peskin}}{2008}]{2008arXiv0801.1928%
P}
{Peskin}, M.~E. (2008).
\newblock {Supersymmetry in Elementary Particle Physics}.
\newblock arXiv:0801.1928.

\bibitem[\protect\citeauthoryear{{Pier}, {Munn}, {Hindsley}, {Hennessy},
  {Kent}, {Lupton}, and {Ivezi{\'c}}}{{Pier}
  et~al.}{2003}]{2003AJ....125.1559P}
{Pier}, J.~R., J.~A. {Munn}, R.~B. {Hindsley}, G.~S. {Hennessy}, S.~M. {Kent},
  R.~H. {Lupton}, and {\v Z}.~{Ivezi{\'c}} (2003).
\newblock {Astrometric Calibration of the Sloan Digital Sky Survey}.
\newblock \aj~125, 1559--1579.

\bibitem[\protect\citeauthoryear{{Press} and {Schechter}}{{Press} and
  {Schechter}}{1974}]{1974ApJ...187..425P}
{Press}, W.~H. and P.~{Schechter} (1974).
\newblock {Formation of Galaxies and Clusters of Galaxies by Self-Similar
  Gravitational Condensation}.
\newblock \apj~187, 425--438.

\bibitem[\protect\citeauthoryear{{Press}, {Teukolsky}, {Vetterling}, and
  {Flannery}}{{Press} et~al.}{1992}]{1992nrca.book.....P}
{Press}, W.~H., S.~A. {Teukolsky}, W.~T. {Vetterling}, and B.~P. {Flannery}
\newblock  (1992){Numerical recipes in C. The art of scientific computing} (2nd
  ed.).
\newblock New York: Cambridge University Press.

\bibitem[\protect\citeauthoryear{{Rassat}, {Land}, {Lahav}, and
  {Abdalla}}{{Rassat} et~al.}{2007}]{2007MNRAS.377.1085R}
{Rassat}, A., K.~{Land}, O.~{Lahav}, and F.~B. {Abdalla} (2007).
\newblock {Cross-correlation of 2MASS and WMAP 3: implications for the
  integrated Sachs-Wolfe effect}.
\newblock \mnras~377, 1085--1094.

\bibitem[\protect\citeauthoryear{{Reid} and {Spergel}}{{Reid} and
  {Spergel}}{2006}]{2006ApJ...651..643R}
{Reid}, B.~A. and D.~N. {Spergel} (2006).
\newblock {Sunyaev-Zel'dovich Effect Signals in Cluster Models}.
\newblock \apj~651, 643--657.

\bibitem[\protect\citeauthoryear{{Riess} et~al.}{{Riess}
  et~al.}{1998}]{1998AJ....116.1009R}
{Riess}, A.~G. et~al. (1998).
\newblock {Observational Evidence from Supernovae for an Accelerating Universe
  and a Cosmological Constant}.
\newblock \aj~116, 1009--1038.

\bibitem[\protect\citeauthoryear{{Romeo}, {Agertz}, {Moore}, and
  {Stadel}}{{Romeo} et~al.}{2008}]{2008arXiv0804.0294R}
{Romeo}, A.~B., O.~{Agertz}, B.~{Moore}, and J.~{Stadel} (2008).
\newblock {Discreteness Effects in Lambda Cold Dark Matter Simulations: A
  Wavelet-Statistical View}.
\newblock arXiv:0804.0294.

\bibitem[\protect\citeauthoryear{{Ryazhskaya}, {Volkova}, and
  {Saavedra}}{{Ryazhskaya} et~al.}{2002}]{2002NuPhS.110..531R}
{Ryazhskaya}, O.~G., L.~V. {Volkova}, and O.~{Saavedra} (2002).
\newblock {A limit on charm produced cosmic ray muon and atmospheric neutrino
  fluxes}.
\newblock Nucl. Phys. Proc. Suppl.~110, 531--533.

\bibitem[\protect\citeauthoryear{{S{\'a}nchez}, {Baugh}, {Percival}, {Peacock},
  {Padilla}, {Cole}, {Frenk}, and {Norberg}}{{S{\'a}nchez}
  et~al.}{2006}]{2006MNRAS.366..189S}
{S{\'a}nchez}, A.~G., C.~M. {Baugh}, W.~J. {Percival}, J.~A. {Peacock}, N.~D.
  {Padilla}, S.~{Cole}, C.~S. {Frenk}, and P.~{Norberg} (2006).
\newblock {Cosmological parameters from cosmic microwave background
  measurements and the final 2dF Galaxy Redshift Survey power spectrum}.
\newblock \mnras~366, 189--207.

\bibitem[\protect\citeauthoryear{{Sandage}}{{Sandage}}{1970}]{1970PhT....23b..%
34S}
{Sandage}, A.~R. (1970).
\newblock {Cosmology: a search for two numbers.}
\newblock Physics Today~23, 34--41.

\bibitem[\protect\citeauthoryear{{Sapienza}}{{Sapienza}}{2006}]{2006astro.ph.1%
1105S}
{Sapienza}, P. (2006).
\newblock {Status of the NEMO Project}.
\newblock astro-ph/0611105.

\bibitem[\protect\citeauthoryear{{Saunders} et~al.}{{Saunders}
  et~al.}{2000}]{2000MNRAS.317...55S}
{Saunders}, W. et~al. (2000).
\newblock {The PSCz catalogue}.
\newblock \mnras~317, 55--63.

\bibitem[\protect\citeauthoryear{{Schechter}}{{Schechter}}{1976}]{1976ApJ...20%
3..297S}
{Schechter}, P. (1976).
\newblock {An analytic expression for the luminosity function for galaxies.}
\newblock \apj~203, 297--306.

\bibitem[\protect\citeauthoryear{{Schlegel}, {Finkbeiner}, and
  {Davis}}{{Schlegel} et~al.}{1998}]{1998ApJ...500..525S}
{Schlegel}, D.~J., D.~P. {Finkbeiner}, and M.~{Davis} (1998).
\newblock {Maps of Dust Infrared Emission for Use in Estimation of Reddening
  and Cosmic Microwave Background Radiation Foregrounds}.
\newblock \apj~500, 525--+.

\bibitem[\protect\citeauthoryear{{Schlickeiser}}{{Schlickeiser}}{2002}]{2002cr%
a..book.....S}
{Schlickeiser}, R. (2002).
\newblock {Cosmic Ray Astrophysics}.
\newblock Berlin: Springer.

\bibitem[\protect\citeauthoryear{{Scoccimarro}, {Sheth}, {Hui}, and
  {Jain}}{{Scoccimarro} et~al.}{2001}]{2001ApJ...546...20S}
{Scoccimarro}, R., R.~K. {Sheth}, L.~{Hui}, and B.~{Jain} (2001).
\newblock {How Many Galaxies Fit in a Halo? Constraints on Galaxy Formation
  Efficiency from Spatial Clustering}.
\newblock \apj~546, 20--34.

\bibitem[\protect\citeauthoryear{{Scott}, {Shane}, and {Swanson}}{{Scott}
  et~al.}{1954}]{1954ApJ...119...91S}
{Scott}, E.~L., C.~D. {Shane}, and M.~D. {Swanson} (1954).
\newblock {Comparison of the Synthetic and Actual Distribution of Galaxies on a
  Photographic Plate.}
\newblock \apj~119, 91--+.

\bibitem[\protect\citeauthoryear{{Scranton}}{{Scranton}}{2003}]{2003MNRAS.339.%
.410S}
{Scranton}, R. (2003).
\newblock {Testing the halo model against the SDSS photometric survey}.
\newblock \mnras~339, 410--426.

\bibitem[\protect\citeauthoryear{{SDSS Collaboration}}{{SDSS
  Collaboration}}{2007}]{boss}
{SDSS Collaboration} (2007).
\newblock {After SDSS-II White Paper}.
\newblock \url{http://cosmology.lbl.gov/BOSS/as2_proposal.pdf}.

\bibitem[\protect\citeauthoryear{{Sefusatti} and {Scoccimarro}}{{Sefusatti} and
  {Scoccimarro}}{2005}]{2005PhRvD..71f3001S}
{Sefusatti}, E. and R.~{Scoccimarro} (2005).
\newblock {Galaxy bias and halo-occupation numbers from large-scale
  clustering}.
\newblock \prd~71(6), 063001--+.

\bibitem[\protect\citeauthoryear{{Seljak}}{{Seljak}}{2000}]{2000MNRAS.318..203%
S}
{Seljak}, U. (2000).
\newblock {Analytic model for galaxy and dark matter clustering}.
\newblock \mnras~318, 203--213.

\bibitem[\protect\citeauthoryear{{Seljak}}{{Seljak}}{2001}]{2001MNRAS.325.1359%
S}
{Seljak}, U. (2001).
\newblock {Redshift-space bias and {$\beta$} from the halo model}.
\newblock \mnras~325, 1359--1364.

\bibitem[\protect\citeauthoryear{{Seljak}, {Makarov}, {Mandelbaum}, {Hirata},
  {Padmanabhan}, {McDonald}, {Blanton}, {Tegmark}, {Bahcall}, and
  {Brinkmann}}{{Seljak} et~al.}{2005}]{2005PhRvD..71d3511S}
{Seljak}, U., A.~{Makarov}, R.~{Mandelbaum}, C.~M. {Hirata}, N.~{Padmanabhan},
  P.~{McDonald}, M.~R. {Blanton}, M.~{Tegmark}, N.~A. {Bahcall}, and
  J.~{Brinkmann} (2005).
\newblock {SDSS galaxy bias from halo mass-bias relation and its cosmological
  implications}.
\newblock \prd~71(4), 043511--+.

\bibitem[\protect\citeauthoryear{{Seljak} and {Warren}}{{Seljak} and
  {Warren}}{2004}]{2004MNRAS.355..129S}
{Seljak}, U. and M.~S. {Warren} (2004).
\newblock {Large-scale bias and stochasticity of haloes and dark matter}.
\newblock \mnras~355, 129--136.

\bibitem[\protect\citeauthoryear{{Seljak} and {Zaldarriaga}}{{Seljak} and
  {Zaldarriaga}}{1996}]{1996ApJ...469..437S}
{Seljak}, U. and M.~{Zaldarriaga} (1996).
\newblock {A Line-of-Sight Integration Approach to Cosmic Microwave Background
  Anisotropies}.
\newblock \apj~469, 437--+.

\bibitem[\protect\citeauthoryear{{Shectman}, {Landy}, {Oemler}, {Tucker},
  {Lin}, {Kirshner}, and {Schechter}}{{Shectman}
  et~al.}{1996}]{1996ApJ...470..172S}
{Shectman}, S.~A., S.~D. {Landy}, A.~{Oemler}, D.~L. {Tucker}, H.~{Lin}, R.~P.
  {Kirshner}, and P.~L. {Schechter} (1996).
\newblock {The Las Campanas Redshift Survey}.
\newblock \apj~470, 172--+.

\bibitem[\protect\citeauthoryear{{Shen} et~al.}{{Shen}
  et~al.}{2007}]{2007AJ....133.2222S}
{Shen}, Y. et~al. (2007).
\newblock {Clustering of High-Redshift (z {$\ge$} 2.9) Quasars from the Sloan
  Digital Sky Survey}.
\newblock \aj~133, 2222--2241.

\bibitem[\protect\citeauthoryear{{Sheth}, {Connolly}, and {Skibba}}{{Sheth}
  et~al.}{2005}]{2005astro.ph.11773S}
{Sheth}, R.~K., A.~J. {Connolly}, and R.~{Skibba} (2005).
\newblock {Marked correlations in galaxy formation models}.
\newblock astro-ph/0511773.

\bibitem[\protect\citeauthoryear{{Sheth} and {Diaferio}}{{Sheth} and
  {Diaferio}}{2001}]{2001MNRAS.322..901S}
{Sheth}, R.~K. and A.~{Diaferio} (2001).
\newblock {Peculiar velocities of galaxies and clusters}.
\newblock \mnras~322, 901--917.

\bibitem[\protect\citeauthoryear{{Sheth}, {Mo}, and {Tormen}}{{Sheth}
  et~al.}{2001}]{2001MNRAS.323....1S}
{Sheth}, R.~K., H.~J. {Mo}, and G.~{Tormen} (2001).
\newblock {Ellipsoidal collapse and an improved model for the number and
  spatial distribution of dark matter haloes}.
\newblock \mnras~323, 1--12.

\bibitem[\protect\citeauthoryear{{Sheth} and {Tormen}}{{Sheth} and
  {Tormen}}{1999}]{1999MNRAS.308..119S}
{Sheth}, R.~K. and G.~{Tormen} (1999).
\newblock {Large-scale bias and the peak background split}.
\newblock \mnras~308, 119--126.

\bibitem[\protect\citeauthoryear{{Simon}}{{Simon}}{2005}]{2005A&A...430..827S}
{Simon}, P. (2005).
\newblock {Time evolution of the stochastic linear bias of interacting galaxies
  on linear scales}.
\newblock \aap~430, 827--842.

\bibitem[\protect\citeauthoryear{{Simon}, {Hetterscheidt}, {Schirmer}, {Erben},
  {Schneider}, {Wolf}, and {Meisenheimer}}{{Simon}
  et~al.}{2007}]{2007A&A...461..861S}
{Simon}, P., M.~{Hetterscheidt}, M.~{Schirmer}, T.~{Erben}, P.~{Schneider},
  C.~{Wolf}, and K.~{Meisenheimer} (2007).
\newblock {GaBoDS: The Garching-Bonn Deep Survey. VI. Probing galaxy bias using
  weak gravitational lensing}.
\newblock \aap~461, 861--879.

\bibitem[\protect\citeauthoryear{{Skibba}, {Sheth}, {Connolly}, and
  {Scranton}}{{Skibba} et~al.}{2006}]{2006MNRAS.369...68S}
{Skibba}, R., R.~K. {Sheth}, A.~J. {Connolly}, and R.~{Scranton} (2006).
\newblock {The luminosity-weighted or `marked' correlation function}.
\newblock \mnras~369, 68--76.

\bibitem[\protect\citeauthoryear{{Slosar} and {Hobson}}{{Slosar} and
  {Hobson}}{2003}]{2003astro.ph..7219S}
{Slosar}, A. and M.~{Hobson} (2003).
\newblock {An improved Markov-chain Monte Carlo sampler for the estimation of
  cosmological parameters from CMB data}.
\newblock astro-ph/0307219.

\bibitem[\protect\citeauthoryear{{Smith} et~al.}{{Smith}
  et~al.}{2002}]{2002AJ....123.2121S}
{Smith}, J.~A. et~al. (2002).
\newblock {The u'g'r'i'z' Standard-Star System}.
\newblock \aj~123, 2121--2144.

\bibitem[\protect\citeauthoryear{{Smith}, {Peacock}, {Jenkins}, {White},
  {Frenk}, {Pearce}, {Thomas}, {Efstathiou}, and {Couchman}}{{Smith}
  et~al.}{2003}]{2003MNRAS.341.1311S}
{Smith}, R.~E., J.~A. {Peacock}, A.~{Jenkins}, S.~D.~M. {White}, C.~S. {Frenk},
  F.~R. {Pearce}, P.~A. {Thomas}, G.~{Efstathiou}, and H.~M.~P. {Couchman}
  (2003).
\newblock {Stable clustering, the halo model and non-linear cosmological power
  spectra}.
\newblock \mnras~341, 1311--1332.

\bibitem[\protect\citeauthoryear{{Smith}, {Scoccimarro}, and {Sheth}}{{Smith}
  et~al.}{2007}]{2007PhRvD..75f3512S}
{Smith}, R.~E., R.~{Scoccimarro}, and R.~K. {Sheth} (2007).
\newblock {Scale dependence of halo and galaxy bias: Effects in real space}.
\newblock \prd~75(6), 063512--+.

\bibitem[\protect\citeauthoryear{{Smith}, {Scoccimarro}, and {Sheth}}{{Smith}
  et~al.}{2008}]{2008PhRvD..77d3525S}
{Smith}, R.~E., R.~{Scoccimarro}, and R.~K. {Sheth} (2008).
\newblock {Motion of the acoustic peak in the correlation function}.
\newblock \prd~77(4), 043525--+.

\bibitem[\protect\citeauthoryear{{Sokolsky} and {Thomson}}{{Sokolsky} and
  {Thomson}}{2007}]{2007JPhG...34..401S}
{Sokolsky}, P. and G.~B. {Thomson} (2007).
\newblock {TOPICAL REVIEW: Highest energy cosmic-rays and results from the
  HiRes experiment}.
\newblock Journal of Physics G Nuclear Physics~34, 401--+.

\bibitem[\protect\citeauthoryear{{Somerville}, {Lemson}, {Sigad}, {Dekel},
  {Kauffmann}, and {White}}{{Somerville} et~al.}{2001}]{2001MNRAS.320..289S}
{Somerville}, R.~S., G.~{Lemson}, Y.~{Sigad}, A.~{Dekel}, G.~{Kauffmann}, and
  S.~D.~M. {White} (2001).
\newblock {Non-linear stochastic galaxy biasing in cosmological simulations}.
\newblock \mnras~320, 289--306.

\bibitem[\protect\citeauthoryear{{Spergel} et~al.}{{Spergel}
  et~al.}{2003}]{2003ApJS..148..175S}
{Spergel}, D.~N. et~al. (2003).
\newblock {First-Year Wilkinson Microwave Anisotropy Probe (WMAP) Observations:
  Determination of Cosmological Parameters}.
\newblock \apjs~148, 175--194.

\bibitem[\protect\citeauthoryear{{Spergel} et~al.}{{Spergel}
  et~al.}{2007}]{2007ApJS..170..377S}
{Spergel}, D.~N. et~al. (2007).
\newblock {Three-Year Wilkinson Microwave Anisotropy Probe (WMAP) Observations:
  Implications for Cosmology}.
\newblock \apjs~170, 377--408.

\bibitem[\protect\citeauthoryear{{Springel} et~al.}{{Springel}
  et~al.}{2005}]{2005Natur.435..629S}
{Springel}, V. et~al. (2005).
\newblock {Simulations of the formation, evolution and clustering of galaxies
  and quasars}.
\newblock \nat~435, 629--636.

\bibitem[\protect\citeauthoryear{{Sta{\'s}to}}{{Sta{\'s}to}}{2004}]{2004IJMPA.%
.19..317S}
{Sta{\'s}to}, A.~M. (2004).
\newblock {Ultrahigh Energy Neutrino Physics}.
\newblock Int. J. Mod. Phys. A~19, 317--340.

\bibitem[\protect\citeauthoryear{{Stecker} and {Salamon}}{{Stecker} and
  {Salamon}}{1996}]{Stecker:1995th}
{Stecker}, F.~W. and M.~H. {Salamon} (1996).
\newblock {High Energy Neutrinos from Quasars}.
\newblock \ssr~75, 341--355.

\bibitem[\protect\citeauthoryear{{Steigman}}{{Steigman}}{2007}]{2007ARNPS..57.%
.463S}
{Steigman}, G. (2007).
\newblock {Primordial Nucleosynthesis in the Precision Cosmology Era}.
\newblock Annual Review of Nuclear and Particle Science~57, 463--491.

\bibitem[\protect\citeauthoryear{{Stoughton} et~al.}{{Stoughton}
  et~al.}{2002}]{2002AJ....123..485S}
{Stoughton}, C. et~al. (2002).
\newblock {Sloan Digital Sky Survey: Early Data Release}.
\newblock \aj~123, 485--548.

\bibitem[\protect\citeauthoryear{{Strauss}, {Huchra}, {Davis}, {Yahil},
  {Fisher}, and {Tonry}}{{Strauss} et~al.}{1992}]{1992ApJS...83...29S}
{Strauss}, M.~A., J.~P. {Huchra}, M.~{Davis}, A.~{Yahil}, K.~B. {Fisher}, and
  J.~{Tonry} (1992).
\newblock {A redshift survey of IRAS galaxies. VII - The infrared and redshift
  data for the 1.936 Jansky sample}.
\newblock \apjs~83, 29--63.

\bibitem[\protect\citeauthoryear{{Stubbs}, {Sweeney}, {Tyson}, and {the LSST
  Collaboration}}{{Stubbs} et~al.}{2004}]{2004AAS...20510802S}
{Stubbs}, C.~W., D.~{Sweeney}, J.~A. {Tyson}, and {the LSST Collaboration}
  (2004).
\newblock {An Overview of the Large Synoptic Survey Telescope (LSST) System}.
\newblock In Bulletin of the American Astronomical Society, Volume~36,
  1527--+.

\bibitem[\protect\citeauthoryear{{Swanson} et~al.}{{Swanson}
  et~al.}{2006}]{2006ApJ...652..206S}
{Swanson}, M.~E.~C. et~al. (2006).
\newblock {Search for Diffuse Astrophysical Neutrino Flux Using
  Ultra-High-Energy Upward-going Muons in Super-Kamiokande I}.
\newblock \apj~652, 206--215.

\bibitem[\protect\citeauthoryear{{Swanson}, {Tegmark}, {Blanton}, and
  {Zehavi}}{{Swanson} et~al.}{2008}]{2008MNRAS.385.1635S}
{Swanson}, M.~E.~C., M.~{Tegmark}, M.~{Blanton}, and I.~{Zehavi} (2008).
\newblock {SDSS galaxy clustering: luminosity and colour dependence and
  stochasticity}.
\newblock \mnras~385, 1635--1655.

\bibitem[\protect\citeauthoryear{{Swanson}, {Tegmark}, {Hamilton}, and
  {Hill}}{{Swanson} et~al.}{2007}]{2007arXiv0711.4352S}
{Swanson}, M.~E.~C., M.~{Tegmark}, A.~J.~S. {Hamilton}, and J.~C. {Hill}
  (2007).
\newblock {Methods for Rapidly Processing Angular Masks of Next-Generation
  Galaxy Surveys}.
\newblock arXiv:0711.4352.

\bibitem[\protect\citeauthoryear{{Szapudi}, {Prunet}, and {Colombi}}{{Szapudi}
  et~al.}{2001}]{2001ApJ...561L..11S}
{Szapudi}, I., S.~{Prunet}, and S.~{Colombi} (2001).
\newblock {Fast Analysis of Inhomogenous Megapixel Cosmic Microwave Background
  Maps}.
\newblock \apjl~561, L11--L14.

\bibitem[\protect\citeauthoryear{{Taruya} and {Hiramatsu}}{{Taruya} and
  {Hiramatsu}}{2008}]{2008ApJ...674..617T}
{Taruya}, A. and T.~{Hiramatsu} (2008).
\newblock {A Closure Theory for Nonlinear Evolution of Cosmological Power
  Spectra}.
\newblock \apj~674, 617--635.

\bibitem[\protect\citeauthoryear{{Tasitsiomi}, {Kravtsov}, {Wechsler}, and
  {Primack}}{{Tasitsiomi} et~al.}{2004}]{2004ApJ...614..533T}
{Tasitsiomi}, A., A.~V. {Kravtsov}, R.~H. {Wechsler}, and J.~R. {Primack}
  (2004).
\newblock {Modeling Galaxy-Mass Correlations in Dissipationless Simulations}.
\newblock \apj~614, 533--546.

\bibitem[\protect\citeauthoryear{{Tegmark}}{{Tegmark}}{1997}]{1997PhRvL..79.38%
06T}
{Tegmark}, M. (1997).
\newblock {Measuring Cosmological Parameters with Galaxy Surveys}.
\newblock Physical Review Letters~79, 3806--3809.

\bibitem[\protect\citeauthoryear{{Tegmark}}{{Tegmark}}{1999}]{1999ApJ...519..5%
13T}
{Tegmark}, M. (1999).
\newblock {Comparing and Combining Cosmic Microwave Background Data Sets}.
\newblock \apj~519, 513--517.

\bibitem[\protect\citeauthoryear{{Tegmark} et~al.}{{Tegmark}
  et~al.}{2004a}]{2004PhRvD..69j3501T}
{Tegmark}, M. et~al. (2004a).
\newblock {Cosmological parameters from SDSS and WMAP}.
\newblock \prd~69(10), 103501--+.

\bibitem[\protect\citeauthoryear{{Tegmark} et~al.}{{Tegmark}
  et~al.}{2004b}]{2004ApJ...606..702T}
{Tegmark}, M. et~al. (2004b).
\newblock {The Three-Dimensional Power Spectrum of Galaxies from the Sloan
  Digital Sky Survey}.
\newblock \apj~606, 702--740.

\bibitem[\protect\citeauthoryear{{Tegmark} et~al.}{{Tegmark}
  et~al.}{2006}]{2006PhRvD..74l3507T}
{Tegmark}, M. et~al. (2006).
\newblock {Cosmological constraints from the SDSS luminous red galaxies}.
\newblock \prd~74(12), 123507--+.

\bibitem[\protect\citeauthoryear{{Tegmark} and {Bromley}}{{Tegmark} and
  {Bromley}}{1999}]{1999ApJ...518L..69T}
{Tegmark}, M. and B.~C. {Bromley} (1999).
\newblock {Observational Evidence for Stochastic Biasing}.
\newblock \apjl~518, L69--L72.

\bibitem[\protect\citeauthoryear{{Tegmark}, {Hamilton}, and {Xu}}{{Tegmark}
  et~al.}{2002}]{2002MNRAS.335..887T}
{Tegmark}, M., A.~J.~S. {Hamilton}, and Y.~{Xu} (2002).
\newblock {The power spectrum of galaxies in the 2dF 100k redshift survey}.
\newblock \mnras~335, 887--908.

\bibitem[\protect\citeauthoryear{{Tegmark} and {Peebles}}{{Tegmark} and
  {Peebles}}{1998}]{1998ApJ...500L..79T}
{Tegmark}, M. and P.~J.~E. {Peebles} (1998).
\newblock {The Time Evolution of Bias}.
\newblock \apjl~500, L79+.

\bibitem[\protect\citeauthoryear{{Tegmark} and {Rees}}{{Tegmark} and
  {Rees}}{1998}]{1998ApJ...499..526T}
{Tegmark}, M. and M.~J. {Rees} (1998).
\newblock {Why is the Cosmic Microwave Background Fluctuation Level $10^{-5}$?}
\newblock \apj~499, 526--+.

\bibitem[\protect\citeauthoryear{{Tegmark}, {Silk}, and {Blanchard}}{{Tegmark}
  et~al.}{1994}]{1994ApJ...420..484T}
{Tegmark}, M., J.~{Silk}, and A.~{Blanchard} (1994).
\newblock {On the inevitability of reionization: Implications for cosmic
  microwave background fluctuations}.
\newblock \apj~420, 484--496.

\bibitem[\protect\citeauthoryear{{Tegmark}, {Vilenkin}, and
  {Pogosian}}{{Tegmark} et~al.}{2005}]{2005PhRvD..71j3523T}
{Tegmark}, M., A.~{Vilenkin}, and L.~{Pogosian} (2005).
\newblock {Anthropic predictions for neutrino masses}.
\newblock \prd~71(10), 103523--+.

\bibitem[\protect\citeauthoryear{{Terzian} and {Lazio}}{{Terzian} and
  {Lazio}}{2006}]{2006SPIE.6267E..76T}
{Terzian}, Y. and J.~{Lazio} (2006).
\newblock {The Square Kilometre Array}.
\newblock Volume 6267 of Proceedings of the SPIE.

\bibitem[\protect\citeauthoryear{{Thunman}, {Ingelman}, and
  {Gondolo}}{{Thunman} et~al.}{1996}]{1996APh.....5..309T}
{Thunman}, M., G.~{Ingelman}, and P.~{Gondolo} (1996).
\newblock {Charm production and high energy atmospheric muon and neutrino
  fluxes}.
\newblock Astropart. Phys.~5, 309--332.

\bibitem[\protect\citeauthoryear{{Tinker}, {Conroy}, {Norberg}, {Patiri},
  {Weinberg}, and {Warren}}{{Tinker} et~al.}{2007}]{2007arXiv0707.3445T}
{Tinker}, J.~L., C.~{Conroy}, P.~{Norberg}, S.~G. {Patiri}, D.~H. {Weinberg},
  and M.~S. {Warren} (2007).
\newblock {Void Statistics in Large Galaxy Redshift Surveys: Does Halo
  Occupation of Field Galaxies Depend on Environment?}
\newblock arXiv:0707.3445.

\bibitem[\protect\citeauthoryear{{Tinker}, {Norberg}, {Weinberg}, and
  {Warren}}{{Tinker} et~al.}{2007}]{2007ApJ...659..877T}
{Tinker}, J.~L., P.~{Norberg}, D.~H. {Weinberg}, and M.~S. {Warren} (2007).
\newblock {On the Luminosity Dependence of the Galaxy Pairwise Velocity
  Dispersion}.
\newblock \apj~659, 877--889.

\bibitem[\protect\citeauthoryear{{Tinker}, {Weinberg}, and {Warren}}{{Tinker}
  et~al.}{2006}]{2006ApJ...647..737T}
{Tinker}, J.~L., D.~H. {Weinberg}, and M.~S. {Warren} (2006).
\newblock {Cosmic Voids and Galaxy Bias in the Halo Occupation Framework}.
\newblock \apj~647, 737--752.

\bibitem[\protect\citeauthoryear{{Tinker}, {Weinberg}, {Zheng}, and
  {Zehavi}}{{Tinker} et~al.}{2005}]{2005ApJ...631...41T}
{Tinker}, J.~L., D.~H. {Weinberg}, Z.~{Zheng}, and I.~{Zehavi} (2005).
\newblock {On the Mass-to-Light Ratio of Large-Scale Structure}.
\newblock \apj~631, 41--58.

\bibitem[\protect\citeauthoryear{{Tucker} et~al.}{{Tucker}
  et~al.}{2006}]{2006AN....327..821T}
{Tucker}, D.~L. et~al. (2006).
\newblock {The Sloan Digital Sky Survey monitor telescope pipeline}.
\newblock Astronomische Nachrichten~327, 821--+.

\bibitem[\protect\citeauthoryear{{Tyson}}{{Tyson}}{2002}]{2002SPIE.4836...10T}
{Tyson}, J.~A. (2002).
\newblock {Large Synoptic Survey Telescope: Overview}.
\newblock In J.~A. {Tyson} and S.~{Wolff} (Eds.), Survey and Other Telescope
  Technologies and Discoveries, Volume 4836 of Proceedings of the SPIE,
  10--20.

\bibitem[\protect\citeauthoryear{{Tyson}}{{Tyson}}{2006}]{2006AIPC..870...44T}
{Tyson}, J.~A. (2006).
\newblock {Precision Studies of Dark Energy with LSST}.
\newblock In T.~M. {Liss} (Ed.), Intersections of Particle and Nuclear Physics:
  9th Conference CIPAN2006, Volume 870 of American Institute of Physics
  Conference Series,  44--52.

\bibitem[\protect\citeauthoryear{{Tytler}, {Fan}, and {Burles}}{{Tytler}
  et~al.}{1996}]{1996Natur.381..207T}
{Tytler}, D., X.-M. {Fan}, and S.~{Burles} (1996).
\newblock {Cosmological baryon density derived from the deuterium abundance at
  redshift z = 3.57}.
\newblock \nat~381, 207--209.

\bibitem[\protect\citeauthoryear{{Uomoto}, {Smee}, and {Barkhouser}}{{Uomoto}
  et~al.}{2004}]{2004SPIE.5492.1411U}
{Uomoto}, A., S.~A. {Smee}, and R.~H. {Barkhouser} (2004).
\newblock {A high-efficiency near-infrared spectrograph for the Apache Point
  3.5 m telescope}.
\newblock In A.~F.~M. {Moorwood} and M.~{Iye} (Eds.), {Ground-based
  Instrumentation for Astronomy}, Volume 5492 of Proceedings of the SPIE,
  1411--1422.

\bibitem[\protect\citeauthoryear{{van den Bosch}, {Yang}, and {Mo}}{{van den
  Bosch} et~al.}{2003}]{2003MNRAS.340..771V}
{van den Bosch}, F.~C., X.~{Yang}, and H.~J. {Mo} (2003).
\newblock {Linking early- and late-type galaxies to their dark matter haloes}.
\newblock \mnras~340, 771--792.

\bibitem[\protect\citeauthoryear{{Verde} et~al.}{{Verde}
  et~al.}{2003}]{2003ApJS..148..195V}
{Verde}, L. et~al. (2003).
\newblock {First-Year Wilkinson Microwave Anisotropy Probe (WMAP) Observations:
  Parameter Estimation Methodology}.
\newblock \apjs~148, 195--211.

\bibitem[\protect\citeauthoryear{{Volkova}}{{Volkova}}{1980}]{1980SvJNP..31..7%
84V}
{Volkova}, L.~V. (1980).
\newblock {Energy Spectra and Angular Distributions of Atmospheric Neutrinos}.
\newblock Sov. J. Nucl. Phys.~31, 784--+.

\bibitem[\protect\citeauthoryear{{Wandelt}, {Hivon}, and {Gorski}}{{Wandelt}
  et~al.}{1998}]{1998astro.ph..3317W}
{Wandelt}, B.~D., E.~{Hivon}, and K.~M. {Gorski} (1998).
\newblock {Topological Analysis of High-Resolution CMB Maps}.
\newblock astro-ph/9803317.

\bibitem[\protect\citeauthoryear{{Wandelt}, {Hivon}, and
  {G{\'o}rski}}{{Wandelt} et~al.}{2001}]{2001PhRvD..64h3003W}
{Wandelt}, B.~D., E.~{Hivon}, and K.~M. {G{\'o}rski} (2001).
\newblock {Cosmic microwave background anisotropy power spectrum statistics for
  high precision cosmology}.
\newblock \prd~64(8), 083003--+.

\bibitem[\protect\citeauthoryear{{Wang}, {Yang}, {Mo}, and {van den
  Bosch}}{{Wang} et~al.}{2007}]{2007ApJ...664..608W}
{Wang}, Y., X.~{Yang}, H.~J. {Mo}, and F.~C. {van den Bosch} (2007).
\newblock {The Cross-Correlation between Galaxies of Different Luminosities and
  Colors}.
\newblock \apj~664, 608--632.

\bibitem[\protect\citeauthoryear{{Waxman} and {Bahcall}}{{Waxman} and
  {Bahcall}}{1999}]{1999PhRvD..59b3002W}
{Waxman}, E. and J.~{Bahcall} (1999).
\newblock {High energy neutrinos from astrophysical sources: An upper bound}.
\newblock \prd~59(2), 023002--+.

\bibitem[\protect\citeauthoryear{{Weinberg}, {Dav{\'e}}, {Katz}, and
  {Kollmeier}}{{Weinberg} et~al.}{2003}]{2003AIPC..666..157W}
{Weinberg}, D.~H., R.~{Dav{\'e}}, N.~{Katz}, and J.~A. {Kollmeier} (2003).
\newblock {The Lyman-{$\alpha$} Forest as a Cosmological Tool}.
\newblock In S.~H. {Holt} and C.~S. {Reynolds} (Eds.), The Emergence of Cosmic
  Structure, Volume 666 of American Institute of Physics Conference Series,
  157--169.

\bibitem[\protect\citeauthoryear{{Weinmann}, {van den Bosch}, {Yang}, and
  {Mo}}{{Weinmann} et~al.}{2006}]{2006MNRAS.366....2W}
{Weinmann}, S.~M., F.~C. {van den Bosch}, X.~{Yang}, and H.~J. {Mo} (2006).
\newblock {Properties of galaxy groups in the Sloan Digital Sky Survey - I. The
  dependence of colour, star formation and morphology on halo mass}.
\newblock \mnras~366, 2--28.

\bibitem[\protect\citeauthoryear{{White}}{{White}}{2007}]{2007RPPh...70..883W}
{White}, S.~D.~M. (2007).
\newblock {Fundamentalist physics: why Dark Energy is bad for astronomy}.
\newblock Reports of Progress in Physics~70, 883--897.

\bibitem[\protect\citeauthoryear{{White}, {Efstathiou}, and {Frenk}}{{White}
  et~al.}{1993}]{1993MNRAS.262.1023W}
{White}, S.~D.~M., G.~{Efstathiou}, and C.~S. {Frenk} (1993).
\newblock {The amplitude of mass fluctuations in the universe}.
\newblock \mnras~262, 1023--1028.

\bibitem[\protect\citeauthoryear{{White}, {Tully}, and {Davis}}{{White}
  et~al.}{1988}]{1988ApJ...333L..45W}
{White}, S.~D.~M., R.~B. {Tully}, and M.~{Davis} (1988).
\newblock {Clustering bias in the nearby galaxies catalog and in cold dark
  matter models}.
\newblock \apjl~333, L45--L49.

\bibitem[\protect\citeauthoryear{{Wild} et~al.}{{Wild}
  et~al.}{2005}]{2005MNRAS.356..247W}
{Wild}, V. et~al. (2005).
\newblock {The 2dF Galaxy Redshift Survey: stochastic relative biasing between
  galaxy populations}.
\newblock \mnras~356, 247--269.

\bibitem[\protect\citeauthoryear{{Willmer}, {da Costa}, and
  {Pellegrini}}{{Willmer} et~al.}{1998}]{1998AJ....115..869W}
{Willmer}, C.~N.~A., L.~N. {da Costa}, and P.~S. {Pellegrini} (1998).
\newblock {Southern Sky Redshift Survey: Clustering of Local Galaxies}.
\newblock \aj~115, 869--884.

\bibitem[\protect\citeauthoryear{{Yadav}, {Bharadwaj}, {Pandey}, and
  {Seshadri}}{{Yadav} et~al.}{2005}]{2005MNRAS.364..601Y}
{Yadav}, J., S.~{Bharadwaj}, B.~{Pandey}, and T.~R. {Seshadri} (2005).
\newblock {Testing homogeneity on large scales in the Sloan Digital Sky Survey
  Data Release One}.
\newblock \mnras~364, 601--606.

\bibitem[\protect\citeauthoryear{{Yamamoto}, {Bassett}, {Nichol}, {Suto}, and
  {Yahata}}{{Yamamoto} et~al.}{2006}]{2006PhRvD..74f3525Y}
{Yamamoto}, K., B.~A. {Bassett}, R.~C. {Nichol}, Y.~{Suto}, and K.~{Yahata}
  (2006).
\newblock {Searching for modified gravity with baryon oscillations: From SDSS
  to wide field multiobject spectroscopy (WFMOS)}.
\newblock \prd~74(6), 063525--+.

\bibitem[\protect\citeauthoryear{{Yang}, {Mo}, {Jing}, and {van den
  Bosch}}{{Yang} et~al.}{2005}]{2005MNRAS.358..217Y}
{Yang}, X., H.~J. {Mo}, Y.~P. {Jing}, and F.~C. {van den Bosch} (2005).
\newblock {Galaxy occupation statistics of dark matter haloes: observational
  results}.
\newblock \mnras~358, 217--232.

\bibitem[\protect\citeauthoryear{{Yang}, {Mo}, and {van den Bosch}}{{Yang}
  et~al.}{2003}]{2003MNRAS.339.1057Y}
{Yang}, X., H.~J. {Mo}, and F.~C. {van den Bosch} (2003).
\newblock {Constraining galaxy formation and cosmology with the conditional
  luminosity function of galaxies}.
\newblock \mnras~339, 1057--1080.

\bibitem[\protect\citeauthoryear{{Yang}, {Mo}, and {van den Bosch}}{{Yang}
  et~al.}{2006}]{2006ApJ...638L..55Y}
{Yang}, X., H.~J. {Mo}, and F.~C. {van den Bosch} (2006).
\newblock {Observational Evidence for an Age Dependence of Halo Bias}.
\newblock \apjl~638, L55--L58.

\bibitem[\protect\citeauthoryear{{Yang}, {Mo}, and {van den Bosch}}{{Yang}
  et~al.}{2008}]{2008ApJ...676..248Y}
{Yang}, X., H.~J. {Mo}, and F.~C. {van den Bosch} (2008).
\newblock {Galaxy Groups in the SDSS DR4. II. Halo Occupation Statistics}.
\newblock \apj~676, 248--261.

\bibitem[\protect\citeauthoryear{{York} et~al.}{{York}
  et~al.}{2000}]{2000AJ....120.1579Y}
{York}, D.~G. et~al. (2000).
\newblock {The Sloan Digital Sky Survey: Technical Summary}.
\newblock \aj~120, 1579--1587.

\bibitem[\protect\citeauthoryear{{Zas}, {Halzen}, and {V{\'a}zquez}}{{Zas}
  et~al.}{1993}]{1993APh.....1..297Z}
{Zas}, E., F.~{Halzen}, and R.~A. {V{\'a}zquez} (1993).
\newblock {High energy neutrino astronomy: Horizontal air shower arrays versus
  underground detectors}.
\newblock Astropart. Phys.~1, 297--315.

\bibitem[\protect\citeauthoryear{{Zatsepin} and {Kuzmin}}{{Zatsepin} and
  {Kuzmin}}{1966}]{GZK2}
{Zatsepin}, G.~T. and V.~A. {Kuzmin} (1966).
\newblock {Upper Limit of the Spectrum of Cosmic Rays}.
\newblock JTEP Letters~4, 79.

\bibitem[\protect\citeauthoryear{{Zehavi} et~al.}{{Zehavi}
  et~al.}{2005}]{2005ApJ...630....1Z}
{Zehavi}, I. et~al. (2005).
\newblock {The Luminosity and Color Dependence of the Galaxy Correlation
  Function}.
\newblock \apj~630, 1--27.

\bibitem[\protect\citeauthoryear{{Zentner}}{{Zentner}}{2007}]{2007IJMPD..16..7%
63Z}
{Zentner}, A.~R. (2007).
\newblock {The Excursion Set Theory of Halo Mass Functions, Halo Clustering,
  and Halo Growth}.
\newblock International Journal of Modern Physics D~16, 763--815.

\bibitem[\protect\citeauthoryear{{Zheng}, {Berlind}, {Weinberg}, {Benson},
  {Baugh}, {Cole}, {Dav{\'e}}, {Frenk}, {Katz}, and {Lacey}}{{Zheng}
  et~al.}{2005}]{2005ApJ...633..791Z}
{Zheng}, Z., A.~A. {Berlind}, D.~H. {Weinberg}, A.~J. {Benson}, C.~M. {Baugh},
  S.~{Cole}, R.~{Dav{\'e}}, C.~S. {Frenk}, N.~{Katz}, and C.~G. {Lacey} (2005).
\newblock {Theoretical Models of the Halo Occupation Distribution: Separating
  Central and Satellite Galaxies}.
\newblock \apj~633, 791--809.

\bibitem[\protect\citeauthoryear{{Zheng} and {Weinberg}}{{Zheng} and
  {Weinberg}}{2007}]{2007ApJ...659....1Z}
{Zheng}, Z. and D.~H. {Weinberg} (2007).
\newblock {Breaking the Degeneracies between Cosmology and Galaxy Bias}.
\newblock \apj~659, 1--28.

\bibitem[\protect\citeauthoryear{{Zwicky}}{{Zwicky}}{1952}]{1952PASP...64..247%
Z}
{Zwicky}, F. (1952).
\newblock {Dispersion in the Large-Scale Distribution of Galaxies}.
\newblock \pasp~64, 247--+.

\end{thebibliography}
\bibliographystyle{chicago}

\end{document}